\documentclass[12pt, a4paper, twoside, openany]{book}
\usepackage{color}
\usepackage[dvipsnames]{xcolor}
\usepackage{layout}
\usepackage[bookmarksnumbered]{hyperref}
\hypersetup{
    hidelinks, 
    colorlinks=false, 
    linkcolor=OrangeRed,
    urlcolor=ForestGreen,
    citecolor=Aquamarine
}
\usepackage{tabstackengine}
\usepackage[titletoc]{appendix}
\usepackage{comment}

\usepackage{amsmath}
\usepackage{amssymb}
\usepackage{graphicx}
\usepackage[font=small, width=0.9\textwidth]{caption}
\usepackage{lipsum}
\usepackage{mathrsfs}
\usepackage[shortcuts]{extdash}
\pdfoutput=1

\usepackage{mathtools}
\usepackage{environ}
\usepackage{xcolor}
\usepackage{pgfplots}
\pgfplotsset{compat=1.15}
\usepackage{tikz}
\usepackage[tikz]{bclogo}
\usepackage{tkz-euclide}
\usetikzlibrary{calc,arrows,decorations}
\usepackage{dsfont}
\usepackage{fancyvrb}

\def\ket#1{|{#1}\rangle}
\def\bra#1{\langle{#1}|}
\def\Tr{{\rm Tr}}

\newcommand{\Dslash}{\nabla \! \! \! \!  / \, } 
\newcommand{\DDslash}{D \! \! \! \!  / \, } 
\newcommand{\Aslash}{A \! \! \!  /} 
\newcommand{\Bslash}{B \! \! \! \! / \, } 
\newcommand{\bpsi}{\overline{\psi}}
\newcommand{\blambda}{\overline{\lambda}}
\newcommand{\brho}{\overline{\rho}}

\def\la{\langle}
\def\ra{\rangle}
\def\tr{{\rm tr}}

\let\svthefootnote\thefootnote
\newcommand\freefootnote[1]{%
  \let\thefootnote\relax%
  \footnotetext{#1}%
  \let\thefootnote\svthefootnote%
}

\usepackage[german,english]{babel}
\usepackage[utf8]{inputenc}
\usepackage[T1]{fontenc}
\usepackage[babel]{csquotes}

\usepackage{fancyhdr}
\makeatletter 
\newcommand{\rightorleftmark}{%
  \begingroup\protected@edef\x{\rightmark}%
  \ifx\x\@empty
    \endgroup\leftmark
  \else
    \endgroup\rightmark
  \fi}
\makeatother

\fancypagestyle{main}{%
    \fancyhead{} 

\fancyhf{}
\fancyhead[LE]{\textsc\leftmark} 
\fancyhead[RO]{\textsc\rightorleftmark} 
\fancyfoot[C]{\thepage}
}

\fancypagestyle{appendices}{%
    \fancyhead{} 

\fancyhf{}
\fancyhead[LE]{\textsc\leftmark} 
\fancyhead[RO]{\textsc\rightorleftmark} 
\fancyfoot[C]{\thepage}
}

\usepackage[backend=biber, style=alphabetic, giveninits,date=year, doi=false,eprint=true,url=false,maxbibnames=4]{biblatex}
\title{Aspects of Conformal Field Theory}
\author{Matteo Broccoli}

\begin{document}

\frontmatter
\begin{titlepage}
\begin{center}
    {\LARGE \sc Aspects of Conformal Field Theory}
\end{center}
\vspace{1ex}
\begin{center}
    Dissertation zur Erlangung des akademischen Grades\\Doctor rerum naturalium\\(Dr.~rer.~nat.)
\end{center}
\begin{center}
    im Fach: Physik\\Spezialisierung: theoretische Physik
\end{center}
\begin{center}
    eingereicht an der\\Mathematisch-Naturwissenschaftlichen Fakult\"at\\der\\Humboldt-Universit\"at zu Berlin\\von
\end{center}
\vspace{1ex}
\begin{center}
    {\large \sc Matteo Broccoli}
\end{center}
\vspace{1ex}
\begin{center}
    Pr\"asidentin der Humboldt-Universit\"at zu Berlin:\\Prof.~Dr.~Julia von Blumenthal
\end{center}
\begin{center}
    Dekanin der Mathematisch-Naturwissenschaftlichen Fakult\"at:\\Prof.~Dr.~Caren Tischendorf
\end{center}
\vspace{1ex}
%
%
\textcolor{black}{\rule{\textwidth}{0.01mm}}
\vspace{1ex}
\begin{center}\textcolor{black}{
    \begin{minipage}[t]{0.4\textwidth}
    Gutachter/innen: 
    \end{minipage}
    \begin{minipage}[t]{0.4\textwidth}
    1.~Dr.~Valentina Forini\\%
    2.~Prof.~Dr.~Stefan Fredenhagen\\%
    3.~Prof.~Dr.~Stefan Theisen%
    \end{minipage}
}
\end{center}
\vspace{1ex}
\begin{center}\textcolor{black}{
    \begin{minipage}[t]{0.4\textwidth}
    Tag der m\"undlichen Pr\"ufung: 
    \end{minipage}
    \begin{minipage}[t]{0.4\textwidth}
    28.~November 2022
    \end{minipage}
    }
\end{center}
\end{titlepage}

\clearpage\shipout\null 
\stepcounter{page} 

\thispagestyle{empty}
\begin{center}
    {\LARGE \sc Aspects of Conformal Field Theory}
\end{center}
\vspace{3ex}
\begin{center}
    {\large \sc Matteo Broccoli}
\end{center}
\vspace{3ex}
\begin{center}
    \begin{figure}[!h]
    \centering
    \begin{minipage}[t]{0.3\textwidth}
        \centering\includegraphics[height=3.5cm]{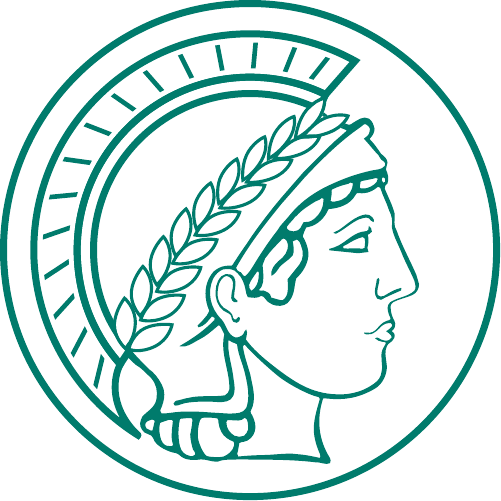}
    \end{minipage}
    \begin{minipage}[t]{0.3\textwidth}
        \centering\includegraphics[trim={1cm 0.75cm 0.7cm 0.62cm},height=3.5cm]{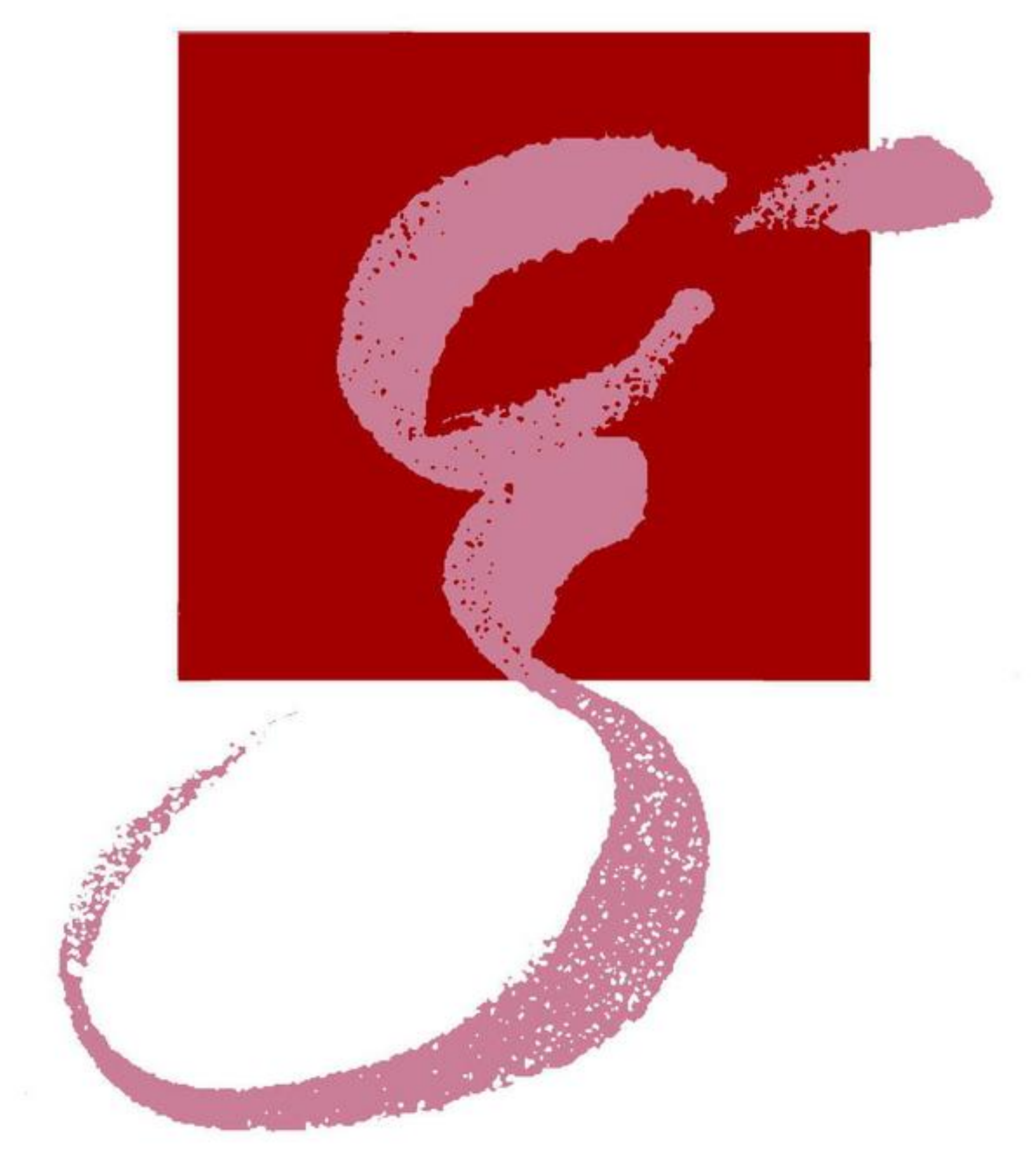}
    \end{minipage}
    \end{figure}
    Max-Planck-Institut f\"ur Gravitationsphysik\\Albert-Einstein-Institut
\end{center}
\vspace{1ex}
\begin{center}
    and
\end{center}
\vspace{1ex}
\begin{center}
    Humboldt-Universit\"at zu Berlin\\Mathematisch-Naturwissenschaftlichen Fakult\"at
\end{center}
\vspace{9ex}
\begin{center}
    June 2022
\end{center}

\clearpage\shipout\null 
\stepcounter{page} 

\newpage\thispagestyle{empty}
\noindent I declare that I have completed the thesis independently using only the aids and tools specified.
I have not applied for a doctor’s degree in the doctoral subject elsewhere and do not hold a corresponding doctor’s degree.
I have taken due note of the Faculty of Mathematics and Natural Sciences PhD Regulations, published in the Official Gazette of Humboldt-Universit\"at zu Berlin no.~42/2018 on 11/07/2018.

\vspace{7mm}

\emph{Matteo Broccoli}

\pagestyle{main}
\clearpage\thispagestyle{empty}
\tableofcontents
\selectlanguage{german}
\chapter{Zusammenfassung}

Konforme Feldtheorien (CFTs) sind eine der am meisten untersuchten Feldtheorien und eine bemerkenswerte Spielwiese in der modernen theoretischen Physik.
In dieser Dissertation analysieren wir drei Aspekte von CFTs in verschiedenen Dimensionen.

Erstens betrachten wir Korrelationsfunktionen von sekundären Zuständen in zweidimensionalen CFTs.
Wir diskutieren eine rekursive Formel zu ihrer Berechnung und erstellen eine Computerimplementierung dieser Formel.
Damit können wir jede Korrelationsfunktion von Sekundärzuständen des Vakuums erhalten und für Nicht-Vakuum-Sekundärzustände den Korrelator als Differentialoperator, der auf den jeweiligen primären Korrelator wirkt, ausdrücken.
Mit diesem Code untersuchen wir dann einige Verschränkungs- und Unterscheidbarkeitsmaße zwischen Sekundärzuständen, nämlich die R\'enyi-Entropie, den Spurquadratabstand und die Sandwich-R\'enyi-Divergenz.
Mit unseren Ergebnissen können wir die R\'enyi Quanten-Null-Energie-Bedingung testen und stellen neue Werkzeuge zur Analyse der holographischen Beschreibung von Sekundärzuständen bereit.

Zweitens untersuchen wir vierdimensionale Weyl-Fermionen auf verschiedenen Hintergründen.
Unser Interesse gilt ihrer Spuranomalie, insbesondere der Frage, ob die Pontryagin-Dichte auftritt.
Um diese Möglichkeit festzustellen, berechnen wir die Anomalien von Dirac-Fermionen, die an vektorielle und axiale nicht-abelsche Eichfelder gekoppelt sind, und dann auf einem metrisch-axialen Tensor Hintergrund.
Geeignete Grenzwerte der Hintergründe erlauben es dann, die Anomalien von Weyl-Fermionen, die an nicht-abelsche Eichfelder gekoppelt sind, und in einer gekrümmten Raumzeit zu berechnen.
In beiden Fällen bestätigen wir das Fehlen der Pontryagin-Dichte in den Spuranomalien.

Drittens liefern wir die holographische Beschreibung einer vierdimensionalen CFT mit einem irrelevanten Operator.
Wenn der Operator eine ganzzahlige konforme Dimension hat, modifiziert sein Vorhandensein in der CFT die Weyl\-/Transformation der Metrik, was wiederum die Spuranomalie ändert.
Unter Ausnutzung der Äquivalenz zwischen Diffeomorphismen im Inneren und Weyl-Transformationen auf dem Rand, berechnen wir diese Modifikationen mithilfe der dualen Gravitationstheorie.
Unsere Ergebnisse repräsentieren einen weiteren Test der AdS/CFT-Korrespondenz.

\selectlanguage{english}
\chapter{Abstract}

Conformal field theories (CFTs) are amongst the most studied field theories and they offer a remarkable playground in modern theoretical physics.
In this thesis we analyse three aspects of CFTs in different dimensions.

First, we consider correlation functions of descendant states in two-dimensional CFTs.
We discuss a recursive formula to calculate them and provide a computer implementation of it.
This allows us to obtain any correlation function of vacuum descendants, and for non-vacuum descendants to express the correlator as a differential operator acting on the respective primary correlator.
With this code, we then study some entanglement and distinguishability measures between descendant states, namely the R\'enyi entropy, trace square distance and sandwiched R\'enyi divergence.
With our results we can test the R\'enyi Quantum Null Energy Condition and provide new tools to analyse the holographic description of descendant states.

Second, we study four-dimensional Weyl fermions on different backgrounds.
Our interest is in their trace anomaly, where the Pontryagin density has been claimed to appear.
To ascertain this possibility, we compute the anomalies of Dirac fermions coupled to vector and axial non-abelian gauge fields and then in a metric-axial-tensor background.
Appropriate limits of the backgrounds allow then to recover the anomalies of Weyl fermions coupled to non-abelian gauge fields and in a curved spacetime.
In both cases, we confirm the absence of the Pontryagin density in the trace anomalies.

Third, we provide the holographic description of a four-dimensional CFT with an irrelevant operator.
When the operator has integer conformal dimension, its presence in the CFT modifies the Weyl transformation of the metric, which in turns modifies the trace anomaly.
Exploiting the equivalence between bulk diffeomorphisms and boundary Weyl transformations, we compute these modifications from the dual gravity theory.
Our results represent an additional test of the AdS/CFT correspondence.

\chapter{List of publications}

This thesis is based on the following papers, which were published during the PhD:
\begin{refsection}
\nocite{*}
\newrefcontext[sorting=ydnt]
\sloppy\printbibliography[heading=none, keyword = {thesis}]
\end{refsection}
The following paper was published during the PhD but will not be discussed in this thesis:
\begin{refsection}
\nocite{*}
\newrefcontext[sorting=ydnt]
\sloppy\printbibliography[heading=none, keyword = {phd}]
\end{refsection}
The full list of publications by the author include:
\begin{refsection}
\nocite{*}
\newrefcontext[sorting=ydnt]
\sloppy\printbibliography[heading=none, keyword = {older}]
\end{refsection}

\chapter{Acknowledgements}

This thesis and the research presented here benefited substantially from Stefan Theisen, whose supervision I gratefully acknowledge.
His guidance and openness towards discussing have been crucial to progress in my research.
His advice about museums and concert halls (wherever I would travel to) has also been always appreciated.

I am grateful to my collaborators Fiorenzo Bastianelli, Enrico Brehm, Sadik Deger and Adriano Vigan\`o for tackling problems together in various stages of my studies and for their patience in answering my questions. 

During my PhD I was supported by the International Max Planck Research School for Mathematical and Physical Aspects of Gravitation, Cosmology and Quantum Field Theory.
I sincerely thank Hadi Godazgar, Axel Kleinschmidt and Hermann Nicolai for managing and coordinating it.

I am thankful to Olaf Hohm for being my supervisor at Humboldt University of Berlin, and to the doctoral committee members for accepting to take part in it.

My time at the Albert Einstein Institute was made precious and fruitful by numerous wonderful people.
In addition to the ones already named, these include: Riccardo, Hugo, Lorenzo, Roberto, Alice, Mehregan, Penelope, Jan, Serena, Caroline, Johannes, Benedikt, Lars, Hannes, Aditya, Darya, Alejandro, Stefano, Tung.
 
I am thankful to Marcello, Serena, Paris, Pedro and Nicole, with whom I have shared enthusiasm and difficulties with physics since the University days.

Special thanks go to Dale and Regina, for the birding adventures, and for the discussions over many cups of tea;
to Hanna and Hans-Gerd, for never running out of beers and hospitality, and for sharing the natural beauty of Golm and Brandenburg with us.
They made me feel at home.

Notwithstanding the distance, Cristiano and Irene have always felt close.
I am thankful to have such wonderful friends and I thank them for always being there.

Finally, I am grateful to my family:
to my parents for their unconditional love and support, to my sister for bringing me down to earth, to my brother-in-law for the warm communion, and to my better half for not having all the answers but figuring them out together.
I am fortunate to have them and I thank them for their love.

\newpage \thispagestyle{empty}
\begin{flushright}\null \vspace{\stretch{1}}
\textit{
e per\`o, prima che tu pi\`u t'inlei,\\
rimira in gi\`u, e vedi quanto mondo\\
sotto li piedi gi\`a esser ti fei\freefootnote{therefore, before thou go farther in, \textbackslash\textbackslash~look down once more, and see how vast a world \textbackslash\textbackslash~thou hast already put beneath thy feet}\\}
\vspace{1ex}
\textsc{Dante Alighieri}, Paradiso XXII%
\vspace{\stretch{3}}\null
\end{flushright}

\clearpage\shipout\null 

\renewcommand{\cleardoublepage}{\newpage} 
\mainmatter
\pagestyle{main}
\chapter{Introduction}\label{chap:intro}

Symmetries play a crucial role in modern physics, and they are often guidelines in the construction of a successful description of natural systems.
A particular example of symmetry is scale invariance, which is realised when enlarging or shrinking the distance between points of a system does not change its features.
Such systems have no preferred scale and they look the same at all distances.

Scale (or dilation) symmetry is found in statistical systems at critical points.
In such systems, it is possible to define the correlation length as the typical distance over which physical observables are statistically correlated.
The correlation length usually depends on external parameters, like the temperature.
The critical points are special values of these parameters at which the correlation length becomes infinite and scale invariance emerges.
Such situation appears in many physical systems, including the vapor-liquid one at the critical point and the ferromagnet-paramagnet transition.
A detailed analysis of critical phenomena can be found in~\cite{Parisi:1988nd}.

It is usually the case that scale invariant theories describing critical points actually possess full conformal symmetry~\cite{Polyakov:1970xd}.
Conformal transformations are local transformations that preserve angles and change the line element by a spacetime dependent scale factor.
Therefore, they are an extension of scale transformation, although in general scale invariance does not imply conformal invariance (the conditions for equivalence between scale and conformal invariance in two and four dimensions are examined in~\cite{Polchinski:1987dy,Dymarsky:2013pqa} and reviewed in~\cite{Nakayama:2013is}).

Besides statistical systems, conformal symmetry is important in many other contexts.
The most developed framework we have to describe known interactions is provided by quantum field theory (QFT) and
conformal symmetry is intrinsic to QFT via the renormalization group (RG) flow~\cite{Wilson:1974mb}.
Indeed, it is usually possible to think of a UV-complete QFT as an RG flow between two fixed points, where the theory is conformal invariant.
This is possible only if the couplings of the operators in the QFT depend on the energy of the process under consideration.
If the coupling grows as the energy is decreased, then the operator is called relevant, while it is called irrelevant if the coupling decreases; if the coupling is constant, the operator is called marginal.
From the RG flow viewpoint, conformal field theories (CFTs) appear as landmarks in the space of QFTs, and their study leads to a better understanding of QFTs.

CFTs are also at the heart of string theory, a candidate theory for the unification of all known interactions (see the textbook~\cite{Blumenhagen:2013fgp} and references therein).
In perturbative string theory, all spacetime fields emerge from a CFT on the worldsheet of the string moving in some background spacetime.
Another important application of CFTs is found in the AdS/CFT correspondence, according to which particular CFTs are conjectured to describe the same physics as particular gravitational theories formulated in anti-de Sitter (AdS) space.
Since the CFT is on the boundary of an AdS space, this correspondence is holographic and fields in the bulk theory are dual to operators of the boundary CFT.
This correspondence has applications ranging from elementary particle physics to condensed matter physics and is a very active area of research (for an introduction and further references see~\cite{Ammon:2015wua}).
Thus, CFTs are a crucial ingredient in modern theoretical physics.


A peculiar aspect of conformal transformations is that they are specified by a finite number of parameters in spacetime with generic $d$ dimensions, except for the case $d=2$. 
In this case, the algebra associated with conformal transformations is infinite dimensional.
This richness of conformal symmetry makes two-dimensional CFTs among the best understood and most studied QFTs.
In some cases they can even be solved exactly \cite{Belavin:1984vu} and under certain conditions all possible CFTs have been classified \cite{Cappelli:1986hf}.

The huge amount of symmetry of two-dimensional CFTs allows to explicitly compute partition and correlation functions.
When dealing with a quantum system, an important example of correlation between the states of the system is called entanglement.
It plays a crucial role in quantum information theory and beyond that provides ways to characterise quantum fluctuations. 
It can reveal whether a system is close to criticality and help classifying its quantum phases~\cite{Amico:2007ag}. 
Therefore measures of entanglement of quantum states play a crucial role in describing the structure of state spaces. 

Another standard way to understand these structures is the development of methods to compare different states. 
Even if the microscopic realization of two states is quite different their meso- or macroscopic features might be very similar. 
One can also consider the opposite scenario. 
Given two states with macroscopically very similar features, e.g.~with the same energy, it would be desirable to understand how distinguishable they are at the microscopic level.

Mathematical measures of distinguishability can attach a lot of structure to the space of states. 
Ideally this structure has physical significance, i.e.~it helps to explain physical phenomena. 
For instance, distinguishability measures help to put the Eigenstate Thermalization Hypothesis, which provides a criterion for a quantum system to reach thermal equilibrium at late times,
on a more quantitative footing~\cite{Srednicki:1995pt};
as another example, these measures should govern the `indistinguishability' of black hole microstates in AdS \cite{Strominger:1996sh,Strominger:1997eq}.
It is therefore relevant to study measures of entanglement and distinguishability, in order to better understand the space of states of quantum systems.

A peculiar feature of a quantum system is that
it may not have the same symmetries present at the classical level.
A classical symmetry which is broken by quantum effects is said to be anomalous.
Since symmetries are extremely important in the structure of a theory, anomalies are extremely important as well.
In particular, two types of anomalies can be distinguished, namely anomalies in rigid or gauge symmetries.
The latter afflict symmetries which are necessary to renormalize the theory, and are thus a tool to spot ill-defined QFTs.
For example, the Standard Model, which is our best description available for the interactions of known particles, has no gauge anomaly; in string theory, the cancellation of anomalies imposes a condition on the spacetime dimensions where the string propagates.

An example of anomaly which will be important in the rest of this thesis is the Weyl (or trace) anomaly.
This anomaly concerns theories which are classically Weyl invariant, i.e.~invariant under local rescaling of the fields and the metric.
This anomaly is of particular interest to CFTs, because it is usually the case that a CFT in flat space can be coupled to a curved metric in a Weyl invariant way, at least at the classical level; conversely, a classical Weyl invariant field theory is a CFT in flat space~\cite{Zumino:1970tu}.
The trace anomaly was first discovered in~\cite{Capper:1974ic} and since then systematically studied, for example, in~\cite{Duff:1977ay,Bonora:1985cq,Deser:1993yx}.
Further review can be found in~\cite{Duff:1993wm,Duff:2020dqb} and a discussion of the history of anomalies is in~\cite{Bastianelli:2006rx}.

The trace anomaly satisfies a consistency condition, called the Wess-Zumino (WZ) consistency condition~\cite{Wess:1971yu}.
This condition alone fixes to a good extent the possible form of the trace anomaly~\cite{Bonora:1985cq}.
In particular, in four dimensions the only solutions to the WZ condition, which cannot be removed with the addition of local counterterms, are the Euler density, the square of the Weyl tensor and the Pontryagin density.

The Pontryagin term, although consistent with the WZ condition, never emerged in the trace anomalies of known Weyl invariant theories.
It was then conjectured to be a possibility for chiral theories in~\cite{Nakayama:2012gu} and it first explicitly appeared in the case of a Weyl fermion in a curved background in~\cite{Bonora:2014qla}, where it is shown that it enters the anomaly with an imaginary coefficient.
Its presence breaks CP symmetry and it was suggested that it might provide a new mechanism for baryogenesis in the early Universe.

The trace anomaly found an important application also in the context of the AdS/CFT correspondence.
Indeed, the computation of the Weyl anomaly from the gravitational bulk theory has been one of the very first non-trivial tests of the AdS/CFT correspondence~\cite{Henningson:1998gx}.
Its success relies on the equivalence between bulk diffeomorphisms and Weyl transformations at the boundary of AdS space, which, since then, has been widely studied.

In~\cite{Henningson:1998gx} Einstein's equations with a negative cosmological constant are solved in terms of the boundary metric and it is shown that the on-shell action is divergent at the boundary.
The divergences can be cancelled with the addition of local counterterms, but the regularisation spoils the conformal symmetry of the boundary theory and gives rise to a holographic Weyl anomaly.

Holographic renormalisation has been further developed in~\cite{deHaro:2000vlm}, where scalar fields coupled to gravity in the bulk and their contributions to the Weyl anomaly are also considered.
When scalar fields are added on top of a dynamical background, they induce a backreaction, and the method presented in~\cite{deHaro:2000vlm} is consistent when the bulk scalars are dual to relevant or marginal operators of the boundary CFT (see also~\cite{Witten:1998qj}).

Scalars that are dual to irrelevant operators induce a stronger backreaction onto the gravitational background, and in~\cite{vanRees:hr_irr_op} the method of holographic renormalisation has been extended to account for such backreaction, when there is no anomaly in the CFT.
This is the case if the irrelevant operators are of non-integer conformal dimension.
Irrelevant operators with integer conformal dimensions are considered in~\cite{vanRees:CS_eq-w_anomalies}, where the conformal anomaly in the three-point function of irrelevant operators is computed in the CFT and derived from holography.
However, in this case no backreaction has been considered and the gravitational background has been taken to be unperturbed by the presence of the scalar fields.
In other words, the boundary field theory is a CFT.

Nonetheless, CFTs with integer-dimensional irrelevant operators have peculiar properties when formulated on a curved background~\cite{Schwimmer:2019efk}.
In order to have a solution of the WZ consistency condition for the Weyl anomaly, the presence of these operators requires a modification of the usual Weyl transformation.
A metric beta-function, which depends on the sources of the irrelevant operators, has to be introduced in the Weyl transformation of the metric.
As a consequence, the Weyl anomaly is deformed by the metric beta-function, i.e.~the solution of the WZ consistency condition is different in the presence of the beta-function.
The geometry, however, is not subject to an RG flow, since correlation functions of irrelevant operators in~\cite{Schwimmer:2019efk} are computed in the undeformed CFT.

\section{Results of this work}

The original results presented in this thesis contribute to three different aspects of CFTs.
We summerise here the problems we analysed, how we tackled them and the results we obtained.

\subsubsection*{Correlators and quantum measures of descendant states}

In the context of two-dimensional CFTs, we put our focus on descendant states, i.e.~states excited by the generators of the conformal algebra.
In particular, we consider an infinite cylinder and a quantum system on a spatial circle of length $L$, and we assume that its physics is described by a two-dimensional CFT on the cylinder.
We further take a segment of this circle of size $l < L$ onto which we reduce the density matrix of the full system and look at reduced density matrices associated with descendant states.
For this kind of construction, it is in principle possible to compute algebraic expressions for entanglement and distinguishability measures for any descendant by relating them to correlation functions of descendant states.
How to do it was shown in \cite{Palmai:2014jqa,Taddia:2016dbm}. 
Correlators of descendant states can then be computed in terms of correlators of the respective primary states by employing the conformal algebra recursively.
However, in practice this recursion soon yields cumbersome expressions which make the computation difficult to handle.

To overcome this issue, we implement a recursive algorithm to analytically compute generic correlators of descendants on a computer.
In case of vacuum descendants it results in an analytic expression of the insertion points and the central charge of the theory. In case of descendants of arbitrary primary states the function returns a differential operator acting on the respective primary correlator.
We use Mathematica for our computations and explicitly display important parts of our code in the appendices. 
The notebooks with the remaining code are openly accessible. 

With this tool at hand, we are able to compute measures of entanglement, like the R\'enyi entropy, and measures of distinguishability, like the sandwiched R\'enyi divergence (SRD) and the trace squared distance (TSD).
The SRD and TSD have not been computed for descendant states before, while in the case of the R\'enyi entropy we can expand on existing results. 
The outcomes for the SRD for example allow us to test a generalisation of the Quantum Null Energy Condition suggested in \cite{Lashkari:2018nsl}. 
Results that we compute for vacuum descendants are universal and, in particular, can be studied at large central charge, i.e.~the regime where two dimensional conformal field theories may have a semi-classical gravitational dual in three-dimensional AdS space.
We will show results for vacuum descendant states in this limit.

These results are presented in chap.~\ref{chap:desc_corr} and were first published in~\cite{Brehm:2020zri}.

\subsubsection*{Weyl fermions and trace anomalies}
The presence of a Pontryagin term indicated in~\cite{Bonora:2014qla} in the four-dimensional trace anomaly of a Weyl fermion is surprising.
Indeed, by CPT symmetry a left-handed fermion has a right-handed antiparticle, which is expected to give an opposite contribution to any chiral imbalance in the coupling to gravity.
To see this, one can cast the theory of Weyl fermion as that of a Majorana fermion; the functional determinant that arises in a path integral quantization can be regulated in such a way to keep it manifestly real, so that the appearance of an imaginary term in the anomaly is to be excluded~\cite{Alvarez-Gaume:1983ihn}.

An independent computation showed indeed the absence of the Pontryagin term \cite{Bastianelli:2016nuf} and called for additional studies to clarify the issue.
Additional analysis eventually appeared, confirming the absence of the Pontryagin term~\cite{Godazgar:2018boc,Frob:2019dgf}.
However, some of the authors of~\cite{Bonora:2014qla} advocated again for the presence of the Pontryagin term~\cite{Bonora:2017gzz}.

To further clarify the issue, we compute the four-dimensional trace anomaly of a Weyl fermion coupled to a non-abelian gauge background~\cite{Bastianelli:2019fot}.
This is analogous to the case of the Weyl fermion in a gravitational background, as a contribution from the parity-odd Chern-Pontryagin density of the gauge background can appear in the trace anomaly.
To compute it, we consider Bardeen's method that embeds the Weyl theory into the theory of Dirac fermions coupled to vector and axial non-abelian gauge fields~\cite{Bardeen:1969md}.
Using a Pauli-Villars (PV) regularization we calculate its trace anomaly, which was not known before, and as an aside we rederive the well-known non-abelian chiral  anomaly to check the consistency of our methods.
A suitable limit on the background produces the searched-for anomalies of the non-abelian Weyl fermions.
Our results show no presence of the Chern-Pontryagin density, confirming previous results in the case of an abelian gauge background~\cite{Bastianelli:2018osv}.

Bardeen's method can be generalized to curved space by considering an axial extension to the ordinary metric~\cite{Bonora:2017gzz,Bonora:2018obr}.
A theory of Dirac fermions in this metric-axial-tensor (MAT) background reproduces that of Weyl fermions coupled to ordinary gravity in a suitable limit of the background field.
We thus extend the analysis of~\cite{Bastianelli:2016nuf} by employing this background, as it gives more freedom for studying chiral couplings to gravity.
To check the consistency of our method, we again compute the known anomalies for the Dirac fermion, and then find the anomalies for the Weyl fermion by taking limits on the background.
Our final results support the absence of the Pontryagin term in the trace anomaly~\cite{Bastianelli:2019zrq}.

Our results for the trace anomalies of Weyl fermions are presented in chap.~\ref{chap:weyl_fermions} and were published in~\cite{Bastianelli:2019fot,Bastianelli:2019zrq}.

\subsubsection*{Irrelevant operators and their holographic anomalies}

In the context of the holographic Weyl anomaly, we compute the contribution of a bulk scalar field to the anomaly when the scalar field sources an integer-dimensional operator on the boundary CFT.
In particular, we consider the case of a four-dimensional boundary CFT with an operator with conformal dimension five, thus providing the holographic description of the four-dimensional CFT studied in~\cite{Schwimmer:2019efk}.
Therefore, we also generalise the analysis of~\cite{vanRees:CS_eq-w_anomalies} to include the case in which the bulk scalar fields are coupled to a dynamical background.
However, to make contact with~\cite{Schwimmer:2019efk}, we will be interested in describing a boundary theory that is not deformed by the irrelevant operators.

To do so, the tool that we find most convenient to use is that of Penrose-Brown-Henneaux (PBH) transformations~\cite{Imbimbo:1999bj}.
These are a particular class of bulk diffeomorphisms that reduces to Weyl transformation on the boundary. 
They consist of a general transformation rule for the bulk metric, and, as such, they do not require solving any equation of motion.
An action evaluated on a metric that is a solution of the PBH transformation allows to study the Weyl anomaly of the boundary theory.
Thus, we extend the analysis of~\cite{Imbimbo:1999bj} to include massive scalar fields in the bulk and provide a holographic description of the results obtained in~\cite{Schwimmer:2019efk}.
As we will show, the condition that the irrelevant operators at the boundary do not deform the CFT requires that we are off-shell in the bulk, thus making the PBH transformation an ideal framework for the present analysis.

Irrelevant deformations of CFTs have received attention in particular in the form of $T \bar T$ deformations.
It is proposed that a two-dimensional $T \bar T$ deformed CFT is dual to a three dimensional AdS space with a sharp cutoff~\cite{McGough:2016lol}.
This conjecture has then been further analysed and extended to higher dimensions~\cite{Guica:2019nzm,Taylor:2018xcy,Hartman:2018tkw}.
Although we are not considering a $T\bar T$ deformation, the scalar field that we add in the bulk is sourcing an integer-dimensional irrelevant operator on the boundary.
As we will see, this addition still has the effect of moving the boundary into the bulk, but, since we are interested in describing an undeformed boundary theory, we have to move the cutoff back to the AdS boundary where the undeformed CFT lives.
In so doing, the solutions of the PBH transformations go off-shell, in the sense that they don't match anymore the solution of the equations of motion of a scalar field coupled to a dynamical background.
Nonetheless, once this is done we precisely recover the physics described in~\cite{Schwimmer:2019efk}.
We will thus see that the Weyl transformation of the metric is no longer the usual one, but it acquires a beta-function, and we are able to study the modified anomaly in our holographic set-up.
In the end, this analysis will sharpen the understanding of the AdS/CFT duality in the presence of a backreaction in the bulk theory.

These results are presented in chap.~\ref{chap:holography} and were first published in~\cite{Broccoli:2021icm}.

\section{Outline}
The rest of the thesis is organised as follows.

In chap.~\ref{chap:intro_tech} we provide a technical introduction to the topics on which later chapters are based.
In sec.~\ref{sec:conf_symm} we discuss conformal symmetry and provides more details about two-dimensional CFTs in sec.~\ref{sec:2d_cft}.
Sec.~\ref{sec:weyl_symm} is devoted to Weyl symmetry and anomalies; in particular, we will discuss the technique that we will use to compute anomalies for the fermionic theories.
Sec.~\ref{sec:holography} provides an introduction to holography and the AdS/CFT correspondence; we will also present the PBH transformations and compute the holographic Weyl anomaly as a test of the correspondence.

In chap.~\ref{chap:desc_corr} we present the results of~\cite{Brehm:2020zri}, namely we will discuss a recursive formula to compute correlation function of descendant states in terms of the correlator of the respective primaries and compute some quantum measures for descendant states.
In sec.~\ref{sec:CFTtec} we derive the recursive formula. 
In the following sec.~\ref{sec:qmeasures} we discuss the quantum measures that we want to compute, i.e.~the R\'enyi entanglement entropy as a measure of entanglement, and the sandwiched R\'enyi divergence and the trace square distance as measures of distinguishability between states reduced to a subsystem. 
In sec.~\ref{sec:universal} we focus on results for descendants of the vacuum; these will apply to all theories with a unique vacuum and, hence, we call them universal. 
In sec.~\ref{sec:nonuniversal} we show the results for descendants of generic primary states, which depend on the primary correlators that are theory dependent;
we compute results in two explicit models, namely the critical Ising model and the three-state Potts model.


In chap.~\ref{chap:weyl_fermions} we present the results of~\cite{Bastianelli:2019fot,Bastianelli:2019zrq} for the anomalies of Weyl fermions.
In sec.~\ref{sec:weyl_fermions_gauge} we discuss the Bardeen model for the Dirac fermion and use it to compute the chiral and trace anomalies for a Weyl fermion.
In sec.~\ref{sec:weyl_mat} we consider the theory of a Dirac fermion in a MAT background and compute related anomalies; taking a suitable limit on the background we will derive the corresponding anomalies for a Weyl fermion coupled to gravity.

In chap.~\ref{chap:holography} we present the results of~\cite{Broccoli:2021icm}.
In sec.~\ref{sec:pbhwscalar} we extend the PBH transformations to describe a  scalar field coupled to gravity; we show that, by choosing the scalar field to be dual to an integer-dimensional irrelevant operator, the transformation of the boundary metric is no longer the usual Weyl transformation.
We derive the modified Weyl anomaly of the four-dimensional boundary theory in sec.~\ref{sec:pbhwscalar_ano}.

We conclude in chap.~\ref{chap:conclusions} and then provide relevant notations, conventions and additional details in the appendices.
\chapter{Basics of Conformal Field Theory and Holography}\label{chap:intro_tech}

In this chapter we introduce the concepts and mathematical tools needed in the rest of the thesis.
In sec.~\ref{sec:conf_symm} we define conformal transformations and give a brief overview of the implications of conformal symmetry in field theories.
We then provide more details about two-dimensional CFTs in sec.~\ref{sec:2d_cft}.
In sec.~\ref{sec:weyl_symm} we define Weyl transformations and anomalies.
Finally, we discuss the importance of CFTs and anomalies in connection with the AdS/CFT correspondence in sec.~\ref{sec:holography}.

Reviews and textbooks have been written on these topics, which are still very active areas of research.
We are not interested in providing a complete overview of them, which would be by far out of our scope.
We will rather briefly review some general aspects which are essential in the following chapters, keeping a bird's-eye view that provides a unifying picture of the different research threads developed in the rest of the thesis.
We will refer to relevant literature for more details.

\section{Conformal symmetry}\label{sec:conf_symm}
We start by presenting conformal transformations and discuss their peculiarities in spacetimes with dimensions higher than two or equal to two, and provide some consequences of conformal symmetry in field theory.
Comprehensive and popular introductions to these topics are~\cite{DiFrancesco:1997nk,Blumenhagen:2009zz};
shorter and self-contained introductions can also be found in~\cite{Blumenhagen:2013fgp,Ammon:2015wua}.

We define conformal transformations as those coordinate transformations that leave angles invariant.
A conformal transformation changes the line element by a spacetime dependent scale factor:
\begin{equation}\label{line_elem}
    x \to x'(x) \, , \qquad\qquad ds'^{ 2} = e^{2\sigma(x)} ds^2 \, .
\end{equation}
In the case of a flat spacetime with metric $\eta_{\mu\nu}$, the definition~\eqref{line_elem} implies that an infinitesimal transformation $x'^\mu = x^\mu + \epsilon^\mu(x)$ has to satisfy
\begin{equation}
    2\sigma \eta_{\mu\nu} = \partial_\mu \epsilon_\nu + \partial_\nu \epsilon_\mu \, .
\end{equation}
After some manipulations, this condition can be recast as
\begin{equation}\label{conf_condition}
    \left( \eta_{\mu\nu} \Box + (d-2)\partial_\mu\partial_\nu \right) \partial \epsilon = 0 \,, \quad\quad \sigma = \frac{\partial \epsilon}{d} \,,
\end{equation}
with $\Box = \partial^\mu\partial_\mu$ and $\partial \epsilon = \partial_\mu \epsilon^\mu$.
We see that setting $d=2$ simplifies this condition and indeed it has dramatic consequences.
Thus, we will treat separately the cases $d=2$ and $d\geq 3$.\footnote{We will not discuss the one-dimensional case, but refer the interested reader to~\cite{deAlfaro:1976vlx} for details about one-dimensional field theory invariant under the conformal group and to~\cite{Cadoni:1999ja,Bianchi:2021piu} and references therein for studies of the holographic duality, correlation functions and corresponding Mellin amplitude in one-dimensional CFT.
One-dimensional statistical models are discussed in~\cite{Mussardo:2020rxh}.}

\subsection{\texorpdfstring{$d\geq 3$}{}}
In $d\geq 3$, eq.~\eqref{conf_condition} is solved by
\begin{equation}\label{epsilon3d}
    \epsilon^\mu(x) = a^\mu + \omega^{\mu}{}_{\nu} x^\nu + \lambda x^\mu + b^\mu x^2 -2 b_\nu x^\nu x^\mu \, , \qquad \sigma(x) = \lambda -2 b_\mu x^\mu \, ,
\end{equation}
where $\omega_{\mu\nu}$ is antisymmetric.
The parameters $a^\mu$ and $\omega_{\mu\nu}$ are related respectively to translations and Lorentz transformations;
$\lambda$ and $b^\mu$ are instead peculiar to conformal transformations and they are related to dilations and special conformal transformations.
Notice that, since these parameters have a finite number of components, the conformal algebra associated with the conformal transformations in $d \geq 3$ is finite dimensional.
Denoting by $P_\mu$ and $L_{\mu\nu}$ the generators of translations and rotations, they satisfy the commutation relations of the Poincar\'e group, familiar from standard classical relativistic field theory.
With $D$ and $K_\mu$ we denote the generators of dilations and special conformal transformations.
Explicitly, these generators read on spin-less fields
\begin{equation}
\begin{aligned}
    P_\mu &= -i \partial_\mu \, , & \qquad\qquad L_{\mu\nu} &= i \left(x_\mu \partial_\nu - x_\nu \partial_\mu \right) \, , \\
    D &= -i x^\mu\partial_\mu \, , &  K_\mu &= -i \left(2 x_\mu x^\nu \partial_\nu - x^\nu x_\nu \partial_\mu\right) \, .
\end{aligned}
\end{equation}
Then, we may define
\begin{equation}
\begin{aligned}
    J_{\mu\nu} &:= L_{\mu\nu} \, , & \qquad\qquad J_{(-1)\mu} &:= \frac12(P_\mu - K_\mu) \, , \\
    J_{(-1)(0)} &:= D \, , & J_{(0)\mu} &:= \frac12 (P_{\mu} + K_{\mu})
\end{aligned}
\end{equation}
and the $J_{mn}$ with $m,n = -1,0,1,\ldots,(d-1)$ satisfy the following commutation relations
\begin{equation}
    [J_{mn}, J_{rs}] = i \left(\eta_{ms} J_{nr} + \eta_{nr} J_{ms} - \eta_{mr} J_{ns}  - \eta_{ns} J_{mr}\right) \, .
\end{equation}
For a $d$-dimensional Minkowski space, $\eta_{mn} = \text{diag}(-1,-1,1,\ldots,1)$ so that we identify the conformal group with $SO(d,2)$.
Interestingly, this is also the isometry group of a $(d+1)$-dimensional Anti-de Sitter spacetime; we will come back to this in sec.~\ref{sec:holography}.

In a field theory, a field $\phi$ is called primary if under a conformal transformation $x^\mu \to x'^\mu$ it transforms as
\begin{equation}\label{prim_highd}
    \phi(x) \to \phi'(x) = \left| \frac{\partial x'}{\partial x} \right|^{\Delta/d} \phi(x') \,,
\end{equation}
where $|\partial x'/\partial x|$ is the Jacobian of the transformation and $\Delta$ is the scaling dimension of $\phi$, as can be seen by restricting the conformal transformation to a scale transformation.
In a CFT, due to the transformation~\eqref{prim_highd}, correlation functions of primary fields are determined to some extent by the conformal symmetry.
In particular, the two- and three-point functions are fixed to be
\begin{align}
    \left\langle \phi_1(x_1)\phi_2(x_2) \right\rangle &= \frac{c_{12}}{\left((x_1-x_2)^2 \right)^{\Delta_1}} \delta_{\Delta_1,\Delta_2} \, ,\\
     \left\langle \phi_1(x_1)\phi_2(x_2) \phi_3(x_3) \right\rangle &= \frac{c_{123}}{x_{12}^{\Delta_1+\Delta_2-\Delta_3} x_{23}^{\Delta_2+\Delta_3-\Delta_1} x_{13}^{\Delta_1+\Delta_3-\Delta_2}} \, , \quad x_{ij} = | x_i-x_j | \,,
\end{align}
where $c_{12}$ is a normalisation factor while $c_{123}$ is characteristic of the CFT.
A comprehensive discussion with more details about conformal two- and three-point functions may be found in~\cite{Osborn:1993cr,Erdmenger:1996yc}.
Higher point functions are less constrained by the symmetry.
This is because out of at least four points it is possible to construct the so-called cross ratios (e.g.~$x_{12}x_{34}/x_{13}x_{24}$) which are invariant under conformal transformations and the dependency of higher point functions on the cross ratios cannot be fixed by symmetry arguments alone.
However, the product of fields appearing in the correlator is associative and for Grassmann-even fields the correlator does not depend on the order of the fields.
These two properties provide further constraints on the correlation functions, but we will not elaborate more on this direction.
Similar results about correlation functions are also obtained in two-dimensional CFTs.

An important field in CFTs is the symmetric and conserved stress-energy tensor $T_{\mu\nu}$.
We recall here some of its properties, but discuss it again in sec.~\ref{sec:weyl_symm} in the context of Lagrangian field theories.
The stress tensor is defined through Noether's theorem when the theory is invariant under translations.
Since any conformally invariant theory is also invariant under translation, the stress-energy tensor is defined in every CFT.
In particular, when a theory is invariant under the conformal transformation $x^\mu \to x^\mu + \epsilon^\mu(x)$ with $\epsilon(x)$ as in~\eqref{epsilon3d} we have a conserved current given by $T_{\mu\nu} \epsilon^\nu$.
From translation invariance, we find that the stress-energy tensor is conserved
\begin{equation}\label{conserved_T}
    \partial_\mu T^{\mu\nu} = 0  \, .
\end{equation}
In a CFT, the conservation of the current $T_{\mu\nu} \epsilon^\nu$ further implies that $T_{\mu\nu}$ is traceless
\begin{equation}\label{traceless_T}
    T^\mu{}_\mu = 0 \, .
\end{equation}
In two-dimensional CFTs the tracelesness condition has further important implications for the stress tensor, and we will discuss them in sec.~\ref{sec:fields_2dcft}.

\subsection{\texorpdfstring{$d=2$}{d=2}}
In two dimensions, the condition~\eqref{conf_condition} is equivalent to the Cauchy-Riemann equations.
Thus, the infinitesimal generator of conformal transformations in two dimensions is a (anti-)holomorphic function and in Euclidean signature it is more convenient to introduce complex coordinates $(z, \bar z)$ so that $\epsilon(z)$ is holomorphic and $\bar\epsilon(\bar z)$ is anti-holomorphic.
If we perform a Laurent expansion of $\epsilon(z)$ and $\bar\epsilon(\bar z)$, we then identify the generators of a conformal transformation as
\begin{equation}
    l_n = - z^{n+1} \partial_z \, , \quad \bar l_n = - \bar z^{n+1} \partial_z \, , \quad n \in \mathbb{Z} \, .
\end{equation}
The generators satisfy the following commutation relations
\begin{equation}
    [l_m, l_n] = (m-n) l_{m+n} \, , \quad [\bar l_m, \bar l_n] = (m-n) \bar l_{m+n} \, , \quad [l_m, \bar l_n] =0 \,,
\end{equation}
which consist of two commuting copies of the Witt algebra.
Notice that, since $n$ is integer, the number of independent infinitesimal conformal transformations is infinite in two dimensions.
However, these transformations are not everywhere well defined, i.e.~they do not always generate an invertible map.
Each of these two infinite-dimensional algebras contains a finite subalgebra generated by the operators $\{l_{-1},l_0,l_1\}$, and their respective anti-holomorphic copies.
This is the subalgebra associated with the global conformal group $SL(2,\mathbb{C})/\mathbb{Z}_2$, which is everywhere well defined and isomorphic to the M\"obius group.
In particular, $l_{-1}$ generates translations, $l_0$ generates scale transformations and rotations and $l_{+1}$ generates special conformal transformations.
The linear combinations $l_0 + \bar l_0$ and $i(l_0 - \bar l_0)$ generate scale transformations and rotations, respectively.

The Witt algebra admits a central extension, whose elements we will call $L_n$ and $\bar L_n$ with $n \in \mathbb{Z}$.
They enjoy the commutation relation
\begin{equation}\label{virasoro}
    [L_m , L_n] = (m-n) L_{m+n} + \frac{c}{12} (m^3 - m) \delta_{m+n,0} \,,
\end{equation}
where $c \in \mathbb{R}$ is called the central charge.
A similar relation holds for the commuting $\bar L_n$ with central charge $\bar c$.
This algebra is called the Virasoro algebra and it is the quantum symmetry algebra of CFTs, while the Witt algebra is the classical one.
Indeed, the term proportional to the central charge in~\eqref{virasoro} physically arises as a quantum effect and it can be related to an anomaly.
We will focus on anomalies in sec.~\ref{sec:weyl_symm}.

Notice that the central term vanishes for $m,n=-1,0,1$, so that $\{L_{-1},L_0,L_1\}$ are generators of $SL(2,\mathbb{C})/\mathbb{Z}_2$ and the linear combinations $L_0 + \bar L_0$ and $i(L_0 - \bar L_0)$ are associated with the operators which generate dilations and rotations respectively.
At the quantum level, we will then employ radial quantization to define two-dimensional CFTs on the complex plane, which means that the plane is foliated into slices of equal radius and the dilaton operator moves from one slice to the other. Products of fields will be radially ordered, so that when we write $\phi_1(z_1)\phi_2(z_2)$ we will implicitly assume that $|z_1|>|z_2|$.

\section{More about two-dimensional CFTs}\label{sec:2d_cft}
We focus here on two-dimensional CFTs, and provide more details that will be useful in chap.~\ref{chap:desc_corr}.
The seminal paper about CFTs in two dimension is~\cite{Belavin:1984vu}.
We refer to the textbooks~\cite{DiFrancesco:1997nk,Blumenhagen:2009zz}.

\subsection{Fields}\label{sec:fields_2dcft}
The basic objects of a two-dimensional CFT are fields $\phi(z,\bar z)$.
Fields which only depend on the (anti-)holomorphic coordinate will be called (anti-)holomorphic.
We will be interested in computing how different fields inserted at different points correlate, which is captured in correlation functions
\begin{equation}
    \langle \phi_1(z_1,\bar z_1) \phi_2(z_2,\bar z_2) \ldots \phi_n(z_n,\bar z_n) \rangle \, .
\end{equation}
Correlation functions are related to measurable quantities.
They are usually singular when fields are inserted at coincident points.
In a quantum field theory, the computation of correlation functions typically requires the notion of an action or a functional integral where the fields explicitly appear.
In a CFT, instead, conformal invariance is usually enough to define a theory and extract information about correlation functions to some extent without reference to an action.

As in the higher-dimensional case, a special class of fields is that of primary fields.
Under a conformal transformation $z \to z'$, $\bar z \to \bar z'$ a field $\phi$ is primary if it transforms as
\begin{equation}\label{prim_field}
    \phi(z,\bar z) \to \phi^\prime(z,\bar z) = \left( \frac{\partial z'}{\partial z} \right)^h \left( \frac{\partial \bar z'}{\partial \bar z} \right)^{\bar h} \phi\left(z',\bar z'\right) \,,
\end{equation}
where $(h,\bar h)$ are called the conformal weights of the field and we will call $\Delta= h + \bar h$ its conformal dimension.
A field for which~\eqref{prim_field} holds only under Möbius transformation is called quasi-primary field.
It is clear that a primary field is also a quasi-primary field.
A field which is neither primary nor quasi-primary is called secondary field.

In two dimension, the stress tensor is a quasi-primary field.
It is conserved and traceless as we argued in~\eqref{conserved_T} and~\eqref{traceless_T}.
Moreover, in two dimensions and on the complex plane, from the tracelessness condition one finds that the stress-energy tensor has only two non-vanishing components, which are holomorphic and anti-holomorphic due to the conservation of $T_{\mu\nu}$.
Thus, we define the holomorphic and anti-holomorphic components of the stress-energy tensor as
\begin{equation}
    T(z) := T_{zz}(z) \, , \quad\quad \bar T(\bar z) := T_{\bar z \bar z} (\bar z) \, .
\end{equation}
Since $T(z)$ is holomorphic, we can Laurent-expand it as
\begin{equation}\label{T_laur_exp}
    T(z) = \sum_n z^{-n-2} L_n \quad \text{s.t.} \quad  L_n = \oint \frac{dz}{2\pi i} z^{n+1} T(z) \,,
\end{equation}
where the $L_n$'s are the Virasoro generators, and similarly for $\bar T(z)$.

\subsection{Correlation functions of quasi-primary fields}
Given the transformation~\eqref{prim_field} of quasi-primary fields under a conformal transformation, in a CFT it is possible to fix their correlation functions to some extent.
Here we present some results about correlation functions of quasi-primary fields on the plane, which follow from employing the $SL(2,\mathbb{C})/\mathbb{Z}_2$ symmetry of the theory.

Due to translation invariance, the $n$-point function can only depend on the difference between the insertion points.
This means that the one-point function has to be a constant, and in particular it vanishes for quasi-primary fields because of dilation invariance.
Thus, the one-point function reads
\begin{equation}
    \langle \phi_i(z,\bar z) \rangle = \delta_{i0} \,,
\end{equation}
where the label 0 stands for the unit operator.
Notice that this result is not necessarily true on different geometries, e.g.~for CFTs on geometries with boundaries.

The global conformal symmetry fixes the two-point function to be
\begin{equation}
    \langle \phi_1(z_1,\bar z_1)\phi_2(z_2,\bar z_2) \rangle = \frac{d_{12} \, \delta_{h_1,h_2} \, \delta_{\bar h_1, \bar h_2}}{(z_1 - z_2)^{2h_1} (\bar z_1 -\bar z_2)^{2 \bar h_1}} \,,
\end{equation}
while the three-point function reads
\begin{equation}
    \langle \phi_1(z_1,\bar z_1)\phi_2(z_2,\bar z_2) \phi_3(z_3,\bar z_3) \rangle = \frac{C_{123}}{z_{12}^{(123)} z_{23}^{(231)} z_{13}^{(132)} \bar z_{12}^{\overline{(123)}} \bar z_{23}^{\overline{(231)}} \bar z_{13}^{\overline{(132)}}} \,,
\end{equation}
where $z_{ij} = z_i - z_j$ and $(ijk) = h_i+h_j-h_k$.
Notice that $d_{12}$ is simply a normalisation factor, while $C_{123}$ is characteristic of the CFT.

Higher-point functions have more complicated structures and they are no longer determined up to a constant.
This is because out of four points it is possible to form cross ratios
\begin{equation}
    \eta_{12}^{34} = \frac{z_{12}z_{34}}{z_{13}z_{24}} \,,
\end{equation}
which are invariant under global conformal transformations.
The general structure for the $n$-point function is
\begin{equation}
     \langle \phi_1(z_1,\bar z_1)\ldots\phi_n(z_n,\bar z_n) \rangle = f\left(\eta,\bar\eta \right) \prod_{i<j} z_{ij}^{-\gamma_{ij}} \bar z_{ij}^{-\bar \gamma_{ij}} \,,
\end{equation}
where the symmetric $\gamma_{ij}$ and $\bar \gamma_{ij}$ are solutions of the sets of equations
\begin{equation}
    \sum_{j\neq i} \gamma_{ij} = 2h_i \, , \quad\quad \sum_{j\neq i} \bar\gamma_{ij} = 2\bar h_i \,,
\end{equation}
and $f$ is a function of the cross ratios which is undetermined by global conformal invariance.
Similar results are also obtained for $n$-point functions in $d>2$ dimensions.
In chap.~\ref{chap:desc_corr} we will see explicit examples of four-point functions in two-dimensional CFTs.

\subsection{Ward identities and OPE}
We just saw that conformal symmetry is very powerful in constraining the form of correlation functions.
Quantum manifestations of the symmetries of a theory on correlation functions are encoded in the so-called Ward identities.
We present here only one example, namely the conformal Ward identity which follows from a conformal transformation on the $n$-point function of primary fields.
Its holomorphic version reads
\begin{equation}\label{wardid}
    \langle T(z) \phi_1(z_1,\bar z_1) \ldots \phi_n(z_n,\bar z_n)\rangle = \sum_{j=1}^n \left( \frac{h_j}{(z-z_j)^2} + \frac{\partial_{z_j}}{z-z_j} \right) \langle \phi_1(z_1,\bar z_1) \ldots \phi_n(z_n,\bar z_n)\rangle
\end{equation}
and gives important information about the singularities of the correlator as $z\to z_j$.
In particular, from the Ward identity we see that we can expand the product of the stress-energy tensor with a primary field of conformal weight $h$ as
\begin{equation}\label{ope_Tphi}
     T(z) \phi(w,\bar w) = \frac{h}{(z-w)^2} \phi(w,\bar w) + \frac{1}{z-w} \partial_w \phi(w,\bar w) + \ldots \,,
\end{equation}
where $\ldots$ means regular terms which are not captured by the Ward identity, since they vanish when integrated over closed contour integrals.
This is our first example of an operator product expansion (OPE) between two fields. Notice that the OPE holds inside correlators, but for simplicity we will not write the correlator brackets.

The idea of the OPE is that two fields at nearby points can be expanded as
\begin{equation}
    O_i(z,\bar z) O_j(w,\bar w) = \sum_k C_{ij}^k(z-w,\bar z - \bar w) O_k(w,\bar w) \,,
\end{equation}
where $O_k$ is a set of local fields and $C_{ij}^k$ are functions which can only depend on the difference of the insertion points.
Thus, we can write the OPE~\eqref{ope_Tphi} as
\begin{equation}
    T(z) \phi(w,\bar w) = \sum_{k=0}^\infty (z-w)^{k-2} \phi^{(k)} (w,\bar w) \,,
\end{equation}
where $\phi^{(0)} = h \phi$ and $\phi^{(1)} = \partial \phi$, while the fields $\phi^{(k)}$ for $k\geq 2$ appear in the regular terms of the OPE~\eqref{ope_Tphi} and we define
\begin{equation}\label{def_desc_field}
    \hat L_{-k} \phi(w) := \phi^{(k)}(w) = \oint_{\gamma_w} \frac{dz}{2\pi i} (z-w)^{1-k} T(z) \phi(w) \,,
\end{equation}
where $\gamma_w$ is a closed path around $w$.
The fields $\phi^{(k)}$ for $k\geq 1$ are called secondary fields or descendants of the primary field $\phi^{(0)}$.
A primary field $\phi$ thus defines an infinite family of fields, which is called the conformal family.

Another important OPE is that of the stress-energy tensor with itself, which for the holomorphic part reads
\begin{equation}\label{opeTT}
    T(z)T(w) = \frac{c/2}{(z-w)^4} + \frac{2}{(z-w)^2} T(w) + \frac{1}{z-w} \partial_w T(w) + \ldots \,,
\end{equation}
from which we see that the stress-energy tensor is not a primary field and its conformal weight is $h=2$.
From the transformation of the stress tensor under an infinitesimal conformal transformation it can be shown that the stress tensor is a quasi-primary field.
Notice that the singular terms of this OPE can be shown to be equivalent to the Virasoro algebra for the $L_n$'s modes, upon using the Laurent expansion~\eqref{T_laur_exp}.
Moreover, $T(z)$ is the second descendant field in the conformal family of the identity operator
\begin{equation}
    \mathbb{I}^{(2)}(z) = \hat L_{-2} \mathbb{I}(z) = \oint_{\gamma_z} \frac{dw}{2\pi i} \frac{1}{w-z} T(w) \mathbb{I}(z) = T(z) 
\end{equation}
and it is present in every CFT, since the identity operator itself is.
Notice that from conformal symmetry and the OPE~\eqref{opeTT} we find the stress-energy tensor two-point function
\begin{equation}\label{<TT>}
    \left\langle T(z)T(w) \right\rangle = \frac{c/2}{(z-w)^4} \, .
\end{equation}

\subsection{State-operator correspondence and the Hilbert space}
Let us discuss now the space of states and some representation theory of CFTs.
We denote the vacuum state of a CFT as $\ket{0}$ and we define it by
\begin{equation}\label{CFT_vac}
    L_n \ket{0} =0 \, \quad\quad \text{for} \quad\quad n\geq -1 \,,
\end{equation}
so that $T(z)\ket{0}$ is regular at the origin of the plane $z=0$.
Since the $L_n$'s with $n\geq -1$ include the generators of global conformal transformation, the vacuum defined in eq.~\eqref{CFT_vac} is invariant under the conformal group.
The states of the theory are generated by acting with fields on the vacuum.
In particular, we define the state $\ket{\Delta}$ corresponding to a primary field $\phi(z)$ with conformal dimension $\Delta$ as
\begin{equation}
    \ket{\Delta} = \phi(0) \ket{0} \,,
\end{equation}
with
\begin{equation}\label{hws}
   L_0 \ket{\Delta} = h \ket{\Delta}\, , \quad\quad L_n \ket{\Delta} = 0 \, \quad\quad \text{for} \quad\quad n>0 \, .
\end{equation}
The $L_{-n}$'s with $n\geq 0$ raise the eigenvalue of $L_0$ as
\begin{equation}
    L_0 \left( L_{-n} \ket{\Delta} \right) = (n+h)\left( L_{-n} \ket{\Delta} \right) \, \quad\quad \text{for} \quad\quad n>0 \, .
\end{equation}
Descendant states are obtained by applying secondary fields to the vacuum, as
\begin{equation}
    \phi^{(-n)}(0)\ket{0} = L_{-n}\ket{\Delta} \, \quad\quad \text{for} \quad\quad n>0 \, .
\end{equation}
Therefore, every field can be related to a corresponding state in a one-to-one relation, which is called state-operator correspondence.
In the following, given a state $\ket{s}$ we will denote the corresponding field as $\phi_{\ket{s}}$.

The dual vacuum $\bra{0}$ is defined as
\begin{equation}
    \bra{0} L_n = 0 \, \quad\quad \text{for} \quad\quad n \leq 1 \,,
\end{equation}
so that the generators of $SL(2,\mathbb{C})$ annihilate both $\ket{0}$ and its dual.
The dual state $\bra{\Delta}$ is defined by the action of $\phi$ at infinity on the dual vacuum and we have that
\begin{equation}
    \bra{\Delta} L_0 = \bra{\Delta} h \, , \quad \bra{\Delta}L_n = 0 \, \quad \text{for} \quad n<0 \, .
\end{equation}
A similar construction holds for the $\bar L_n$'s as well.

A state which satisfies the conditions~\eqref{hws} is called a highest weight state.
Notice that the vacuum $\ket{0}$ is a highest weight state itself, and in a unitary CFT it is the one with the lowest eigenvalue of $L_0$.
Descendant states are generated by acting with the raising operators $L_{-n}$ with $n>0$ on the highest weight states.
In particular, given a primary state $\ket{\Delta}$, we write the descendant states and corresponding descendant fields as
\begin{equation}
    \ket{\Delta,\{(m_i,n_i)\}} = \prod_i L^{n_i}_{-m_i} \ket{\Delta} \quad \leftrightarrow \quad  \phi_{ \ket{\Delta,\{(m_i,n_i)\}}} = \prod_i \hat{L}_{-m_i}^{n_i} \phi_{\ket{\Delta}} \,,
\end{equation}
where $\hat{L}_{-m} \phi_{\ket{\Delta}}$ is as in~\eqref{def_desc_field} and $\sum_i m_i n_i$ is called the level of the state.
A whole tower of raising operators can be applied to a highest weight state, and with the Virasoro algebra it is possible to order the modes so that $m_i \geq m_{i+1}$ and we obtain linear independent states in the Hilbert space.
The collection of linear independent descendant states constructed out of a highest weight state is called the Verma module.
Due to the state-operator correspondence, the set of states in the Verma module of a primary state $\ket{\Delta}$ corresponds to the set of fields in the conformal family of the primary field $\phi$ with conformal dimension $\Delta$.
A similar construction can be done also for the Verma modules associated with the anti-holomorphic generators $\bar L_n$.
The complete Hilbert space is then obtained as the sum of the tensor product of the Verma modules.

The duality structure of the Hilbert space is fixed by the definitions $L_{-n}^\dagger = L_{n}$ and $\bra{\Delta}\Delta'\rangle = \delta_{\Delta,\Delta'}$. This structure needs to be recovered from the two-point function of the respective fields when the two points coincide, namely 
\begin{equation}
    \bra{s}s'\rangle \equiv \lim_{z\to 0}\left\langle  \phi_{\bra{s}}(\bar{z},z) \phi_{\ket{s'}}(0,0) \right\rangle \,,
\end{equation}
where we denote the field dual to $\phi_{\ket{s}}(\bar{z},z)$ by
\begin{equation}
    \phi_{\bra{s}}(\bar{z},z) := \left(\phi_{\ket{s}}(z,\bar{z})\right)^\dagger\,.
\end{equation}
In chap.~\ref{chap:desc_corr} we will discuss how to explicitly construct the dual field.

In general, a Verma module can contain states with vanishing or negative norm, depending on the combination of the conformal weights of primary states and the central charge of the theory.
Vanishing norm states can be removed from the Verma module, and what we are left with is an irreducible representation of the Virasoro algebra, also called highest weight representation.
Negative norm states should be removed as well in order for the representation to be unitary, and in the following we will refer to the remaining positive norm states as the physical states of a CFT.
Unitarity translates into conditions on the primaries' conformal weights and $c$, which have to be non-negative.
In particular, unitary representations can exist for $c\geq 1$ and non-negative conformal weights, while for $c<1$ only the values
\begin{align}
    c = 1- \frac{6}{m(m+1)} \quad &\text{with} \quad m=3,4,\ldots \\
    h_{p,q}(m) = \frac{\left( (m+1)p-mq \right)^2 -1}{4m (m+1)} \quad &\text{with} \quad 1\leq p\leq m-1 \text{ and } 1\leq q \leq m
\end{align}
allow for unitary representations.
Unitary CFTs with a finite number of highest weight representations are known only for rational values of $c$, and are thus called Rational CFTs.
In chap.~\ref{chap:desc_corr} we will see a few examples of them.

\subsection{Correlation functions of descendant fields}\label{sec:intro_corr_desc}
We have seen that the correlation functions of (quasi-)primary field are determined to a great extent by conformal symmetry.
However, in a CFT there is a whole tower of descendant fields that can be constructed starting from primary fields.
It is then natural to ask what correlators with descendant fields look like.
It turns out that this kind of correlators can always be expressed as a differential operator acting on the correlator of the respective primary fields.
Indeed, consider for example the correlator
\begin{equation}
    \left\langle \hat L_{-n} \phi(z) \phi_1(z_1) \ldots \phi_N (z_N) \right\rangle \,,
\end{equation}
where we have inserted the descendant field $\hat L_{-n} \phi(z)$ as defined in~\eqref{def_desc_field}.
We will focus here on holomorphic fields, but the extension to anti-holomorphic fields follows straightforwardly.
Deforming the integration contour in~\eqref{def_desc_field}, we apply the Ward identity~\eqref{wardid} to find
\begin{equation}
    \left\langle \hat L_{-n} \phi(z) \phi_1(z_1) \ldots \phi_N (z_N) \right\rangle = \mathcal{L}_{-n} \left\langle \phi(z) \phi_1(z_1) \ldots \phi_N (z_N) \right\rangle \,,
\end{equation}
where the differential operator reads
\begin{equation}
    \mathcal{L}_{-n} = \sum_{i=1}^N \left( \frac{(n-1)}{(z_i -z)^n} h_i - \frac{1}{(z_i -z)^{n-1}} \partial_{z_i} \right) \, .
\end{equation}
Thus, as we anticipated, a correlator involving a descendant field reduces to the correlator of the respective primary fields on which a differential operator act.
This is true even for more complicated descendants, for instance
\begin{equation}
    \left\langle \left(\hat L_{-n_1} \ldots \hat L_{-n_k} \phi(z) \right) \phi_1(z_1) \ldots \phi_N (z_N) \right\rangle = \mathcal{L}_{-n_1} \ldots \mathcal{L}_{-n_k} \left\langle \phi(z) \phi_1(z_1) \ldots \phi_N (z_N) \right\rangle \, .
\end{equation}
Of course, we can also insert more than one descendant in the correlator and still apply Ward identities to reduce it to a differential operator acting on the correlator of primaries.
However, this is a more laborious procedure and we will show how to efficiently compute this kind of correlators in chap.~\ref{chap:desc_corr}.

\section{Weyl symmetry and anomalies}\label{sec:weyl_symm}
So far we discussed conformal transformations, which are a particular class of coordinate change, and the properties of CFTs in flat spacetimes.
Closely related to conformal transformations are Weyl transformations, which are defined as a local scaling of the curved metric and matter field as
\begin{equation}
    g_{\mu\nu}(x) \to e^{2\sigma(x)} g_{\mu\nu}(x) \, ,\qquad \phi(x) \to e^{-\Delta \sigma(x)} \phi(x) \,,
\end{equation}
where $\sigma(x)$ is some function and $\Delta$ is the conformal dimension associated to the field $\phi$.
Thus, we can think of conformal transformations as diffeomorphisms which coincide with a Weyl transformation of the metric and it is indeed true that at the classical level a Weyl invariant theory on a curved background is a CFT when formulated in flat spacetime. Conversely, it is usually the case that a CFT can be coupled to a curved background in a Weyl invariant way, at least at the classical level~\cite{Zumino:1970tu}.

A curved metric $g_{\mu\nu}$ can be regarded as the source for the stress-energy tensor.
We already introduced it in the context of CFTs, but we will now present some of its properties which hold for generic (non-flat) spacetime.
In particular, we now define the stress tensor through an action as
\begin{equation}
    T^{\mu\nu} = \frac{2}{\sqrt{g}} \frac{\delta S}{\delta g_{\mu\nu}} \,,
\end{equation}
where the action $S=S[g,\phi^i]$ is a functional of the metric and might include matter fields $\phi^i$ coupled to it; for simplicity, we write the absolute value of the metric determinant as $g$.
From its definition, we see that the stress tensor is symmetric.
Additional properties follow from the classical symmetries of the action.
Under a diffeomorphism $x^\mu\to x^\mu + \xi^\mu(x)$ the infinitesimal change for the metric is $\delta g_{\mu\nu} = \nabla_{\mu} \xi_\nu + \nabla_{\nu}\xi_\mu$, so that from invariance of the action under diffeomorphisms follows
\begin{equation}
    0 = \delta S = \int d^d x \sqrt{g} \, \xi_\nu (x) \nabla_\mu T^{\mu\nu} \,,
\end{equation}
where we have integrated by parts and discarded terms proportional to the equations of motion $\delta S / \delta\phi^i$ for the matter fields.
Thus, on the shell of the $\phi^i$ equations of motion, diffeomorphism invariance implies that the stress-energy tensor is conserved
\begin{equation}
    \nabla_\mu T^{\mu\nu} = 0 \, .
\end{equation}
If the action is also Weyl invariant, then when the matter fields are on-shell we also find that
\begin{equation}
    0= \delta S = \int d^d x \sqrt{g} \sigma(x) T^\mu{}_\mu \,,
\end{equation}
i.e.~the stress-energy tensor is traceless
\begin{equation}
    T^\mu{}_\mu = 0 \, .
\end{equation}
As Weyl invariance implies conformal symmetry in flat space, in the flat space limit we recover the conservation and tracelessness properties of the stress tensor that we stated in sec.~\ref{sec:2d_cft}.
Notice that if the action is invariant only under rigid Weyl transformations, i.e.~for constant $\sigma$, then the trace of the stress tensor does not have to vanish but vanishes up to a total derivative.

\subsection{Trace anomaly}\label{sec:intro_trace_ano}
Classical symmetries do not necessarily hold at the quantum level.
A quantum field theory usually needs to be regulated in order to make sense out of diverging Feynman diagrams.
The regularisation scheme may not preserve all the symmetries of the classical action, and in this case the classical symmetries which are broken at the quantum level are said to be anomalous.
It may still happen that the addition of local counterterms to the effective action restores the classical symmetry; if this is not the case, than there is a true anomaly.

Anomalies made their first clear appearance in the study of the axial-vector vertex in quantum electrodynamics~\cite{Adler:1969gk,Bell:1969ts}, related to the quantum violation of the classical chiral symmetry.
Here, we will focus on the anomaly related to the Weyl symmetry, namely the trace anomaly, which was first discovered in~\cite{Capper:1974ic}.

In the context of the path integral formulation of a quantum field theory, anomalies arise from the non-invariance of the path integral measure under the classical symmetry transformation~\cite{Fujikawa:1979ay,Fujikawa:1980vr}.
Given a classical action $S[\phi]$, which contains up to first derivatives of $\phi$, the quantum theory is defined by the path integral
\begin{equation}
    Z = e^{i \Gamma} = \int \mathcal{D}\phi \, e^{i S[\phi]} \,,
\end{equation}
where $\Gamma$ is the effective action.
Consider an infinitesimal field transformation
\begin{equation}
    \phi \to \phi'=\phi + \alpha \, \delta_\alpha\phi \,,
\end{equation}
parametrised by an infinitesimal constant parameter $\alpha$, which leaves the classical action invariant.
If we promote the constant $\alpha$ to an arbitrary function $\alpha(x)$, then we read the Noether current $J_\mu$ associated to the symmetry from
\begin{equation}
     \delta_{\alpha(x)} S[\phi] = \int d^d x \, J^\mu \partial_\mu \alpha(x) \,,
\end{equation}
where terms proportional to undifferentiated $\alpha(x)$ cannot appear, as for constant parameter the symmetry has to be recovered.
On-shell $\delta S = 0 $ for arbitrary variations and we thus deduce that the current is classically conserved
\begin{equation}
    \partial_\mu J^\mu =0 \, .
\end{equation}
The path integral is invariant under a dummy change of integration variable
\begin{equation}
    \int \mathcal{D}\phi' \, e^{i S[\phi']} = \int \mathcal{D}\phi \, e^{i S[\phi]} \,,
\end{equation}
and the Jacobian of the transformation reads
\begin{equation}
    \det \frac{\partial \phi'}{\partial \phi} = 1 + \Tr J \, ,
\end{equation}
where the functional trace and $J$ are explicitly given by
\begin{equation}\label{functional_trace}
    \Tr J = \int d^4x \int d^4 y \, J(x,y) \delta^4(x-y) \, , \quad \text{with } \quad J(x,y) = \frac{\delta \left( \delta_{\alpha(x)} \phi(x)\right)}{\delta \phi(y)} \,.
\end{equation}
Invariance of the path integral then implies
\begin{equation}
    \left\langle \Tr J + i \delta_{\alpha(x)} S \right\rangle = 0 \,,
\end{equation}
which, written in terms of the Noether current yields
\begin{equation}\label{fuji}
    i \int d^d x \, \alpha(x) \partial_\mu \left\langle J^\mu \right\rangle = \Tr J \,,
\end{equation}
where brackets $\la ...\ra$ denote normalized correlation functions and we assumed that $J$ is independent of the quantum fields.
Thus, if the path integral measure transforms with a non-trivial Jacobian, then the Noether current is not conserved at the quantum level
\begin{equation}
    \partial_\mu \left\langle J^\mu \right\rangle \neq 0 \,,
\end{equation}
and the Ward identity corresponding to the classical symmetry is violated.
Still, the part which violates the Ward identity has to satisfy certain consistency conditions~\cite{Wess:1971yu}, since the anomaly originates from the variation of the effective action $ \delta \Gamma = \langle \delta S \rangle$.
Thus, this anomaly is usually called `consistent' anomaly.\footnote{The consistent anomaly may not transform covariantly under gauge transformations. However, it is usually possible to bring it into a gauge invariant form, as we will show in some examples in chap.~\ref{chap:weyl_fermions}.}

Let us now consider a quantum field theory coupled to a curved background, where the metric is sourcing the stress-energy tensor whose expectation value is given through the effective action as
\begin{equation}
    \left\langle T^{\mu\nu} \right\rangle = \frac{2}{\sqrt{g}} \frac{\delta \Gamma}{\delta g_{\mu\nu}} \, .
\end{equation}
If the theory is classically Weyl invariant, then the stress-energy tensor is traceless.
In the quantum theory, however, we might expect the corresponding Ward identity to be anomalous if Weyl invariance is spoiled at the quantum level by the regularisation, so that under a Weyl variation
\begin{equation}\label{ano_trace}
    0 \neq \delta \Gamma = \langle \delta S \rangle =  \int d^d x \sqrt{g} \, \sigma(x) \left\langle T^\mu{}_{\mu} \right\rangle \,,
\end{equation}
from which we conclude the violation of the Ward identity related to the classical symmetry
\begin{equation}
    \left\langle T^\mu{}_{\mu} \right\rangle \neq 0 \, .
\end{equation}
Since the Weyl transformation is abelian, two Weyl transformations commute, so that the effective action should obey the following consistency condition
\begin{equation}
    \delta_2 \delta_1 \Gamma - \delta_1 \delta_2 \Gamma =0 \,,
\end{equation}
which is actually a condition on the trace anomaly
\begin{equation}
    \delta_2 \int d^d x \sqrt{g} \, \sigma_1(x) \left\langle T^\mu{}_{\mu} \right\rangle = \delta_1 \int d^d x \sqrt{g} \, \sigma_2(x) \left\langle T^\mu{}_{\mu} \right\rangle \, .
\end{equation}
This is called the Wess-Zumino consistency condition~\cite{Wess:1971yu} and is a powerful condition to fix to a good extent the possible form of the trace anomaly~\cite{Bonora:1985cq}.

In two dimensions, the only solution for the trace anomaly to the WZ condition consistent with dimensional arguments is given by the Ricci scalar $R$.
Thus, in two dimensions we can parametrize the trace anomaly as
\begin{equation}\label{<Tt>2d}
    \left\langle T^\mu{}_{\mu} \right\rangle = \frac{c}{24\pi} R = \frac{c}{24\pi} E_2 \,,
\end{equation}
where $E_2$ is the two-dimensional Euler density and the coefficient $c$ is indeed the central charge of the Virasoro algebra appearing in the two-point function of the stress tensor~\eqref{<TT>}.
Notice that in the geometrical classifications of trace anomalies of~\cite{Deser:1993yx} this is a type A anomaly.

In four dimensions the trace anomaly consistent with the WZ condition is given by a linear combination of the topological Euler density $E_4$, the square of the Weyl tensor $C^2 = C^{\mu\nu\rho\sigma}C_{\mu\nu\rho\sigma}$ and the Pontryagin term $P_4$; respectively, they read
\begin{align}
    E_4 &= R^{\mu\nu\rho\sigma}R_{\mu\nu\rho\sigma} - 4 R^{\mu\nu}R_{\mu\nu} + R^2 \,, \label{euler_4}\\
    C^2 &= R^{\mu\nu\rho\sigma}R_{\mu\nu\rho\sigma} - 2 R^{\mu\nu}R_{\mu\nu} + \frac13 R^2 \,, \label{weyl_sqr}\\
    P_4 &= \varepsilon_{\mu\nu\rho\sigma} R^{\mu\nu\alpha\beta} R^{\rho\sigma}{}_{\alpha\beta} \,, \label{pontryagin_4}
\end{align}
with the four-dimensional Levi-Civita tensor $\varepsilon_{\mu\nu\rho\sigma}$. 
A term proportional to $\Box R$ also solves the WZ condition, but it can be removed with the addition of a local counterterm to the effective action.
Notice that, even though $P_4$ is a consistent term, it has never appeared in the trace anomaly of known CFTs until the claim of~\cite{Bonora:2014qla} for the case of a chiral fermion.
We will address this issue in details in chap.~\ref{chap:weyl_fermions}.
For the scope of the present analysis, it is enough to parametrise the four-dimensional trace anomaly as
\begin{equation}\label{<Tt>4d}
    \left\langle T^\mu{}_{\mu} \right\rangle = \frac{1}{180(4\pi)^2} \left( - a \, E_4 + b \, \Box R + c \, C^2 \right) \,,
\end{equation}
where we stress once again that $b$ can take any value with the addition of a local counterterm, while $a$ and $c$ are theory-dependent coefficients and respectively parametrise the type A and B anomalies in the classification of~\cite{Deser:1993yx}. 

Even though the trace anomalies~\eqref{<Tt>2d} and~\eqref{<Tt>4d} vanish in flat space, they have implications for correlation functions of CFTs in flat space.
Taking a variation of~\eqref{<Tt>2d} with respect to the metric indeed produces the expression for $\left\langle T^\mu{}_\mu T_{\rho\sigma} \right\rangle$, which does not vanish in flat space.
Relating the resulting expression to the stress tensor two-point function~\eqref{<TT>} and assuming that the stress tensor is conserved at the quantum level would then show that the $c$ appearing in~\eqref{<Tt>2d} is the central charge of the Virasoro algebra.\footnote{It is possible to quantize the theory in such a way that the stress tensor conservation is anomalous but the trace vanishes; however, no quantization allows to respect both symmetries unless $c=0$.}
From~\eqref{<Tt>4d} we have to take two variations with respect to the metric to obtain a non-vanishing result in flat space. 
The resulting expression gives $\left\langle T^\mu{}_\mu T_{\rho\sigma} T_{\alpha\beta} \right\rangle$, which can again be related to the three-point function of the stress energy tensor in flat space constrained by conformal symmetry.
We are not interested in presenting the explicit expressions here (which are analysed in detail in~\cite{Osborn:1993cr}), but rather we notice that in this way the coefficients of the trace anomaly can be related to the correlation functions in flat space.
Thus, knowing the coefficients of trace anomalies provides information about correlation functions of CFTs in flat space, and, vice versa, computing the flat space correlation functions of stress-energy tensor provides information about the quantum theory on curved backgrounds.

In the following we will discuss a method to compute anomalies that relies on the path integral and that we will use in chap.~\ref{chap:weyl_fermions}; we will then conclude this section with a sample calculation.

\subsection{Regulators and consistent anomalies}\label{sec:regulators}
In order to compute the anomalies from the Jacobian of the path integral measure, it would be desirable to fully specify the path integral measure and compute the path integral.
In practice, this is only doable for Gaussian path integrals.
Nonetheless, it was shown by Fujikawa~\cite{Fujikawa:1979ay,Fujikawa:1980vr} that it is still possible to obtain one-loop anomalies by regulating the trace in~\eqref{fuji} with a positive-definite operator $\mathcal{R}$ as
\begin{equation}\label{fuji_jaco}
    \Tr J \quad \to \quad \lim_{M\to \infty} \Tr \left[ J \, e^{- \frac{\mathcal{R}}{M^2}} \right] = \lim_{M\to\infty} \int d^4x \int d^4 y \, J(x,y) e^{- \frac{\mathcal{R}(x)}{M^2}} \delta^4(x-y) \, .
\end{equation}
In the Fujikawa scheme the regulator remains to be specified and it is not a priori clear which regulator yields which anomaly.
Thus, in the following we will use a well-defined algorithm to determine the regulators which produce consistent anomalies, following~\cite{Diaz:1989nx,Hatsuda:1989qy}.
In particular, we first work in a Pauli-Villars (PV) regularisation scheme, where the anomaly follows from the non-invariance of the mass term of the PV field.
Once the mass term is specified, the regulators and Jacobians to be employed in the Fujikawa's scheme are determined and anomalies are computed by making use of heat kernel formulae.

We review now the scheme of~\cite{Diaz:1989nx}, which was also discussed in~\cite{Bastianelli:2016nuf,Bastianelli:2018osv,Bastianelli:2019fot,Bastianelli:2019zrq} and whose presentation we will closely follow.
Consider an action for a field~$\varphi$ with lagrangian\footnote{The field $\varphi$ might carry additional indices, but we suppress them for simplicity of notation.}
\begin{equation}
\mathcal{L} = \frac12 \varphi^T T \mathcal{O} \varphi \,,
\end{equation}
where $\mathcal{O}$ is a differential operator and $T$ will be defined shortly; we will refer to $T {\cal O}$ as the kinetic operator, which may depend on background fields.
The action is invariant under a linear symmetry 
\begin{equation}
\delta \varphi = K\varphi 
\end{equation}
that generically acts also on the kinetic operator.
The one-loop effective action is regulated by subtracting a loop  of a massive PV field~$\phi$ with lagrangian 
\begin{equation} 
\mathcal{L}_{\scriptscriptstyle PV} =  \frac12 \phi^T T  \mathcal{O} \phi +\frac12 M \phi^T T \phi \,,
\label{PV-l}
\end{equation} 
where $M$ is a real parameter\footnote{To be precise, one should employ a set of PV fields with mass $M_i$ and relative weight $c_i$ in the loop to be able to regulate and cancel all possible one-loop divergences \cite{Pauli:1949zm}.
For simplicity, we consider only one PV field with relative weight $c=-1$, as this is enough  for our purposes.
The weight $c=-1$  means that we are subtracting one massive PV loop from the original one.} and the last term corresponds to the mass of the PV fields. 
The mass term defines the mass operator $T$, that in turn allows to find the operator ${\cal O}$.
The latter may act as a regulator in the final formula for the anomaly; in fermionic theories, where it is usually a first order differential operator, the regulator is rather given by $\mathcal{O}^2$.
The invariance of the original action extends to an invariance of the massless part of the PV action by defining 
\begin{equation}
\delta \phi = K \phi
\label{PV-t}
\end{equation}
so that only the mass term may break the symmetry 
\begin{equation}
\delta\mathcal{L}_{\scriptscriptstyle PV} =  \frac12 M \phi^T (T K +K^T T +\delta T) \phi =
M \phi^T (T K +\frac12 \delta T) \phi \,,
\end{equation}
where we assumed $T=T^T$.
If it is possible to introduce a mass term which is invariant under the transformation~\eqref{PV-t}, then the symmetry is anomaly free.
The path integral $Z$ and the one-loop effective action $\Gamma$ are regulated by the PV field
\begin{equation}
Z=e^{i \Gamma} =\int \mathcal{D}\varphi\; e^{i S}  \quad \to \quad Z=e^{i \Gamma} =
\int \mathcal{D}\varphi \mathcal{D}\phi\;  e^{i (S + S_{\scriptscriptstyle PV})} \,,
\end{equation}
where the limit $M\to \infty$ is understood.
The anomalous response of the path integral under a symmetry is now due to the PV mass term only, as the measure of the PV field can be defined so that the measure of the whole path integral is invariant~\cite{Diaz:1989nx}. 
Computing the symmetry variation of the regulated path integral we obtain
\begin{align}\label{reg_var}
i\delta \Gamma =i \left\langle\delta S\right\rangle  &=  \lim_{M \to \infty}  i M \langle \phi^T (TK +\frac12 \delta T) \phi \rangle \nonumber\\
&=  \lim_{M \to \infty}  
\Tr \left[ \left(K + \frac12 T^{-1} \delta T \right) \left( 
1+ \frac{\cal O}{M} \right)^{-1} \right] \,,
\end{align}
where we considered a scalar field theory with PV propagator
\begin{equation}
\la \phi \phi^T\ra = \frac{-i}{T {\cal O} + T M} \,,
\end{equation}
and assumed $T$ to be invertible.
In the limit $M\to\infty$, we can replace $\left( 
1+ {\cal O}/M \right)^{-1}$ with $e^{-\mathcal{O}/M}$ and the anomaly calculation is cast into the Fujikawa calculation~\eqref{fuji_jaco}
\begin{equation}
i \la\delta S\ra  = \lim_{M \to \infty}   \Tr \left[J e^{ -\frac{\mathcal{R}}{M}}\right]  \,,
\label{tra_scalar}
\end{equation}
by identifying the Jacobian and the regulator as
\begin{equation}\label{scalar_jaco}
    J = K + \frac12 T^{-1} \delta T \, , \quad \quad \mathcal{R} = \mathcal{O} \,,
\end{equation}
assuming that $\mathcal{O}$ is positive definite after Wick rotation to Euclidean space.
This is enough for many cases when $\mathcal{O}$ is a second-order differential operator, e.g.~for scalar theories.

For fermionic theories, $\mathcal{O}$ is typically a first-order differential operator, so that further manipulations are required.
Inserting the identity $ 1 = (1 - \frac{\cal O}{M}) (1 - \frac{\cal O}{M})^{-1}$ and using 
\begin{equation}
 \varphi^T \left ( T {\cal O} K +\frac12 \delta T {\cal O} +\frac12 T \delta  {\cal O} \right ) \varphi =0 \,,
\end{equation}
which follows from the invariance of the massless action,
we obtain the equivalent form~\cite{Hatsuda:1989qy}
 \begin{equation}
i \la\delta S\ra  =
 \lim_{M \to \infty}  
\Tr \left[\left(K + \frac12 T^{-1} \delta T + \frac12 \frac{\delta {\cal O}}{M} \right) 
\left( 1- \frac{{\cal O}^2}{M^2} \right)^{-1} \right] \, .
\label{3.8}
\end{equation}
In the limit $M\to \infty$ the regulating term $ ( 1- \frac{{\cal O}^2}{M^2})^{-1}$ inside~\eqref{3.8} can be replaced by $e^{{\cal O}^2/M^2}$, assuming that ${\cal O}^2$ is negative definite after a Wick rotation to Euclidean space.
The connection to heat kernel formulae is now explicit.
Indeed, denoting 
\begin{equation}
J=K + \frac12 T^{-1} \delta T + \frac12 \frac{\delta {\cal O}}{M} \;, \qquad { \cal R}=-{\cal O}^2 \,,
\label{jac}
\end{equation}
the anomaly is related to the trace of the heat kernel of the regulator  ${\cal R}$ with an insertion of $J$
\begin{equation}
i \la\delta S\ra  = \lim_{M \to \infty}   \Tr \left[J e^{ -\frac{\mathcal{R}}{M^2}}\right] \,,
\label{tra}
\end{equation}
where the $M\to\infty$ limit extracts only the mass independent term (negative powers of the mass vanish in the limit, while positive (diverging) powers are made to cancel by using additional PV fields).
This has the same form that appears in the original Fujikawa's method for computing anomalies~\cite{Fujikawa:1979ay,Fujikawa:1980vr}, where in the Fujikawa scheme $J$ is the infinitesimal part of the Jacobian arising from a change of the path integral variables under a symmetry transformation, and $\mathcal{R}$ is the regulator. 
As we anticipated, in the PV scheme that we adopt the regulator is directly read from the PV lagrangian.
Moreover, the PV scheme guarantees that the regulator $\mathcal{R}$ together with $J$ produces consistent anomalies, since we are computing directly the variation of the effective action.

\subsection{Heat kernel}\label{sec:heat_kernel}
To compute the functional trace in~\eqref{tra} we use heat kernel techniques.
Consider the heat equation with Euclidean time $\beta$
\begin{equation}
    -\frac{\partial}{\partial \beta} K(\beta,x,y) = H K(\beta,x,y) \,,
\end{equation}
where $H$ is a second order elliptic differential operator.
The heat equation can be obtained from the non-relativistic Schr\"odinger equation by performing a Wick rotation to the Euclidean time $\beta$.
Introducing eigenstates for the position operator $\ket{x}$, its formal solution in operator formalism is then given by the heat kernel
\begin{equation}
    K(\beta,x,y) = \bra{y} e^{-\beta H} \ket{x} \,,
\end{equation}
with boundary condition
\begin{equation}
    K(0,x,y) = \delta(x-y) \, .
\end{equation}
The connection between the heat kernel and anomalies is now clear, since we write the functional trace as
\begin{equation}
    \Tr \left[ J e^{- \beta H} \right] = \int d^d x \, \tr \left[ J(x) \bra{x} e^{-\beta H} \ket{x} \right] \,,
\end{equation}
where the remaining trace is over discrete (internal) indices and $J(x)$ is an arbitrary matrix function.
Thus, knowing the kernel of the operator $H$ allows us to study the effective action $\Gamma$ and its anomalies.

We will be interested in operators $H$ defined on manifolds without boundaries of the form
\begin{equation}\label{opH}
    H = - \Box + V \,,
\end{equation}
where $V$ is a matrix potential and $\Box = \nabla^\mu\nabla_\mu$ is constructed with a gravitational and gauge covariant derivative $\nabla_\mu$ that satisfies
\begin{equation}
    [\nabla_\mu, \nabla_\nu] \phi = \mathcal{F}_{\mu\nu} \phi \, , \quad\quad [\nabla_\mu, \nabla_\nu] A^\lambda = R_{\mu\nu}{}^\lambda{}_{\rho} A^\rho \,,
\end{equation}
with $\phi$ a charged scalar field and $A^\mu$ a real (uncharged) vector.
We further define the Ricci tensor and Ricci scalar as
\begin{equation}
    R_{\mu\nu} = R_{\mu\lambda\nu}{}^\lambda \, , \quad \quad R = R^\mu{}_\mu \, .
\end{equation}
Then, we can perturbatively expand the trace of the corresponding heat kernel for small $\beta$ as
\begin{equation}
    \Tr \left[ J e^{- \beta H} \right] = \int\frac{d^d x \sqrt{g}}{(2\pi \beta)^{d/2}} \sum_{n=0}^{\infty} \tr \left[ J(x)  a_n(x,H) \beta^n \right] \,,
\end{equation}
where the $a_n$ are the heat kernel, or Seeley-DeWitt, coefficients~\cite{DeWitt:1964mxt,DeWitt:1984sjp}.
We list here the first few of them:
\begin{align}
    a_0(x,H) &= \mathbb{I} \,,\\
    a_1(x,H) &= \frac{1}{6} R - V \,,\label{hkc1}\\
    a_2(x,H) &= \frac{1}{180} \left( R_{\mu\nu\lambda\rho} R^{\mu\nu\lambda\rho} - R_{\mu\nu}R^{\mu\nu} \right) + \frac12 \left(\frac16 R -V\right)^2 + \frac16 \Box \left( \frac15 R -V \right) \nonumber\\
    &\phantom{=} + \frac{1}{12} \mathcal{F}_{\mu\nu}\mathcal{F}^{\mu\nu} \,, \label{hkc2}
\end{align}
where $\mathbb{I}$ is the identity in the internal space.
In the following we will mostly use $a_2$ which is related to the trace anomaly of four-dimensional CFTs, but higher order coefficients are also known and relevant for higher-dimensional theories (see for instance~\cite{Vassilevich:2003xt} and references therein).

To make contact with eq.~\eqref{tra}, we write the functional trace in Lorentzian time as
\begin{equation}\label{hkt}
{\rm Tr} \left [ J  e^{-i s H} \right ]  = \int \frac{d^dx\, i\sqrt{g}}{(4 \pi i s)^{d/2}} \sum_{n=0}^\infty \tr [ J(x)a_n(x,H)] (is)^n \, .
\end{equation}
When computing anomalies the role of the operator $H$ is played by the regulator $\mathcal{R}$, while $is$ is proportional to a negative power of the PV mass.
In $d$ dimensions, the coefficient $a_{d/2}$ is thus the one which produces the anomalies, and it is non-vanishing only for even $d$ \cite{Vassilevich:2003xt}.
Henceforth we will work in a Minkowskian set-up and we justify heat kernel formulae by Wick rotating to an Euclidean time and back, when necessary.

\subsection{Case study: conformally coupled scalar field}
As an example application of the scheme we just introduced, let us apply it to compute the trace anomaly of a scalar field conformally coupled to gravity~\cite{Bastianelli:2016nuf}.
In $d$ dimensions, its action is
\begin{equation}
    S = \frac12 \int d^d x \sqrt{g}\left( g^{\mu\nu} \partial_\mu \varphi \partial_\nu \varphi + \xi_d R \varphi^2 \right) \, , \quad \xi_d = \frac{d-2}{4(d-1)} \, .
\end{equation}
The action is Weyl invariant for the following Weyl variations of the scalar and metric field
\begin{equation}
    \delta \varphi = -\frac{d-2}{2} \sigma(x) \varphi \, , \quad\quad \delta g_{\mu\nu} = 2\sigma(x) g_{\mu\nu} \, .
\end{equation}
Thus, classically the stress tensor is traceless, but it develops an anomaly at the quantum level.
We now compute it with the technique we introduced in sec.~\ref{sec:regulators} and~\ref{sec:heat_kernel}.

We introduce a massive PV field with action
\begin{equation}
    S_{\scriptscriptstyle PV} = \frac12 \int d^d x \sqrt{g} \phi \left( - \Box + \xi_d R + M^2 \right) \phi \,,
\end{equation}
where we integrated by parts to bring it in the symmetric form~\eqref{PV-l} which allows us to recognise
\begin{equation}
    T = \sqrt{g} \, , \quad\quad \mathcal{O} = - \Box + \xi_d R \,,
\end{equation}
and $\mathcal{O}$ identifies the regulator $\mathcal{R} = \mathcal{O}$.
Notice that the mass term preserves general coordinate invariance, so that this symmetry remains free from anomalies.
From the field transformation we identify $K=-(d-2)\sigma/2$, so that the Jacobian in~\eqref{scalar_jaco} is $J=\sigma$ and from eq.~\eqref{tra_scalar} and~\eqref{hkt} we then find
\begin{equation}
    \langle T^\mu{}_{\mu} \rangle = \frac{1}{(4\pi)^{\frac{d}{2}}} a_{\frac{d}{2}} (x, \mathcal{R}) \, .
\end{equation}
As a side result, this formula also tells that in spacetime without boundaries there are no trace anomalies in odd dimensions, since the corresponding heat kernel coefficients vanish.
In two dimensions, $\xi_2 = 0$ and the trace anomaly is proportional to the heat kernel coefficient~\eqref{hkc1} with $V = 0$
\begin{equation}
    \langle T^\mu{}_{\mu} \rangle = \frac{1}{24 \pi} R \, .
\end{equation}
In four dimensions, we need the heat kernel coefficient~\eqref{hkc2} with $V = R/6$ and $\mathcal{F}_{\mu\nu} =0$ since we are not considering background gauge fields;
then, the trace anomaly is
\begin{align}
    \langle T^\mu{}_{\mu} \rangle &= \frac{1}{180(4\pi)^2} \left( R_{\mu\nu\lambda\rho} R^{\mu\nu\lambda\rho} - R_{\mu\nu}R^{\mu\nu} + \Box R \right) \nonumber\\
    &= \frac{1}{180(4\pi)^2} \left( \frac32 C^2 - \frac12 E_4 + \Box R \right) \, .
\end{align}
As we already stated, we notice that adding the counterterm
\begin{equation}
    \Gamma_{\text{ct}} = \frac{1}{180(4\pi)^2} \frac{1}{12} \int d^4x \sqrt{g} R^2
\end{equation}
to the effective action would cancel the $\Box R$ term in the trace anomaly.

\section{Holography}\label{sec:holography}
CFTs play an important role in the context of the gauge/gravity duality, which is a realisation of the holographic principle.
Amongst these dualities, CFTs are conjectured to be dual to gravitational theories in anti-de Sitter (AdS) spacetime.
The AdS/CFT correpondence is a very active area of research, which has led to insights into a vast area of physics ranging from particle physics to condensed matter systems.
In this thesis, we are interested in possible tests of the AdS/CFT correspondence, which consist in computing observables on both side of the duality to find agreement between the results.
In particular, we will focus on the trace anomaly, which as we saw is a peculiar feature of a CFT, and we will show how to compute it holographically.
Before doing so, we will present the holographic principle and motivate the AdS/CFT correspondence.
A complete and comprehensive introduction to this topic is by far out of the scope of the present section, and can be found in reviews and books such as~\cite{Aharony:1999ti,DHoker:2002nbb,Casalderrey-Solana:2011dxg,Ammon:2015wua,Nastase:2015wjb}.
Here we will cover the essential points needed for the computation of the holographic anomaly.

\subsection{The holographic principle}
The holographic principle states that the information content of a quantum theory of gravity in a certain volume of space is encoded at the boundary surface of this volume~\cite{tHooft:1993dmi}.
Thus, a gravitational theory in a given space should be equivalent to a non-gravitational theory defined at the boundary of this space.
This principle originates from black hole thermodynamics, according to which the entropy of a black hole is proportional to its horizon area, where it is encoded~\cite{Bekenstein:1973ur,Hawking:1976ra}.
The entropy measures the information inside the horizon, which is inaccessible to an exterior observer, thus realising a hologram.

We provide now an intuitive motivation for a gauge/gravity duality, following~\cite{Casalderrey-Solana:2011dxg}.
Consider a QFT in a $d$-dimensional Minkowski spacetime with coordinates~$x^i$ and with a short-distance cutoff~$\epsilon$ to avoid UV divergences.
We assume that there is no correlation between degrees of freedom at largely separated points.
If we are interested in the physics at large length scales $z \gg \epsilon$ we can integrate out the short-distance degrees of freedom and obtain an effective theory at length scale~$z$.
Similarly, we would use an effective theory at scale $z'\gg z$ if we are interested in the physics at even larger scale~$z'$. 
Iterating this procedure defines an RG flow~\cite{Wilson:1974mb} and gives rise to a continuous family of $d$-dimensional effective field theories labelled by the RG scale~$z$.
Pictorially, we can think of this continuous family of $d$-dimensional theories as a single theory in a $(d+1)$-dimensional spacetime with coordinates $X^\mu=(z,x^i)$ where the additional coordinate is the RG scale $z$, which foliates the $(d+1)$-dimensional spacetime into $d$-dimensional Minkowski spacetime slices.
By construction, the $(d+1)$-dimensional theory should be invariant under reparametrization of the $z$-coordinate, since the physics of the original theory is invariant under reparametrization of the RG scale.
Moreover, the resulting theory does not have more degrees of freedom than the $d$-dimensional theory, and in particular the effective theory at $z=z_0$ can describe all the physics for $z\geq z_0$.
Together with the holographic principle, this means that the $(d+1)$-dimensional theory is a gravitational theory, whose hologram is the original $d$-dimensional non-gravitational theory.

\subsection{The AdS/CFT correspondence}
To realise a concrete example of gauge/gravity duality we need to specify the theories involved.
Consider a $d$-dimensional CFT with flat metric $\eta_{ij}$ and assume that it is the hologram of a $(d+1)$-dimensional gravitational theory.
Demanding that the hologram is conformally invariant allows us to reconstruct the gravitational theory to some extent.
Indeed, the most general $(d+1)$-dimensional metric consistent with $d$-dimensional Poincar\'e symmetry is given by
\begin{equation}\label{gen_grav}
    ds^2 = \Omega^2(u) \left( du^2 + \eta_{ij} dx^i dx^j \right) \,,
\end{equation}
where $u$ is the extra spatial direction and the warp factor $\Omega$ can only depend on $u$ to have translational symmetries in the $x^i$ directions.
Moreover, since the $d$-dimensional theory is a CFT, it is also in particular invariant under scaling of the $x^i$ coordinates $x^i \to \lambda x^i$.
As we argued in the previously, $u$ represents a length scale in the boundary theory, and thus it also has to transform as $u \to \lambda u$ under scaling of the $x^i$ coordinates.
Assuming that the gravity theory describes such a field theory, then the metric~\eqref{gen_grav} should respect the scaling symmetry $(u,x^i)\to\lambda(u,x^i)$ and
the warp factor of the $(d+1)$-dimensional theory is thus fixed to be 
\begin{equation}
    \Omega(u) = \frac{\ell}{u}
\end{equation}
for some constant $\ell$.
The metric~\eqref{gen_grav} becomes then
\begin{equation}\label{ads_poincare}
    ds^2 = \frac{\ell^2}{u^2} \left( du^2 + \eta_{ij} dx^i dx^j \right) \, .
\end{equation}
This is the metric of a $(d+1)$-dimensional AdS spacetime with radius $\ell$ in Poincar\'e coordinates.
Thus, assuming the gauge/gravity duality, we conclude that a $d$-dimensional CFT, or $\text{CFT}_d$, is dual to a gravitational theory formulated in a $(d+1)$-dimensional AdS spacetime, or $\text{AdS}_{d+1}$.

AdS is a maximally symmetric spacetime with negative cosmological constant.
The isometry group of $\text{AdS}_{d+1}$ is $SO(d,2)$, the same symmetry group of a $\text{CFT}_d$ in Minkowski spacetime.
This symmetry matching provides additional support to the conjecture.
Although AdS spacetime has no boundary in the topological sense, the notion of confomal boundary can be defined.\footnote{Henceforth we will refer to the conformal boundary simply as `boundary' for short.}
Indeed, consider the metric $\hat\eta_{MN}=\text{diag}(-1,1,\ldots,1,-1)$ in $\mathbb{R}^{d,2}$ with coordinates $y^M$.
Then $\text{AdS}_{d+1}$ can be embedded into $\mathbb{R}^{d,2}$, where it is given by the hypersurface
\begin{equation}\label{ads_hyper}
    \hat\eta_{MN} y^M y^N = -\ell^2 \,.
\end{equation}
Consider now solutions of~\eqref{ads_hyper} of the form $\hat\eta_{MN} \tilde{y}^M \tilde{y}^N = -\ell^2/R^2$.
In the limit $R\to\infty$ the hyperboloid becomes $\hat\eta_{MN} \tilde{y}^M \tilde{y}^N = 0$, i.e.~the light-cone in $\mathbb{R}^{d,2}$.
But this is true even if we scale out $\lambda R$ instead of $R$.
Thus, considering the equivalence classes of points which satisfy $\hat\eta_{MN} {y}^M {y}^N = 0$, the boundary of $\text{AdS}_{d+1}$ is defined as any section of the light-cone in $\mathbb{R}^{d,2}$ which is pierced once be each null ray.
It can be verified that the boundary of $\text{AdS}_{d+1}$ corresponds to a compactification of $d$-dimensional Minkowski spacetime, and e.g.~in Poincar\'e coordinates~\eqref{ads_poincare} the boundary is at $u=0$.

If we define a new radial coordinate $z = u^2$, then the AdS line element becomes
\begin{equation}\label{pure_FG}
    ds^2 = \ell^2 \left( \frac{dz^2}{4z^2} + \frac{1}{z} \eta_{ij} dx^i dx^j \right) \,,
\end{equation}
and the boundary is at $z=0$.
We will refer to these coordinates as the Fefferman-Graham (FG) coordinates.

The discussion so far provided a first hint for the AdS/CFT correspondence.
Another evidence that a QFT admits a dual description as a gravitational theory comes from the large $N$ expansion of a non-abelian gauge theory, as suggested by 't~Hooft~\cite{tHooft:1973alw}.
We will state here some remarkable results, without providing any derivation.
In particular, we are interested in the vacuum-to-vacuum amplitude of the Yang-Mills theory with gauge group $U(N)$, which can be expanded in $1/N$ as
\begin{equation}\label{thooft_exp}
    \mathcal{A}_{\text{YM}} = \sum_{k=0}^\infty N^{2-2k} f_k(\lambda) \,,
\end{equation}
where $\lambda$ is the 't~Hooft coupling which is related to the gauge coupling $g_{\text{YM}}$ and to $N$ by
\begin{equation}
    \lambda = g_{\text{YM}}^2 N \,,
\end{equation}
and the $f_k(\lambda)$ are functions of the 't~Hooft coupling only.
In particular, $f_k(\lambda)$ includes the contributions of all Feynman diagrams which can be drawn on a two-dimensional compact surface with genus $k$ without crossing any lines.
Thus, the expansion in~\eqref{thooft_exp} can be thought of as an expansion in terms of the topology of the two-dimensional surface.
A similar expression arises in the context of string theory (see e.g.~\cite{Blumenhagen:2013fgp}).
The worldsheet of a closed string is a two-dimensional compact surface and the string perturbative expansion is again given by a sum over the worldsheet's topologies.
The vacuum-to-vacuum amplitude in this case can be written as
\begin{equation}\label{string_exp}
    \mathcal{A}_{\text{string}} = \sum_{k=0}^{\infty} g_s^{2k-2} F_k(\alpha') \,,
\end{equation}
where $g_s$ is the string coupling and $\alpha'$ is related to the inverse of the string tension.
The $F_k(\alpha')$ are the contributions to the amplitude of two-dimensional surfaces with genus~$k$.
Comparing~\eqref{thooft_exp} with~\eqref{string_exp}, it is thus tempting to relate
\begin{equation}
    f_k(\lambda) \leftrightarrow F_k(\alpha') \,,
\end{equation}
so that the 't~Hooft expansion is interpreted as a closed string expansion with
\begin{equation}\label{duality_par}
    g_s \sim 1/N \, , \quad \quad \alpha' = \alpha'(\lambda) \,,
\end{equation}
providing further motivation for a gauge/gravity duality, in this case between a non-abelian gauge theory and a theory of closed string.
What this argument does not specify is what particular string theory is taking part in the duality.
A specific example of AdS/CFT correspondence is the famous duality between $\mathcal{N} = 4$ super Yang-Mills theory and type IIB string theory on an $\text{AdS}_5\times S^5$ background~\cite{Maldacena:1997re}.
In this case, the relations~\eqref{duality_par} between the free parameters of the theories take the explicit forms
\begin{equation}\label{duality_mald}
    g_{\text{YM}}^2 = 2\pi g_s \, , \quad \quad 2 g_{\text{YM}}^2 N = \ell^4/(\alpha')^2 \,,
\end{equation}
where $\ell$ is the radius of $\text{AdS}_5$ and of the five-dimensional sphere.

\subsection{The dictionary}\label{sec:dictionary}
Once established, the AdS/CFT duality becomes a useful tool if it is known how to map one theory to the other.
This set of relations is usually called the holographic dictionary.
A very important entry of the dictionary is the relation between fields in the bulk and operators in the boundary.
In general, given a bulk field $\Phi_i$ it is possible to solve its equation of motion with prescribed boundary value $\phi_i$.
The $\phi_i$ is then interpreted as the source for an operator $\mathcal{O}_i$ in the CFT.
This establishes a one-to-one correspondence between bulk fields and boundary operators
\begin{equation}
    \Phi_i \leftrightarrow \mathcal{O}_i \, .
\end{equation}

Consider for example a bulk scalar field $\Phi$ with mass $m$ coupled to a background metric $G_{\mu\nu}$.\footnote{When dealing with holographic systems (i.e.~in the remaining of this section and in chap.~\ref{chap:holography}), we use Greek letters for $(d+1)$-dimensional bulk indices, while Latin letters for $d$-dimensional boundary indices.}
Its bulk action is
\begin{equation}
    S = \frac12 \int d^{d+1} X \sqrt{G} \left( G^{\mu\nu} \partial_\mu \Phi \partial_\nu \Phi + m^2 \Phi^2 \right) \,,
\end{equation}
from which we derive the Klein-Gordon equation of motion
\begin{equation}
    \left(\Box_G - m^2 \right) \Phi = 0 \, , \quad \quad \Box_G \Phi = \frac{1}{\sqrt{G}} \partial_\mu \left( \sqrt{G} \, G^{\mu\nu} \partial_\nu \Phi \right) \, .
\end{equation}
By writing the equation of motion in FG coordinates~\eqref{pure_FG}, then the solution near the boundary $z\to 0$ has the following power series form
\begin{equation}
    \Phi(z,x) = z^{\frac12 \Delta_-} \left( \phi(x) + \ldots \right) + z^{\frac12 \Delta_+} \left( \varphi(x) + \ldots \right) \,,
\end{equation}
where we indicated only the leading terms and $\Delta_\pm$ are the roots of $(m\ell)^2 = \Delta(\Delta-d)$.
Explicitly,
\begin{equation}
    \Delta_\pm = \frac{d}{2} \pm \sqrt{\frac{d^2}{4} + \left(m\ell\right)^2} \, .
\end{equation}
Notice that $\Delta_+ \geq \Delta_-$ and $\Delta_- = d - \Delta_+$.
Defining $\Delta := \Delta_+$, one can identify $\varphi$ as the vacuum expectation value of a scalar operator $\mathcal{O}$ of conformal dimension $\Delta$ in the dual field theory, and $\phi$ as its source~\cite{Gubser:1998bc,Witten:1998qj}.
The relation between the source of the boundary operator and the bulk field is then simply
\begin{equation}
    \phi (x) = \lim_{z\to 0} \Phi(z,x) z^{\frac12(\Delta-d)} \, .
\end{equation}
The roots $\Delta_\pm$ are real as long as the mass satisfies the Breitenlohner-Freedman bound $(m\ell)^2 \geq -d^2/4$ \cite{Breitenlohner:1982bm}.
When $\Delta > d$, the dual operator is irrelevant and $m>0$.
For $\Delta =d$, the dual operator is marginal and the bulk field is massless $m=0$.
When the bulk field has a mass that satisfies $0> (m\ell)^2 \geq -d^2/4$, then the dual operator is relevant with $d/2 \leq \Delta < d$.
When $\Delta< d/2$ the identification of the source and vev of the boundary operator $\mathcal{O}$ can be exchanged and we can describe dual operators of dimension $(d-2)/2 \leq \Delta < d/2 $ while still respecting the mass bound~\cite{Klebanov:1999tb}.

Another important entry in the dictionary is the identification of the bulk and boundary partition functions
\begin{equation}
    Z_{\text{bulk}} (\phi_i) = Z_{\text{CFT}}(\phi_i) \, .
\end{equation}
Together with the relation between bulk fields and boundary operators, the identification of the partition functions is very powerful since it allows to perform non-trivial tests of the AdS/CFT correspondence.
The first non-trivial tests consisted in the holographic computation of CFT correlation functions (see e.g.~\cite{Lee:1998bxa,Freedman:1998tz}) and trace anomaly~\cite{Henningson:1998gx}.
Here we will focus on the trace anomaly, which we will derive with the help of a particular class of bulk diffeomorphisms which reduce to Weyl tranformations at the boundary, first constructed in~\cite{Imbimbo:1999bj} (see also~\cite{Schwimmer:2000cu,Schwimmer:2003eq,Schwimmer:2008yh,Schwimmer:2008zz}).
These diffeomorphisms are called Penrose-Brown-Henneaux (PBH) transformations and we first briefly review them before computing the holographic anomaly, mainly following the presentation in~\cite{Broccoli:2021icm} to set our notation.

\subsection{PBH transformations}\label{sec:pbh}
Consider a theory of gravity in a $(d+1)$-dimensional spacetime with coordinates $(z,x^i)$.
It is possible to bring the bulk metric in the FG form
\begin{equation}\label{fg}
    ds^2 = G_{\mu\nu} dX^\mu dX^\nu = \ell^2 \left( \frac{dz^2}{4z^2} + \frac{1}{z}  g_{ij}(z,x) dx^i dx^j \right) \, ,
\end{equation}
where $g_{ij}$ depends on the bulk coordinates and $g_{ij}(z=0,x)=g^{(0)}_{ij}(x)$ is the boundary metric.
For $g^{(0)}_{ij} = \eta_{ij}$ we recover the AdS metric~\eqref{pure_FG} with radius $\ell$,\footnote{For simplicity, in the following we will set $\ell = 1$, but we will reintroduce it when necessary.}
but the bulk metric~\eqref{fg} also allows for a curved boundary where $g^{(0)}_{ij}$ sources the stress tensor of the CFT.

In order to study the Weyl anomaly of the dual CFT, we first study the behaviour of $G_{\mu\nu}$ under diffeomorphisms.
Under a general coordinate transformation
\begin{equation}\label{diffeo}
    X^\mu = X'^\mu + \xi^\mu (X') \, ,
\end{equation}
the bulk metric $G_{\mu\nu}$ transforms as
\begin{equation}\label{Gtransf}
    \delta G_{\mu\nu} = G_{\mu\rho} \partial_\nu \xi^\rho + G_{\nu\rho} \partial_\mu \xi^\rho + \xi^\rho \partial_\rho G_{\mu\nu} \, ,
\end{equation}
with $\delta G_{\mu\nu} = \mathcal{L}_\xi G_{\mu\nu} = G'_{\mu\nu}(z,x) - G_{\mu\nu}(z,x)$
and we require that $\xi^\mu$ is such that \eqref{fg} is form invariant under~\eqref{diffeo}, i.e.~$\mathcal{L}_\xi G_{zz} = 0 = \mathcal{L}_\xi G_{zi}$.
The solution is given by
\begin{equation}\label{a}
    \xi^z = -2 z \sigma(x) \, , \quad \xi^i = a^i(z,x) = \frac12 \partial_j \sigma(x) \int_0^z dz' \,   g^{ij}(z',x) \, ,
\end{equation}
where $\sigma(x)$ is an arbitrary function, the $a^i$ are infinitesimal and we will work to order $\mathcal{O}(\sigma, a^i)$.
The lower end of the integration is chosen so that there are no residual diffeomorphisms at the boundary $z=0$.
In other words, the boundary condition $a^i(z=0,x) =0$ holds.
Then, performing the diffeomorphism~\eqref{diffeo} with~\eqref{a}, $g_{ij}$ transforms as
\begin{equation}\label{g}
    \delta g_{ij} = 2\sigma (1-z\partial_z) g_{ij} + \nabla_i a_j + \nabla_j a_i \, ,
\end{equation}
where indices are lowered with (and derivatives are covariant w.r.t.) $g_{ij}$.
The equations~\eqref{a} and~\eqref{g} define the PBH transformations.

We show now that this particular class of bulk diffeomorphisms reduces to a Weyl transformation at the boundary.
Indeed, the commutator of two diffeomorphisms~\eqref{Gtransf} is again a diffeomorphism
\begin{equation}\label{commdiffeo}
    [\delta_2,\delta_1] G_{\mu\nu} = G_{\mu\rho} \partial_\nu \hat\xi^\rho + G_{\nu\rho} \partial_\mu \hat\xi^\rho + \hat\xi^\rho \partial_\rho G_{\mu\nu} \, ,
\end{equation}
where $\hat\xi^\rho$ is defined as
\begin{equation}\label{xihat}
    \hat\xi^\rho = \xi^\sigma_1 \partial_\sigma \xi^\rho_2 - \xi^\sigma_2 \partial_\sigma \xi^\rho_1 + \delta_2 \xi^\rho_1 - \delta_1 \xi^\rho_2 \,,
\end{equation}
and the last two terms are non-vanishing if we allow $\xi^\mu$ to be field dependent.
If the diffeomorphism is a PBH, then it is possible to derive that $\hat\xi^\mu = 0$, which is called the PBH group property \cite{Schwimmer:2008yh,Fiorucci:2020xto}, so that $[\delta_2,\delta_1] G_{\mu\nu} = 0$.
Since the PBH transformations do not act on coordinates but on fields, it follows that \mbox{$[\delta_2,\delta_1] g_{ij} = 0$} and the PBH transformations reduce to a Weyl transformation on the boundary metric.
Indeed, from~\eqref{g} at $z=0$ it follows that
\begin{equation}
    \delta g^{(0)}_{ij} = 2 \sigma g^{(0)}_{ij} \, .
\end{equation}

Close to the boundary, $g_{ij}$ has a FG expansion\footnote{If $d$ is an even integer, the expansion of the metric contains also logarithmic terms.
Here we work in generic dimensions, and we do not need to include such terms.} \cite{fefferman-graham}
\begin{equation}
    g_{ij}(z,x) = \sum _{n=0}^\infty g^{(n)}_{ij}(x) z^n \,,  \label{gansatz}
\end{equation}
which induces also the following expansion\footnote{The label $(n)$ indicating the $n$th term in the radial expansion will be always equivalently used as a super- or subscript.}
\begin{equation}
    a^i(z,x) = \sum_{n=1}^\infty a^i_{(n)}(x) z^n \,. \label{aansatz}
\end{equation}
Using the PBH equation~\eqref{g}, we can determine the coefficient in the expansion of the metric in terms of covariant tensors built from the boundary metric as follows.
First, we compute the $a_{(n)}$ in terms of the $g_{(n)}$, and for the first few terms we find
\begin{align}
    a^i_{(1)} &= \frac12 g_{(0)}^{ij}\partial_j \sigma \,, \label{a1}\\
    a^i_{(2)} &= -\frac14 g_{(1)}^{ij}\partial_j \sigma \label{a2} \, .
\end{align}
Indices are now lowered (raised) with the (inverse of) $g^{(0)}_{ij}$; curvatures and covariant derivatives will be w.r.t.~$g^{(0)}_{ij}$.\footnote{We will stress this by adding a label $(0)$ on top of curvature tensors and covariant derivatives.}
Then, combining the expansions and the $a_{(n)}$ into~\eqref{g} we find the variation of $g_{(n)}$ as
\begin{align}
    \delta g^{(0)}_{ij} =& \, 2\sigma g^{(0)}_{ij} \,, \label{dg0}\\
    \vdots & \nonumber\\
    \delta g^{(n)}_{ij} =& \, 2\sigma (1-n) g^{(n)}_{ij} + \sum_{m=1}^{n} \left( g^{(n-m)}_{ik} \partial_j a^k_{(m)} + g^{(n-m)}_{jk} \partial_i a^k_{(m)} + a^k_{(m)} \partial_ k g^{(n-m)}_{ij} \right) \,, \label{dgn}
\end{align}
and we see that in general $g_{(n)}$ contains $2n$ derivatives.
Thus, to compute it we make the most general Ansatz for a symmetric and covariant tensor built from $g^{(0)}_{ij}$ with two indices and with $2n$ derivatives, take its variation according to~\eqref{dg0} and impose that it satisfies the PBH equation~\eqref{dgn}.
In this way, for example for the first two terms we find
\begin{align}\label{g1}
    g^{(1)}_{ij} &= - \frac{1}{d-2} \bigg( \overset{(0)}{R}_{ij} -\frac{1}{2(d-1)}\overset{(0)}{R} g^{(0)}_{ij} \bigg) \,, \displaybreak[0] \\
    g^{(2)}_{ij} &= \frac{1}{d-4} \Bigg( \frac{1}{8(d-1)}\overset{(0)}{\nabla}_i \overset{(0)}{\nabla}_j \overset{(0)}{R} - \frac{1}{4(d-2)} \overset{(0)}{\Box} \overset{(0)}{R}_{ij} + \frac{1}{8(d-2)(d-1)}g_{(0)ij} \overset{(0)}{\Box}\overset{(0)}{ R}  \nonumber\\
    &\phantom{=} - \frac{1}{2(d-2)} \overset{(0)}{R}{}^{kl}\overset{(0)}{R}_{ikjl} + \frac{d-4}{2(d-2)^2}\overset{(0)}{R}_{ik}\overset{(0)}{R}_j{}^k + \frac{1}{(d-2)^2(d-1)} \overset{(0)}{R}_{ij} \overset{(0)}{R}  \nonumber \\
    & \phantom{=} + \frac{1}{4(d-2)^2} \overset{(0)}{R}_{kl}\overset{(0)}{R}{}^{kl} g^{(0)}_{ij} - \frac{3d}{16(d-2)^2(d-1)^2} g^{(0)}_{ij} \overset{(0)}{R}{}^{2} \Bigg) \nonumber \\
    &\phantom{=} + c_1 \, \overset{(0)}{C}{}^{2} g^{(0)}_{ij} + c_2 \, \overset{(0)}{C}_{iklm}\overset{(0)}{C}_j{}^{klm} \,, \label{g2}
\end{align}
where $C_{ijkl}$ is the Weyl tensor.
Starting from $g_{(2)}$, the solutions will have free coefficients that are not fixed by the PBH equation.
On the other hand, if one perturbatively solves Einstein's equations of motion for the metric~\eqref{fg} using the expansion~\eqref{gansatz}, $g_{(2)}$ is completely determined in generic dimensions.
The free coefficients in the PBH solutions are thus fixed on-shell given an action.

\subsection{Effective boundary action and Weyl anomalies}\label{sec:PBHanomaly}
Consider now an action
\begin{equation} \label{puregravity}
    S = \int_M d^{d+1} X \sqrt{G} f(R(G)) \, ,
\end{equation}
where $f$ is a local function of the curvature and its covariant derivatives and we require that $f(R)$ is such that the equations of motion are solved by $\text{AdS}_{d+1}$ in order to have a CFT at the boundary.
Under a bulk diffeomorphism, the action $S$ is invariant up to a boundary term
\begin{equation}\label{deltaS}
    \delta S = \int_M d^{d+1} X \, \partial_\mu \left( \xi^\mu \mathcal{L} \right) \, , \quad \mathcal{L} = \sqrt{G} f(R(G))\,,
\end{equation}
from which we read the transformation $\delta\mathcal{L} = \partial_\mu \left( \xi^\mu \mathcal{L} \right)$. If the diffeomorphism is a PBH, then one can show that $[\delta_2, \delta_1]\mathcal{L}=0$ upon using the PBH group property \cite{Schwimmer:2008yh}.
Taking the FG form for the bulk metric, then the FG expansion~\eqref{gansatz} induces a power series expansion for $\mathcal{L}$ as well
\begin{equation}\label{lg_FG}
    \mathcal{L} = \sqrt{g_{(0)}} z^{-d/2-1} \mathcal{L}_g \, , \quad  \mathcal{L}_g = \sum_{n=0}^\infty \mathcal{L}^{(n)}_g (x) z^n \, .
\end{equation}
Then, by virtue of $[\delta_2, \delta_1]\mathcal{L}=0$, it is possible to show that $\mathcal{L}_g$ satisfies a WZ condition
\begin{equation}
    \int_{\partial M} d^d x \sqrt{g_{(0)}} \left( \sigma_2(x) \delta_1 \mathcal{L}_g - \sigma_1(x) \delta_2 \mathcal{L}_g \right) = 0 \, ,
\end{equation}
which means that $\mathcal{L}_g$ is a candidate for the anomaly of the boundary CFT.
To make the connection precise, from~\eqref{deltaS} we have
\begin{equation}\label{deltaSbdy}
    \delta S = \int_{\partial M} d^d x \, \xi^z \mathcal{L} |_{z=0} = -2 \int_{\partial M} d^d x \, z \, \sigma \, \mathcal{L} |_{z=0}
\end{equation}
restricting the diffeomorphism to a PBH.
Using the holographic dictionary, we interpret the variation of the bulk action as the variation of the generating functional of the CFT correlators, so that the finite piece in~\eqref{deltaSbdy} gives the holographic Weyl anomaly.\footnote{Divergent terms can be cancelled with the addition of counterterms to the bulk action, as first shown for the Einstein-Hilbert action in~\cite{Henningson:1998gx} and then further elaborated upon and proven for a generic action in~\cite{deHaro:2000vlm,Papadimitriou:2004ap,Papadimitriou:2005ii,Andrade:2006pg}.}
In particular, recall that the Weyl variation of the effective action produces the anomalous trace of the stress tensor and comparing~\eqref{ano_trace} with~\eqref{deltaSbdy} we find the relation
\begin{equation}
    g^{(0)}_{ij} \langle T^{ij} \rangle = - 2 \mathcal{L}_g^{d/2} \,,
\end{equation}
which means that $\mathcal{L}^{(n)}_g(x)$ gives the trace anomaly of the $d=2n$ dimensional CFT.

Consider now as an example
\begin{equation}\label{EHwriem}
    2\kappa^2 f(R(G)) = \Lambda - R(G) + \gamma \left(R_{\mu\nu\rho\sigma}R^{\mu\nu\rho\sigma}\right) \left( G \right) \, , \quad \Lambda = -d (d-1) -2d (d-3)\gamma \, ,
\end{equation}
where $\gamma$ is a dimensionless parameter and $\Lambda$ is such that $\text{AdS}_{d+1}$ with radius $\ell =1$ is a solution of the equations of motion.\footnote{For simplicity, in the following we will take $2\kappa^2 = 16\pi G^{(d+1)}_N =1$ but reintroduce it when necessary.}
By writing the action in the FG coordinates~\eqref{fg} we find (a prime denotes a derivative w.r.t.~$z$)
\begin{align}
    \mathcal{L}_g &= d (1 +4 \gamma ) + \frac12 z (1+4\gamma) \left[ 2(1 - d) g^{ij}g'_{ij} - R(g) \right]  + \frac12 z^2 \big[ 4(1 + 4 \gamma ) g^{ij} g''_{ij}\nonumber\\
    &\phantom{=} - (3 - 4 (d-5) \gamma ) g^{ik}g^{jl}g'_{ij} g'_{kl} + (1 + 8 \gamma ) \left( g^{ij} g'_{ij} \right)^2  + \gamma (R_{ijkl} R^{ijkl})(g) \nonumber\\
    &\phantom{=} + 8 \gamma g'_{ij} R^{ij}(g) \big] + \gamma z^3 \big[ 4 g^{im}g^{jn}g^{kl} g'_{ik} g'_{mn} g'_{jl}  - 4 g^{ij} g^{kl} g^{mn} g'_{ij} g'_{km} g'_{ln} \nonumber\\
    &\phantom{=} -2  g'_{ij} g'_{kl} R^{ikjl}(g)  - 4 g^{im}g^{jn}\nabla_{j}g'_{ik}\nabla^{k}g'_{mn} + 4 g^{im}g^{jn} \nabla_{k}g'_{ij} \nabla^{k}g'_{mn} \big]  \nonumber\\
    &\phantom{=} + \gamma z^4 \big[ g^{im}g^{jn}g^{kp}g^{lq} g'_{ik} g'_{mn} g'_{jl} g'_{pq} + (g^{ik}g^{jl}g'_{ij} g'_{kl})^2 + 8 g^{ik} g^{jl} g''_{ij} g''_{kl}  \nonumber\\
    &\phantom{=} - 8 g^{im}g^{jn}g^{kl} g'_{ik} g'_{mn} g''_{jl} \big] \,, \label{Og}
\end{align}
and expanding the metric according to~\eqref{gansatz}, we identify for instance the terms
\begingroup
\allowdisplaybreaks[1]
\begin{align}
    \mathcal{L}^{(1)}_g &= - \frac{1}{2} (1+4\gamma) \overset{(0)}{R} + \frac12 \left(2 - d \right)(1+4\gamma) g^{(1)i}{}_{i} \,, \label{Og1}\\
    \mathcal{L}^{(2)}_g &= \frac{1}{2} (1+12\gamma) \overset{(0)}{R}{}^{ij} g^{(1)}{}_{ij}  - \frac{1}{4} (1+4\gamma) \overset{(0)}{R} g^{(1)i}{}_{i} + \frac32 (4 -  d) (1+4\gamma) g^{(2)i}{}_{i}  \nonumber\\*
    &\phantom{=} + \left(\frac18 (8  - 3 d) +\frac32 \gamma (4-d) \right) g^{(1)i}{}_{i} g^{(1)j}{}_{j} - \frac{1}{2}(1+4\gamma) \overset{(0)}{\nabla}_{j}\overset{(0)}{\nabla}_{i}g^{(1)ij}   \nonumber\\*
    &\phantom{=} + \left(\frac14 ( 3 d - 10 ) +\gamma(5d-14) \right) g^{(1)}{}_{ij} g^{(1)ij}  + \frac{1}{2}(1+4\gamma) \overset{(0)}{\Box}g^{(1)i}{}_{i} + \frac12 \gamma \overset{(0)}{R}_{ijkl}\overset{(0)}{R}{}^{ijkl} \,. \label{Og2}
\end{align}
\endgroup
On the PBH solutions~\eqref{g1},~\eqref{g2} we find the trace anomaly in $d=2,4$ respectively:
\begin{align}
    \mathcal{L}^{(1)}_g &= -\frac12 (1+4\gamma) \overset{(0)}{R} = -\frac12 a \overset{(0)}{E}{}_2 \,, \\
    \mathcal{L}^{(2)}_g &= -\frac18 (1+12\gamma) \overset{(0)}{R}{}_{ij} \overset{(0)}{R}{}^{ij} + \frac{1}{24} (1+8\gamma) \overset{(0)}{R}{}^{2} + \frac12 \gamma \overset{(0)}{R}_{ijkl}\overset{(0)}{R}{}^{ijkl} = -\frac{1}{16} \left( c \, \overset{(0)}{C}{}^2 - a \, \overset{(0)}{E}{}_4 \right) 
\end{align}
with $c=1-4\gamma$ and $a=1+4\gamma$.
${E}{}_{2n}$ is the Euler density in $d=2n$, and explicitly ${E}{}_4$ is as in~\eqref{euler_4} while ${C}{}^2$ as in~\eqref{weyl_sqr} here with curvature w.r.t.~the boundary metric.
Following the classification of~\cite{Deser:1993yx}, we notice that in $d=2$ the holographic computation yields an anomaly which is entirely type A, while in $d=4$ there is also a type B as expected from the CFT analysis.
Moreover, for $\gamma =0$ we have $a=c$, while in the presence of the quadratic term in the curvature in~\eqref{EHwriem} then $a-c\neq 0$.
It can be verified that including $R^2$ and $R_{\mu\nu}R^{\mu\nu}$ terms in the action would also change the values of $a$ and $c$, but not their difference~\cite{Blumenhagen:2013fgp}, so that \eqref{EHwriem} produces the minimal bulk action which allows to distinguish between type A and B anomalies in the pure gravity case.
Having $a\neq c$ will be useful in chap.~\ref{chap:holography} when we include scalar fields in the bulk.
As a final comment, notice that $g^{(n)}_{ij}$ does not contribute to $\mathcal{L}^{(n)}_g$ in $d=2n$ \cite{Schwimmer:2003eq,Schwimmer:2008yh,Miao:2013nfa}.

Consider the case $\gamma =0$ and $d=4$.
Then, the action that follows from~\eqref{EHwriem} corresponds to the action of type~IIB supergravity on $\text{AdS}_5 \times S^5$ after integration over the five-dimensional sphere and setting to zero all fields except for the metric.
The five-dimensional Newton constant can be expressed as
\begin{equation}\label{newton5}
    G^{(5)}_N = \frac{G^{(10)}_N}{\text{Vol}(S^5)} \quad \text{with} \quad 16\pi G^{(10)}_N = (2\pi)^7 g_s^2 (\alpha')^4 \, .
\end{equation}
The dual CFT is $\mathcal{N}=4$ super Yang-Mills with gauge group $SU(N)$ and we can check that the holographic computation for the anomaly reproduces the anomaly of this CFT.
Indeed, the trace anomaly computed from holography reads
\begin{equation}
    g^{(0)}_{ij} \langle T^{ij} \rangle = \frac{\ell^3}{64\pi G_N^{(5)}} \left( \overset{(0)}{R}{}_{ij} \overset{(0)}{R}{}^{ij} - \frac13 \overset{(0)}{R}{}^{2} \right) = \frac{N^2}{32\pi^2} \left( \overset{(0)}{R}{}_{ij} \overset{(0)}{R}{}^{ij} - \frac13 \overset{(0)}{R}{}^{2} \right) \,,
\end{equation}
where we reintroduced the $2\kappa^2$ factor and the dependence on the AdS radius $\ell$; the second equality follows from~\eqref{newton5} and~\eqref{duality_mald}.
This coincides with the trace anomaly of $\mathcal{N}=4$ super Yang-Mills with gauge group $SU(N)$ at large $N$ and thus provides a test for the AdS/CFT correspondence.

In chap.~\ref{chap:holography}, we will modify the PBH transformations by including a scalar field in the bulk, focusing on the case in which the scalar field sources an integer-dimensional irrelevant operator on the boundary.
The trace anomaly of a CFT with integer-dimensional irrelevant operators is computed in~\cite{Schwimmer:2019efk}.
We will compute it again from holography and thus provide an additional test of the AdS/CFT correspondence.
\chapter{Correlators and quantum measures of descendant states}\label{chap:desc_corr}

In sec.~\ref{sec:intro_corr_desc} we stated that correlation functions of descendant fields in two\-/dimensional CFTs can be written as a differential operator acting on the correlator of the respective primary fields.
As an example, we showed the case of the correlator of primary fields with one descendant field, and argued that computing the correlator with more than one descendant is a more laborious task.
We now come back to that issue and in sec.~\ref{sec:CFTtec} we derive a recursive formula valid in a two-dimensional CFT to calculate correlation functions of an arbitrary number of descendants, whose result will depend on the correlator of the respective primaries.
In sec.~\ref{sec:qmeasures} we review some entanglement and distinguishability measures between quantum states, in particular the R\'enyi entropy, trace square distance and sandwiched R\'enyi divergence, that are related to correlation function of states.
With our recursive formula, we will then be able to compute these measures for descendant states.
In sec.~\ref{sec:universal} we show some results for descendants of the vacuum; since these are valid in all theories with a unique vacuum, we call them universal.
In sec.~\ref{sec:nonuniversal} we focus on descendants of non-vacuum primary fields; these are theory dependent, and we compute the quantum measures in the critical Ising model and the three-state Potts model.
As an application of our results, we will provide a test of the conjectured R\'enyi Quantum Null Energy Condition (QNEC).
More details and expressions are presented in app.~\ref{app:descendant}; there, we also report a Mathematica code that implements the recursive formula.
The content of the present chapter was first published in~\cite{Brehm:2020zri}, whose presentation we closely follow.

\section{Computing correlators of descendant fields on the plane}\label{sec:CFTtec}
Recall from sec.~\ref{sec:2d_cft} that descendant states in two-dimensional CFTs are generated by acting with the Virasoro generators $L_{-n}$ with $n>0$ on the primary states.
Since the state-operator correspondence holds, we can assign a field to every state, and for descendant states we write the states and corresponding fields as
\begin{equation}
   \ket{\Delta,\{(m_i,n_i)\}} = \prod_i L^{n_i}_{-m_i} \ket{\Delta} \quad \leftrightarrow \quad f_{ \ket{\Delta,\{(m_i,n_i)\}}} = \prod_i \hat{L}_{-m_i}^{n_i} f_{\ket{\Delta}}\,,
\end{equation}
where $\ket{\Delta}$ is a primary state with conformal dimension $\Delta=h+\bar h$, with holomorphic and anti-holomorphic conformal weights $h$ and $\bar h$, and with
\begin{equation}\label{eq:desfield}
    \hat{L}_{-m} g(w) := \oint_{\gamma_w} \frac{dz}{2\pi i} \frac{1}{(z-w)^{m-1}} T(z) g(w) 
\end{equation}
for any field $g$; $\gamma_w$ is a closed path surrounding $w$. $\hat{L}_{-m} g(w)$ is the $m$th `expansion coefficient' in the OPE of the stress tensor $T$ with the field $g$. 
A similar construction holds for the anti-holomorphic copy of the Virasoro generators.
In what follows we will focus on spinless states $h-\bar h =0$

Recall also that we denote the field dual to $f_{\ket{s}}(z,\bar{z})$ by
\begin{equation}
    f_{\bra{s}}(\bar{z},z) := \left(f_{\ket{s}}(z,\bar{z})\right)^\dagger\,.
\end{equation}
The duality structure of the Hilbert space is fixed by the definitions $L_{-n}^\dagger = L_{n}$ and $\bra{\Delta}\Delta'\rangle = \delta_{\Delta,\Delta'}$, and has to be recovered from the two-point function of the corresponding fields at coincident points
\begin{equation}\label{eq:contraint0}
    \bra{s}s'\rangle \equiv \lim_{z\to w}\left\langle  f_{\bra{s}}(\bar{z},z) f_{\ket{s'}}(w,\bar{w}) \right\rangle\,.
\end{equation}

\noindent
To achieve this one chooses radial quantization around the second insertion point $w$ and defines the dual field $f_{\bra{s}}(\bar{z},z)$ as the outcome of the transformation $G(z)=\frac{1}{z-w}+w$ of the field $f_{\ket{s}}(z,\bar{z})$ at the unit circle surrounding $w$.
With the help of the transformation rules that we define in the following sec.~\ref{sec:trafo}, we can write
\begin{equation}\label{eq:DualFeld}
    f_{\bra{s}}(\bar{z},z) = f_{\Gamma_{G} \ket{s}}\left(\frac1{z-w}+w,\frac1{\bar{z}-\bar{w}}+\bar{w}\right)\,,
\end{equation}
where the action $\Gamma_G$ on the local Hilber space takes the simple form 
\begin{equation}
    \Gamma_G = \left(-\frac{1}{(z-w)^2}\right)^{L_0}\left(-\frac{1}{(\bar{z}-\bar{w})^2}\right)^{\bar{L}_0} \exp\left(\frac{L_1}{w-z}+\frac{\bar{L}_1}{\bar{w}-\bar{z}}\right)\,.
\end{equation}

\noindent
In what follows we will use radial quantization around the origin of the complex plane, i.e.~we will choose $w=0$. Note, that \eqref{eq:DualFeld} gives \eqref{eq:contraint0} up to a phase factor $(-1)^{h-\bar h}$, which will not be important in our analysis.

\subsection{Transformation of states and fields}\label{sec:trafo}
The transformation rule for arbitrary holomorphic fields was first presented in \cite{Gaberdiel:1994fs}. We will, however, use the (equivalent) method introduced in \cite{frenkel2004vertex}. 

There is a natural action $M(G)$ of a conformal transformation $G$ on any Virasoro module and, hence, on the full space of states. For a field $f_{\ket{s}}(w)$ we need to know how the transformation acts locally around $w$ and transform the field accordingly. It works as follows.

Consider a conformal transformation $G$ and choose local coordinates around the insertion point $w$ and the point $G(w)$. The induced local coordinate change can be written as $\mathfrak{g}(z) = \sum_{k=1}^\infty a_k z^k$, where $z$ are the local coordinates around $w$ that are mapped to the local coordinates $\mathfrak{g}(z)$ around $G(w)$. Now solve the equation 
\begin{equation}
    v_0 \exp\left(\sum_{j=1}^\infty v_j t^{j+1}\partial_t\right)t = \mathfrak{g}(t)
\end{equation}
for the coefficients $v_j$ order by order in $t$. The local action of $G$ on the module is then given by $M(G) := \exp\left(-\sum_{j=1}^\infty v_j L_j\right) v_0^{-L_0}$. The inverse, that we will rather use, is then given by
\begin{equation}
    \Gamma := M(G)^{-1} = v_0^{L_0} \exp\left(\sum_{j=1}^\infty v_j L_j\right)\,,
\end{equation}
such that we can write
\begin{equation}
    f_{\ket{s}}(G(w)) = f_{\ket{s'} = \Gamma \ket{s}}(w)\,.
\end{equation}

\noindent
Note that for a descendant at level $k$ we only need the coefficients $v_j$ up to $j=k$. A Mathematica code to obtain the relation between the coefficients $v_j$ and $a_k$ is given in appendix~\ref{app:matv}.

\subsection{The recursive formula}\label{sec:rec_formula}

We will be interested in computing correlation functions 
\begin{equation}
    \langle \prod_{i=1}^N f_{\ket{s_i}}(z_i) \rangle \, ,
\end{equation}
where $\ket{s_i}$ are some descendant states. 

To get a handle on them we use Ward identities in a particular way. Therefore, consider a meromorphic function $\rho(z)$ that has singularities at most at $z\in \left\{z_i\right\}\cup \{0,\infty\} $, where $z_i$ are the insertion points.
Let us make the particular choice
\begin{equation}
    \rho(z) = \prod_{i=1}^N (z-z_i)^{a_i}
\end{equation}
for $a_i\in\mathbb{Z}$, which is in particular regular at $0$. Now, consider the integral identity
\begin{equation}
    \sum_{i=1}^N \oint_{\gamma_{z_i}} \frac{dz}{2\pi i} \rho(z) \left\langle T(z) g_i(z_i)  \prod_{j\neq i} g_j(z_j) \right\rangle 
    = - \oint_{\gamma_\infty} \frac{dz}{2\pi i} \rho(z) \left\langle T(z) \prod_{j=1}^N g_j(z_j) \right\rangle\,,  
\end{equation}
where $g_j$ are arbitrary fields, e.g.~descendant fields. The latter identity simply follows from deforming the integral contour accordingly. The r.h.s.~vanishes for $\sum_{i=1}^N a_i \le2$. Next, we consider the functions
\begin{equation}
    \rho_i(z) := \prod_{j\neq i} (z-z_j)^{a_j} = \frac{\rho(z)}{(z-z_i)^{a_i}}
\end{equation}
for which we need the expansion around $z_i$,
\begin{equation}
    \rho_i(z) \equiv \sum_{n=0}^\infty \rho_i^{(n)} \, (z-z_i)^n\,.
\end{equation}

\noindent
Note, that the expansion coefficients $\rho_i^{(n)}$ are some rational expressions that depend on all $z_j\neq z_i$ and $a_j$. 
Now, using the definition of $\hat{L}_m$ in~\eqref{eq:desfield}, and the latter expansion we obtain
\begin{equation}
    \sum_{i=1}^N \sum_{n=0}^\infty \rho_i^{(n)} \left\langle\left(\hat{L}_{a_i+n-1}g_i(z_i)\right)   \prod_{j\neq i} g_j(z_j)\right\rangle = 0\,\label{eq:WardIdNpt}
\end{equation}
for $\sum a_i\leq 2$. Note that, even if not written explicitly, the sums over $n$ do always terminate for descendant fields $g_i$. Note further that these relations among correlation functions depend on the choice of $a_i$ but the correlators that can be computed from these relations are unique.

One very immediate choice is $a_i = 1-m$ and $a_{j\neq i}=0$ which gives the relation
\begin{align}
    \left\langle \left(\hat{L}_{-m}g_i(z_i)\right) \prod_{j\neq i}g_j(z_j)\right\rangle = - \sum_{j\neq i} \sum_{n=0}^{\text{lvl}(g_j)+1} \rho_j^{(n)} \left\langle \left(\hat{L}_{n-1} g_j(z_j)\right) \prod_{k\neq j}g_k(z_k) \right\rangle \label{eq:rec1}
\end{align}
with 
\begin{equation}
    \rho_j^{(n)} = (-1)^{n}\binom{n+m-2}{n} (z_j-z_i)^{1-m-n}\,.
\end{equation}

\noindent
For $m>1$ we see that the total level of each correlator on the r.h.s., i.e.~the sum over all levels of fields appearing in the correlation functions, is lower than the one on the l.h.s.~of the equation. We, hence, can express correlation functions of higher total level by correlators of lower total level. One way of computing correlation functions of descendants is using the above formula recursively until there are only $L_{-1}$ left. These simply act as derivative operators on the respective primary. 

The Mathematica code that uses above equation recursively and computes arbitrary correlation functions of vacuum descendants is given in appendix \ref{app:VacDesCorr}. It produces an algebraic expression of the insertion points and the central charge $c$. The Mathematica code to compute correlation function for descendants of generic primary fields is given in appendix \ref{app:PrimDesCorr}. It produces a derivative operator that acts on the respective primary correlator, which in general is theory dependent. 

\section{Review of some quantum measures in CFT}\label{sec:qmeasures}

We want to consider an isolated quantum system living on a circle of length $L$ whose (low-energy) physics is governed by a (1+1)-dimensional effective field theory. We assume that at some critical value of its couplings the theory becomes conformal. Then, the system is in some pure state of a (1+1)-dimensional CFT, associated with a density matrix $\rho = \ket{s}\bra{s}$. 

Let us further consider a spatial bipartition into a region $A$ of size $l<L$ and its complement $\overline{A}$. Assume a situation where one has no access to the complement, i.e.~all measurements are restricted to the subregion $A$. Our ignorance of the complement means that the state in the region we have access to can be  reduced to the density matrix
\begin{equation}
    \rho_A = \Tr_{\overline{A}}\rho\,,
\end{equation}
where $\Tr_{\overline{A}}$ is the partial trace over the degrees of freedom of the complement.

Our focus of interest lies in reduced density matrices that originate from descendant states of the full system. In particular, we want to study their entanglement and measures of distinguishability between them. 

\subsection{Entanglement measure: R\'enyi entropy} \label{sec:Renyi}

The $n$th R\'enyi entropy \cite{renyi2012probability,nielsen_chuang_2010} is defined as
\begin{equation}
    S_n(A) = \frac{1}{1-n} \log \Tr_A \rho_A^n\,.
\end{equation}

\noindent 
For $n\to 1$ it converges to the (von Neumann) entanglement entropy $S(A) = -\Tr \rho_A \log\rho_A$ which is the most common entanglement measure \cite{nielsen_chuang_2010}. However, particularly in field theories, there exist alluring analytical tools that make it much easier to compute R\'enyi entropies for $n>1$ than the entanglement entropy. Additionally, many key properties of the entanglement entropy, such as the proportionality of ground state entanglement to the central charge in critical systems and the area law of gapped states, hold for R\'enyi entropies too. In principle, the knowledge of the  R\'enyi entropy for all $n\in \mathbb{N}$ allows to determine all eigenvalues of the reduced density matrix $\rho_A$. 

In the present case, the full system can be described by a CFT on the Euclidean spacetime manifold of an infinite cylinder for which we choose complex coordinates $u = x + i \tau$ with $\tau\in\mathbb{R}$ and $x + L \equiv x \in \left(-\frac L2 ,\frac L2\right]$. The variable $\tau$ is regarded as the time coordinate and $x$ is the spatial coordinate. As subsystem $A$ we choose the spatial interval $\left(-\frac{l}2,\frac{l}{2}\right)$\,. In two-dimensional CFTs, the trace over the $n$th power of the reduced density matrix $\rho_A = \Tr_{\overline{A}}\ket{s}\bra{s}$ is equivalent to a $2n$-point function on the so-called replica manifold which is given by $n$ copies of the cylinder glued together cyclically across branch cuts along the subsystem $A$ at $\tau =0$ \cite{Holzhey:1994we,Calabrese:2009qy}. The exponential map $z(u) = \exp\left(2\pi i u/L\right)$ maps the latter manifold to the $n$-sheeted plane $\Sigma_n$, where the branch cut now extends between $\exp\left(\pm i \pi \frac{l}{L}\right)$\,. The $2n$ fields are those that correspond to the state $\ket{s}$ and its dual $\bra{s}$, where one of each is inserted at the origin of each sheet:
\begin{align}
    \Tr_A \rho_A^n  
                    &= \mathcal{N}_n \left\langle \prod_{k=1}^n f_{\bra{s}}(0_k)f_{\ket{s}}(0_k)\right\rangle_{\Sigma_n}\\
                    &= \mathcal{N}_n \left\langle \prod_{k=1}^n f_{\Gamma_{-1/z}\ket{s}}(\infty_k)f_{\ket{s}}(0_k)\right\rangle_{\Sigma_n}\,.
\end{align}

\noindent
The constant $\mathcal{N}_n =  Z(\Sigma_n)/Z(\mathbb{C})^n = \left(\frac{L}{\pi a} \sin\left(\frac{\pi l}{L}\right)\right)^{\frac{c}{3} \left(n-\frac{1}{n}\right)}$, $Z$ being the partition function on the respective manifold, ensures the normalization $\Tr_A \rho_A =1$, with some UV regulator $a$ (for example some lattice spacing). In the second line we use the definition of the dual state. 

One way to compute the above correlation function is to use a uniformization map from $\Sigma_n$ to the complex plane. It is given by composing a Möbius transformation with the $n$th root,
\begin{equation}\label{eq:uniformization}
     w(z) = \left(\frac{z e^{ -i\pi \frac{l}{L}} - 1}{z -  e^{-i\pi\frac{l}{L}} }\right)^{\frac1n}\,.
\end{equation}

\noindent 
The $2n$ fields are mapped to the insertion points 
\begin{equation}
\begin{aligned}
        w(0_k) &= \exp\left(\frac{i \pi l}{ n L}+\frac{2\pi i(k-1)}{n} \right)\label{eq:InsPoints}\\
        w(\infty_k) &= \exp\left(-\frac{i \pi l}{ n L}+\frac{2\pi i (k-1)}{n}\right)
\end{aligned}
\end{equation}
on the unite circle, and the fields have to transform as described in section \ref{sec:trafo}. The change of local coordinates is given in \ref{app:uniformization}. The local action is denoted by $\Gamma_{w(z)} \equiv \Gamma_{k,l}$ and for the dual fields we get $\Gamma_{w(1/z)} = \Gamma_{w(z)} \Gamma_{1/z} \equiv \Gamma_{k,-l}$.

Putting all together we see that computing the $n$th R\'enyi entropy is basically equivalent to computing a $2n$-point function of particularly transformed fields: 

\begin{align}\label{eq:RFE}
    e^{(1-n)S_n(A)} = \Tr_A \rho_A^n \equiv \mathcal{N}_n \left\langle \prod_{k=1}^n f_{\Gamma_{k,l} \ket{s}}\left(w(0_k)\right)f_{\Gamma_{k,-l} \ket{s}}\left(w(\infty_k)\right) \right\rangle_{\mathbb{C}}=: \mathcal{N}_n F_{\ket{s}}^{(n)} \,. 
\end{align}

\noindent 
See also \cite{Palmai:2014jqa,Taddia:2016dbm} for derivations of the latter formula. Other computations of the entanglement entropy of excited states (not necessarily descendants) can also be found in \cite{Alcaraz:2011tn,Berganza:2011mh,Mosaffa:2012mz,Bhattacharya:2012mi,Taddia_2013}.

\subsection{Distance measures}

Distance and other similarity measures between density matrices provide quantitative methods to evaluate how distinguishable they are, where distinguishability refers to the outcome of generic measurements in the different states. There is not a single best measure and not even agreement upon criteria to evaluate different distance measures. Most of them are designed such that they provide the space of (not necessarily pure) states with some additional structure that ideally allows to draw some physically relevant conclusions about the system under consideration. In case of reduced density matrices, distance measures quantify how distinguishable they are by measurements confined to the subregion~$A$.  

We want to consider two of these measurements for reduced density matrices in two-dimensional CFT, namely the sandwiched R\'enyi divergence (which is a generalization of the relative entropy) and the trace square distance. Let us denote the reduced density matrices as $\rho_i = \Tr_{\overline{A}} \ket{s_i}\bra{s_i}$, with $\rho_0 \equiv \Tr_{\overline{A}} \ket{0}\bra{0}$ the reduce density matrix of the vacuum.

\subsubsection{Relative entropy}

The relative entropy between two reduced density matrices $\rho_{1}$ and $\rho_{2}$ is given by 
\begin{equation}\label{eq:RelEntropy}
    S(\rho_{1},\rho_{2}) = \Tr  ( \rho_{1} \log \rho_{1}) - \Tr  ( \rho_{1} \log \rho_{2}) \,.
\end{equation}

\noindent 
It is free from UV divergences, positive definite and one of the most commonly used distance measures in quantum information, particularly because several other important quantum information quantities are special cases of it, e.g.~the quantum mutual information and quantum conditional entropy. The relative entropy also shows to be useful in high energy application when e.g.~coupling theories to (semiclassical) gravity. It allows a precise formulation of the Bekenstein bound \cite{Casini_2008}, a proof of the generalized second law \cite{Wall_2010,Wall:2011hj} and the quantum Bousso bound \cite{Bousso:2014sda,Bousso:2014uxa}. It also appears in the context of holography where it can be used to formulate important bulk energy conditions (see e.g.~\cite{Lin:2014hva,Lashkari:2014kda,Lashkari:2015hha}). 

However, as in the case of the entanglement entropy there exist no direct analytical tools to compute the relative entropy in generic two-dimensional CFTs.  There exist several R\'enyi type generalisations (see e.g.~\cite{Lashkari:2014yva,Lashkari:2015dia}) that are more straight forward to compute. We here want to focus on a quite common one called the sandwiched R\'enyi divergence. 

\subsubsection*{Sandwiched R\'enyi divergence}\label{sec:SRD}

The sandwiched R\'enyi divergence (SRD) between two density matrices $\rho_1$ and $\rho_2$ is given by 
\begin{equation}
    \mathcal{S}_n(\rho_{1},\rho_{2}) = \frac{1}{n-1} \log \Tr \left(\rho_1^{\frac{1-n}{2n}} \, \rho_2 \, \rho_1^{\frac{1-n}{2n}}\right)^n\,.\label{eq:SRD}
\end{equation} 

\noindent 
It is a possible one-parameter generalization of the relative entropy \eqref{eq:RelEntropy}, with the parameter $n \in[\frac12,\infty)$ and $S(\rho_{1},\rho_{2}) \equiv \mathcal{S}_{n\to1} (\rho_{1},\rho_{2})$\,. The SRD by itself has been shown to enjoy important properties of a measure of distinguishability of quantum states. It is, in particular, positive for all states, unitarily invariant, and decreases under tracing out degrees of freedom \cite{Mueller-Lennert:2013,Wilde:2014eda,Frank_2013,Beigi_2013}.

Due to the negative fractional power of $\rho_1$, there is no general method known to compute the SRD for arbitrary states in a CFT. However, if $\rho_1$ is the reduced density matrix of the theory's vacuum, then there is a technique introduced in \cite{Lashkari:2018nsl} to express it in terms of correlation functions. We now briefly review it, and refer to~\cite{Lashkari:2018nsl} for more details.

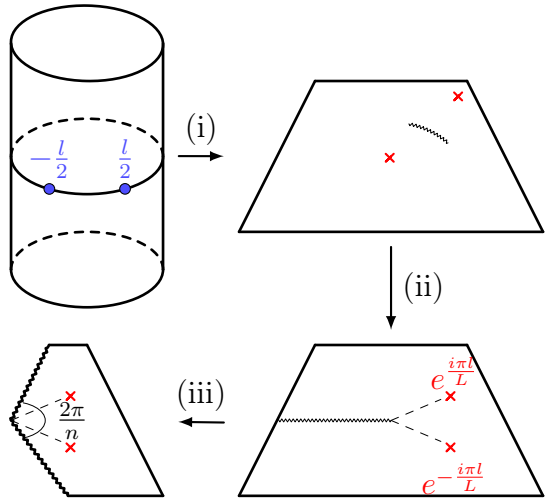
\begin{figure}[t]
\centering
\definecolor{ududff}{rgb}{0.30196078431372547,0.30196078431372547,1}
\begin{tikzpicture}[line cap=round,line join=round,>=triangle 45,x=1cm,y=1cm,scale=1]

\begin{scope}[local bounding box = cylinder]
\draw [rotate around={178.65449714894862:(1,3)},line width=1pt] (1,3) ellipse (1.0008289996875666cm and 0.4998965358776981cm);
\draw [line width=1pt] (2,3)-- (2,0);
\draw [line width=1pt] (2,0) arc(0:-180:1cm and 0.5cm);
\draw [dashed, line width=1pt] (2,0) arc(0:180:1cm and 0.5cm);
\draw [line width=1pt] (2,1.5) arc(0:-180:1cm and 0.5cm);
\draw [dashed, line width=1pt] (2,1.5) arc(0:180:1cm and 0.5cm);
\draw [line width=1pt] (0,3)-- (0,0);
\draw [fill=ududff] (0.5040500631758853,1.068897766899791) circle (2pt) node[anchor=south,color=ududff] {$-\frac{l}{2}$};
\draw [fill=ududff] (1.5,1.07) circle (2pt) node[anchor=south,color=ududff] {$\frac{l}{2}$};
\end{scope}

\begin{scope}[local bounding box = plane, shift={(5,1.5)}]
\draw[decoration = {zigzag,segment length = 0.5mm, amplitude = 0.2mm},decorate,line width=0.25pt] (0.75,0.18) arc(49.03:79.29:1.1cm);
\draw [line width=1pt] (-2,-1)-- (-1,1);
\draw [line width=1pt] (-1,1)-- (1,1);
\draw [line width=1pt] (1,1)-- (2,-1);
\draw [line width=1pt] (-2,-1)-- (2,-1);
\draw [color=red,thick] (0,0) -- ++(-2pt,-2pt) -- ++(3pt,3pt) ++(-3pt,0) -- ++(3pt,-3pt);
\draw [color=red,thick] (0.9,0.8133333333333327) -- ++(-2pt,-2pt) -- ++(3pt,3pt) ++(-3pt,0) -- ++(3pt,-3pt);
\end{scope}

\draw[shorten >=2mm,shorten <=2mm, thick,-latex] (cylinder) -- node[above]{(i)} (plane);

\begin{scope}[local bounding box = moebius, shift={(5,-2)}]
\draw [line width=1pt] (-2,-1)-- (-1,1);
\draw [line width=1pt] (-1,1)-- (1,1);
\draw [line width=1pt] (1,1)-- (2,-1);
\draw [line width=1pt] (-2,-1)-- (2,-1);
\draw[decoration = {zigzag,segment length = 0.5mm, amplitude = 0.2mm},decorate,line width=0.25pt] (0,0) -- (-1.5,0);
\node at (0.8,-0.34) (a) {};
\node at (0.8,0.34) (aa) {};
\draw [color=red,thick] (0.8,-0.34) -- ++(-2pt,-2pt) -- ++(3pt,3pt) ++(-3pt,0) -- ++(3pt,-3pt) node[anchor=north, color=red] {$e^{-\frac{i \pi l}{L}}$};
\draw[dashed, line width=0.25pt] (0,0) -- (a);
\draw [color=red,thick] (0.8,0.34) -- ++(-2pt,-2pt) -- ++(3pt,3pt) ++(-3pt,0) -- ++(3pt,-3pt) node[anchor=south, color=red] {$e^{\frac{i \pi l}{L}}$};
\draw[dashed, line width=0.25pt] (0,0) -- (aa);
\end{scope}

\draw[shorten >=2mm,shorten <=2mm, thick,-latex] (plane) --node[right]{(ii)} (moebius);

\begin{scope}[local bounding box = moebius_negpow, shift={(0,-2)}]
\draw [line width=1pt] (0.5,1)-- (1,1);
\draw [line width=1pt] (1,1)-- (2,-1);
\draw [line width=1pt] (0.75,-1)-- (2,-1);
\draw[decoration = {zigzag,segment length = 0.8mm, amplitude = 0.2mm},decorate,line width=1pt] (0,0) -- (0.5,1);
\draw[decoration = {zigzag,segment length = 0.8mm, amplitude = 0.2mm},decorate,line width=1pt] (0,0) -- (0.75,-1);
\node at (0.8,-0.34) (a) {};
\node at (0.8,0.34) (aa) {};
\draw [color=red,thick] (0.8,-0.34) -- ++(-2pt,-2pt) -- ++(3pt,3pt) ++(-3pt,0) -- ++(3pt,-3pt);
\draw[dashed, line width=0.25pt] (0,0) -- (a);
\draw [color=red,thick] (0.8,0.34) -- ++(-2pt,-2pt) -- ++(3pt,3pt) ++(-3pt,0) -- ++(3pt,-3pt);
\draw[dashed, line width=0.25pt] (0,0) -- (aa);
\begin{scope}
    \clip (0.17,-0.25) rectangle (0.5,0.25);
    \draw[line width=0.25pt] (0,0) ellipse (0.45cm and 0.25cm);
\end{scope}
\node[anchor=west] at (0.45,0) {$\frac{2\pi}{n}$}; 
\end{scope}

\draw[shorten >=2mm,shorten <=2mm, thick, -latex] (moebius) -- node[above]{(iii)} (moebius_negpow);

\end{tikzpicture}
    
\caption{Pictorial representation of the geometric setting for the SRD. (i) The reduced density matrix is represented by the sheet with respective operator insertions (red crosses) at 0 and $\infty$. (ii) A M\"obius transformation maps the insertion points to $e^{\pm \frac{i\pi l}{L}}$ and the branch cut to the negative real line. (iii) The multiplication by negative fractional powers of the reduced vacuum states is given cutting out respective parts of the sheet.}
\label{fig:SRD}
\end{figure}

Let us remind that the reduced density matrix for a sub-system on the cylinder is represented by a sheet of the complex plane with a brunch cut along some fraction of the unit circle with the respective operator insertions at the origin and at infinity of that sheet. In case of the vacuum the corresponding operator is the identity and, hence, we regard it as no operator insertion. Multiplication of reduced density matrices is represented by gluing them along the branch cut. Now, let us consider the M\"obius transformation
\begin{equation}\label{eq:Moebius}
   w(z) =  \frac{z e^{ -i\pi \frac{l}{L}} - 1}{z- e^{-i\pi\frac{l}{L}} }\,,
\end{equation}
which maps the two insertions points $0$ and $\infty$ of a sheet to $e^{\pm \frac{i\pi l}{L} }$ and the cut to the negative real axis on every sheet. Now, the reduced density operators can be regarded as operators acting on states defined on the negative real axis by rotating them by $2\pi$ and exciting them by locally acting with the respective operators at $e^{\pm \frac{i\pi l}{L} }$. In case of the vacuum reduced density matrix this now allows to define fractional powers by rotating by a fractional angle and even negative powers by rotating by negative angles, which basically means removing a portion of the previous sheet. The latter is, however, only possible if no operator insertion is removed. In the present case, the negative power $\frac{1-n}{2n}$ corresponds to an angle $-\pi + \frac{\pi}{n}$. Hence, this construction only makes sense for $\frac{l}{L}< \frac{1}{n}$.\footnote{In \cite{Moosa:2020jwt} the interested reader can find arguments why this is not simply an artifact of the CFT construction but holds generally when one assumes that the state is prepared from a Euclidean path integral.} If this requirement holds then $\rho_0^{\frac{1-n}{2n}} \rho_2 \rho_0^{\frac{1-n}{2n}}$ can be interpreted as a part of the complex plane between angles~$\pm \frac{\pi}{n}$ with operator insertions at angles $\pm\frac{\pi l}{L}$. This procedure is pictorially presented in figure \ref{fig:SRD}. Finally, taking the cyclic trace of $n$ copies of it means gluing $n$ of these regions onto each other which results in a $2n$-point function on the complex plane
\begin{align}
   \mathcal{F}^{(n)}_{\ket{s}} := \Tr\left(\rho_0^{\frac{1-n}{2n}} \rho_2 \rho_0^{\frac{1-n}{2n}}\right)^n = \left\langle \prod\limits_{k=0}^{n-1} f_{\Gamma_{k,l} \ket{s}}\left(e^{\frac{i\pi l}{L} + \frac{2\pi i k}{n}}\right)f_{\Gamma_{k,-l} \ket{s}}\left(e^{-\frac{i\pi l}{L} + \frac{2\pi i k}{n}}\right) \right\rangle_\mathbb{C} \label{eq:SRDcorr}
\end{align}
where, in contrast to the previous and following section, $\Gamma_{k,l}$ is the local action of the above M\"obius transformation $w(z)$ followed by a rotation $e^{\frac{2\pi i k}{n}}$ to obtain the correct gluing. As before, for the dual field one has to consider $w(1/z)$ which is done by replacing $l\to-l$\,. 

Let us highlight the connection between rather formal definitions of distinguishability measures and physical features of a theory. The latter is the Quantum Null Energy Condition (QNEC) which follows from the so-called Quantum Focusing Conjecture \cite{Bousso:2015mna}. The QNEC gives a lower bound on the stress-energy tensor in a relativistic quantum field theory that depends on the second variation of entanglement of a subregion. The QNEC can also be formulated solely in terms of quantum information theoretical quantities and has been shown to be equivalent to positivity of the second variation of relative entropies \cite{Leichenauer:2018obf}. After the QNEC has been proven in free and holographic theories \cite{Bousso:2015wca,Koeller:2015qmn,Malik:2019dpg}, it has since been shown to hold quite generally in the context of Tomita-Takesaki modular theory \cite{Balakrishnan:2017bjg,Ceyhan:2018zfg}. A generalized version of QNEC has been suggested in \cite{Lashkari:2018nsl} and proven to be true in free theories in dimensions larger than two \cite{Moosa:2020jwt}. This generalization may be called `R\'enyi QNEC' and is formulated as the positivity of the second variation of sandwiched R\'enyi entropies. The diagonal part of the second variation is simply given by the second derivative of the SRD with respect to the subsystem size. Hence, the R\'enyi QNEC can only be true in a theory if any SRD is a convex function of the subsystem size. We will explicitly check if this is true in our results in secs.~\ref{sec:universal} and~\ref{sec:nonuniversal}. 

\subsubsection{Trace square distance}

The trace square distance (TSD) between two reduced density matrices is given by 
\begin{equation}
    T^{(2)}(\rho_{1},\rho_{2}) := \frac{\Tr|\rho_{1} -\rho_{2}|^2}{\Tr \rho_{0}^2} = \frac{\Tr \rho_1^2 + \Tr \rho_2^2 -2\Tr\rho_1\rho_2}{\Tr \rho_{0}^2}\,,
\end{equation}
where the factor $\Tr \rho_{0}^2$ in particular removes any UV divergences and allows to directly express the TSD in terms of four-point functions on the two-sheeted surface $\Sigma_2$ (see also \cite{Sarosi:2016oks}),
\begin{align}
     T^{(2)}(\rho_{1},\rho_{2}) \equiv& \left\langle f_{\bra{1}}(0_1)f_{\ket{1}}(0_1)f_{\bra{1}}(0_2)f_{\ket{1}}(0_2)\right\rangle_{\Sigma_2}  + \left\langle f_{\bra{2}}(0_1)f_{\ket{2}}(0_1)f_{\bra{2}}(0_2)f_{\ket{2}}(0_2)\right\rangle_{\Sigma_2} \nonumber\\
     & - 2  \left\langle f_{\bra{1}}(0_1)f_{\ket{1}}(0_1)f_{\bra{2}}(0_2)f_{\ket{2}}(0_2)\right\rangle_{\Sigma_2}\,.
\end{align}
Using the uniformization map \eqref{eq:uniformization} with $n=2$ we can express it in terms of four-point functions on the complex plane,
\begin{align} \label{eq:TSDcorr}
    T^{(2)}(\rho_{1},\rho_{2}) \equiv& \left\langle f_{\Gamma_{1,-l}\ket{1}}\left(e^{-\frac{i\pi l}{2L}}\right)f_{\Gamma_{1,l}\ket{1}}\left(e^{\frac{i\pi l}{2L}}\right)f_{\Gamma_{2,-l}\ket{1}}\left(-e^{-\frac{i\pi l}{2L}}\right)f_{\Gamma_{2,l}\ket{1}}\left(-e^{\frac{i\pi l}{2L}}\right)\right\rangle_{\mathbb{C}} \nonumber\\  
     & + \left\langle f_{\Gamma_{1,-l}\ket{2}}\left(e^{-\frac{i\pi l}{2L}}\right)f_{\Gamma_{1,l}\ket{2}}\left(e^{\frac{i\pi l}{2L}}\right)f_{\Gamma_{2,-l}\ket{2}}\left(-e^{-\frac{i\pi l}{2L}}\right)f_{\Gamma_{2,l}\ket{2}}\left(-e^{\frac{i\pi l}{2L}}\right)\right\rangle_{\mathbb{C}} \nonumber\\
     & - 2  \left\langle f_{\Gamma_{1,-l}\ket{1}}\left(e^{-\frac{i\pi l}{2L}}\right)f_{\Gamma_{1,l}\ket{1}}\left(e^{\frac{i\pi l}{2L}}\right)f_{\Gamma_{2,-l}\ket{2}}\left(-e^{-\frac{i\pi l}{2L}}\right)f_{\Gamma_{2,l}\ket{2}}\left(-e^{\frac{i\pi l}{2L}}\right)\right\rangle_{\mathbb{C}}.
\end{align}
The trace square distance is manifestly positive and has the great advantage that we can compute it directly in terms of four-point correlators, i.e.~there is no need to consider higher-sheeted replica manifolds and we do not need to take any analytic continuations. 
Different trace distances between (not necessarily descendant) states in two-dimensional CFTs have been considered in e.g.~\cite{Sarosi:2016oks,Zhang:2019itb,Zhang:2019wqo}.

\section{Universal results from the vacuum representation} \label{sec:universal}

Most physically interesting conformal field theories contain a unique vacuum that naturally corresponds to the identity field. For the vacuum all the above correlation functions to compute the quantum measures become trivial. However, the theories also contain the whole vacuum representation which for example consists of the state $L_{-2}\ket{0}$ that corresponds to the holomorphic part of the stress tensor, $T(z)$. Correlation functions of  vacuum descendant fields generically depend on the central charge of the theory and can be computed explicitly using the Ward identities \eqref{eq:WardIdNpt} or \eqref{eq:rec1} recursively. 
Since all quantities discussed in section \ref{sec:qmeasures} can be expressed in terms of correlators, we can in principle compute all of them as closed form expressions, too. 
However, since we use computer algebra to perform the transformations and compute the correlation functions, computer resources are the biggest limiting factor. 
Here we mostly present results for all descendants up to conformal weight five, and in some cases up to descendants at level 10. In particular, we want to check how the measures depend on the conformal weights of the states and if states at the same conformal weight can be regarded as similar. 

\subsection{R\'enyi entanglement entropy}\label{sec:renyivac}

Only for the first few excited states in the identity tower, the expressions \eqref{eq:RFE} to compute the second R\'enyi entanglement entropy are compact enough to display them explicitly. In case of the first descendant $L_{-2}\ket{0}$, i.e.~the state that corresponds to the stress tensor, we get
\begin{align}\label{eq:RFE[-2]}
    F^{(2)}_{L_{-2}\ket{0}} &= \frac{c^2 \sin ^8(\pi x)}{1024}+\frac{c \sin ^4(\pi x) (\cos (2 \pi x)+7)^2}{1024}+\frac{\sin ^4(\pi x) (\cos (2 \pi x)+7)}{16
   c}\nonumber\\
   &\phantom{=} +\frac{16200 \cos (2 \pi x)-228 \cos (4 \pi x)+120 \cos (6 \pi x)+\cos (8 \pi x)+16675}{32768}\,,
\end{align}
where we defined $x=l/L$\,. The results for the states $L_{-n} \ket{0}$ with $n=3,4,5$ are given in \ref{app:REresultsVac}. The results here agree with those in \cite{Taddia:2016dbm} when present.

One important case is the limit of small subsystem size, i.e.~when $x\ll 1$. In this limit to leading order any of the above $2n$-point functions \eqref{eq:RFE} decouple into $n$ $2$-point functions. This is because the operator product of a field and its conjugate includes the identity. Then, in the limit $x\to 0$ the respective identity block dominates and takes the form of a product of $n$ 2-point functions. Those two point functions are, however, given by the transition amplitude from the state to its dual on the $k$th sheet that decouples in the limit $x\to 0$ from all other sheets. The latter is simply given by the squared norm of the state, i.e.~it gives one for normalized states. Hence, we can write
\begin{align}
    \lim_{x\to 0} F_{\ket{s}}^{(n)} &= \prod\limits_{k=1}^n \lim_{x\to 0} \langle f_{\Gamma_{k,l} \ket{s}}\left(w(0_k)\right)f_{\Gamma_{k,-l} \ket{s}}\left(w(\infty_k)\right) \rangle_\mathbb{C}\\ 
    &=\prod\limits_{k=1}^n \bra{s}s\rangle = 1\,.
\end{align}

\noindent
Hence, to order $x^0$ the descendant does not play any role at all. For the next-to-leading-order result there are expectations from primary excitations and the change of the entanglement entropy computed from holography. E.g.~in \cite{Bhattacharya:2012mi} it is shown that the change should be proportional to the excitation energy and, in particular, should be independent from $c$. Expanding the results \eqref{eq:RFE[-2]}, \eqref{eq:RFE[-3]}-\eqref{eq:RFE[-5]} we obtain 
\begin{equation}
    F_{L_{-n}\ket{0}}^{(2)} = 1 - \frac{n}{2}\left(\pi x\right)^2 + \mathcal{O}\!\left(x^4\right)\,, \quad\quad \text{for} ~ n = 2,3,4,5\,,\label{eq:RElowx}
\end{equation}
which is in agreement with all above expectations. 

\begin{figure}[t]
    \centering
    \begin{tikzpicture}
    \node at (0,0) {\includegraphics[width=.45\textwidth]{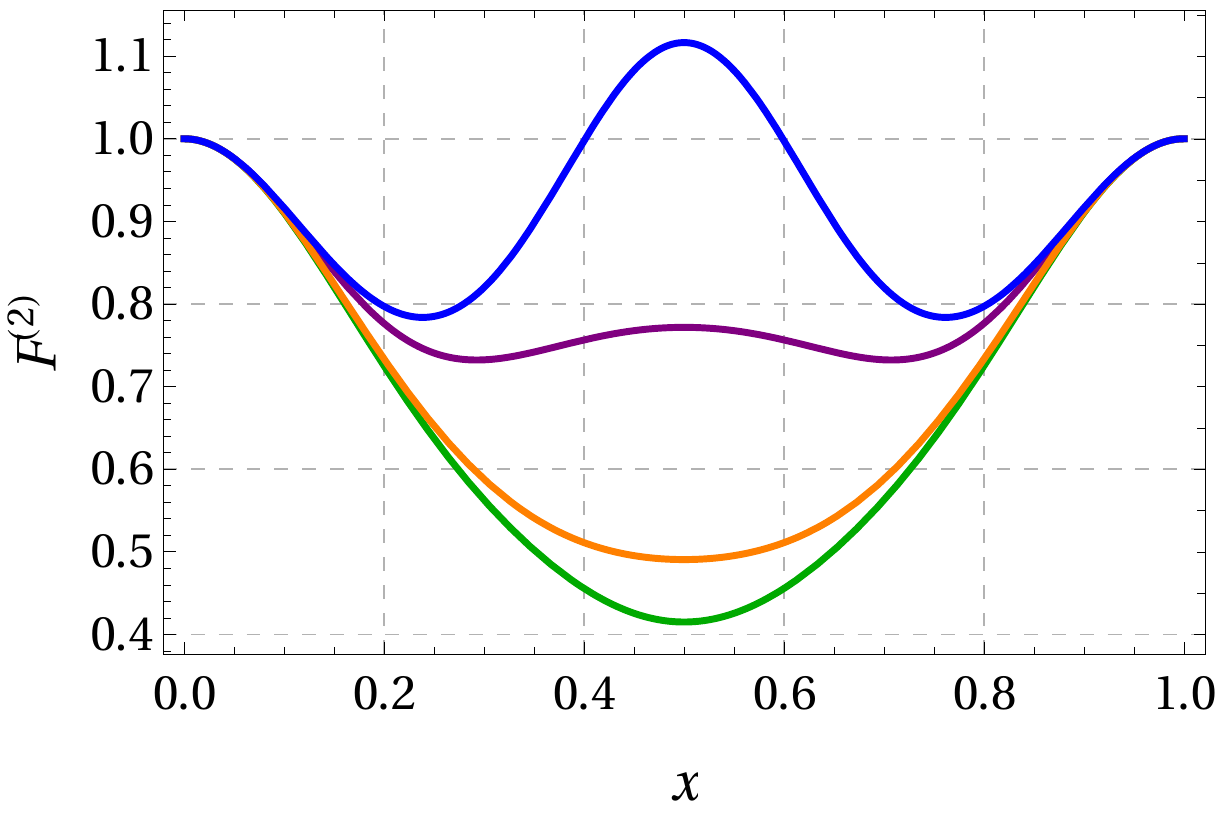}};
    \node at (.47\textwidth,0)  {\includegraphics[width=.45\textwidth]{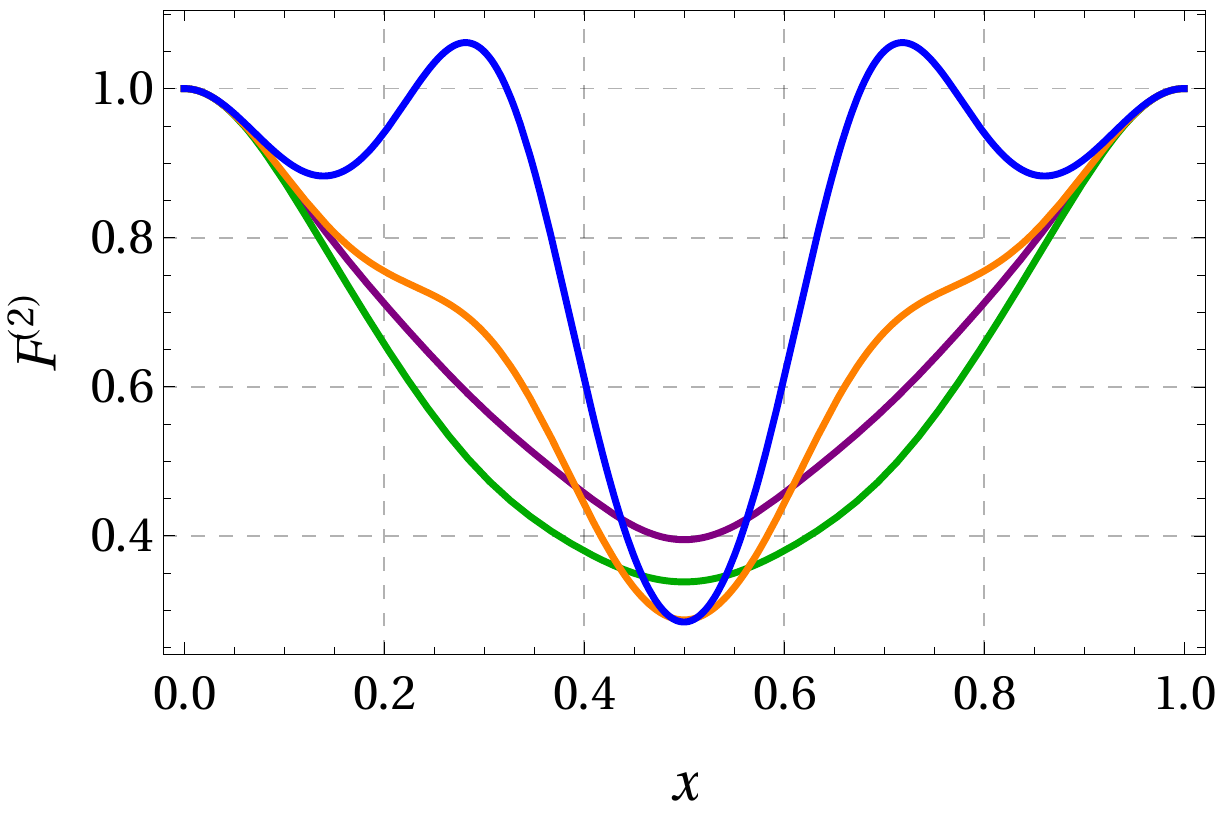}};
    \node at (0,-4.7) {\includegraphics[width=.45\textwidth]{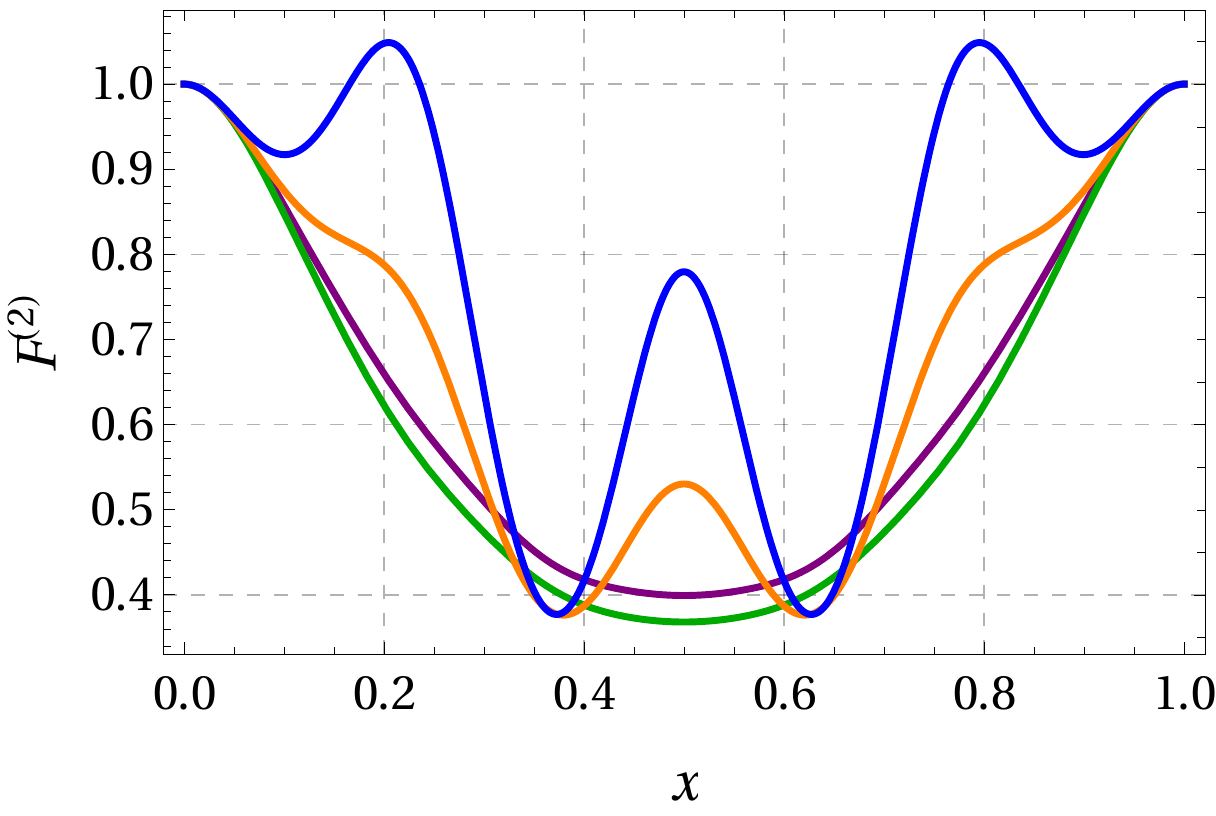}};
    \node at (.47\textwidth,-4.7)  {\includegraphics[width=.45\textwidth]{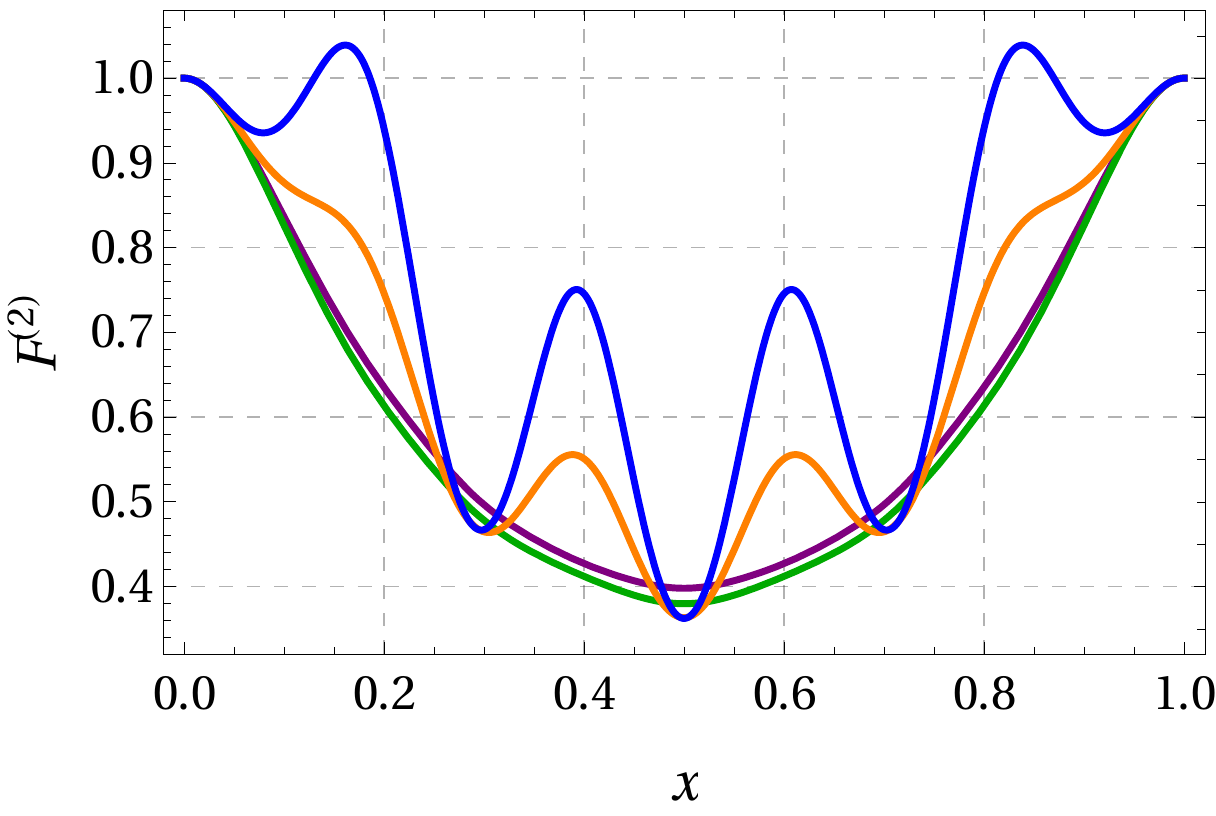}};
    \node at (.7\textwidth,-1.5) {\includegraphics[width=.1\textwidth]{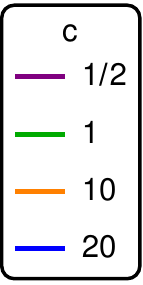}};
    \node at (.4,1.5) {(a)};
    \node at (7.5,1.5) {(b)};
    \node at (.4,-3.2) {(c)};
    \node at (7.5,-3.2) {(d)};
    \end{tikzpicture}
    \caption{The correlator $F^{(2)}_{\ket{s}}$ for (a) $\ket{s} = L_{-2}\ket{0}$, (b) $\ket{s} = L_{-3}\ket{0}$, (c) $\ket{s} = L_{-4}\ket{0}$, (d) $\ket{s} = L_{-5}\ket{0}$, for several values of the central charge.}
    \label{fig:RE2lowA}
\end{figure}

In figure \ref{fig:RE2lowA} we show the results for $F^{(2)}_{\ket{s}}$ for the states $\ket{s} = L_{-n}\ket{0}$, $n=2,3,4,5$\,. The first observation is that at large $c$ the correlator shows an oscillating behaviour with oscillation period proportional to $1/n$. In fact, we can see this also from the explicit results \eqref{eq:RFE[-2]}, \eqref{eq:RFE[-3]}-\eqref{eq:RFE[-5]} where at large central charge the term proportional to $c^2$ dominates. 
Note that the correlator $F^{(n)}$ can become larger than one at large central charge and, hence, its contribution to the R\'enyi entropy $S^{(n)}$ can get negative. For example, in case of $n=2$ and $\ket{s} = L_{-2}\ket{0}$ this happens at $x=1/2$ for $c\gtrsim 18.3745$.  

\begin{figure}[t]
    \centering
    \begin{tikzpicture}
    \node at (0,0) {\includegraphics[width=.45\textwidth]{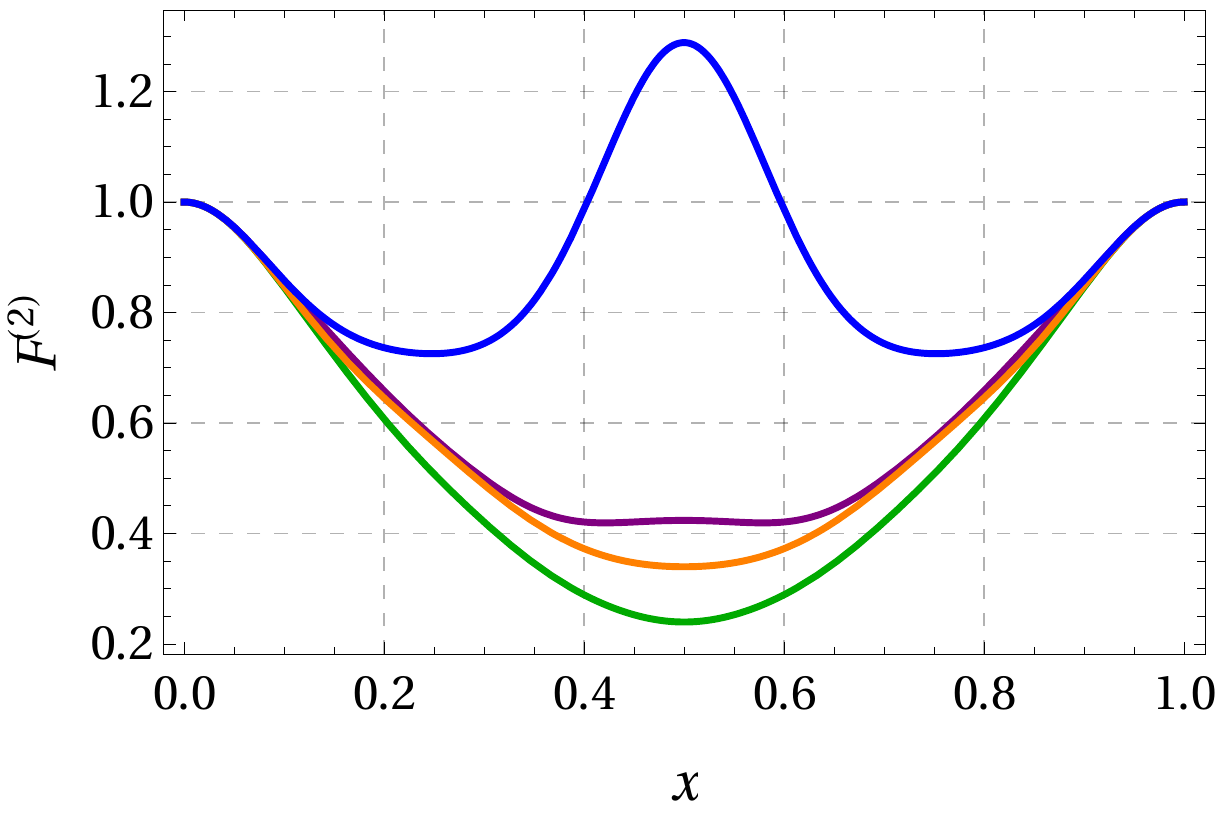}};
    \node at (.47\textwidth,0)  {\includegraphics[width=.45\textwidth]{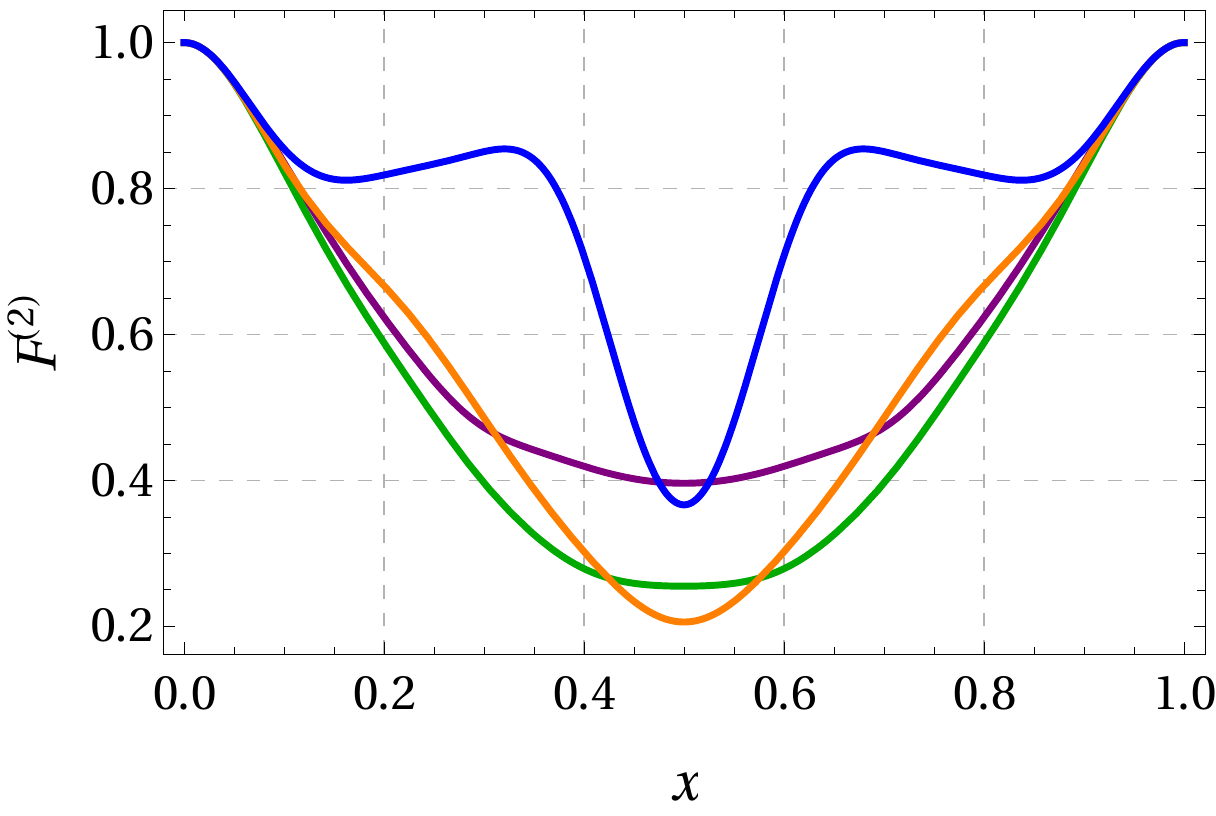}};
    \node at (0,-4.7) {\includegraphics[width=.45\textwidth]{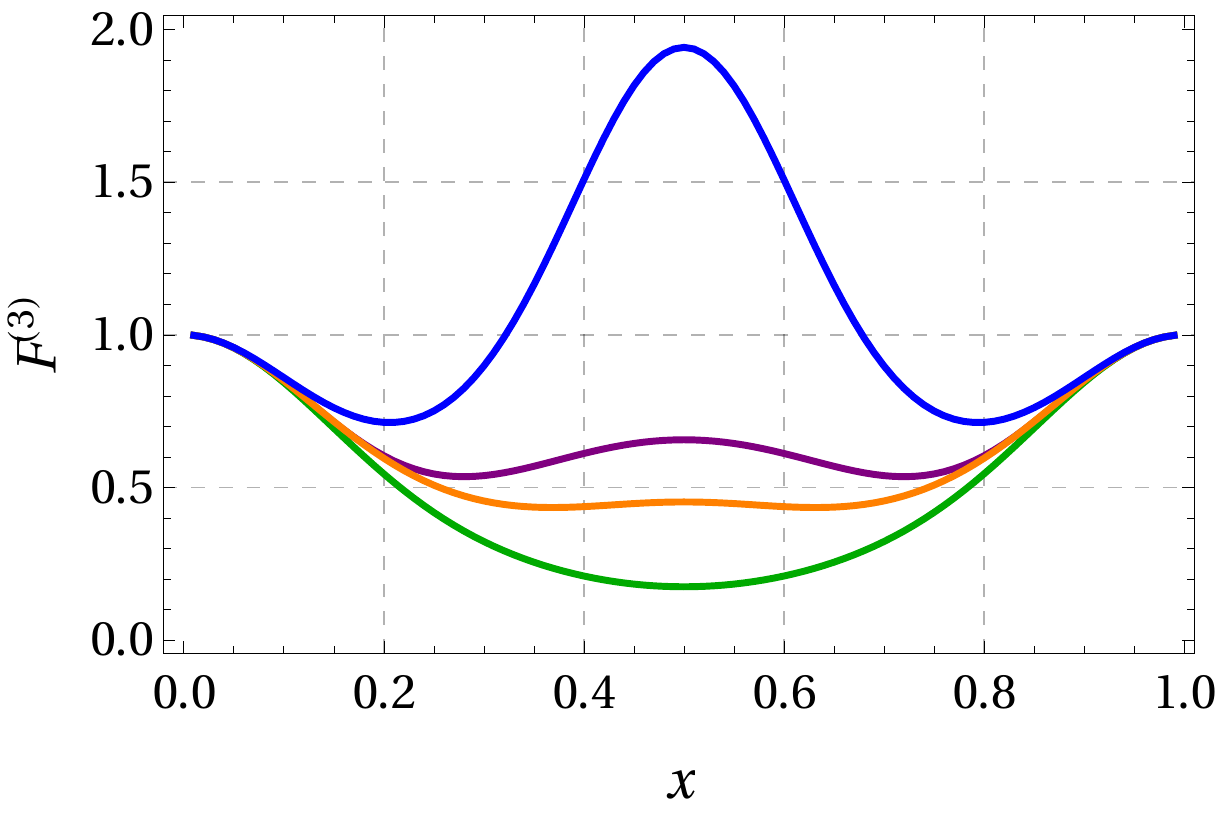}};
    \node at (.47\textwidth,-4.7)  {\includegraphics[width=.45\textwidth]{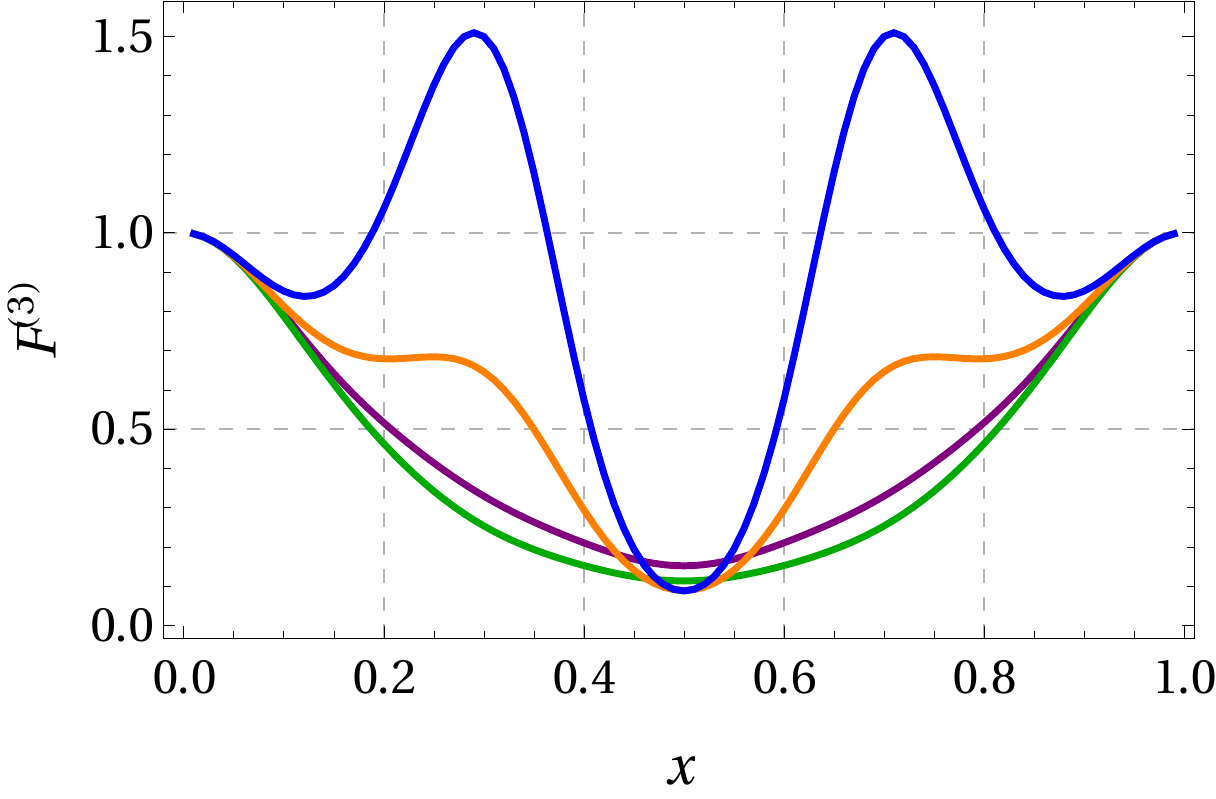}};
    \node at (.25\textwidth,-2.1
    ) {\includegraphics[width=.25\textwidth]{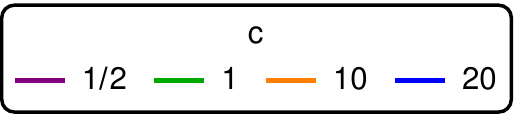}};
    \node at (.4,1.5) {(a)};
    \node at (7.5,1.5) {(b)};
    \node at (.4,-3.2) {(c)};
    \node at (7.5,-3.2) {(d)};
    \end{tikzpicture}  
    \caption{The correlator $F^{(n)}_{\ket{s}}$ for (a) $n=2$, $\ket{s} = L_{-2}^2\ket{0}$ , (b) $n=2$, $\ket{s} = L_{-3}L_{-2}\ket{0}$, (c) $n=3$ $\ket{s} = L_{-2}\ket{0}$, and (d)  $n=3$ $\ket{s} = L_{-3}\ket{0}$ for several values of the central charge.}
    \label{fig:RE2lowB}
\end{figure}

The vacuum module is degenerate at conformal weight $h=4$ and $h=5$. In addition to the states $L_{-4}\ket{0}$ and $L_{-5}\ket{0}$ there are the states $L_{-2}^2\ket{0}$ and $L_{-3}L_{-2}\ket{0}$, respectively. Their correlators $F^{(2)}_{\ket{s}}$ are shown in figure \ref{fig:RE2lowB} (a) and (b) for different values of the central charge. Interestingly, although their small subsystem behaviour is given by \eqref{eq:RElowx} and, hence, it is the same as for $L_{-4}\ket{0}$ and $L_{-5}\ket{0}$, respectively, their general behaviour is rather different at large central charge. Their oscillation period is not proportional to the conformal weight but proportional to the level of the lowest Virasoro generator appearing in it.

Already these two examples show that the behaviour of the R\'enyi entropy and, hence, also of the entanglement entropy of descendant states does not only depend on their conformal weight, i.e.~the energy of the state, but also significantly on their building structure. 
This is particularly true at large central charge, limit in which a CFT is conjectured to be dual to a (semi-)classical gravity theory. It is widely believed that black hole microstates in $\text{AdS}_3$ correspond to typical high conformal dimension states in the CFT. However, a typical state at conformal dimension $\Delta\gg 1$ is a descendant at level $\Delta/c$ of a primary with conformal dimension $\tfrac{c-1}{c}\Delta$ (see e.g.~\cite{Datta:2019jeo}). This means that a typical state will be a descendant at large but finite central charge $c$. The results we present here show that descendants with the same conformal dimension can in fact show very different behaviour when it comes to the entanglement structure.
It will be interesting to further study the large $c$ limit, in particular for non-vacuum descendants, to analyse the holographic effect of these different behaviours.

Finally, in figure \ref{fig:RE2lowB} (c) and (d) we show the correlator $F^{(3)}$ for the first two excited states $L_{-2}\ket{0}$ and $L_{-3}\ket{0}$. They show qualitatively the same behaviour as the respective correlators for $n=2$ (see figure \ref{fig:RE2lowA} (a) and (b)). However, their dependence on the central charge is stronger and the oscillating behaviour starts at lower $c$. For example, $F^{(3)}_{L_{-2}\ket{0}}$ is larger than one at $x=1/2$ for $c\gtrsim14.74945$. 

The stronger dependence on the central charge for larger $n$ is expected. Indeed, we can expand $F^{(n)}_{\ket{s}}$ as 
\begin{equation}
    F^{(n)}_{\ket{s}} = \sum_{k=-n+1}^{n} A_k^{(n)} c^k\,,
\end{equation}
where all the dependence on the state $\ket{s}$ and the  relative subsystem size $x=l/L$ sits in the coefficients $A_k^{(n)}$ . The negative powers of $c$ originate from the normalization of the state, while positive powers of $c$ follow from the Virasoro commutation relations when using the Ward identities. Therefore, at large central charge we get 
\begin{equation}
     \left.F^{(n)}_{\ket{s}}\right|_{c\gg 1} \simeq A_n c^n\,.
\end{equation}

\subsection{Sandwiched R\'enyi divergence}

As argued in section \ref{sec:SRD} it is possible to express the SRD~\eqref{eq:SRD} for integer parameters $n$ in terms of a $2 n$-point function $\mathcal{F}^{(n)}$ \eqref{eq:SRDcorr} if $\rho_1$ is the reduced density matrix of the vacuum.
When $\rho_2$ is associated to the state $L_{-2}\ket{0}$ we obtain
\begin{align}
    \mathcal{F}^{(2)}_{L_{-2}\ket{0}} &= \frac{(\cos (4 \pi x)+7)\sec ^8(\pi x) }{16384 c} [ -512 \cos (4 \pi x)+128 \cos (8 \pi x)+384 \nonumber\\
    &\phantom{=}+ c (847 \cos (4 \pi x)-22
   \cos (8 \pi x)+\cos (12 \pi x)+1222) ] \, ,
\end{align}
where $x = l/L < 1/2$\,. More expressions for different descendants can be found in appendix \ref{app:SRDresultsvac}. 

Again we first want to draw attention to the small subsystem limit. The results for the second SRD between the reduced vacuum state and all states up to conformal weight five show the small subsystem behaviour
\begin{equation}
    \mathcal{S}^{(2)}_{\ket{s}} = \frac{2 h_s^2}{c} \pi^4 x^4 +  \frac{2 h_s^2}{3c} \pi^6 x^6 + \mathcal{O}(x^8)\,. \label{eq:SRDsmallx}
\end{equation}

\noindent
The small subsystem behaviour only depends on the central charge and the conformal weight of the respective state and is independent of the specific structure of the state.

In case of $n=2$, the SRD diverges at $x=1/2$. In the limit $x\to 1/2$ find the behaviour 
\begin{equation}\label{eq:srd_vac_divergence}
 \mathcal{F}^{(2)}_{\ket{s}} = \exp\left(\mathcal{S}^{(2)}_{\ket{s}}\right) = \frac{A_{\ket{s}}}{\pi^{4 h_s}\left(x-\frac12\right)^{4h_s}} + \ldots \, ,
\end{equation}
where the coefficient $A_{\ket{s}}$ depends on the specifics of the state. For states of the form $L_{-n}\ket{0}$ up to $n=10$ it takes the form 
\begin{equation}
    A_{L_{-n}\ket{0}} = \binom{2n-1}{n-2}^2\,.
\end{equation}

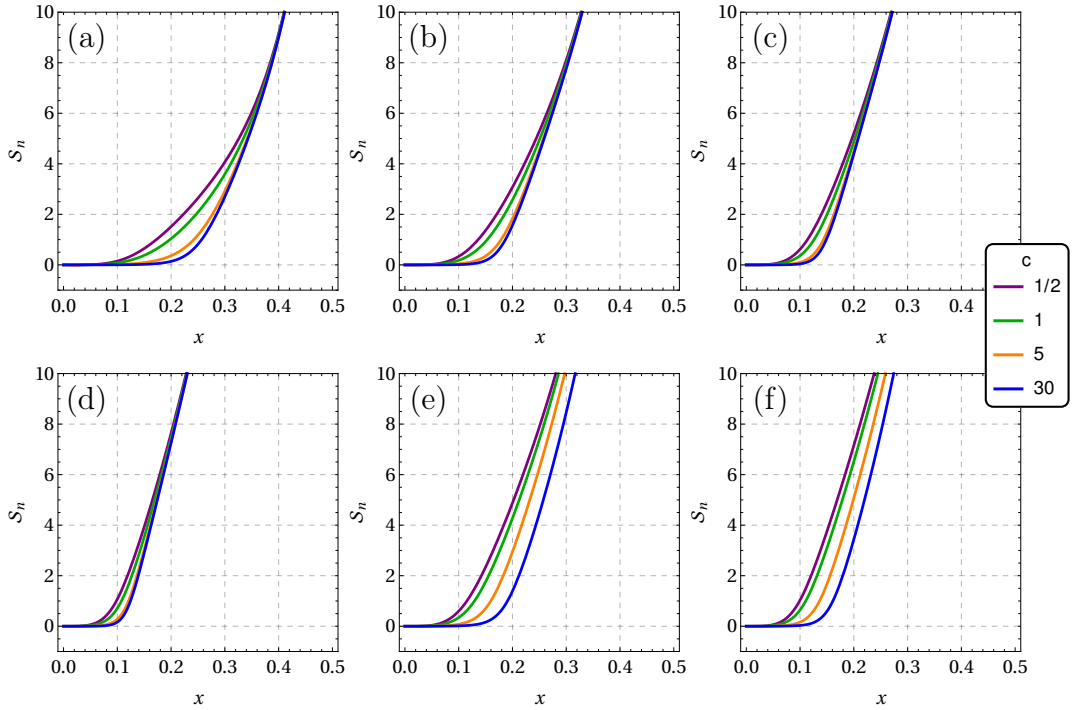
\begin{figure}[t]
    \centering
    \begin{tikzpicture}
    \node at (0,0) {\includegraphics[width=.3\textwidth]{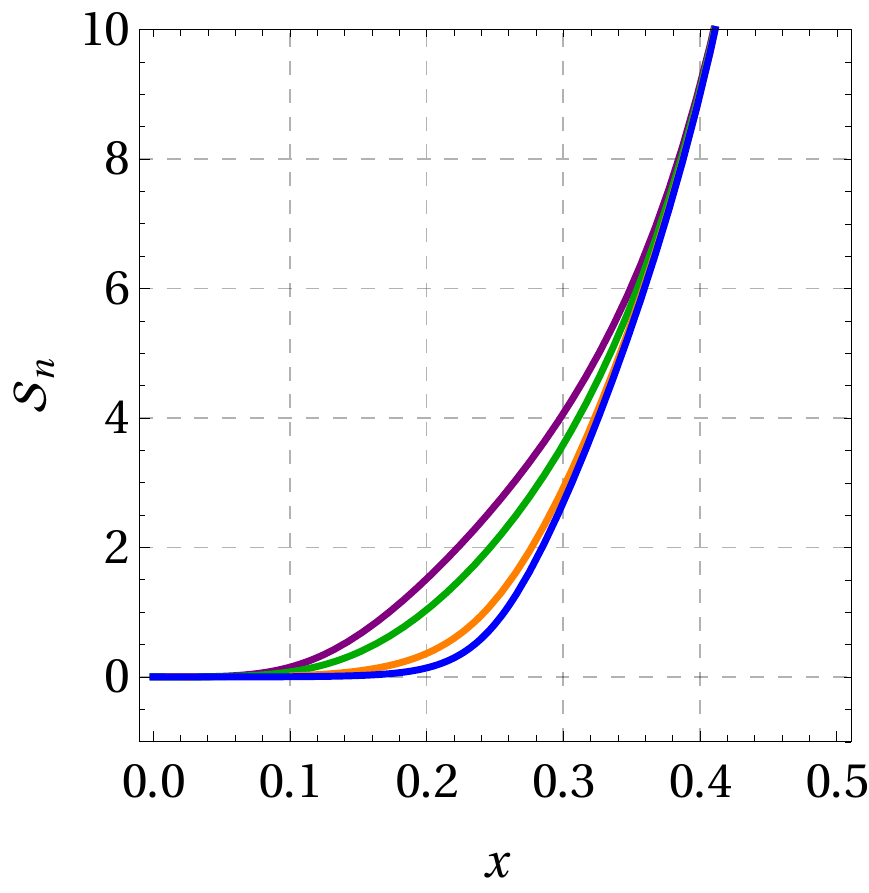}};
    \node at (.3\textwidth,0) {\includegraphics[width=.3\textwidth]{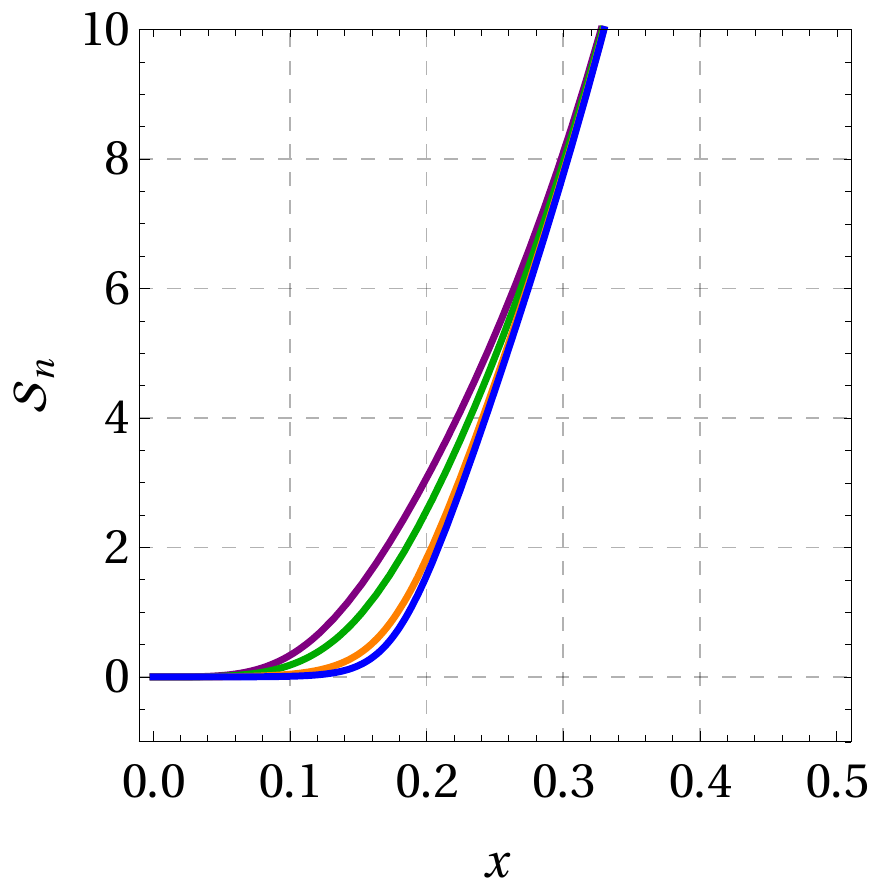}};
    \node at (.6\textwidth,0) {\includegraphics[width=.3\textwidth]{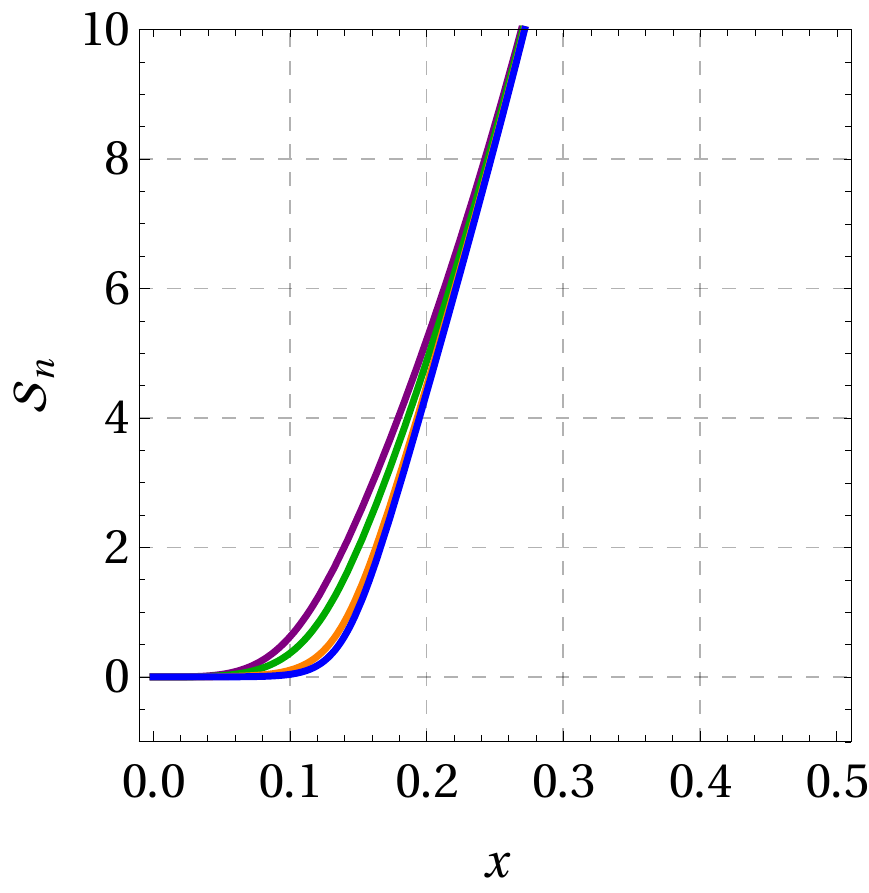}};
    \node at (0,-.32\textwidth) {\includegraphics[width=.3\textwidth]{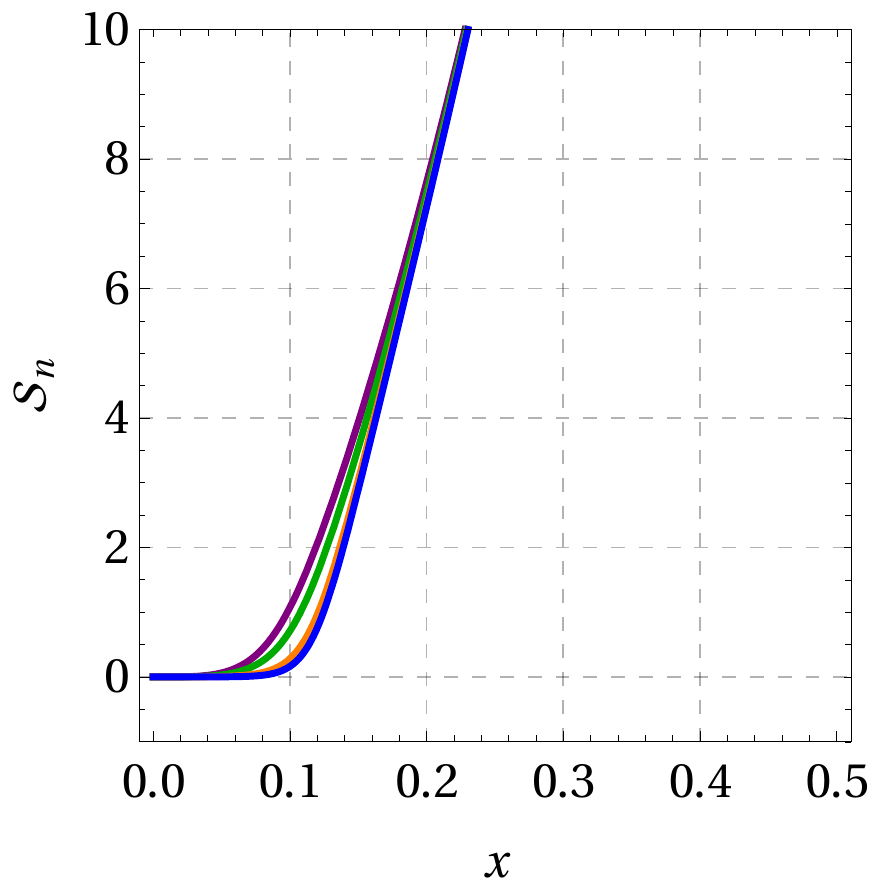}};
    \node at (.3\textwidth,-.32\textwidth) {\includegraphics[width=.3\textwidth]{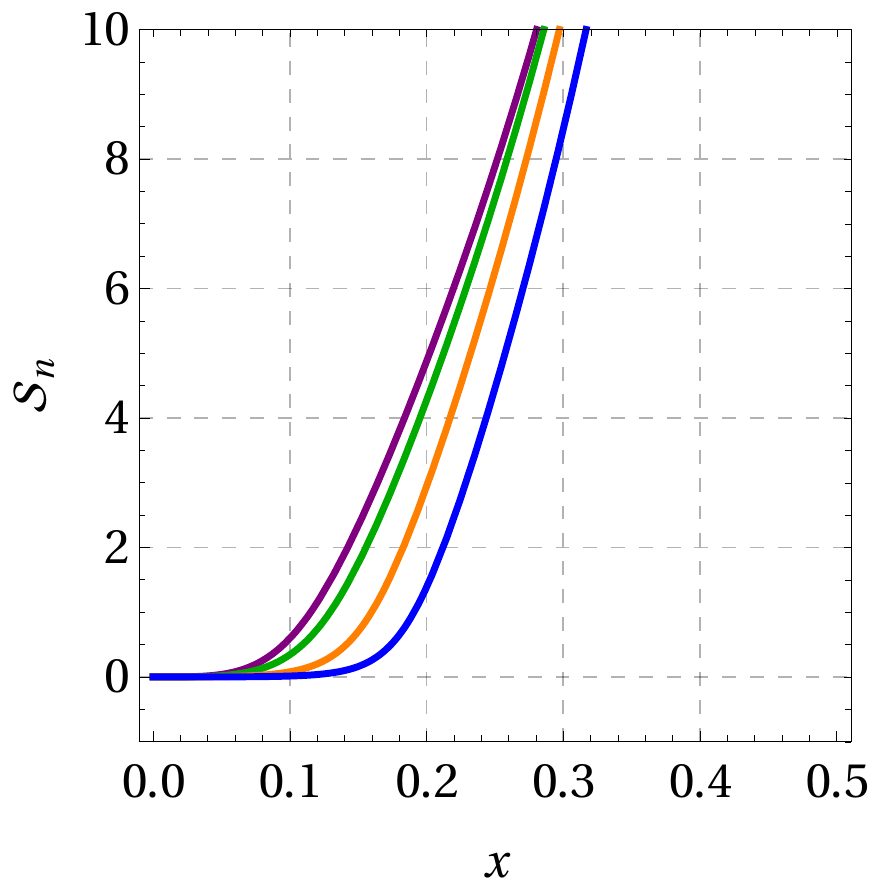}};
    \node at (.6\textwidth,-.32\textwidth) {\includegraphics[width=.3\textwidth]{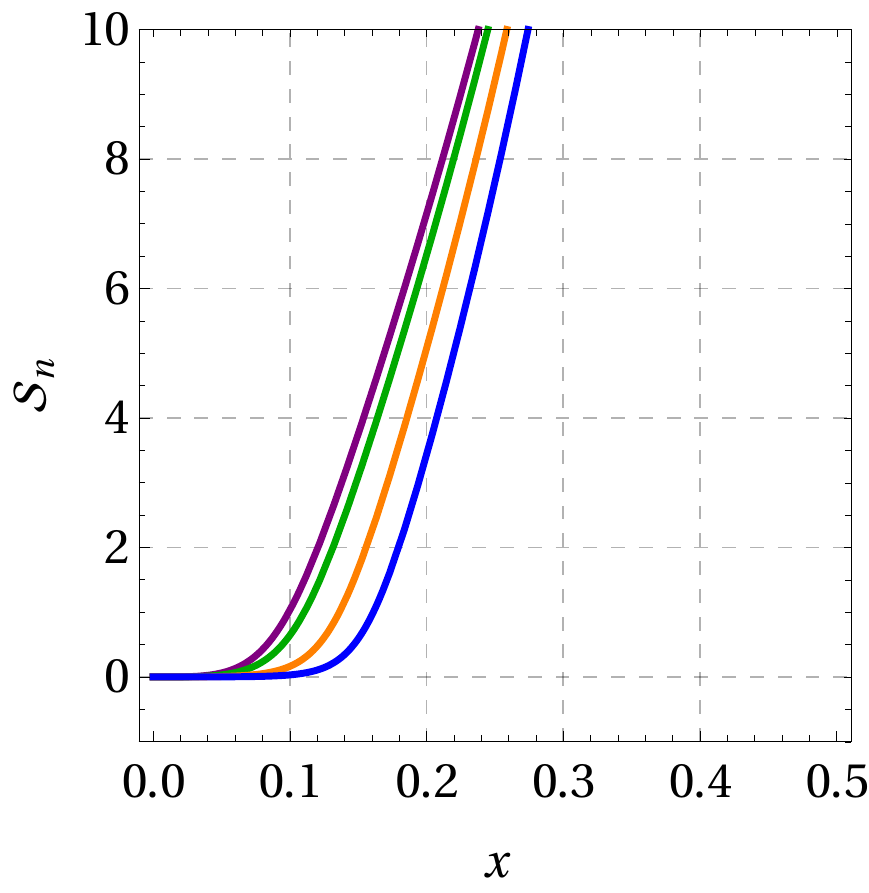}};
    \node at (.75\textwidth,-2) {\includegraphics[width=.075\textwidth]{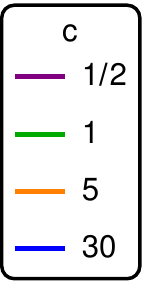}};
    \node at (-1.15,1.8) {(a)};
    \node at (3.35,1.8) {(b)};
    \node at (7.85,1.8) {(c)};
    \node at (-1.15,-3) {(d)};
    \node at (3.35,-3) {(e)};
    \node at (7.85,-3) {(f)};
    \end{tikzpicture}
    \caption{The SRD for $n=2$ between the reduced groundstate and (a) $L_{-2}\ket{0}$, (b) $L_{-3}\ket{0}$, (c) $L_{-4}\ket{0}$, (d) $L_{-5}\ket{0}$, (e) $L_{-2}^2\ket{0}$, (f) $L_{-3}L_{-2}\ket{0}$ for different values of the central charge $c$.}
    \label{fig:SRE2}
\end{figure}

In figure \ref{fig:SRE2} we show the SRD for the first six excited states.
All of them show a plateau at small values of $x$ that increases for larger $c$ and shrinks for higher energy. This is expected from the asymptotic result \eqref{eq:SRDsmallx}. 
It is interesting to compare the second SRD of the states $L_{-2}^2\ket{0}$ and $L_{-3}L_{-2}\ket{0}$ to that of the states $L_{-4}\ket{0}$ and $L_{-5}\ket{0}$ with the same conformal weight.
The plots show that, although in the asymptotic regimes $x\to 0$ and $x\to1/2$ the second SRD behaves similarly for degenerate states, it looks quite different for intermediate regimes of $x$.
In particular, the SRD of the states $L_{-2}^2\ket{0}$ and $L_{-3}L_{-2}\ket{0}$ is more sensible to the central charge.
This shows again that descendant states at the same conformal dimension can behave quite differently, especially at large central charge.

\begin{figure}[th!]
    \centering
    \includegraphics[width=.6\textwidth]{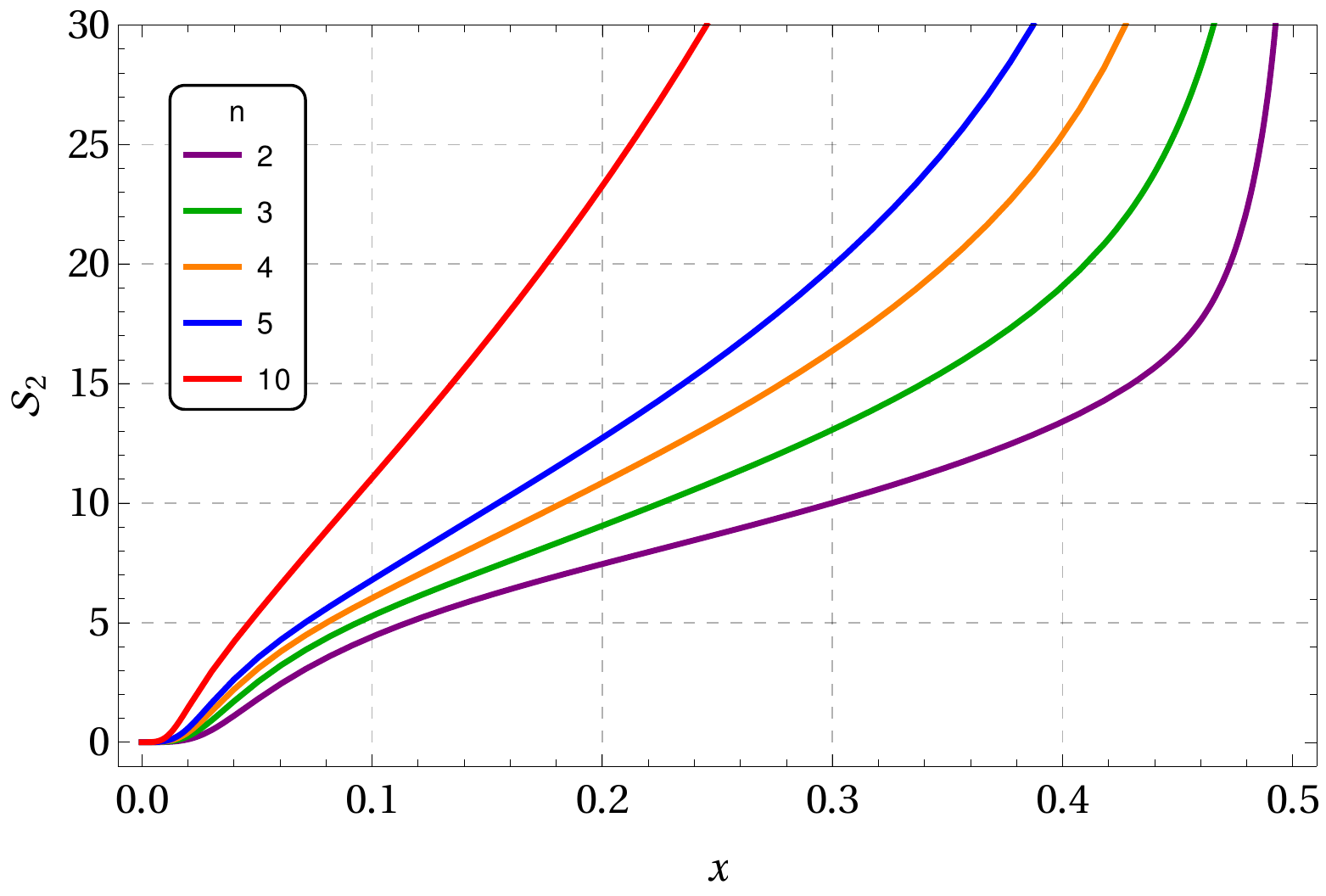}
    \caption{The second SRD between the reduced groundstate and the states $L_{-n}\ket{0}$, $n=2,3,4,5,10$ at $c=1/1000$, showing non-convex behaviour and thus violation of the R\'enyi QNEC conjecture.}
    \label{fig:nonconvex}
\end{figure}

In all plots so far the second SRD is a convex function of the relative subsystem size $x = l/L$. However, in cases of small central charge it is not, i.e.~there are regions of $x$ with $\partial^2 S^{(2)}/\partial x^2 <0$. For example, in case of $\ket{s} = L_{-2}\ket{0}$ the second SRD is not convex for $c\lesssim 0.1098$\,. This shows that there are examples where the generalized version of the QNEC is not true. However, CFTs with central charges smaller than 1/2 are quite unusual. They cannot be part of the ADE classifiation of rational, unitary, modular invariant CFTs \cite{Cappelli:1986hf} but could e.g.~be logarithmic \cite{Nivesvivat:2020gdj}. 
In figure \ref{fig:nonconvex} we show the second SRD for states $L_{-n}\ket{0}$ with $n=2,3,4,5,10$ and $c=1/1000$ to illustrate its non-convexity for all these states.

\subsection{Trace squared distance}\label{sec:vacSRD}

Again only the expressions for the first few excited states are compact enough to display them explicitly. For example, the TSD between the vacuum and the state $L_{-2}\ket{0}$ is given by
\begin{align}\label{eq:TSD21}
    T^{(2)}_{L_{-2}\ket{0},\ket{0}} &= \frac{c^2 \sin ^8(\pi x)}{1024}-\frac{1}{512} c \sin ^6(\pi x) (\cos (2 \pi x)+15)+\frac{\sin ^4(\pi x) (\cos (2 \pi x)+7)}{16 c}\nonumber\\
    &\phantom{=}+\frac{1}{32768} (-32768 \cos(\pi x)+8008 \cos (2 \pi x)-228 \cos (4 \pi x) \nonumber\\
    &\phantom{=} +120 \cos (6 \pi x)+\cos (8 \pi x)+24867)\,,
\end{align}
where we use the abbreviation $x = {l}/{L}$ again. 
Some other explicit expressions can be found in appendix \ref{app:TSDvacResults}. 

In the limit $x\to 0$ the reduced states have no support and, hence, must be trivial. Consequently, the trace square distance vanishes in this limit independently of the original states we choose. We checked the leading order in $x\ll1$ for all states up to conformal weight five and we find the behaviour
\begin{equation}\label{eq:tsd_vacuum_smallx}
    T^{(2)}_{s_1,s_2} =  \frac{2+c}{16 c}  (h_1-h_2)^2 \pi^4 x^4 + \mathcal{O}(x^6)\,.
\end{equation}

\noindent
We can see that to leading order, $x^4$, the TSD depends on the central charge and the difference in conformal weight of the two states. We also see that for large central charge the dependence on $c$ is negligible. 

In case of $h_1 -h_2 =0$ the TSD starts at order $x^8$ for small $x$. We e.g.~obtain
\begin{align}
    T^{(2)}_{L_{-2}^2\ket{0},L_{-4}\ket{0}} & = \frac{ (2 c+1)^2 \left(25 c^3+420 c^2+2444 c+4752\right) \pi ^8 x^8}{1600 c (c+8)^2} + \mathcal{O}(x^{10}) \,, \label{eq:TSDsmallxdeg1}\\
    T^{(2)}_{L_{-3}L_{-2}\ket{0},L_{-5}\ket{0}} & = \frac{9  c \left(25 c^3+420 c^2+2444 c+4752\right) \pi ^8 x^8}{1024 (c+6)^2}+ \mathcal{O}(x^{10})\,.\label{eq:TSDsmallxdeg2}
\end{align}

\noindent
Albeit one common factor, the latter expressions do not seem to show a straightforward dependence on the states. They also show that the large $c$ behaviour is more subtle because the $x^8$ coefficient diverges as $c\to\infty$\,.

In the opposite limit $x\to1$ the TSD can be computed easily because the states become pure. One obtains
\begin{align}
    \lim_{x\to1} T^{(2)}_{\ket{s_1},\ket{s_2}} &= \frac{\Tr(\ket{s_1}\bra{s_1}^2)+\Tr(\ket{s_2}\bra{s_2}^2)-2 \Tr(\ket{s_1}\bra{s_1}\ket{s_2}\bra{s_2})}{\Tr(\ket{0}\bra{0}^2)}\\
    &=2 \left(1- |\bra{s_1}s_2\rangle|^2\right) =: \mathcal{T}\,.
\end{align}

\noindent 
We can see that $0\le \lim_{x\to1} T^{(2)}(\rho_1,\rho_2)\le 2$ where we get the first equal sign if and only if $s_1=s_2$ and the second one if and only if the two states are orthogonal to each other. 

The explicit results up to conformal weight five show that the expansion around $x=1$ is given by
\begin{equation}
    T^{(2)}_{\ket{s_1},\ket{s_2}} = \mathcal{T} \left(1 - \frac{h_1+h_2}{4} \pi^2 (x-1)^2  + \mathcal{O}\!\left((x-1)^4\right) \right)\,.
\end{equation}

\noindent
We can see that the behaviour of the TSD close to $x=1$ depends on the sum of conformal weights $h_1 + h_2$\,. 
This is in contrast to the small $x$ behaviour that depends on the difference. 
Let us, for example, consider the second TSD between the vacuum and $L_{-2}\ket{0}$ in~\eqref{eq:TSD21} and the second TSD between the vacuum and $L_{-3}\ket{0}$ in~\eqref{eq:TSD31}. 
From the difference of conformal weight we get
\begin{equation}
    T^{(2)}_{L_{-2}\ket{0},\ket{0}}(x) < T^{(2)}_{L_{-3}\ket{0},\ket{0}}(x)
\end{equation}
for small $x$. 
However, from the sum of conformal weights we obtain
\begin{equation}
    T^{(2)}_{L_{-2}\ket{0},\ket{0}}(x) > T^{(2)}_{L_{-3}\ket{0},\ket{0}}(x)
\end{equation}
for $x$ close to one. 
We immediately can conclude that there must be an odd number of values $x\in (0,1)$, which in particular means at least one, with
\begin{equation}
    T^{(2)}_{L_{-2}\ket{0},\ket{0}}(x) = T^{(2)}_{L_{-3}\ket{0},\ket{0}}(x) \,.
\end{equation}

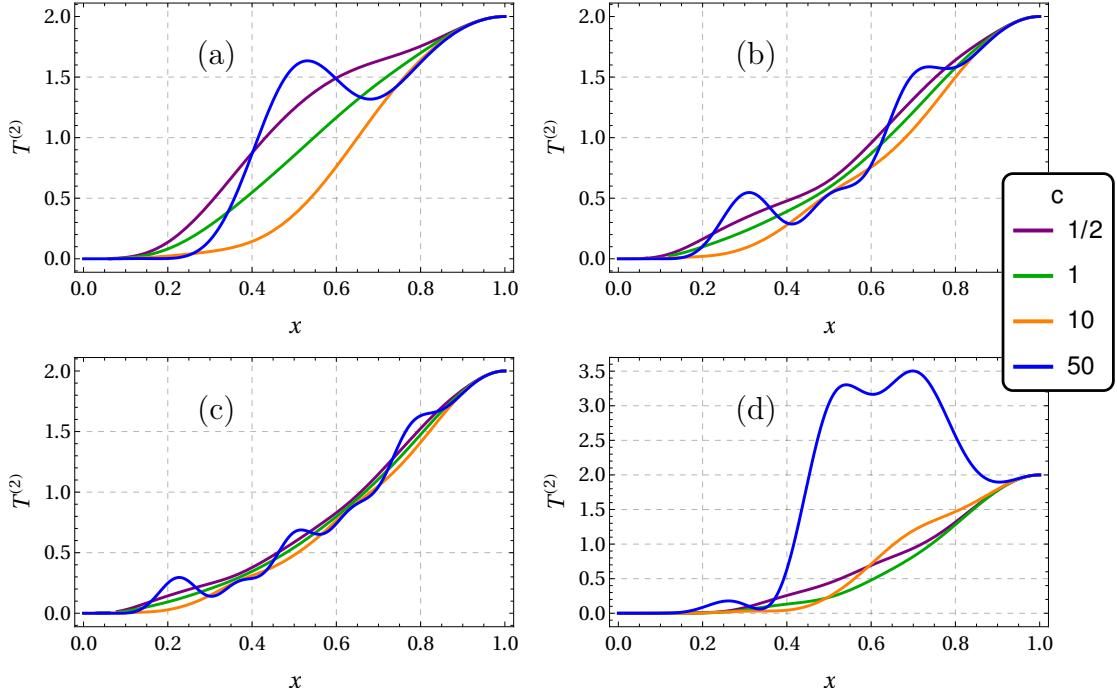
\begin{figure}[t]
    \centering
    \begin{tikzpicture}
    \node at (0,0) {\includegraphics[width=.45\textwidth]{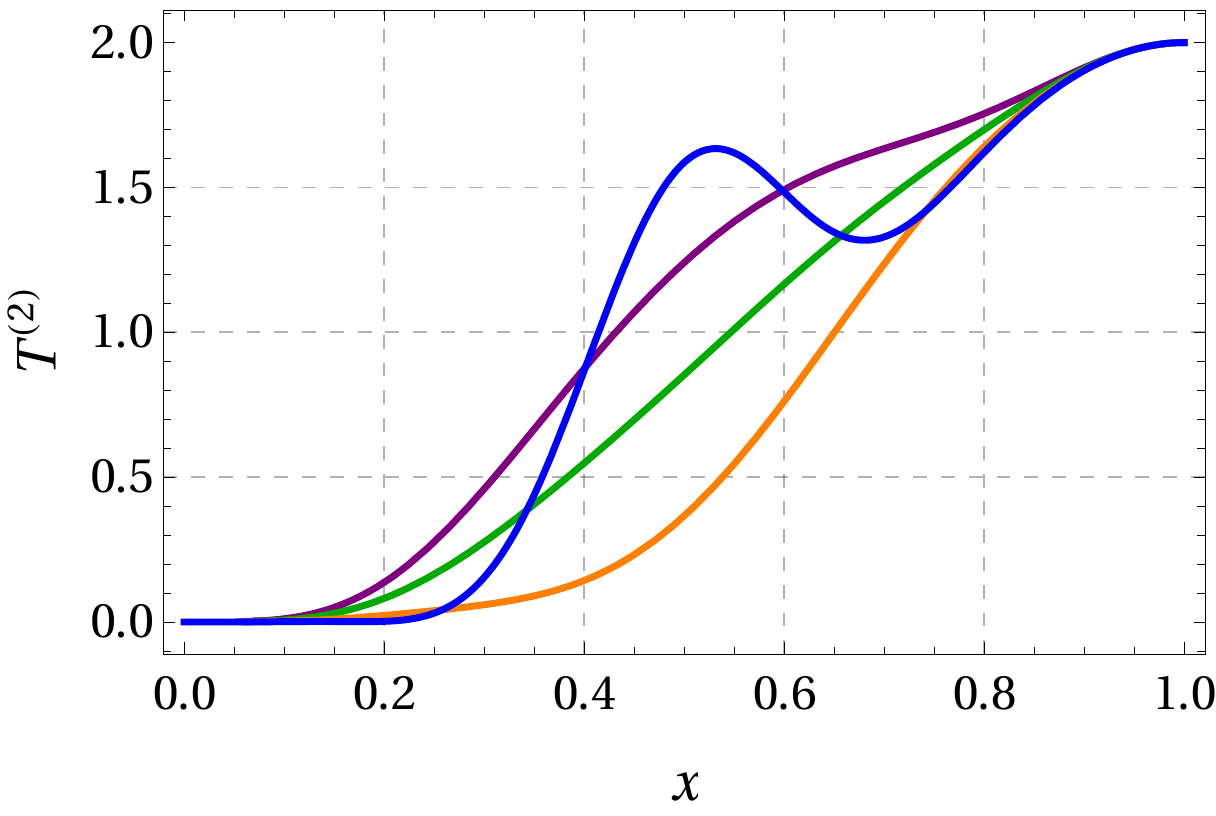}};
    \node at (.47\textwidth,0)  {\includegraphics[width=.45\textwidth]{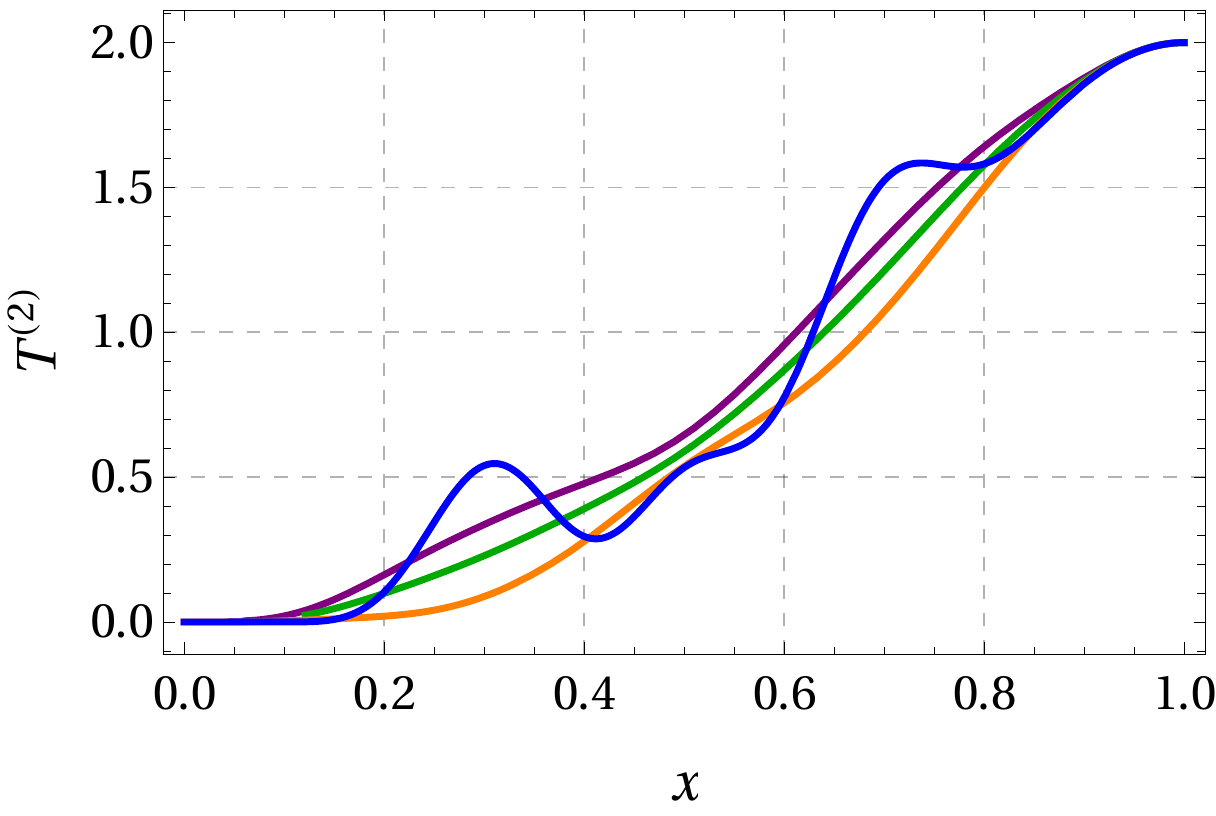}};
    \node at (0,-4.7) {\includegraphics[width=.45\textwidth]{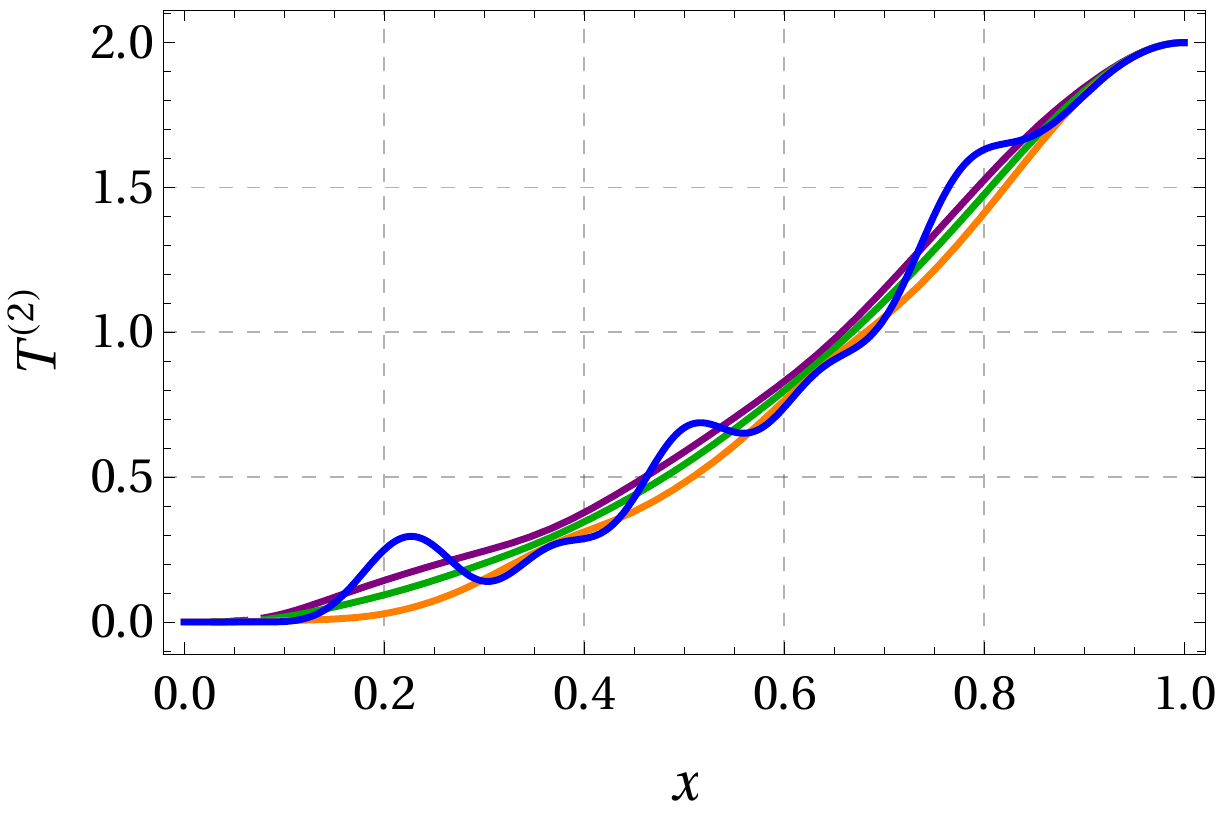}};
    \node at (.47\textwidth,-4.7)  {\includegraphics[width=.45\textwidth]{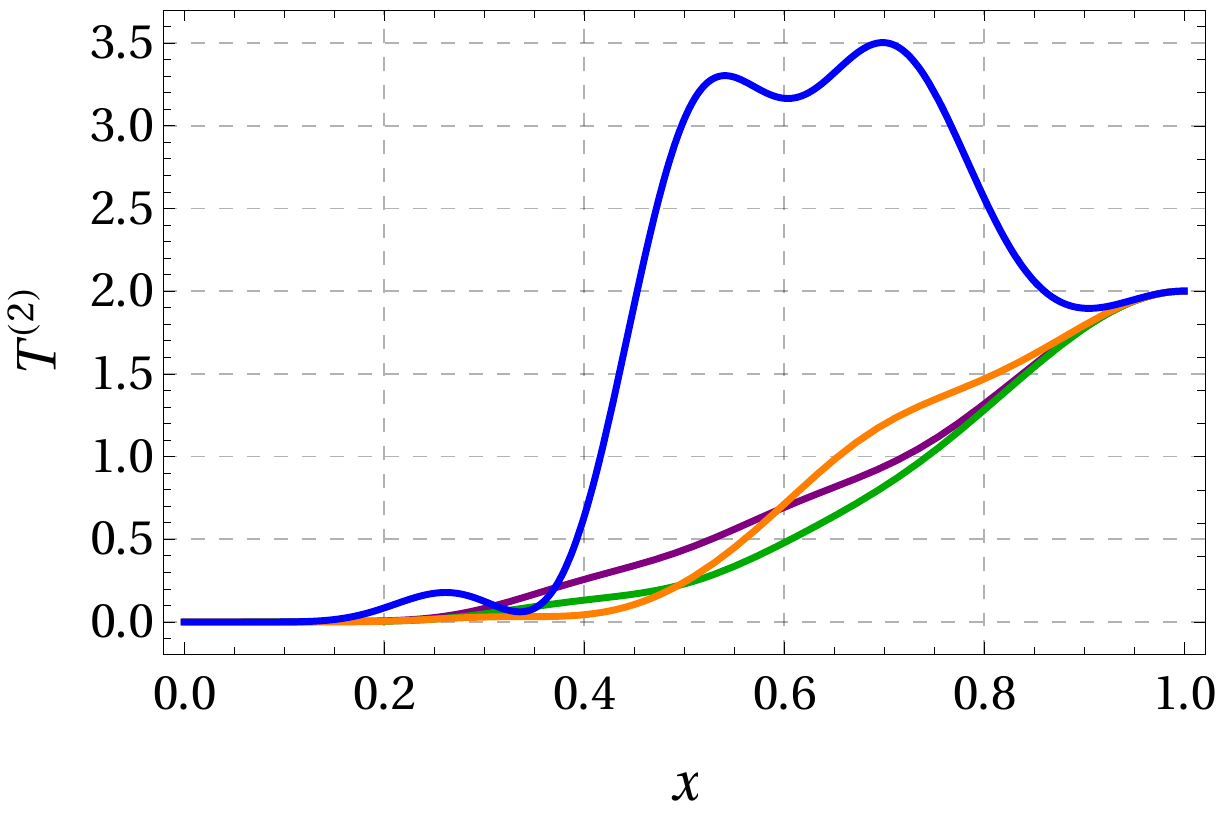}};
    \node at (.7\textwidth,-1.5) {\includegraphics[width=.1\textwidth]{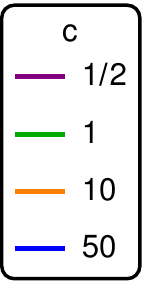}};
    \node at (-.6,1.5) {(a)};
    \node at (6.5,1.5) {(b)};
    \node at (-.6,-3.2) {(c)};
    \node at (6.5,-3.2) {(d)};
    \end{tikzpicture}
    \caption{The TSD between the reduced states of (a) the vacuum and $L_{-2}\ket{0}$, (b) the vacuum and $L_{-3}\ket{0}$, (c) the vacuum and $L_{-4}\ket{0}$, and (d) the states $L_{-3}\ket{0}$ and $L_{-2}\ket{0}$\, for different values of the central charge $c$.}
    \label{fig:TSDA}
\end{figure}

We also visualise some of the results. In figure \ref{fig:TSDA} we show the second TSD between the vacuum $\ket{0}$ and $L_{-n}\ket{0}$ for $n=2,3,4$, and between the first two excited states in the vacuum module, $L_{-2}\ket{0}$ and $L_{-3}\ket{0}$\,. In all these examples only for small enough $c$ the TSD is a monotonic function for $x\in[0,1]$\,. At larger $c$ the function starts to meander and can get even bigger than 2, the maximum value of the TSD between pure states. However, the reduced density matrices are not pure and it is not a contradiction that the TSD behaves like this. Still, it is hard to interpret the quantity as a meaningful measure of distinguishability for large values of $c$ at intermediate values of the relative subsystem size $x=l/L$.  

\begin{figure}[ht!]
    \centering
    \begin{tikzpicture}
    \node at (0,0) {\includegraphics[width=.45\textwidth]{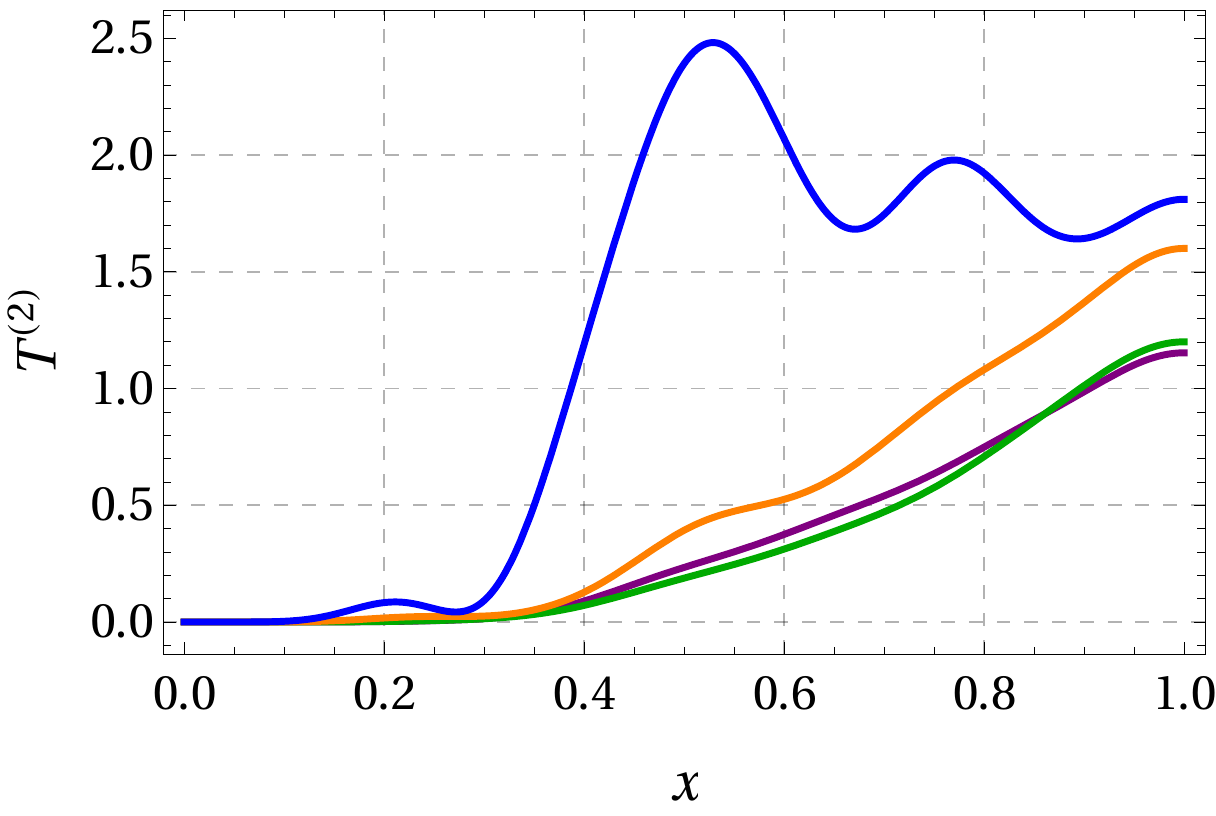}};
    \node at (.47\textwidth,0)  {\includegraphics[width=.45\textwidth]{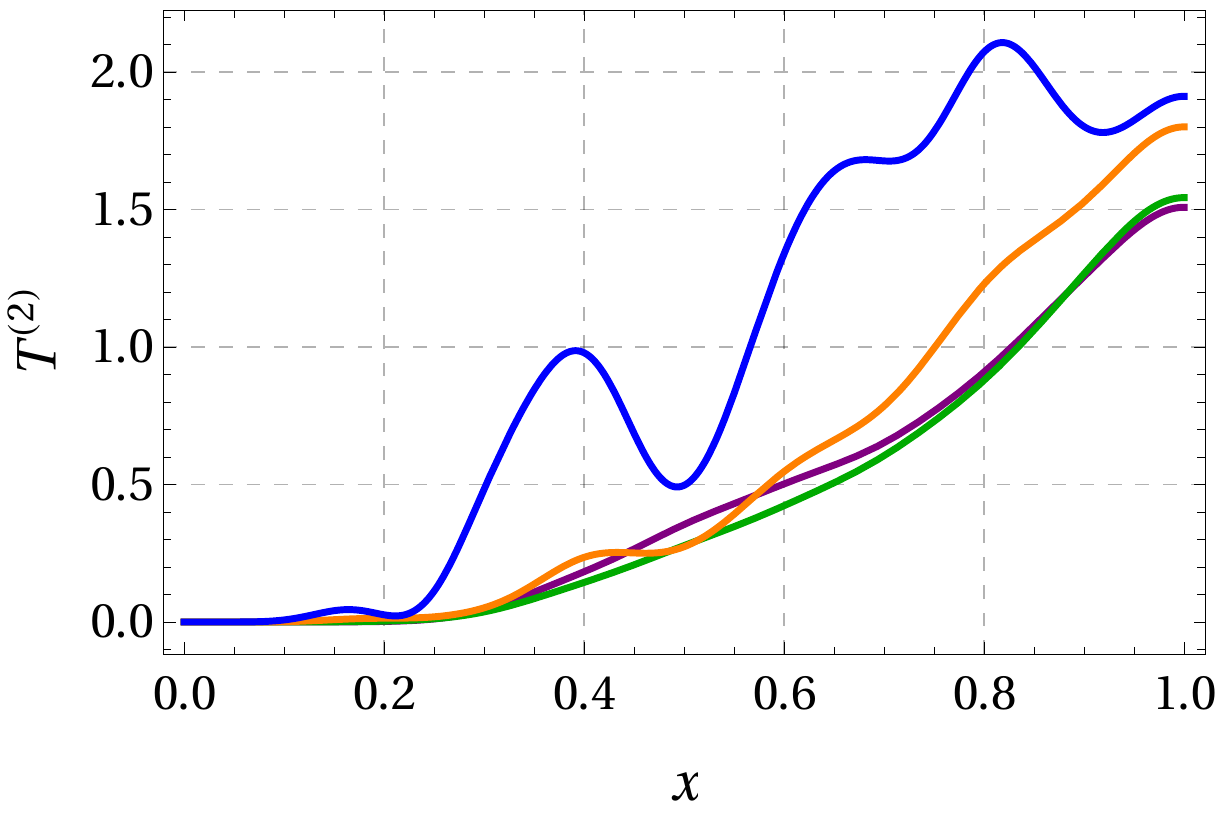}};
    \node at (.25\textwidth,-2.3
    ) {\includegraphics[width=.25\textwidth]{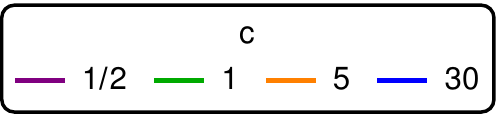}};
    \node at (-.6,1.5) {(a)};
    \node at (6.5,1.5) {(b)};
    \end{tikzpicture}  
    \caption{The TSD between the degenerate states at (a) $h_s= 4$, i.e.~$L_{-4}\ket{0}$ and $L_{-2}^2\ket{0}$, and (b) $h_s = 5$, i.e.~$L_{-5}\ket{0}$ and $L_{-3}L_{-2}\ket{0}$, for different values of $c$. }
    \label{fig:TSDB}
\end{figure}

In figure \ref{fig:TSDB} we show the TSD between the two degenerate states at conformal dimension $h_s=4$ and $h_s=5$ for different values of $c$. As expected from the results \eqref{eq:TSDsmallxdeg1} and \eqref{eq:TSDsmallxdeg2} we see a quite large flat region at small $x$. At $x\to1$ they converge to the TSD of the respective pure states. In the regions in between they show qualitatively the same behaviour as the other TSDs. For larger central charge they start to meander and at very large $c$ the term proportional to $c^2$ dominates, so that the TSD becomes very large, too.

\section{Theory dependent results} \label{sec:nonuniversal}

For non-vacuum descendant states, using the relation \eqref{eq:rec1} recursively allows to express the correlation function of holomorphic descendants $f_{\ket{s_i}}$ as a differential operator acting on the correlation function of the respective  primary fields 
\begin{equation}\label{eq:diff}
    \langle \prod_{i=1}^N f_{\ket{s_i}}(z_i) \rangle = \mathcal{D} \, \langle \prod_{i=1}^N f_{\ket{\Delta_i}}(z_i) \rangle \, .
\end{equation}

\noindent
In general, $\mathcal{D}$ depends on the central charge of the CFT, on the conformal weights of the primary fields, and on the insertion points. As a differential operator it acts on the holomorphic coordinates. 
In appendix~\ref{app:PrimDesCorr} we provide a code to compute it analytically in Mathematica.
If the correlation function of the primaries is known, then it is possible to compute the descendant correlator through~\eqref{eq:diff}.

The correlators in \eqref{eq:RFE}, \eqref{eq:SRDcorr}, and \eqref{eq:TSDcorr} can be written as linear combinations of correlation functions of descendants with coefficients that follow from the respective conformal transformations, i.e.~the uniformization map \eqref{eq:uniformization} in case of the R\'enyi entropy and the TSD, and the  usual M\"obius transformations \eqref{eq:Moebius} followed by a rotation in case of the SRD. Combining this with \eqref{eq:diff} we can write each of the correlators as
\begin{equation}
    D \bar{D} \langle \prod_{i=1}^N f_{\ket{\Delta_i}}(z_i) \rangle\,,
\end{equation}
with differential operators $D,\bar{D}$. Since we only consider holomorphic descendants, $\bar{D}$ is simply given by the anti-holomorphic part of the transformation of primaries,
\begin{equation}
    \bar{D} = \prod_{k=1}^n \bar{v}_{0;(k,l)}^{\bar{h}_k}\bar{v}_{0;(k,-l)}^{\bar{h}_k}\,.
\end{equation}

\noindent
To give an example, for the correlator of the $n$th R\'enyi entropy \eqref{eq:RFE} we simply get $\bar{D} = \sin^{4\bar{h}}(\pi l/L)$ from the uniformization map.

In the following sections we explicitly show the expressions of the differential operators $D\bar{D}$ for the simplest descendant state $L_{-1}\ket{\Delta}$. We will then consider results for higher-level descendants by acting with the operators on particular primary four-point functions in two specific CFTs, the Ising model and the three-state Potts model.

The Ising model is one of the simplest CFTs \cite{DiFrancesco:1997nk}.
It is a unitary minimal model with central charge $c=1/2$ and contains three primary operators: the identity, the energy density $\varepsilon$ and the spin field $\sigma$, whose holomorphic conformal weights are $0$, $1/2$, $1/16$ respectively.
The $2n$-point correlation functions on the plane of the $\varepsilon$ and $\sigma$ operators are known~\cite{DiFrancesco:1997nk} and, in particular, the four-point correlator of the energy density reads
\begin{equation}\label{eq:isingen}
    \left\langle \varepsilon(z_1,\bar{z}_1) \ldots \varepsilon(z_4 ,\bar{z}_4) \right\rangle = \left| \frac{1}{(z_{12}z_{34})^2} + \frac{1}{(z_{13}z_{24})^2} + \frac{1}{(z_{23}z_{14})^2}  \right| \, ,
\end{equation}
while the four-point correlator of the spin is given by
\begin{equation}\label{eq:isingsig}
    \left\langle \sigma(z_1,\bar{z}_1) \ldots \sigma(z_4 ,\bar{z}_4) \right\rangle = \frac{1}{\sqrt{2}} \frac{1}{|z_{14} z_{23}|^{1/4} } \frac{\sqrt{1 + |\eta| + |1-\eta|}}{|\eta|^{1/4}} \,,
\end{equation}
where $z_{ij} = z_i - z_j$ and $\eta = z_{12}z_{34}/z_{13}z_{24}$ is the cross ratio.
Given these expressions, it is possible to study the R\'enyi entanglement entropy  and the quantum measures for various descendants of $\varepsilon$ and $\sigma$.

The three-state Potts model is the unitary minimal model with $c= 4/5$~\cite{DiFrancesco:1997nk}.
It can be realized as a particular class of the more general $N$-state clock model which enjoys $\mathbb{Z}_N$ symmetry:
for $N=2$ one recovers the Ising model, while the case $N=3$ is equivalent to the three-state Potts model~\cite{Fradkin:1980th,Fateev:1985mm,Fateev:1987vh,Dotsenko:1984if}.
Its operator content is richer than that of the Ising model and it includes six spinless primary operators with conformal weight $0$, $2/5$, $7/5$, $3$, $1/15$, and $2/3$.
In particular, the dimensions of the thermal operator $\varepsilon$ and the spin field $\sigma$ are $2/5$ and $1/15$ respectively. 
Again, a number of correlation functions between operators of the three-states Potts model are known (see e.g.~\cite{Fateev:1985mm,Dotsenko:1984if}) and, since we will focus on descendants of the energy operator in the following, we provide here the four-point correlation function of the energy density \cite{Dotsenko:1984if}:
\begin{align}\label{eq:pottsen}
    \left\langle \varepsilon(z_1,\bar{z}_1) \ldots \varepsilon(z_4 ,\bar{z}_4) \right\rangle &=  \frac{1}{|z_{13} z_{24}|^{8/5}} \Bigg[ \frac{1}{| \eta (1-\eta) |^{8/5}} \left| _2F_1\left( -\tfrac85, -\tfrac15; -\tfrac25; \eta \right) \right|^2  \nonumber\\
    & \phantom{=}  -   \frac{ \Gamma\left(-\tfrac25\right)^2 \Gamma\left(\tfrac65\right)\Gamma\left(\tfrac{13}5\right)}{\Gamma\left(\tfrac{12}{5}\right)^2 \Gamma\left(-\tfrac15\right)\Gamma\left(-\tfrac85\right)} |\eta(1-\eta)|^{6/5} \left| _2F_1\left( \tfrac65, \tfrac{13}{5}; \tfrac{12}{5}; \eta \right) \right|^2  \Bigg] \, ,
\end{align}
where $ _2F_1 $ is the hypergeometric function.

\subsection{R\'enyi entanglement entropy}

Let us first consider $F_{\ket{s}}^{(2)}$ with $\ket{s} = L_{-1}\ket{\Delta}$.
As discussed above we can write
\begin{equation}\label{eq:ree1prim}
F_{\ket{s}}^{(2)} = \bar{D}^{F^{(2)}} D_{L_{-1}}^{F{(2)}} \, \left\langle f_{\ket{\Delta}}(e^{-\frac12 i\pi x}) f_{\ket{\Delta}}( e^{\frac12 i\pi x} ) f_{\ket{\Delta}}(- e^{-\frac12 i\pi x}) f_{\ket{\Delta}}( - e^{\frac12 i\pi x} ) \right\rangle_\mathbb{C} \, ,
\end{equation}
with $\bar{D}_{L_{-1}}^{F^{(2)}}= \sin ^{4 \bar{h}}(\pi  x)$ and $D_{L_{-1}}^{F^{(2)}}$ can be computed to be
\begin{align}
    D_{L_{-1}}^{F(2)} =& \, \frac{1}{64} \sin ^{4 h}(\pi x)  \Big[ 4 h^2 (3 \cos (2 \pi  x)+5)^2 + \frac{16 \sin ^4(\pi  x)}{h^2} \partial_1 \partial_2 \partial_3 \partial_4 \nonumber\\
    &+h e^{-\frac{7}{2} i \pi  x} \left(3+e^{2 i \pi  x}\right)^2 \left(-2 e^{2 i \pi  x}+3 e^{4 i \pi  x}-1\right) \left( \partial_2 - \partial_4 \right) \nonumber\\
    &+ h e^{-\frac{9}{2} i \pi  x} \left(1+3 e^{2 i \pi  x}\right)^2 \left(2 e^{2 i \pi  x}+e^{4 i \pi  x}-3\right) \left( \partial_3 - \partial_1 \right) \nonumber\\
    &+ 8 \sin ^2(\pi  x) (3 \cos (2 \pi  x)+5) \left( \partial_1\partial_2 + \partial_3\partial_4 - \partial_2\partial_3 - \partial_1\partial_4 \right)\nonumber\\
    & -e^{-3 i \pi  x} \left(2 e^{2 i \pi  x}+e^{4 i \pi  x}-3\right)^2 \partial_2 \partial_4 -e^{-5 i \pi  x} \left(2 e^{2 i \pi  x}-3 e^{4 i \pi  x}+1\right)^2 \partial_1 \partial_3 \nonumber\\
    & + \frac{1}{h}e^{-\frac{7}{2} i \pi  x} \left(-1+e^{2 i \pi  x}\right)^3 \left(3+e^{2 i \pi  x}\right) \left( \partial_1 \partial_2 \partial_4 - \partial_2 \partial_3\partial_4 \right) \nonumber\\
    & + \frac{1}{h} e^{-\frac{9}{2} i \pi  x} \left(-1+e^{2 i \pi  x}\right)^3 \left(1+3 e^{2 i \pi  x}\right) \left( \partial_1\partial_3\partial_4 - \partial_1\partial_2\partial_3 \right) \Big]\,,
\end{align}
where $\partial_n$ is the partial differentiation w.r.t.~the $n$-th insertion point and $x=l/L$.
Unfortunately, already at level 2 the general expression are too cumbersome to express them here explicitly.

Given the four-point correlation functions~\eqref{eq:isingen}, \eqref{eq:isingsig}, \eqref{eq:pottsen}, we can compute $F_{L_{-1}\ket{\Delta}}^{(2)}$ from eq.~\eqref{eq:ree1prim} for $h= 1/2, \, 1/16$ in the Ising model and $h=2/5$ in the three-states Potts model.
We performed the same computations for descendants up to level 3 and show the results in figure~\ref{fig:REEprim}; some analytic expressions are given in appendix~\ref{app:REEresultsIsing} and~\ref{app:REEresultsPotts}.

\begin{figure}[tb]
    \centering
    \begin{tikzpicture}
        \node at (0,0) {\includegraphics[width=.32\textwidth]{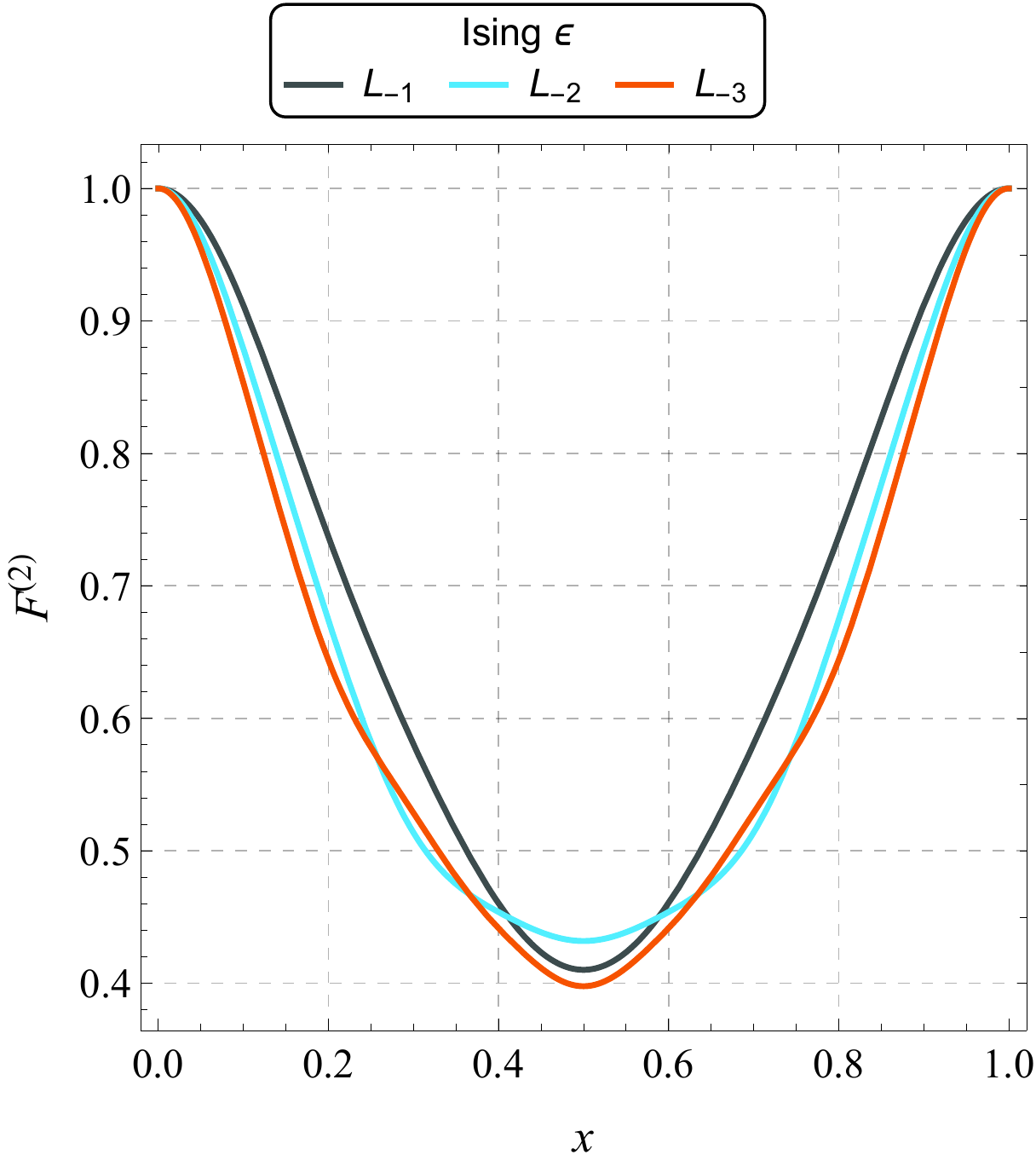}};
    \node at (.33\textwidth,0)  {\includegraphics[width=.32\textwidth]{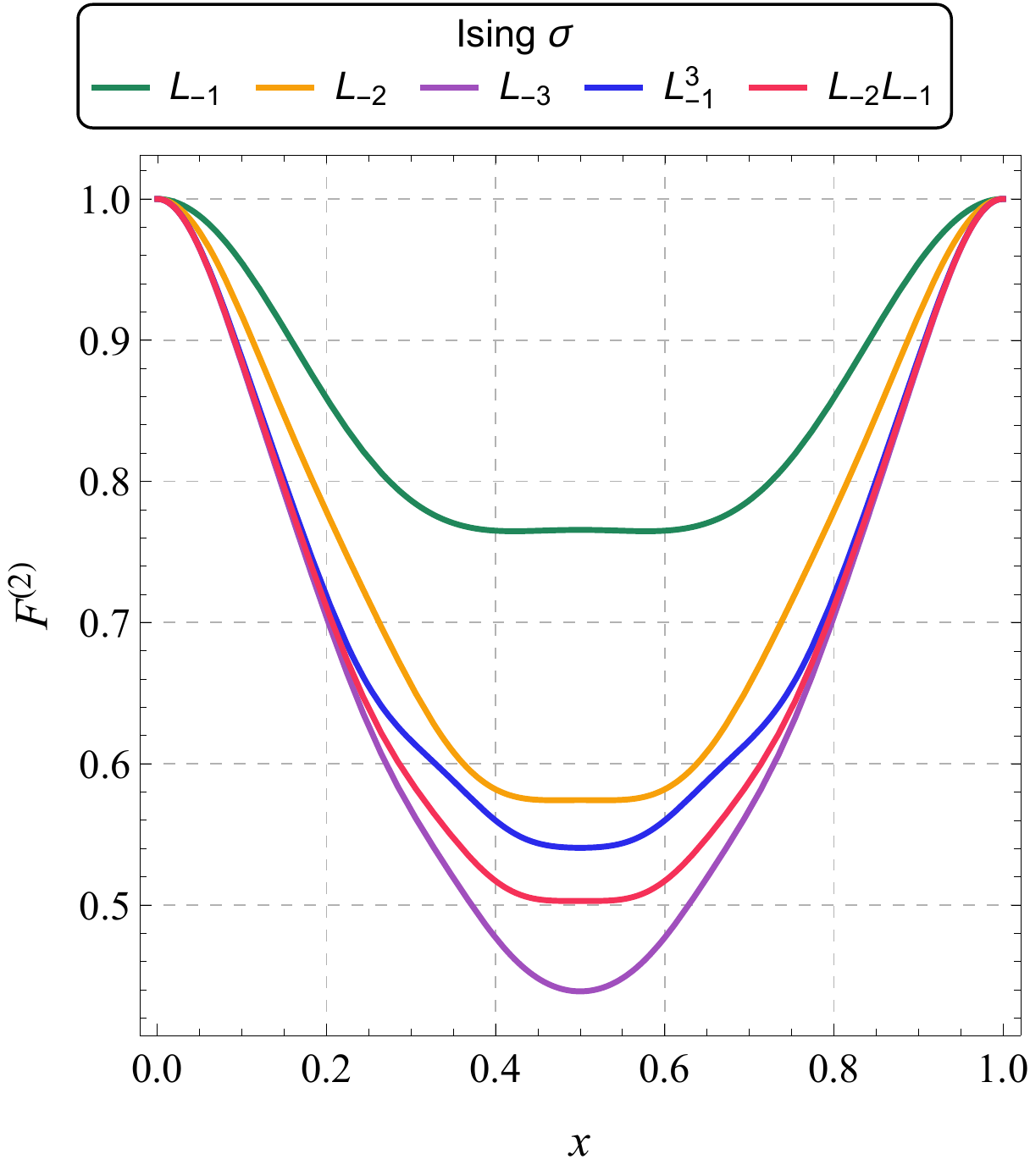}};
    \node at (.66\textwidth,0)  {\includegraphics[width=.32\textwidth]{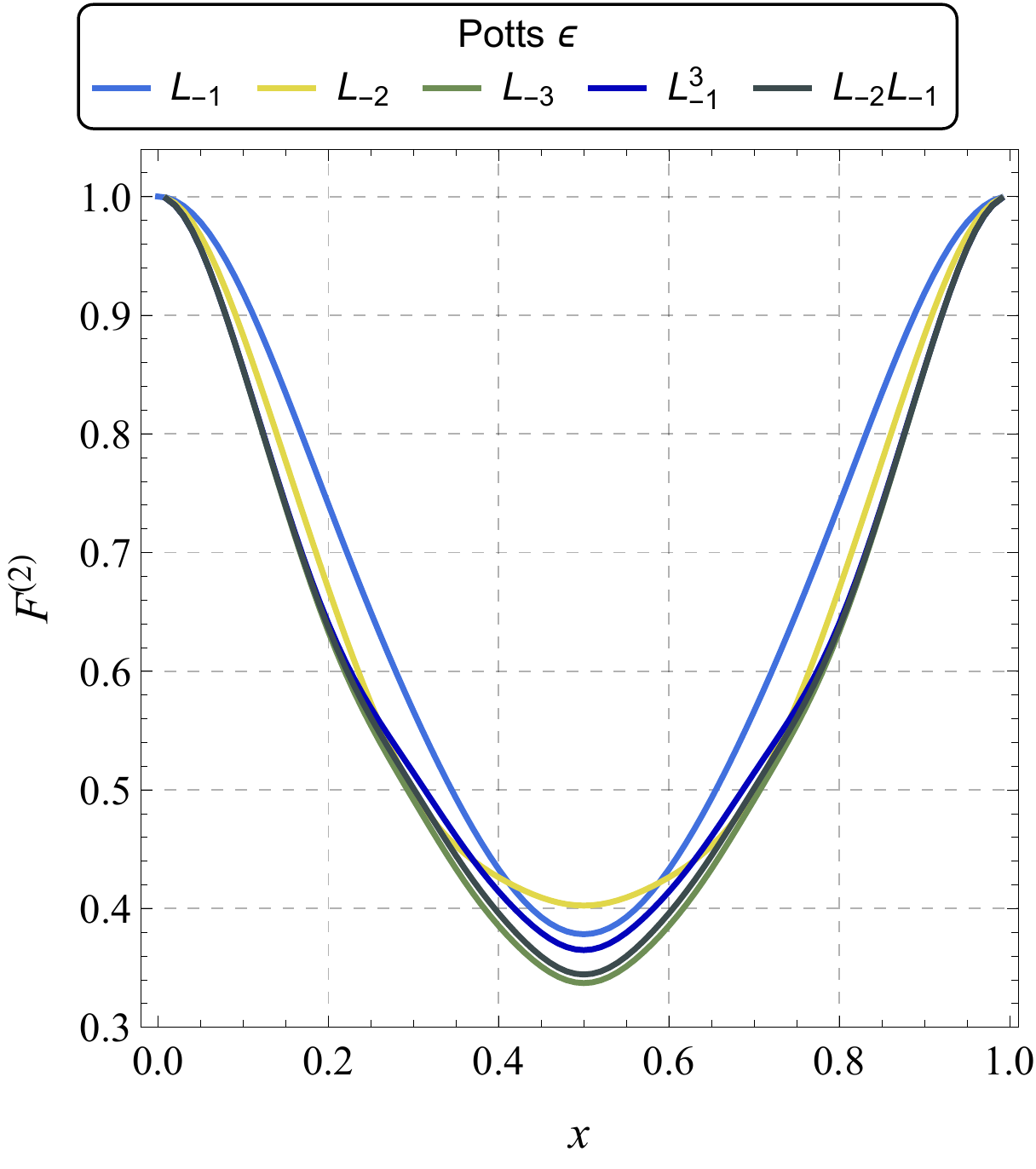}};
    \end{tikzpicture}
    \caption{The correlator $F^{(2)}_{\ket{s}}$ for different descendants of $\ket{\varepsilon}$ and $\ket{\sigma}$ in the Ising model and $\ket{\varepsilon}$ in the Potts model.}
    \label{fig:REEprim}
\end{figure}

In the Ising model, there is only one physical state in the module of the energy operator at each level up to level 3.
A consequence is that $F_{L_{-2}\ket{\varepsilon}}^{(2)} = F_{L_{-1}^2\ket{\varepsilon}}^{(2)}$, even though $D^{F^{(2)}}_{L_{-2}} \neq D^{F^{(2)}}_{L_{-1}^2}$.
The same happens at level 3 for the different descendant states $L_{-3}\ket{\varepsilon}$, $L_{-1}^3\ket{\varepsilon}$ and $L_{-2}L_{-1}\ket{\varepsilon}$.\footnote{With an abuse of notation, we will henceforth denote the primary states related to the energy and spin fields in the Ising and Potts models with $\ket{\varepsilon}$ and $\ket{\sigma}$, instead of labelling them with their conformal dimensions.}
For $\sigma$ descendants, again there is only one physical state at level 2 and $F_{L_{-2}\ket{\sigma}}^{(2)} = F_{L_{-1}^2\ket{\sigma}}^{(2)}$, but at level 3 there are two physical states and $L_{-3}\ket{\sigma}$, $L_{-1}^3\ket{\sigma}$ and $L_{-2}L_{-1}\ket{\sigma}$ produce different REEs as shown in figure~\ref{fig:REEprim}.
Notice that the REEs for the different descendants of $\sigma$ at level 3 have a similar behaviour for small values of $x$, but are clearly distinguishable for $x \sim 1/2$.

For descendants states of the energy density of the three-states Potts model there is again only one physical state at level 2 and two physical states at level 3.
Similarly to the case of descendants of $\sigma$ in Ising, we found that $F_{L_{-2}\ket{\varepsilon}}^{(2)} = F_{L_{-1}^2\ket{\varepsilon}}^{(2)}$ but the different descendants that we considered at level 3 produced different REEs, as plotted in figure~\ref{fig:REEprim}.
Notice that also in Potts the behaviour for small $x$ is given by the level and not by the state configuration, while all the curves are distinguishable for $x\sim 1/2$.
In particular, $F_{L_{-1}^3\ket{\varepsilon}}^{(2)}$ behaves more like $F_{L_{-1}\ket{\varepsilon}}^{(2)}$ than $F_{L_{-3}\ket{\varepsilon}}^{(2)}$ for $x\sim 1/2$, while the plot of $F_{L_{-2}L_{-1}\ket{\varepsilon}}^{(2)}$ is very similar to $F_{L_{-3}\ket{\varepsilon}}^{(2)}$.

If we expand the analytic results for energy descendants in both the Ising and Potts model for small $x$, we find the behaviour
\begin{equation}\label{eq:reesmallx}
    F_{L_{-n}\ket{\varepsilon}}^{(2)} = 1 - \frac{n + 2h_\varepsilon}{2} (\pi x)^2 + \mathcal{O}(x^4) \quad h_\varepsilon= \left\{\begin{matrix}1/2& \text{Ising}\\ 2/5 & \text{Potts}\end{matrix}\right.\,, \quad n=1,2,3\,.
\end{equation}

\noindent
This is in general expected, since for small subsystem size $z_1 \sim z_2$ and $z_3 \sim z_4$ and to first order the four-point function is $(h=\bar{h}=\Delta/2)$
\begin{equation}
    \left\langle f_{\ket{\Delta}}(z_1, \bar{z}_1) f_{\ket{\Delta}}(z_2, \bar{z}_2) f_{\ket{\Delta}}(z_3, \bar{z}_3)  f_{\ket{\Delta}}(z_4, \bar{z}_4) \right\rangle_\mathbb{C} \simeq \frac{1}{| z_{12} z_{34}|^{4 h}}\,.
\end{equation}

\noindent
Then, using this correlation function in~\eqref{eq:ree1prim} as well as in the corresponding equations for higher descendants and taking the small $x$ limit we reproduce precisely eq.~\eqref{eq:reesmallx}, which is the clear generalization of eq.~\eqref{eq:RElowx} in agreement with~\cite{Alcaraz:2011tn}.
However, the leading behaviour of $F^{(2)}_{L_{-n} \ket{\sigma}}$ is different from the one outlined in~\eqref{eq:reesmallx}.
This happens because in the OPE of two Ising spin operator there is an additional contribution, that is absent in the OPE of two energy operators or subleading in the case of Potts.
Indeed, consider in general the OPE between two primary fields 
\begin{equation}\label{eq:ope_light}
    f_{\ket{\Delta_i}} (z_1,\bar{z}_1) f_{\ket{\Delta_i}}(z_2,\bar{z}_2) = \frac{1}{|z_{12}|^{4 h_i}} + \frac{C^k_{ii} \, f_{\ket{\Delta_k}}(z_2,\bar{z}_2)}{|z_{12}|^{4 h_i - 2 h_k}} + \ldots \, ,
\end{equation}
where we included the contribution from the lightest primary field $f_{\ket{\Delta_k}}$ in the module of $f_{\ket{\Delta_i}}$.
Then, to this order the four-point function for $z_1 \sim z_2$ and $z_3 \sim z_4$ becomes
\begin{align}\label{eq:corr_light}
    \left\langle f_{\ket{\Delta_i}}(z_1, \bar{z}_1) \ldots f_{\ket{\Delta_i}}(z_4, \bar{z}_4) \right\rangle_\mathbb{C} &\simeq \frac{1}{| z_{12} z_{34}|^{4 h_i}} + \frac{(C^{k}_{ii})^2}{|z_{12}z_{34}|^{4h_i - 2 h_k}} \frac{1}{|z_{24}|^{4 h_k}} \, ,
\end{align}
so that
\begin{align}\label{eq:reesmallxsubleading}
    F_{L_{-n}\ket{\Delta_i}}^{(2)} &= 1 - \frac{n + 2 h_i}{2} (\pi x)^2 \nonumber\\
    &\phantom{=} + \left(C^{k}_{ii}\right)^2 \left( \frac{ c (n-1)^2  + 4 n h_i  + 2 n^2 (h_k -1) h_k }{ c (n-1)^2 + 4 n h_i }\right)^2 \left( \frac{\pi x}{2} \right)^{4 h_k} + \ldots \, .
\end{align}
The second term is in general a subleading contribution, like in the Potts model $\varepsilon \times \varepsilon = \mathbb{I} + X$ with X having dimension $7/5$.
However, due to the fusion rule $\sigma \times \sigma = \mathbb{I} + \varepsilon$ in Ising, in this case $h_k = 1/2$ with $C_{\sigma\sigma}^\varepsilon = 1/2$~\cite{Mussardo:2020rxh}, and we see that the second term in~\eqref{eq:reesmallxsubleading} contributes to leading order.
Indeed, eq.~\eqref{eq:reesmallxsubleading} correctly predicts the small $x$ behaviour of $F^{(2)}_{L_{-n}\ket{\sigma}}$ for $n=1,2,3$ that we computed and report in app.~\ref{app:REEresultsIsing}.

Some results for the R\'enyi entanglement entropy in the Ising and three-states Potts models were already considered in~\cite{Palmai:2014jqa,Taddia:2016dbm,Taddia_2013}; we checked that our code produces the same analytic expressions studied in these references.

\subsection{Sandwiched R\'enyi divergence}

Consider now the correlator $\mathcal{F}^{(2)}_{\ket{s}}$ related to the SRD as in eq.~\eqref{eq:SRDcorr} with $\ket{s} = L_{-1} \ket{\Delta}$.
Then, we find
\begin{equation}\label{eq:srdprimlvl1}
    \mathcal{F}^{(2)}_{\ket{s}} = \bar{D}^{\mathcal{F}(2)} D_{L_{-1}}^{\mathcal{F}(2)}  \, \left\langle f_{\ket{\Delta}}(e^{- i\pi x}) f_{\ket{\Delta}}( e^{ i\pi x} ) f_{\ket{\Delta}}(- e^{- i\pi x}) f_{\ket{\Delta}}( - e^{ i\pi x} ) \right\rangle_\mathbb{C}\,. 
\end{equation}
From the anti-holomorphic part of the conformal transformation we now obtain
\begin{equation}
    \bar{D}^{\mathcal{F}(2)} = 2^{4 \bar{h}}  \sin ^{4 \bar{h}}(\pi  x)
\end{equation}
and the differential operator acting on the holomorphic coordinates reads
\begin{align}
    D_{L_{-1}}^{\mathcal{F}(2)} &= \frac{2^{4h}}{h^2} e^{-2 i \pi x} \sin ^{4 h}(\pi 
   x) \bigg[ 4 h^4 e^{2 i \pi  (h+1) x} \left(e^{-2 i \pi  x}\right)^h  \nonumber\\
   &\phantom{=}  + 2 h^3 e^{i \pi  x} \left(1 - e^{2 i \pi  x}\right) ( \partial_1 + \partial_4 - \partial_2 - \partial_3 )   \nonumber\\
   &\phantom{=}  + h^2 \left(e^{2 i \pi  x} - 1\right)^2 ( \partial_1\partial_4 + \partial_2\partial_3 -\partial_1\partial_2  -\partial_1\partial_3 -\partial_2\partial_4 - \partial_3\partial_4 )  \nonumber\\
   &\phantom{=}  +4 i h e^{2 i \pi  x} \sin ^3(\pi  x) ( \partial_1\partial_2\partial_3 + \partial_2\partial_3\partial_4 - \partial_1\partial_3\partial_4 - \partial_1\partial_2\partial_4 )  \nonumber\\
   &\phantom{=}  + \frac{1}{4} \left(e^{2 i \pi  x} -1 \right)^4 e^{2 i \pi  (h-1) x} \left(e^{-2 i \pi  x}\right)^h \partial_1\partial_2\partial_3\partial_4 \bigg] \, .
\end{align}

\noindent 
We explicitly study the results for descendants up to level 3. The general expressions for $D$ are, however, again too cumbersome to show them here. 
With the four-point functions \eqref{eq:isingen}, \eqref{eq:isingsig}, \eqref{eq:pottsen} we compute $\mathcal{S}^{(2)}_{\ket{s}}$ for the descendants of the energy and spin primary states in Ising and of the energy state in Potts.
The results are plotted in figure \ref{fig:SRDprim} and some closed expressions are given for descendants of the energy state of Ising in appendix~\ref{app:SRDising}.

\begin{figure}[tb]
    \centering
    \begin{tikzpicture}
        \node at (0,0) {\includegraphics[width=.32\textwidth]{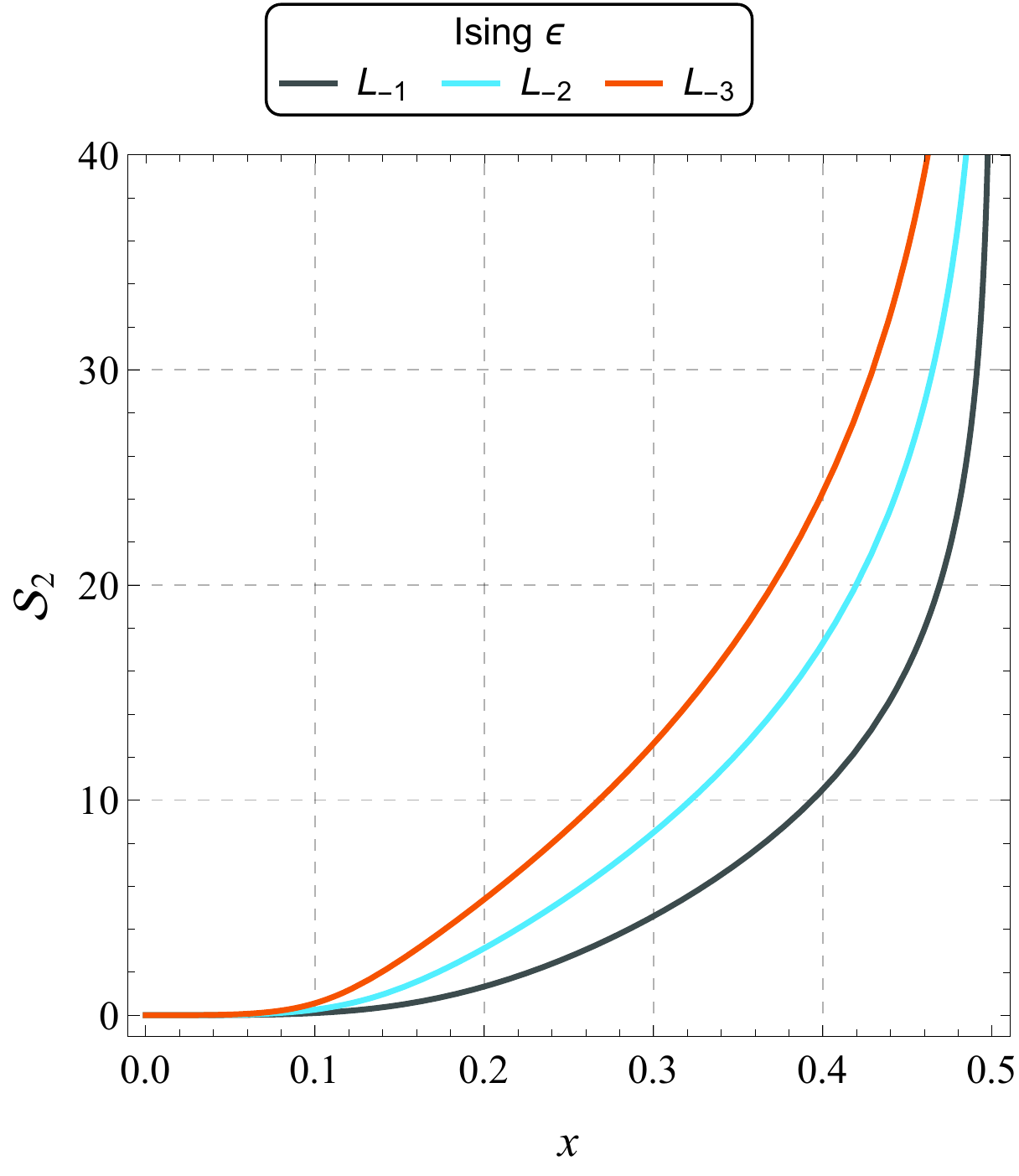}};
    \node at (.33\textwidth,0)  {\includegraphics[width=.32\textwidth]{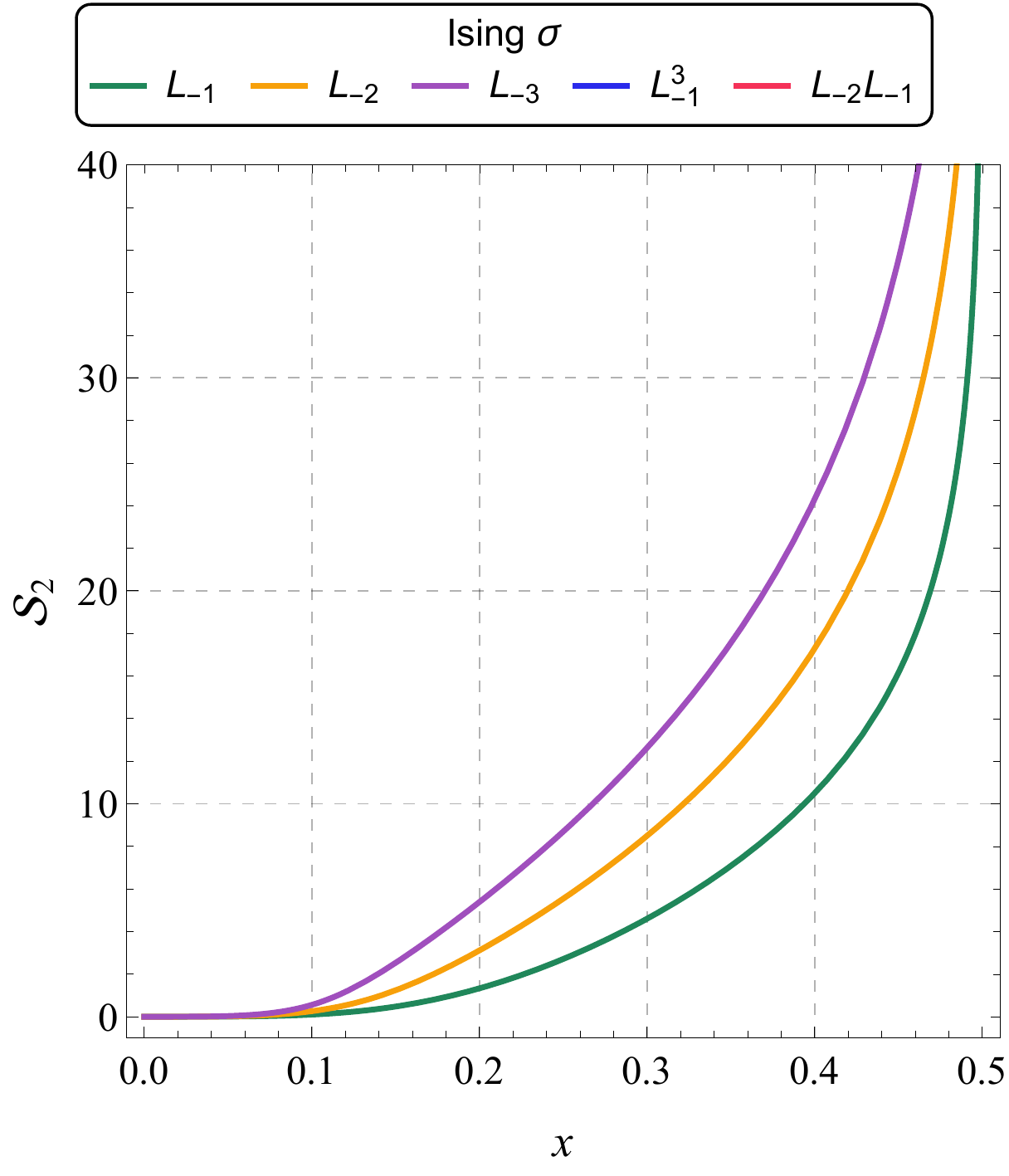}};
    \node at (.66\textwidth,0)  {\includegraphics[width=.32\textwidth]{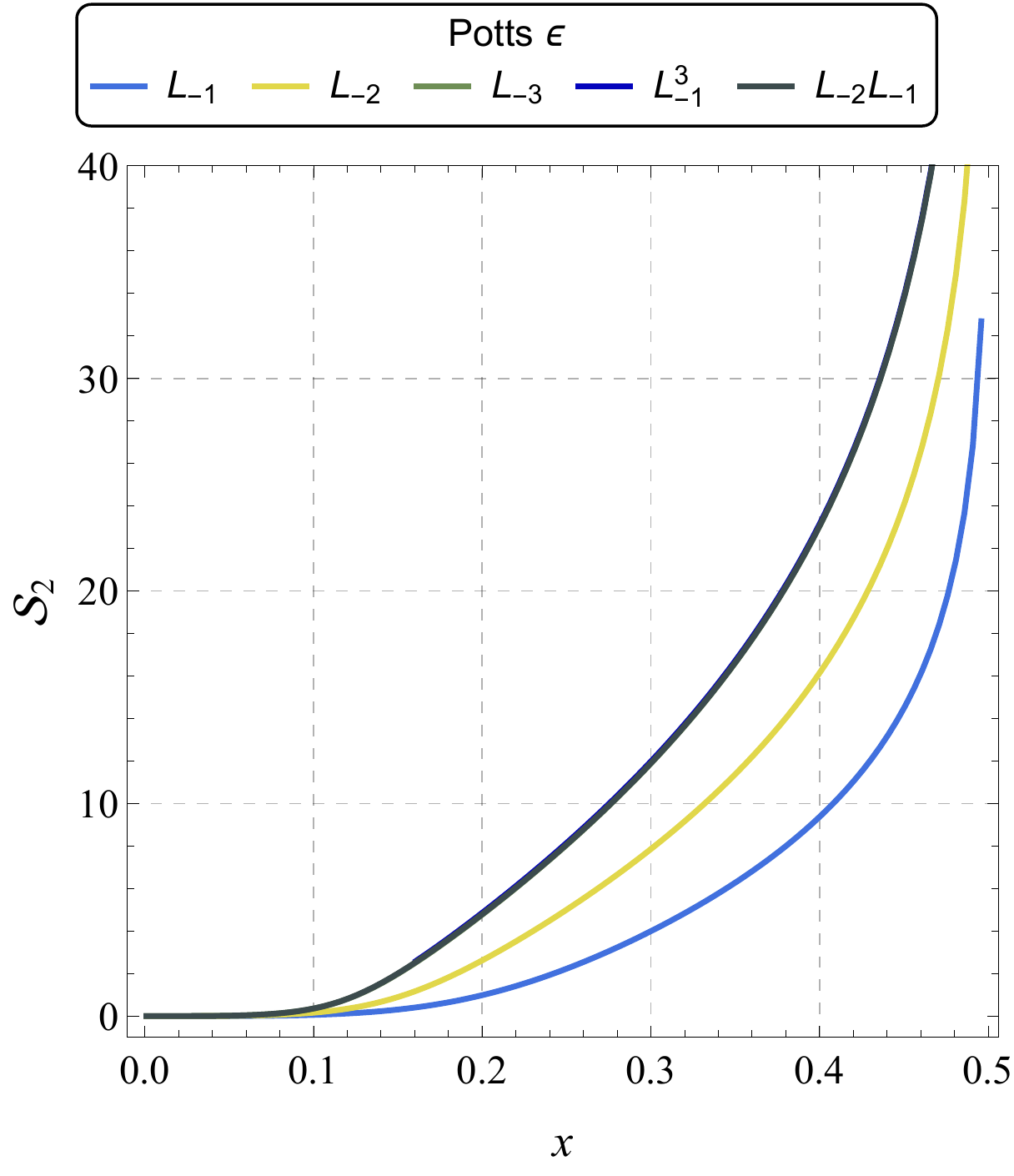}};
    \end{tikzpicture}
    \caption{The SRD between the reduced groundstate and different descendants of $\ket{\varepsilon}$ and $\ket{\sigma}$ in the Ising model and $\ket{\varepsilon}$ in the Potts model.}
    \label{fig:SRDprim}
\end{figure}

As expected, the SRDs start from 0 and diverge at $x=1/2$. 
We also see from the plots that for higher-level descendants the SRD grows more rapidly.
In the Ising model degenerate descendants of $\varepsilon$ at level~2 and~3 produce the same SRDs, while for degenerate descendants of $\sigma$ at level~3 we found three different expressions. 
However, the differences between the plotted results are so small that the three curves at level~3 overlap in figure~\ref{fig:SRDprim}.
The same happens for descendants of $\varepsilon$ in the Potts model.

Now, let us check the limit of small subsystem size.
Consider the OPE between two primary fields ($h = \bar{h} = \Delta/2$)
\begin{equation}\label{eq:ope_srd}
    f_{\Delta} (z_1,\bar{z}_1) f_{\Delta}(z_2,\bar{z}_2) = \frac{1}{|z_{12}|^{4 h}} + \frac{2 h c^{-1} \, T(z_2)}{z_{12}^{2h -2} \bar{z}_{12}^{2 h}} + \frac{2 h c^{-1} \, \bar{T}(\bar{z}_2)}{z_{12}^{2h} \bar{z}_{12}^{2 h - 2}} + \ldots \,,
\end{equation}
where for now we only included the leading contributions from the vacuum module.
Then, if we insert this OPE in the four-point function for $z_1 \sim z_2$ and $z_3\sim z_4$ we obtain
\begin{align}\label{eq:opecorr_srd}
    \left\langle f_{\ket{\Delta}}(z_1, \bar{z}_1) \ldots f_{\ket{\Delta}}(z_4, \bar{z}_4) \right\rangle_\mathbb{C} &\simeq \frac{1}{| z_{12} z_{34}|^{4 h}} + \frac{2 h^2 c^{-1}}{|z_{12}z_{34}|^{4 h}} \frac{z_{12}^2 z_{34}^2}{z_{24}^4}  + \frac{2 h^2 c^{-1}}{|z_{12}z_{34}|^{4 h}} \frac{\bar{z}_{12}^2 \bar{z}_{34}^2}{\bar{z}_{24}^4} \, .
\end{align}

\noindent 
With this expression we can study the limit $x\to 0$ in~\eqref{eq:srdprimlvl1} and similar expressions for higher-level descendants.
We find 
\begin{equation}\label{eq:srdsmallxlaw}
    \mathcal{S}^{(2)}_{L_{-n}\ket{\Delta}} = \frac2c \left( n^2 + 2n h + 2 h^2 \right) ( \pi x)^4 + \ldots \, .
\end{equation}

\noindent 
Expanding our analytic results for descendants of the energy in Ising and Potts for $x\to 0$ we found agreement with eq.~\eqref{eq:srdsmallxlaw}.
For $\sigma$ descendants, however, the leading order contribution to the SRD in the limit $x\to 0$ is different.
Indeed, if we think of the OPE as in~\eqref{eq:ope_light} with the correlator~\eqref{eq:corr_light}, then we find the following leading contribution in the SRD for $n=1,2,3$
\begin{equation}\label{eq:srdsmallxsigma}
    \mathcal{S}^{(2)}_{L_{-n}\ket{\Delta_i}} = \left(C^{k}_{ii}\right)^2 \left( \frac{ c (n-1)^2  + 4 n h_i  + 2 n^2 (h_k -1) h_k }{ c (n-1)^2 + 4 n h_i }\right)^2 \left( \pi x \right)^{4 h_k} + \ldots \, .
\end{equation}

\noindent 
Since $h_k = 1/2$ for $\ket{\Delta_i} = \ket{\sigma}$ in the Ising model, we see that the contribution from the $\varepsilon$ channel dominates over the one from the stress tensor in~\eqref{eq:srdsmallxlaw}.
We checked that~\eqref{eq:srdsmallxsigma} with $C^\varepsilon_{\sigma\sigma} = 1/2$ correctly reproduce the $x\to 0$ limit of our results.

It is interesting to consider also the opposite limit $x\to 1/2$ and see how the SRDs scale with the singularity.
In this case, it is enough to consider the first contribution in the OPE~\eqref{eq:ope_srd}, but making the appropriate changes as with our insertion points $x\to 1/2$ means $z_1 \sim z_4$ and $z_2 \sim z_3$.
Then, for $n=1,2,3$ we find the following expression
\begin{equation}
    \mathcal{F}^{(2)}_{L_{-n}\ket{\Delta}} = \exp \left( \mathcal{S}^{(2)}_{L_{-n}\ket{\Delta}} \right) =  \frac{A_n}{ \pi^{-4 (2h + n)} \left( x -\frac12 \right) ^{-4 (2h + n)}}  + \ldots \, ,
\end{equation}
with
\begin{equation}
    A_n = (-1)^{8 h} \left( \frac{(n-1)(3n-5)(3n-4) \frac{c}{2} + 4 \left( \frac{6^n}{3} -1 \right) h + 2(n+1)^2 h^2 }{ c(n-1)^2 + 4n h } \right)^2 \, .
\end{equation}
Notice that for $h\to 0$ we recover the same scaling as in~\eqref{eq:srd_vac_divergence}.

In all the examples that we considered, the SRD proved to be a convex function of $x$, providing further evidence to the validity of the R\'enyi QNEC in two dimensions \cite{Moosa:2020jwt} for large enough central charge.

\subsection{Trace square distance}

Consider now the trace square distance between a primary state $\ket{\Delta}$ and its first descendants $L_{-1}\ket{\Delta}$. Then
\begin{equation}\label{eq:TSDprim}
T_{L_{-1}\ket{\Delta},\ket{\Delta}}^{(2)} =  \bar{D}^{T(2)} D_{L_{-1}}^{T(2)} \, \left\langle f_{\ket{\Delta}}(e^{-\frac12 i\pi x}) f_{\ket{\Delta}}( e^{\frac12 i\pi x} ) f_{\ket{\Delta}}(- e^{-\frac12 i\pi x}) f_{\ket{\Delta}}( - e^{\frac12 i\pi x} ) \right\rangle_\mathbb{C} \, ,
\end{equation}
where again the differential operator on the anti-holomorphic coordinates is simply given by the transformation factor
\begin{equation}
    \bar{D}^{T(2)} = \sin ^{4 \bar{h}}(\pi  x) \,,
\end{equation}
while the differential operator on the holomorphic coordinates is given by:
\begingroup
\allowdisplaybreaks[1]
\begin{align}
    D_{L_{-1}}^{T(2)} &= \frac{1}{64} \sin ^{4 h}(\pi  x) \bigg[ 4 (3 h \cos (2 \pi  x)+5 h-4)^2  \nonumber\\
    &\phantom{=}  + 2 e^{-\frac{5}{2} i \pi  x} \left(2 e^{2 i \pi  x}-3 e^{4 i \pi  x}+1\right) (3 h \cos (2 \pi  x)+5 h-8) \partial_1  \nonumber\\
    &\phantom{=}  + 2 e^{-\frac{3}{2} i \pi  x} \left(2 e^{2 i \pi  x}+e^{4 i \pi  x}-3\right) (3 h \cos (2 \pi  x)+5 h-8) \partial_2  \nonumber\\
    &\phantom{=}  + h e^{-\frac{9}{2} i \pi  x} \left(1+3 e^{2 i \pi  x}\right)^2 \left(2 e^{2 i \pi  x}+e^{4 i \pi  x}-3\right) \partial_3 \nonumber\\
    &\phantom{=}  - h e^{-\frac{7}{2} i \pi  x} \left(3+e^{2 i \pi  x}\right)^2 \left(-2 e^{2 i \pi  x}+3 e^{4 i \pi  x}-1\right)\partial_4  \nonumber\\
    &\phantom{=}  + 8 \sin ^2(\pi  x) (3 \cos (2 \pi  x)+5) \left( \partial_3\partial_4 - \partial_2\partial_3 - \partial_1\partial_4 \right)  \nonumber\\
    & \phantom{=}  -e^{-3 i \pi  x} \left(2 e^{2 i \pi  x}+e^{4 i \pi  x}-3\right)^2 \partial_2\partial_4 -4 e^{-i \pi  x} (2 i \sin (2 \pi  x)+\cos (2 \pi  x)-1)^2 \partial_1\partial_3  \nonumber\\
    &\phantom{=}  + \frac{8}{h} \sin ^2(\pi  x) (3 h \cos (2 \pi  x)+5 h-8) \partial_1\partial_2  \nonumber\\
    &\phantom{=}  + \frac{16}{h} e^{\frac{i \pi  x}{2}} \sin ^3(\pi  x) (\sin (\pi  x)+2 i \cos (\pi  x)) \partial_2\partial_3\partial_4  \nonumber\\
    &\phantom{=}  + \frac{16}{h} e^{-\frac{1}{2} i \pi  x} \sin ^4(\pi  x) (1-2 i \cot (\pi  x)) \left( \partial_1\partial_3\partial_4 - \partial_1\partial_2\partial_3 \right) \nonumber\\
    &\phantom{=}  \frac{1}{h}e^{-\frac{7}{2} i \pi  x} \left(e^{2 i \pi  x} -1 \right)^3 \left(3+e^{2 i \pi  x}\right) \partial_1\partial_2\partial_4 + \frac{16}{h^2} \sin ^4(\pi  x) \partial_1\partial_2\partial_3\partial_4 \bigg] \,.
\end{align}
\endgroup

\noindent 
Again, we limit ourselves to display this result, which is the simplest, since for higher descendants the expressions become much more involved.
As in the previous cases, we computed $T^{(2)}_{L_{-n}\ket{\Delta},\ket{\Delta}}$ as in~\eqref{eq:TSDprim} for $n=1,2,3$ and for the degenerate states at level~2 and~3.
Then, by using the four-point functions~\eqref{eq:isingen}, \eqref{eq:isingsig}, and \eqref{eq:pottsen} we obtained analytic expressions for the TSD between the primary state and its descendants for the energy and spin operators in the Ising model and for the energy in the three-states Potts model.
Figure~\ref{fig:TSDprim} shows the plots of the results, while in appendix~\ref{app:TSDising} and~\ref{app:TSDpotts} we provide some explicit expressions.

\begin{figure}[t!]
    \centering
    \begin{tikzpicture}
        \node at (0,0) {\includegraphics[width=.32\textwidth]{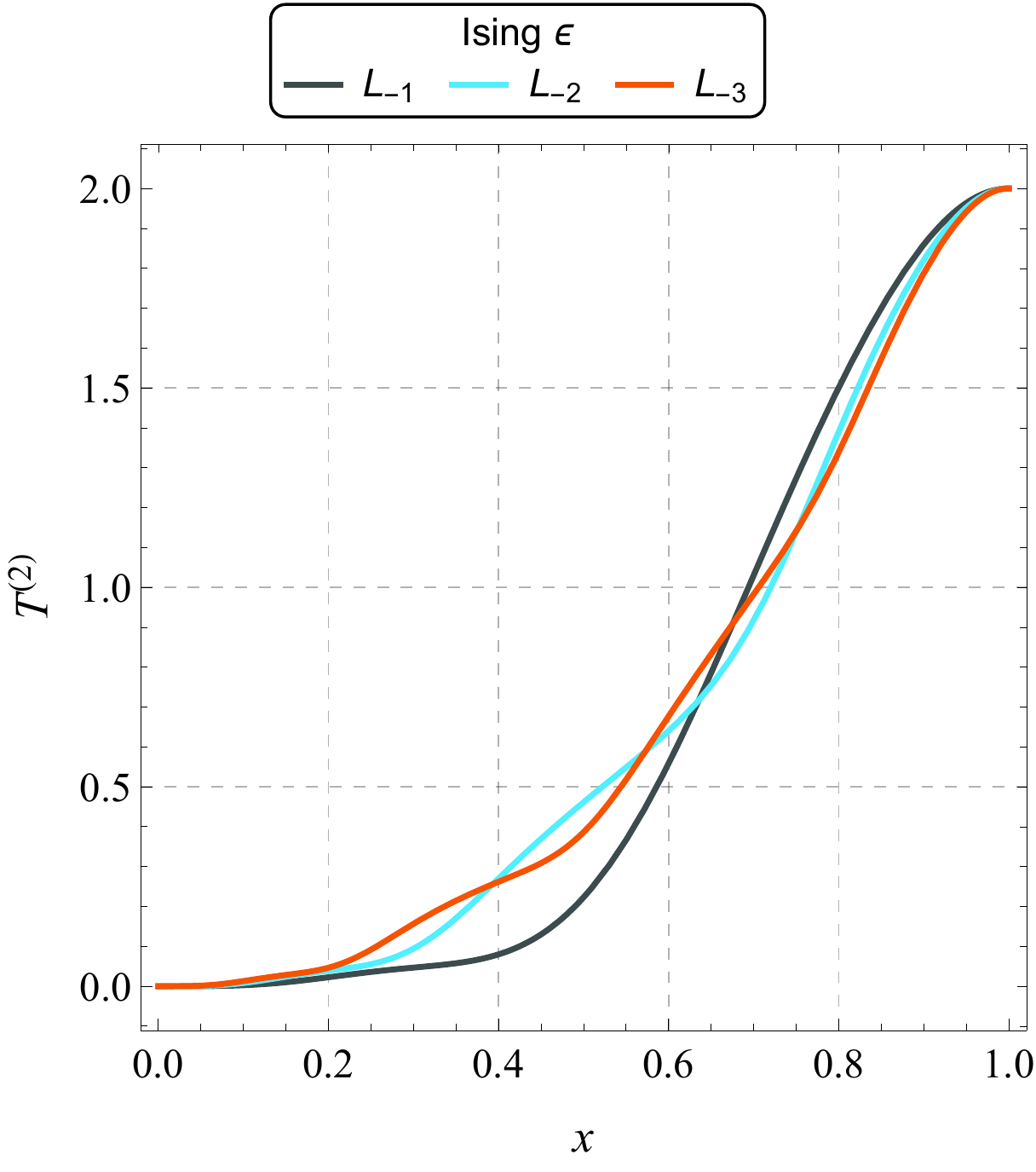}};
    \node at (.33\textwidth,0)  {\includegraphics[width=.32\textwidth]{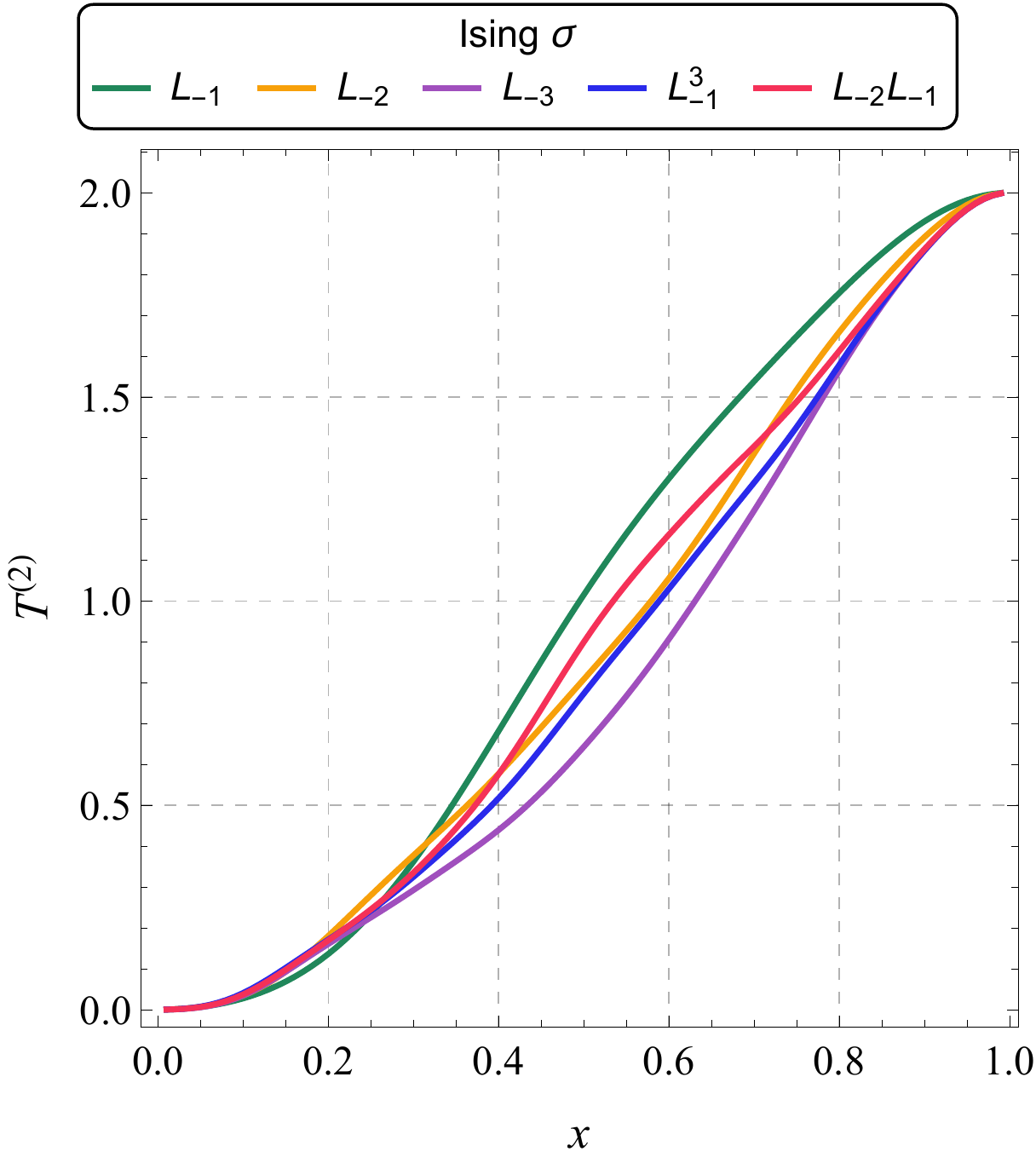}};
    \node at (.66\textwidth,0)  {\includegraphics[width=.32\textwidth]{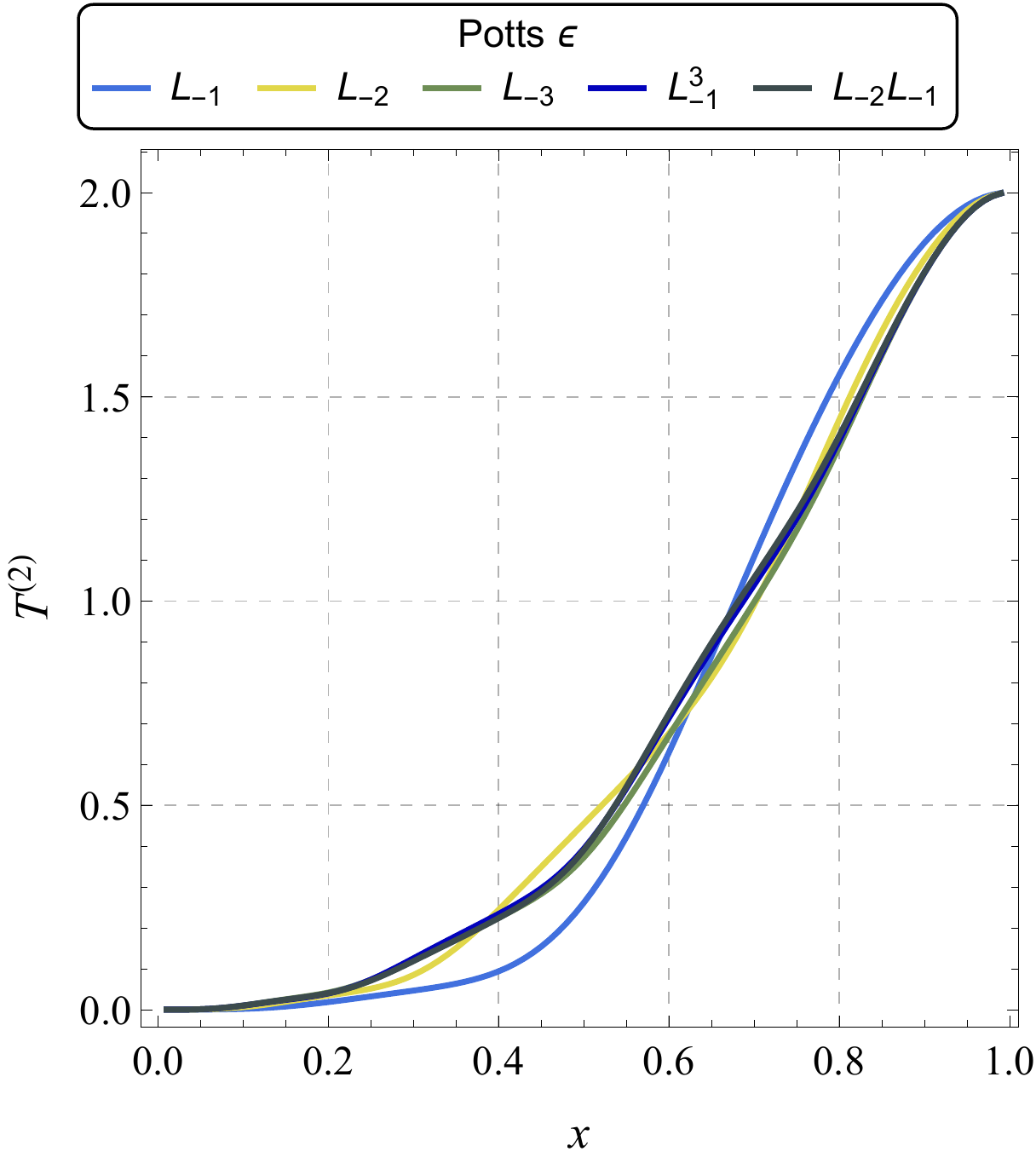}};
    \end{tikzpicture}
    \caption{TSD between different descendants and their primary state $\ket{\varepsilon}$ and  $\ket{\sigma}$ in the Ising model and $\ket{\varepsilon}$ in the Potts model.}
    \label{fig:TSDprim}
\end{figure}

In the Ising model we find that degenerate states of the energy density produce the same TSD w.r.t.~the primary state up to level~3. This again is as expected.
For spin descendants instead this is not true at level~3, with $T^{(2)}_{L_{-3}\ket{\Delta},\ket{\Delta}} \neq T^{(2)}_{L_{-1}^3\ket{\Delta},\ket{\Delta}} \neq T^{(2)}_{L_{-2}L_{-1}\ket{\Delta},\ket{\Delta}}$.
However, in the small and large subsystem size limits we see that these different expressions have the same behaviour, while they differ the most around $x\sim 1/2$.
In the Potts model, TSDs between degenerate states at level~3 and the energy density are again different, but from the plots we see that the difference is barely visible, and in particular for $x\to0$ and $x\to 1$ it is negligible.

If we study the small subsystem size limit, we can generically predict the behaviour of the TSD.
Consider for instance the OPE between two primary states as given by~\eqref{eq:ope_srd} and the correlator as in~\eqref{eq:opecorr_srd}. 
Then, we find the following behaviour in the limit $x\to 0$ for $n=1,2,3$
\begin{equation}\label{eq:tsd_smallx}
    T^{(2)}_{L_{-n}\ket{\Delta},\ket{\Delta}} = \frac{2+c}{16 c} n^2 (\pi x)^4 + \mathcal{O}(x^6)
\end{equation}
in agreement with the vacuum result in~\eqref{eq:tsd_vacuum_smallx} and with the analytic expressions that we found in Ising and Potts models for energy descendants.
However, for $\sigma$ descendants in the Ising model the leading-order contribution as $x\to 0$ does not come from the stress tensor but from the energy field $\varepsilon$ in the OPE.
Indeed, consider again the OPE as in~\eqref{eq:ope_light} with the correlator~\eqref{eq:corr_light}, then the contribution to the TSD as $x\to 0$ for $n=1,2,3$ reads
\begin{align}\label{eq:tsd_smallx_nonvacchannel}
    T^{(2)}_{L_{-n}\ket{\Delta},\ket{\Delta}} = \left(C_{hh}^{h_k}\right)^2 \left( \frac{2 n^2 (h_k-1)h_k }{c (n-1)^2 + 4 n h} \right)^2 \left(\frac{\pi x}{2} \right)^{4 h_k} + \ldots \, .
\end{align}

\noindent
We see that this term dominates over the one one outlined in~\eqref{eq:tsd_smallx} for $h_k < 1$, which is the case for the Ising spin.
We checked that~\eqref{eq:tsd_smallx_nonvacchannel} with $C^\varepsilon_{\sigma\sigma}=1/2$ matches the small $x$ behaviour of the results for $\sigma$ presented in appendix~\ref{app:TSDising}.

Consider now the large subsystem size limit $x \to 1$. 
Then, with our coordinates we have $z_1 \sim z_4$ and $z_2 \sim z_3$ and by taking the OPE similarly as in~\eqref{eq:ope_srd} but with appropriate insertion points we find the behaviour
\begin{equation}
    T^{(2)}_{L_{-n}\ket{\Delta},\ket{\Delta}} = 2 - \left(2h + \frac{n}{2}\right) \pi^2 (x-1)^2 + \ldots \, ,
\end{equation}
which agrees with the $x\to 1$ limit of the explicit results we found for descendants of the energy in Ising and Potts.
Again, for $\sigma$ descendants we need to take into account the contribution from the lightest field in the OPE.
We then find 
\begin{equation}\label{eq:tsd_largex}
    T^{(2)}_{L_{-n}\ket{\Delta},\ket{\Delta}} = 2 \left(C^{h_k}_{hh}\right)^2 C_n \left( \frac{\pi}{2} \right)^{4 h_k} (x-1)^{4 h_k}
\end{equation}
where
\begin{align}
    C_n &= \frac{c^2 (n-1)^4 + 4 c n(n-1) \left( 2(n-1) h + (1 - 2^{n-1}) h_k \right)   }{\left( c(n-1)^2 + 4 n h \right)^2} \nonumber\\
    &\phantom{=} + \frac{(4n)^2 h^2 - (2n)^3 h h_k + 2n^4(h_k -1)^2 h_k^2}{\left( c(n-1)^2 + 4 n h \right)^2} \, .
\end{align}
For $\sigma$ in the Ising model $h_k = 1/2$ and we see that the contribution from the $\varepsilon$ channel sums up with the leading correction in~\eqref{eq:tsd_largex}.
Once this is taken into account, we correctly match the large $x$ limit of the $\sigma$ expressions in appendix~\ref{app:TSDising}.
\chapter{Weyl fermions and trace anomalies}\label{chap:weyl_fermions}

In this chapter we study the trace anomalies of Weyl fermions in four dimensions, first in a gauge background and then in a gravitational background.
Our interest is in the Pontryagin term~\eqref{pontryagin_4} presented in sec.~\ref{sec:intro_trace_ano}, that, as we already stated, has never appeared in the trace anomalies of known CFTs, until its presence in the case of Weyl fermions was first claimed in~\cite{Bonora:2014qla} and then confirmed in~\cite{Bonora:2015nqa,Bonora:2017gzz,Bonora:2018obr}.
However, an independent analysis could not detect this term~\cite{Bastianelli:2016nuf} and claimed for additional studies.
Here we present our contribution to this subject.

In sec.~\ref{sec:weyl_fermions_gauge} we first consider an analogous and simplified example, namely that of a Weyl fermion in a non-abelian gauge background, whose trace anomaly could receive contribution from a parity-odd term proportional to the Chern-Pontryagin density of the gauge background.
To ascertain this possibility, we consider the Bardeen model of Dirac fermions coupled to non-abelian gauge fields~\cite{Bardeen:1969md} and compute all chiral and trace anomalies with the PV regularisation scheme presented in sec.~\ref{sec:regulators}.
The results for the Weyl fermions are then obtained by taking suitable limits on the background.
We reproduce the already known chiral anomalies to test our scheme, and then compute the trace anomaly, finding no parity-odd term in it.
Our result for the trace anomaly generalises to the non-abelian case the expression found in~\cite{Bastianelli:2018osv}.

In sec.~\ref{sec:weyl_mat} we consider the metric-axial-tensor (MAT) background constructed in~\cite{Bonora:2017gzz,Bonora:2018obr} that generalises to curved space the approach used by Bardeen.
We study a Dirac fermion coupled to this background and compute the anomalies again in the PV regularisation scheme.
Suitable limits on the background eventually allow us to obtain the anomalies for the Weyl fermions in a curved spacetime, and again we find that the Pontryagin density does not contribute to the trace anomaly.

We collect in app.~\ref{app:weyl_fermion} our conventions for Dirac and Weyl fermions and more details on the calculations.

The results of this chapter were first published in~\cite{Bastianelli:2019fot,Bastianelli:2019zrq}, whose presentation we follow.

\section{Non-abelian gauge background}
\label{sec:weyl_fermions_gauge}
In this section we compute the chiral and trace anomalies of Weyl fermions coupled to non-abelian gauge fields in four dimensions.
To do so, we consider the Bardeen model~\cite{Bardeen:1969md}, which embeds the Weyl theory into the theory of massless Dirac fermions $\psi$ coupled to vector and axial  
non-abelian gauge fields, $A_a$ and $B_a$.
This model is classically gauge invariant and conformally invariant. We wish to compute systematically the anomalies.
The chiral anomaly was already studied in~\cite{Bardeen:1969md} and we recompute it to test our methods. The main result will be to obtain the trace anomaly. 

\subsection{Bardeen model}
The lagrangian of the Bardeen model is given by
\begin{equation}
{\cal L} = -\bpsi \gamma^a \left( \partial_a + A_a+ B_a\gamma^5 \right) \psi = -\bpsi \gamma^a D_a(A,B) \psi = -\bpsi \DDslash(A,B)\psi  \,,
\label{lag}
\end{equation}
where $D_a(A,B)$ is the covariant derivative for the gauge group $G\times G$. 
We  expand the gauge fields on the generators of the gauge group as $A_a= -i A_a^\alpha T^\alpha$ and $B_a= -i B_a^\alpha T^\alpha$. 
The components  $A_a^\alpha$ and $B_a^\alpha$ are real, 
and $T^\alpha$ denote the hermitian generators in the representation of $G$ chosen for $\psi$.
The  generators satisfy the Lie algebra $ [T^\alpha, T^\beta] = i f^{\alpha \beta \gamma}T^\gamma$ with anti-symmetric structure constants $f^{\alpha \beta \gamma}$.
Taking an appropriate limit on the background, i.e.~by setting $A_a=B_a \to {A_a}/{2}$, a chiral projector arises and we are left with the theory of left-hended Weyl fermions, which is the one we are eventually interested in (see app.~\ref{app:fermion_conventions} for our conventions on Dirac and Weyl fermions).

Let us first review the classical symmetries. The lagrangian is invariant under the $G\times G$ gauge transformations. 
Using infinitesimal parameters $\alpha=-i \alpha^\alpha T^\alpha $ and $\beta=-i \beta^\alpha T^\alpha $, they read
\begin{equation} 
\begin{aligned}
  \delta \psi  &= - \left( \alpha + \beta \gamma^5 \right)\psi \,,  \\
  \delta \bpsi & = \bpsi \left(\alpha -  \beta \gamma^5\right) \,,\\
  \delta \psi_c & =   \left(\alpha^T  -\beta^T \gamma^5\right) \psi_c \,, \\
  \delta  A_a  &=       \partial_a  \alpha + [A_a, \alpha] +[B_a, \beta]  \,, \\
   \delta  B_a  &=   \partial_a  \beta + [A_a, \beta]  +  [B_a, \alpha] \,,
\end{aligned}
\label{gauge-tra}
\end{equation}
where $\psi_c = C^{-1} \bpsi^T$ is the charge conjugated spinor.
The transformations of the gauge fields can be written more compactly as 
\begin{equation}
\delta {\cal A}_a  =   \partial_a  \tilde \alpha + [{\cal A}_a, \tilde \alpha] \,,
\end{equation}
where ${\cal A}_a = A_a +B_a\gamma^5$ and $ \tilde \alpha = \alpha +\beta \gamma^5$.
The corresponding field strength 
\begin{equation}
\mathscr{F}_{ab} = \partial_a  {\cal A}_b -\partial_b  {\cal A}_a  +  [{\cal A}_a , {\cal A}_b] = \hat F_{ab} + \hat G_{ab}\gamma^5 
\end{equation}
contains  the Bardeen curvatures  $\hat F_{ab}$ and $\hat G_{ab}$ 
\begin{equation}
\begin{aligned}
\hat F_{ab} &=  \partial_a  A_b -\partial_b  A_a  +  [A_a , A_b] +   [B_a , B_b] \,, \\
\hat G_{ab} &=  \partial_a  B_b -\partial_b  B_a  +  [A_a , B_b]  +  [B_a , A_b] \,.
 \end{aligned}
 \label{Bardeen-curvatures}
\end{equation} 
In the following we prefer to use the more explicit notation with $\gamma^5$.

We can use $A_a^\alpha$ and $B_a^\alpha$  as sources for the vector 
$J^{a\alpha}= i \bpsi \gamma^a T^\alpha \psi$ and axial $J_5^{a\alpha}= i \bpsi \gamma^a \gamma^5 T^\alpha\psi$
currents, respectively.
These currents are on-shell covariantly conserved, with the conservation law reading  
\begin{equation}
\begin{aligned}
\left(D_a J^a\right)^\alpha = 
\partial_a J^{a\alpha} - i \bpsi [\Aslash+\Bslash \gamma^5, T^\alpha] \psi =0 \,, \\
\left(D_a J^a_5\right)^\alpha = 
\partial_a J_5^{a\alpha} - i \bpsi [\Aslash \gamma^5+\Bslash, T^\alpha] \psi =0 \,,
\end{aligned}
\end{equation}
or, equivalently, as 
\begin{equation}
\begin{aligned}
& \left(D_a J^a\right)^\alpha = 
\partial_a J^{a\alpha} + f^{\alpha\beta \gamma} A_a^\beta J^{a\gamma} + f^{\alpha\beta \gamma} B_a^\beta J_5^{a\gamma} =0 \,,\\
& \left(D_a J^a_5\right)^\alpha=  
\partial_a J_5^{a\alpha} + f^{\alpha\beta \gamma} A_a^\beta J_5^{a\gamma} + f^{\alpha\beta \gamma} B_a^\beta J^{a\gamma} =0 \,.
\label{gauge-conservation}
\end{aligned}
\end{equation}
Indeed, under an infinitesimal gauge variation of the external sources  $A_a^\alpha$ and $B_a^\alpha$, 
the action $S =\int d^4 x\, {\cal L}$ varies as 
\begin{equation}
 \delta^{(A,B)} S = - \int d^4 x \,\left( \alpha^\alpha  \left(D_a J^a\right)^\alpha +   \beta^\alpha  \left(D_a J_5^a\right)^\alpha \right)   \;,
 \label{uno}
\end{equation}
and the gauge symmetries imply that $J^{a\alpha}$ and $J_5^{a\alpha}$ are covariantly conserved on-shell, as stated above.

In order to study the stress tensor, we couple the theory to gravity by introducing
the vierbein $e_\mu{}^a$  and related spin connection (see app.~\ref{app:vielbein}).
Then the new lagrangian becomes
\begin{equation}
{\cal L} = - e\bpsi 
\gamma^a e_a{}^\mu \left( \partial_\mu + A_\mu +B_\mu \gamma^5 +\frac14 \omega_{\mu ab}\gamma^{ab} \right) \psi = - e\bpsi 
\gamma^a e_a{}^\mu \nabla_\mu\psi  \,,
\label{lag2}
\end{equation}
where $e$ is the determinant of the vierbein, $e_a{}^\mu$ its inverse, and 
$\nabla_\mu $ the covariant derivative that, since it acts on the fermion field, also contains the spin connection arising from the vierbein.
One may verify that, on top of the background general coordinates and local Lorentz symmetries, 
the model acquires also a Weyl invariance, which through eq.~\eqref{vielbein} translates into an invariance under 
arbitrary local scalings of the vierbein. 
This suffices to prove conformal invariance in flat space, as we discussed in sec.~\ref{sec:weyl_symm}.
In fermionic theories, the vierbein is used also as an external source for the stress tensor, which we define as
\begin{equation}
T^{\mu a} = \frac{1}{e}\frac{\delta S}{\delta e_{\mu a}} \;.
\end{equation}
The Weyl symmetry implies that the stress tensor is traceless on-shell.
Indeed, an infinitesimal  Weyl transformation with local parameter $\sigma $ is of the form
\begin{equation} 
	\begin{aligned}
	    \delta \psi  &= - \frac{3}{2}\sigma  \psi \,,  \\
        \delta  A_\mu  =  \delta  B_\mu  &=  0 \,,\\
        \delta e_\mu{}^a & =  \sigma e_\mu{}^a \,,
		\end{aligned}
	\label{weyl-tra}
\end{equation}
and  varying the action under an infinitesimal Weyl transformation of the vierbein only produces the trace of the stress tensor
\begin{equation}
\delta^{(e)} S = \int d^4 x e \, \sigma   T^a{}_a \;.
\label{due}
\end{equation}
Then, the full Weyl symmetry of the action implies that the trace vanishes on-shell, $T^a{}_a =0$.
On top of that, on-shell the stress tensor can be shown to be symmetric, once the curved index is made flat 
with the vierbein ($T^{ab}=T^{ba}$),  while satisfying a suitable conservation law.
For completeness, we report the stress tensor's main properties relevant for our model and its explicit expression in app.~\ref{appB}.

\subsection{PV regularization} 

To regulate the one-loop effective action we introduce massive PV fields.  The mass term produces the anomalies, 
which we will compute with heat kernel methods following the scheme outlined in sec.~\ref{sec:regulators}. 

We denote by $\psi$ the PV fields as well (as this should not cause any confusion)
and add  a Dirac mass term 
\begin{equation}
\Delta {\cal L} = - M \bpsi  \psi =   \frac{M}{2} \left(\psi^T_c C \psi +\psi^T C \psi_c\right) 
\label{mass}
\end{equation}
to the massless lagrangian of the PV field which is as in~\eqref{lag}.
The Dirac mass preserves  vector gauge invariance but breaks axial  gauge invariance. Indeed under \eqref{gauge-tra}
the mass term varies as
\begin{equation}
\delta \Delta {\cal L} = 2 M \bpsi \beta \gamma^5 \psi = - M  \left(\psi^T_c \beta C \gamma^5 \psi 
+\psi^T \beta^T C \gamma^5 \psi_c\right) \,,
\end{equation}
where we remind that $\beta =-i\beta^\alpha T^\alpha$, and it shows that the vector gauge symmetry is preserved, leaving room for an  anomaly in the axial gauge symmetry.

The mass term sources also a trace anomaly, as the curved space version of \eqref{mass}
\begin{equation}
\Delta {\cal L} = - e M \bpsi  \psi =   \frac{e M}{2} \left(\psi^T_c C \psi +\psi^T C \psi_c\right)
\label{cov-mass}
\end{equation}
varies under the infinitesimal Weyl transformation \eqref{weyl-tra} as
 \begin{equation}
\delta \Delta {\cal L} = - e \sigma M \bpsi  \psi =   \frac{e \sigma M}{2} \left(\psi^T_c C \psi +\psi^T C \psi_c\right) \;.
\end{equation}
However, the mass term preserves the general coordinate and local Lorentz symmetries.
We conclude that only axial gauge and trace anomalies are to be expected.

Casting the PV lagrangian   ${\cal L}_{\scriptscriptstyle PV} = {\cal L} + \Delta{\cal L}$
in the form 
\begin{equation} 
{\cal L}_{\scriptscriptstyle PV} =  \frac12 \phi^T T  {\cal O} \phi +\frac12 M \phi^T T \phi \,,
\label{PV-lag}
\end{equation} 
where  $ \phi  = \left( \begin{array}{c}    \psi \\  \psi_c    \end{array} \right)  $, allows us to recognize the operators\footnote{We prefer to use $\psi_c = C^{-1} \bpsi^T$ rather than $\bpsi$, as the former has the same index structure of   
 $\psi$, and thus lives in the same spinor space.\label{footnote:psic}}
\begin{equation}
T {\cal O} =
\left( \begin{array}{cc}
  0 &  C \DDslash\left(-A^T,B^T\right)    \\   C \DDslash(A,B)  &0  
\end{array} \right)  \,, \qquad 
T  = \left( \begin{array}{cc}    0& C  \\  C &  0    \end{array} \right) \,, 
\end{equation}
and 
\begin{equation}
{\cal O} =
\left( \begin{array}{cc}
  \DDslash(A,B)   &0  \\  0&  \DDslash\left(-A^T,B^T\right)  
\end{array} \right)  \,, \qquad 
{\cal O}^2 =
\left( \begin{array}{cc}
  \DDslash^2(A,B)   &0  \\  0 &  \DDslash^2\left(-A^T,B^T\right) 
\end{array} \right)  \,.
\label{PVD-Dmass}
\end{equation}
The latter identifies the regulators, as discussed in sec.~\ref{sec:regulators}.
However, when deriving the formulae~\eqref{reg_var}, \eqref{tra_scalar}, \eqref{3.8}, \eqref{tra} for a fermionic theory, an overall minus sign appears in the final results due to the Grassmannian nature of the field.
Thus, the formula that will be relevant in the present context for the anomaly calculations is 
\begin{equation}\label{tra_fermion}
i \la\delta S\ra  =
- \lim_{M \to \infty}   {\rm Tr} \left[J e^{ -\frac{\mathcal R}{M^2}}\right] \,.
\end{equation}
We recall that 
\begin{equation}
J=K + \frac12 T^{-1} \delta T + \frac12 \frac{\delta {\cal O}}{M} \,, \qquad {\mathcal R}=-{\cal O}^2 \,,
\label{jac_fermion}
\end{equation}
where $K$ is read from the field transformation $\delta \phi = K \phi$ which is a symmetry of the massless action. 
Thus, in the case of the Bardeen model, for each of the symmetries we outlined we can construct the corresponding $J$ and then derive the heat kernel traces that compute the anomalies.

To start with, the vector current $J^{a\alpha}$ remains covariantly conserved also at the quantum level, as the PV mass term is invariant under vector gauge transformations.

For the axial current, recalling the transformation laws in \eqref{gauge-tra}, we find 
\begin{equation}
J= \left( \begin{array}{cc}   i\beta^\alpha T^\alpha \gamma^5 
& 0  \\  0&  i\beta^\alpha T^{\alpha T} \gamma^5       \end{array} \right)  
\end{equation}
as $\delta T$ vanishes, while we discarded the contributions from $\delta {\cal O}$ as they would vanish under the Dirac trace.
Here, $T^{\alpha T} $ denotes the  transposed of  $T^{\alpha} $. 
Considering now~\eqref{tra_fermion}, \eqref{uno} and~\eqref{hkt} we find
\begin{equation}
(D_a \la J^a_5 \ra)^\alpha
=   
\frac{i}{(4\pi)^2} \left\{\tr \left[\gamma^5 T^\alpha a_{2} ({R}_\psi)\right] +\tr \left[\gamma^5 T^{\alpha T} a_{2} ({R}_{\psi_c}) \right] \right\} \,,
\label{Bardeen-anomaly}
\end{equation}
where the remaining trace is the finite dimensional one on the gamma matrices and color space.
Here, we  find the Seeley-DeWitt coefficients $a_2({R_i})$ corresponding to the regulators $R_i$ 
associated to  the  fields assembled into $\phi$ that we read from~\eqref{jac_fermion} and~\eqref{PVD-Dmass}, explicitly
\begin{equation}
{R}_\psi   = -  \DDslash^2(A,B) \;,
\qquad
{R}_{\psi_c} = - \DDslash^2\left(-A^T,B^T\right) \;.
\label{regu}
\end{equation}
 The  $a_2$  coefficients are the only ones that survive renormalization and the limit ${M \to \infty}$.

Similarly, for the Weyl symmetry and the transformations in~\eqref{weyl-tra} we find
\begin{equation}
\la T^a{}_a \ra = -\frac{1}{2 (4\pi)^2}  \left\{ \tr \left[ a_{2}\left({R}_\psi\right)\right]  +\tr \left[ a_{2}\left({R}_{\psi_c}\right)\right] \right\} \,,
\label{trace-anomaly}
\end{equation}
where now also $\delta T$ contributes to \eqref{jac_fermion}, while $\delta {\cal O}$  drops out as before.
Again, all remaining traces are in spinor and color spaces. 

Finally, since the mass term is invariant under general coordinate and local Lorentz transformations, no anomalies arise in those symmetries.

\subsection{Anomalies}

We compute now the anomalies produced by the traces of the heat kernel coefficients  $a_2$ in 
\eqref{Bardeen-anomaly} and \eqref{trace-anomaly}, with the  regulators \eqref{regu}. 
We limit ourselves to show the results here, but we report more details of the computations in app.~\ref{appA}.

The vector symmetry is guaranteed to remain anomaly free by the invariance of the mass term. As a check  
one may verify, using the explicit traces given in appendix \ref{appA}, that the would-be anomaly vanishes 
\begin{equation}
\left(D_a \la J^a\ra\right)^\alpha
=   
\frac{i}{(4\pi)^2} \left\{\tr \left[ T^\alpha a_{2} \left({R}_\psi\right)\right]  -  \tr \left[ T^{\alpha T} a_{2} \left({R}_{\psi_c}\right) \right] \right\} =0 \;.
\end{equation}

\subsubsection{Chiral anomaly} 

Performing the spinor traces in~\eqref{Bardeen-anomaly} results in the following expression for the chiral anomaly 
\begin{align}\label{axial-anomaly2}
\left(D_a \la J^a_5 \ra\right)^\alpha &=
 -\frac{1}{(4\pi)^2} \epsilon^{abcd}\, \tr_{_{\!\! YM}}  T^\alpha
\bigg [  \hat F_{ab} \hat F_{cd}  + \frac13 \hat G_{ab} \hat G_{cd}   \nonumber\\
&\phantom{=}-\frac83 \left(\hat F_{ab} B_c B_d + B_a \hat F_{bc} B_d +  B_a B_b\hat F_{cd}\right)
+\frac{32}{3} B_a B_b B_c B_d  \bigg ]  \nonumber\\
&\phantom{=}+ PETs \,,
\end{align}
where the remaining trace is only in color space.
$PETs$ indicate the parity-even terms that take the form
\begin{align}\label{pet}
PETs  &=
\frac{i}{(4\pi)^2} \tr_{_{\!\! YM}}  T^\alpha
\bigg [ \frac43 D^2 D B +
\frac23 [\hat F^{ab},\hat G_{ab}] +\frac83 [D^a \hat F_{ab}, B^b]  \nonumber\\
&\phantom{=}- \frac43 \{B^2, DB \} + 8 B_a DB B^a +\frac83 \{ \{ B^a,B^b\}, D_a B_b \}
\bigg ]  \,,
\end{align}
where $DB=D_a B^a$, $D^2 = D^aD_a$ and $\{\cdot,\cdot\}$ is the anti-commutator.
They are canceled by the chiral gauge variation of a local counterterm
\begin{equation}
\begin{aligned}
\Gamma_{ct} = \int \frac{d^4x}{(4\pi)^2} \, \tr_{_{\!\! YM}} 
&\bigg [ \frac23 \left(D^aB^b\right)\left( D_a B_b\right)  +4F^{ab}(A)B_aB_b
- \frac83 B^4  +\frac43 B^a  B^b  B_a  B_b  \bigg ]
\end{aligned}
\label{ct1}
\end{equation}
and the remaining expression coincides with the famous result obtained by Bardeen \cite{Bardeen:1969md}.

\subsubsection{Trace anomaly}
Evaluation of the spinor traces in~\eqref{trace-anomaly} produces the trace anomaly 
\begin{equation}
\begin{aligned}
\la T^a{}_a \ra &= \frac{1}{(4\pi)^2}  \tr_{_{\!\! YM}} 
 \bigg [ \frac23 \hat F^{ab} \hat F_{ab} +\frac23 \hat G^{ab} \hat G_{ab} \bigg]  
+ CTTs \,,
\end{aligned}
\label{trace-anomaly2}
\end{equation}
where $CTTs$ are the cohomologically trivial terms
\begin{equation}
 CTTs =  -\frac43 \frac{1}{(4\pi)^2} \tr_{_{\!\! YM}}  \left[ D^2 B^2 +DBDB - \left(D^aB^b\right) \left(D_bB_a\right) - 2  F^{ab}(A) B_a B_b \right] \,.
\end{equation}
 They are canceled by the  Weyl variation of the following counterterm 
\begin{equation}
\begin{aligned}
\bar \Gamma_{ct} = \int \frac{d^4x \sqrt{g}}{(4\pi)^2} \, \tr_{_{\!\! YM}} 
&\bigg [ \frac23 \left( D^\mu B^\nu \right)\left( D_\mu B_\nu \right)  +4F^{\mu\nu}(A)B_\mu B_\nu +\frac13 R B^2
\bigg ] \,,
\end{aligned}
\label{ct2}
\end{equation}
where $\mu, \nu$ are curved indices, and $R$ the Ricci scalar. Of course, one restricts to flat space after variation.

The counterterms \eqref{ct1} and  \eqref{ct2} merge  
consistently into a unique counterterm that in curved space reads
\begin{equation}
\Gamma^{tot}_{ct} = \int \frac{d^4x \sqrt{g}}{(4\pi)^2} \, \tr_{_{\!\! YM}} 
\bigg [ \frac23 (D^\mu B^\nu)^2  +4F^{\mu\nu}(A)B_\mu B_\nu +\frac13 R B^2
- \frac83 B^4  +\frac43 B^\mu  B^\nu  B_\mu  B_\nu  
\bigg ]\,.
\end{equation}

We notice that, on top of the complete gauge invariance of the trace anomaly, there is no parity-odd term present.

\subsection{Chiral and trace anomalies of Weyl fermions}

We are now ready to study the chiral limit of the Bardeen model, and identify the chiral and trace anomalies
of Weyl fermions.
We take the limit $A_a=B_a \to A_a/2$, which creates a chiral projector in the coupling to the gauge field.
Then, $\hat F_{ab} = \hat G_{ab} \to  F_{ab}(A) /2$ 
and $J^a=J^a_5 \to J_a= \frac12(J^a+J^a_5) $, so that from  \eqref{axial-anomaly2} and \eqref{trace-anomaly2}  we immediately derive the searched for anomalies for the  
left-handed Weyl fermions  coupled to non-abelian gauge fields
\begin{align}
(D_a \la J^a \ra)^\alpha &=
-\frac{1}{(4\pi)^2} \epsilon^{abcd} \, \tr_{_{\!\! YM}}  T^\alpha \partial_a 
\bigg [ \frac23 A_b\partial_c  A_d  + \frac13 A_b A_c  A_d  \bigg ] \,, 
\\
\la T^a{}_a \ra &= \frac{1}{(4\pi)^2}  \tr_{_{\!\! YM}} 
 \bigg [ \frac13  F^{ab}  F_{ab} \bigg ]  \,,
\end{align}
where we discarded the cohomologically trivial terms.

The chiral anomaly is the standard one, rederived as a check on the methods used here.
The trace anomaly is our new result, that verifies 
the absence of parity-odd terms. It is just half the trace anomaly of non-chiral Dirac fermions.

Thus, we have computed the consistent anomalies for Dirac and Weyl fermions, 
defined as arising from the symmetry
variation of an effective action. In particular, we find that the consistent trace anomaly acquires 
a gauge invariant form.  
The property of gauge invariance of the trace anomaly is not explicitly maintained by
our regulator (as far as the axial gauge symmetry is concerned),
but the breaking terms can be removed by
the variation of a local counterterm, as we have indicated.


\section{Axial gravity background}\label{sec:weyl_mat}

Following the approach by Bardeen to analyse the anomalies of chiral fermions by taking appropriate limits on the gauge background to which Dirac fermions are coupled to,
a metric-axial-tensor (MAT) background has been constructed in~\cite{Bonora:2017gzz, Bonora:2018obr}.
In particular, the MAT background generalizes to curved space the approach used by Bardeen:
Dirac fermions are coupled to a metric-axial-tensor defined as
\begin{equation}\label{axial_metric}
\hat g_{\mu\nu} = g_{\mu\nu} +\gamma^5 f_{\mu\nu}  \,,
\end{equation}
and suitable limits of the MAT background reproduce different theories of (chiral) fermions.
In the present section we will study the classical symmetries of this model and compute its anomalies.

\subsection{The MAT model}

The axial extension of the metric~\eqref{axial_metric} induces similar (i.e~with a $\gamma^5$ component) extensions to the other geometrical quantities, like the vierbein $\hat e_\mu{}^a$
 and the spin connection  $\hat \omega_{\mu ab}$.
 Thus, a massless Dirac fermion coupled to the MAT background has a lagrangian of the form
\begin{equation}
 \mathcal{L} = -   \bpsi \gamma^a \sqrt{\hat g} \hat e_a{}^\mu \left( \partial_\mu + \frac 14 \hat \omega_{\mu ab}\gamma^{ab} \right) \psi = -   \bpsi \gamma^a \sqrt{\hat g} \hat e_a{}^\mu \hat \nabla_\mu \psi \,,
 \label{lag_axial}
\end{equation}
with $\hat \nabla_\mu$ the MAT covariant derivative.
All quantities with a hat  contain an axial extension with $\gamma^5$ and appear
always sandwiched between the Dirac spinors $\bpsi$ and $\psi$.
For a comprehensive presentation of the MAT model we refer to  \cite{Bonora:2017gzz, Bonora:2018obr}. 

For our purposes it is more convenient to split the Dirac fermion $\psi$ into its two independent and Lorentz 
irreducible chiral components $\lambda$ and $\rho$ of opposite chiralities, $\psi = \lambda + \rho$ with $\gamma^5$-eigenvalues $ \gamma^5 \lambda =\lambda$ and $ \gamma^5 \rho =- \rho$.
Then, the lagrangian takes the form
\begin{equation}
 \mathcal{L} = -  \sqrt{g_+}\, \blambda \gamma^a  e^{+\mu}_a \nabla^+_\mu \lambda 
 - \sqrt{g_-}\, \brho \gamma^a e^{-\mu}_a \nabla^-_\mu \rho  \,,
 \label{Dirac-mat}
\end{equation}
where $g^\pm_{\mu\nu} = g_{\mu\nu} \pm f_{\mu\nu} $ are two different effective metrics, with
related compatible vierbeins, spin connections, and covariant derivatives
(which we indicate with the $\pm$ sub/superscripts).
This happens since the $\gamma^5$ matrices acting on chiral fermions  are substituted by 
the corresponding eigenvalues.
One could be more general,  allowing also for the spacetime points to have an axial extension, 
as outlined in \cite{Bonora:2018obr}, but the present formulation   is  sufficient for our purposes.

The limit $f_{\mu\nu}  = 0$  recovers the standard massless Dirac fermion in the metric $g_{\mu\nu}$.
Setting  $g_{\mu\nu}  =  f_{\mu\nu}\to  g_{\mu\nu}/2 $  produces instead a left handed chiral
fermion $\lambda$ coupled to the final metric $g_{\mu\nu}$, while the other chirality is projected out since it would remain coupled to the metric $g^-_{\mu\nu}=0$.
A less singular limit, which keeps a free propagating  right-handed fermion, consists in setting $g^+_{\mu\nu}  = g_{\mu\nu}$ and  $g^-_{\mu\nu}  = \eta_{\mu\nu}$, 
with  $\eta_{\mu\nu}$  the flat Minkowski metric.
  
The Dirac theory in the MAT background has several symmetries which may become anomalous, 
and limits on the background can be used to recover the anomalies of a chiral fermion, as in the Bardeen method.
The classical symmetries of the model are: diffeomorphisms, 
local Lorentz transformation and Weyl rescalings, 
together with their axial extensions, all of which are background symmetries since 
they act on the MAT fields as well. In addition, the model admits
global vector and axial $U(1)$ symmetries, that rotate the spinors by a phase. 
This  global $U(1)_V\times U(1)_A$ symmetry group does 
 not act on the MAT background or any other background, 
 as for simplicity we do not couple the model to the corresponding 
abelian gauge fields, although that could be done as well. 
We will review shortly these symmetries, compute systematically 
all of their anomalies, and then study the chiral limit.

Let us start by considering invariance under diffeomorphisms and their axial extension.
The change of coordinates $x^\mu \to x^\mu -\xi^\mu(x)$ induce the standard transformation law on the fields as generated by the Lie derivative ${\cal L}_\xi$
\begin{equation}
\begin{aligned} 
 \delta e^{\pm a}_\mu = {\cal L}_{\xi} e^{\pm a}_\mu  &= \xi^\nu \partial_\nu e^{\pm a}_\mu + (\partial_\mu \xi^\nu)    e^{\pm a}_\nu \,, \\
 \delta \psi ={\cal L}_\xi \psi &= \xi^\mu \partial_\mu \psi \,,
\end{aligned}
\end{equation}
which are a symmetry of the action defined with the lagrangian~\eqref{lag_axial}.
However, we can also define chiral transformation rules that leave the action invariant.
We may define left infinitesimal diffeomorphisms generated by a vector field $\xi_+^\mu(x)$
\begin{equation}\label{left-diffeo}
\begin{aligned} 
\delta e^{+ a}_\mu &=  {\cal L}_{\xi_+} e^{+ a}_\mu \,,\\
\delta \lambda &=  {\cal L}_{\xi_+} \lambda \,,\\ 
\delta e^{- a}_\mu &=  0 \,,\\
\delta \rho &=  0 \,,
\end{aligned}
\end{equation}
and right infinitesimal diffeomorphisms generated by a vector field $\xi_-^\mu(x)$
\begin{equation}\label{right-diffeo}
\begin{aligned} 
\delta e^{+ a}_\mu &=  0 \,,\\
\delta \lambda &=  0 \,,\\
\delta e^{- a}_\mu &=  {\cal L}_{\xi_-} e^{- a}_\mu \,,\\ 
\delta \rho &=   {\cal L}_{\xi_-} \rho  \,.
\end{aligned}
\end{equation}
It is only the sum with local parameters identified, i.e.~with  $\xi^\mu =\xi_+^\mu =\xi_-^\mu$, that
plays the role of the geometrical transformation induced by the translation of the 
spacetime point $x^\mu$ described above. 
Nevertheless, they are independent symmetries of the action.
They acquire a clear geometrical meaning once the spacetime point $x^\mu$ is extended  to have
an axial partner \cite{Bonora:2018obr}, but we do not need to do that for the scope of the present  
analysis. These symmetries
imply that the stress tensor and its axial partner satisfy suitable covariant conservation laws.
Indeed, by varying the action under the two effective vierbeins  $e^\pm_{\mu a}$ we can define two stress tensors as
\begin{equation}\label{mat_T}
T_\pm^{\mu a}(x) = \frac{1}{\sqrt{g_\pm}}\frac{\delta S} {\delta e^\pm_{\mu a} (x)}\,.
\end{equation}
As we will comment in the next section, our regularization will not spoil these symmetries and no anomaly will arise in the conservation of the stress tensor at the quantum level.

The action is also invariant under the local Lorentz symmetries 
that act independently on the $+$ (left) and $-$ (right) sector. 
On the $+$ sector, the  left-handed  local Lorentz symmetry acts by
   \begin{equation}
  \begin{aligned} 
  \delta e^{+ a}_\mu &=   \omega^{+a}{}_b e^{+ b}_\mu \,, \cr
  \delta \lambda  & =  \frac14 \omega^+_{ab} \gamma^{ab}\lambda \,,\cr
   \delta e^{- a}_\mu &=  0 \,,\cr
  \delta \rho  & = 0 \,,
\label{l-lL}
   \end{aligned}
 \end{equation}
 where $\omega^+_{ab}= - \omega^+_{ba}$ are local parameters. 
  Similarly, on the right sector we have  
\begin{equation}
  \begin{aligned} 
    \delta e^{+ a}_\mu &=  0 \,, \cr
    \delta \lambda  & =  0 \,, \cr
        \delta e^{- a}_\mu &=   \omega^{-a}{}_b e^{- b}_\mu \,, \cr
 \delta \rho  & =  \frac14 \omega^-_{ab} \gamma^{ab}\rho \,.
\label{r-lL}
   \end{aligned}
 \end{equation}
These transformations leave the action invariant, so that the stress tensor and its axial companion are symmetric.

As for the Weyl and axial Weyl symmetries, it is again more convenient to consider their $\pm$ linear combinations, that act separately on the chiral sectors of the theory.
  The infinitesimal Weyl symmetries are defined by
 \begin{equation} 
\begin{aligned}
\delta e_\mu^{\pm a}  & = \sigma^\pm e_\mu^{\pm a} \,,\cr
\delta \lambda  & = -\frac32 \sigma^+ \lambda \,,\cr
\delta \rho  & = -\frac32 \sigma^- \rho \,,
\end{aligned}
\label{Weyl-tra-rules}
\end{equation}
where $\sigma^\pm$ are the two independent Weyl local parameters.
Invariance of the action under these transformations implies tracelessness of the stress tensor and its axial partner.

Finally, for convenience also in the case of the vector and axial $U(1)$ symmetries we study their $\pm$ linear combinations, which produce global $U(1)_L\times U(1)_R$ symmetries.
They have the following transformation rules 
\begin{equation} 
\begin{aligned}
\delta \lambda  & = i\alpha^+ \lambda \,, \cr
\delta \rho  & = i\alpha^- \rho \,, \cr
\end{aligned}
\label{u1-tra-rules}
\end{equation}    
where $\alpha^\pm$ are independent infinitesimal parameters.
By Noether theorem we find the covariantly conserved Noether currents
\begin{equation} \label{u1_currents}
\delta S = \int d^4 x \sqrt{g_+}\, \alpha_+ \nabla^+_\mu J_+^\mu + \int d^4 x \sqrt{g_-}\, 
\alpha_- \nabla^-_\mu J_-^\mu \,,
\end{equation}
where the constants $\alpha_\pm$ in \eqref{u1-tra-rules} are extended to arbitrary functions,
with the currents taking the explicit form
$J_+^\mu= i\blambda \gamma^a  e^{+\mu}_a \lambda$ and  
$J_-^\mu= i\brho \gamma^a  e^{-\mu}_a \rho$.

\subsection{PV regularization}
We regulate the one-loop effective action of the Dirac theory in the MAT background with massive PV fields.
As for Bardeen model, the mass term produces the anomalies, 
which we will compute with heat kernel methods following the scheme outlined in sec.~\ref{sec:regulators}.

In this case we choose PV fields with a lagrangian of the same form as \eqref{Dirac-mat},
but augmented by a Majorana mass coupled to the MAT background as
  \begin{equation}
  \begin{aligned} 
\Delta_{\scriptscriptstyle M}  {\cal L}
= \frac{M}{2}   \sqrt{g_+}\, \left( \lambda^T C \lambda + h.c.\right) +  \frac{M}{2}  \sqrt{g_-}\,\left( \rho^T C \rho + h.c.\right)  \,,
     \label{majorana-mass}
   \end{aligned}
 \end{equation}
where $h.c.$ denotes hermitian conjugation and $C$ is the charge conjugation matrix.
For notational simplicity we use the same symbols for the PV fields and the original variables,
since no confusion should arise in the following.
The advantage of this specific mass term is that it is invariant under the left and right diffeomorphisms~\eqref{left-diffeo} and~\eqref{right-diffeo},  and thus guarantees 
the absence of gravitational anomalies \cite{Alvarez-Gaume:1983ihn} and the covariant conservation of the stress tensor at the quantum level.

Similarly, the mass term is invariant under the local Lorentz symmetries~\eqref{l-lL} and~\eqref{r-lL} that act independently on the $+$ and $-$ sector.
These are full symmetries of the total PV lagrangian, and the invariance of the regulating fields guarantees that the 
stress tensor and its axial companion remain symmetric at the quantum level.
 
 The only possible anomalies appear in the Weyl and axial Weyl symmetries, and in the vector  and axial $U(1)$ symmetries.
 Under the Weyl transformations in~\eqref{Weyl-tra-rules}, the mass term transforms as follows
  \begin{equation}
\delta \Delta_{\scriptscriptstyle M}  {\cal L}
= \sigma^+ \frac{M}{2}   \sqrt{g_+}\, ( \lambda^T C \lambda + h.c.) + \sigma^- \frac{M}{2}  \sqrt{g_-}\,( \rho^T C \rho + h.c.)  \,,
 \end{equation}
so that it breaks the Weyl symmetries causing anomalies to appear.
The mass term also breaks the vector and axial $U(1)$ symmetry of the massless action, since under~\eqref{u1-tra-rules} it transforms as
 \begin{equation}
\delta \Delta_{\scriptscriptstyle M} {\cal L}
= i\alpha^+ M   \sqrt{g_+}\, ( \lambda^T C \lambda - h.c.) + i \alpha^- M \sqrt{g_-}\,( \rho^T C \rho - h.c.)  \,.
 \end{equation}

Before computing the anomalies, let us cast the lagrangian with the Majorana mass term 
using the  Dirac basis of spinors $\psi$ and $\psi_c$,\footnote{See footnote~\ref{footnote:psic}.} so to recognize the operators in  
\eqref{PV-l} and identify our regulator ${\cal O}^2$. 
 The massless lagrangian \eqref{lag_axial} with the addition of the Majorana mass term
\eqref{majorana-mass}  fixes the PV lagrangian
 \begin{equation}
 \mathcal{L}_{PV}   = \frac12 \psi_c^T C  \sqrt{\bar{\hat{g}}} \hat{\Dslash}  \psi
+ \frac12 \psi^T C  \sqrt{\hat{g}} \bar{\hat{\Dslash}} \psi_c
+ \frac{M}{2} \left( \psi^T \sqrt{\hat g} C \psi +\psi_c^T \sqrt{\bar{\hat{g}}}C \psi_c\right) \,,
\end{equation}
 where a bar indicates a sign change in the axial extension 
(e.g.~$\bar {\hat g}_{\mu\nu} = g_{\mu\nu} -\gamma^5 f_{\mu\nu}$)
and $\hat{\Dslash} = \gamma^a  \hat e_a^\mu \hat \nabla_\mu$, 
 so that on the field basis  $ \phi = \left(    \begin{array}{c} \psi\\     \psi_c    \end{array} \right) $ we recognize
\begin{equation}
    T\mathcal{O} = \left(
    \begin{array}{cc}
        0 & C \sqrt{\hat{g}}  \bar{\hat{\Dslash}} \\
        C \sqrt{\bar{\hat{g}}}  {\hat{\Dslash}} & 0
    \end{array}
    \right) 
       \;, \quad
    T = \left(
    \begin{array}{cc}
\sqrt{\hat{g}} C & 0\\ 
0 &      \sqrt{\bar{\hat{g}}} C  
    \end{array}
    \right) 
    \;, \quad
    {\cal O} = \left(
    \begin{array}{cc}
    0 &  \bar{\hat{\Dslash}} \\ 
{\hat{\Dslash}}  &     0
    \end{array}
    \right) \,,
\end{equation}
and the regulator
\begin{equation}
    {\cal O}^2 = \left(
    \begin{array}{cc}
\bar{\hat{\Dslash}} {\hat{\Dslash}}  & 0\\ 
0 &  {\hat{\Dslash}}  \bar{\hat{\Dslash}} 
    \end{array}
    \right) \,, \quad\quad \mathcal{R} = - \mathcal{O}^2  \,,
\end{equation}
which appear in the heat kernel formulae as in~\eqref{tra_fermion}.
The regulator structure is perhaps more transparent 
when the Dirac fermions are split in their  chiral parts 
\begin{equation}
    \phi = \left(
    \begin{array}{c}
    \lambda\\
    \rho \\
    \rho_c\\
    \lambda_c
    \end{array}
    \right) \,,
\end{equation}
and one recognizes a block diagonal operator
\begin{equation}\label{mat_regulators}
    \mathcal{R} 
    = \left(
    \begin{array}{cccc} R_\lambda &0  &0 &0 \\
        0& R_\rho   &0  &0 \\
  0 & 0& R_{\rho_c}
  &0 \\  0& 0 &0 & R_{\lambda_c}
      \end{array}
    \right) \,,
\end{equation}
with entries
\begin{equation}
\begin{aligned}
R_\lambda &=  -\Dslash^2_+ P_+ \,, \qquad  R_{\lambda_c} = - \Dslash^2_+ P_- \,, \\[2mm]
R_\rho  &= - \Dslash^2_- P_- \,, \qquad  R_{\rho _c} = - \Dslash^2_- P_+\,. 
\end{aligned}
\end{equation}  
where we have used  the left and right  chiral  projectors 
$P_+=P_L = {(\mathbb{I}+\gamma^5)}/{2}$ and $ P_-= P_R = {(\mathbb{I}-\gamma^5)}/{2}$,
and denoted by $\Dslash_\pm= \gamma^a  e^{\pm\mu}_a \nabla^\pm_\mu $ 
the Dirac operators coupled to the $\pm$ effective vierbeins.

Before computing the anomalies, let us stress that the choice of the mass term to be used in the PV sector is quite arbitrary, and it is most convenient to choose the one which preserves the maximal number of symmetries at  the quantum level. 
The choice is essentially between the Dirac and Majorana masses, suitably coupled to the MAT background, but the Majorana mass is the one which breaks the least number of symmetries.
This happens because the Majorana mass keeps a split structure for the couplings of the chiral irreducible components of the Dirac fermion
to the effective metrics $g^\pm_{\mu\nu}$, and produces 
anomalies in the Weyl and $U(1)$ symmetries and their axial extensions only.
We provide more details on the regularisation with a Dirac mass in app.~\ref{app:dirac_mass}.
This mass term turns out to be much less symmetric since it destroys the axial symmetries.
It could be employed as well, but calculations become much more cumbersome, producing more anomalies
then necessary, that eventually must be cured by adding countertems to the effective action. 
However, let us recall once more that any choice of the PV mass term is valid, as  local counterterms can
 be added to the effective action to recover the same final result, independently 
 of the regularization scheme adopted.
This arbitrariness is a general feature of the renormalization process of QFTs.

\subsection{Anomalies}
Let us now compute the anomalies. For the Weyl symmetries we get anomalies 
in the traces of the stress tensors, as defined by
varying the action under the two effective vierbeins $e^\pm_{\mu a}$ in~\eqref{mat_T}.
In each chiral sector we use the corresponding chiral metric, and related vierbein, to perform covariant operations and take traces, and the calculation is just  a double copy of the one 
presented in \cite{Bastianelli:2016nuf}.
Recalling~\eqref{tra_fermion} and~\eqref{jac_fermion}, we identify the structure of the breaking term $J$, entirely due  
to the PV mass. The functional trace is then computed by using 
the heat kernel coefficients $a_2(R_i)$ for the regulators $R_i$ induced by the PV fields. 
For the traces of the stress tensors on the MAT background we find
\begin{equation}
\begin{aligned}
\la T_{+}^\mu{}_\mu \ra &= -\frac{1}{2 (4\pi)^2} \left\{ \tr \left[P_+ a_{2}\left(R_\lambda\right)\right]  +  \tr\left[ P_- a_{2}\left(R_{\lambda_c}\right)\right]\right\} \,, \cr
\la T_{-}^\mu{}_\mu \ra &= -\frac{1}{2 (4\pi)^2} \left\{ \tr \left[P_- a_{2}\left(R_\rho\right)\right]  +  \tr\left[ P_+ a_{2}\left(R_{\rho_c}\right)\right]\right\} \,, \cr
\end{aligned}
\label{trace-an}
\end{equation}
where the remaining final dimensional traces are traces on the gamma matrices. 
The projectors in~\eqref{mat_regulators} can be dropped, as they  get absorbed by the explicit projectors
already present in~\eqref{trace-an}.
Thus, we may use  $R_\lambda = R_{\lambda_c}  =-\Dslash^2_+ $ and $R_\rho =  R_{\rho _c} = - \Dslash^2_-$  to simplify the anomaly expressions to 
\begin{equation}
\la T_{\pm}^\mu{}_\mu \ra = -\frac{1}{2 (4\pi)^2} \tr \left[a_{2}\left( - \Dslash^2_\pm\right)\right] \,,   
\label{trace-an-2}
\end{equation}
and we find the following trace anomalies on the MAT background
\begin{equation}
\la T_{\pm}^\mu{}_\mu \ra =
\frac{1}{720 (4 \pi)^2} \left( 7 R_{\mu \nu \lambda \rho} R^{\mu \nu \lambda \rho} +8 R_{\mu \nu} R^{\mu \nu } -5 R^2 +12 \square R \right)\left(g_\pm\right) \,,
\label{tracepm}
\end{equation}
where the functional dependence on  $g_\pm$ reminds that all the geometrical quantities 
and covariant operations are computed using the effective metric $g^\pm_{\mu\nu}$.
 
We now compute the anomalies in the conservation of the $U(1)_L\times U(1)_R$ currents determined in~\eqref{u1_currents}. 
From the PV regularization we find the following expressions
\begin{equation}
\begin{aligned}
& \nabla^+_\mu \la J^\mu_+ \ra =  \frac{ i }{(4\pi)^2}  \left\{\tr  \left[P_+ a_{2}\left(R_\lambda\right)\right]  -\tr \left[P_- a_{2} \left(R_{\lambda_c}\right)\right]\right\} \,,
\cr 
& \nabla^-_\mu \la J^\mu_- \ra =  \frac{ i }{(4\pi)^2}  \left\{\tr  \left[P_- a_{2}\left(R_\rho\right)\right]  -\tr \left[P_+ a_{2} \left(R_{\rho_c}\right)\right]\right\} \,,
\end{aligned}
\label{chir-an}
\end{equation}
that once more can be simplified to  
\begin{equation}
 \nabla^\pm_\mu \la J^\mu_\pm \ra =  \pm \frac{ i }{(4\pi)^2}  \tr  \left[\gamma^5\, a_{2}\left(-\Dslash^2_\pm\right)\right]  \;.
\end{equation}
Their evaluation in terms of the heat kernel coefficients  produces anomalies proportional 
to the Pontryagin density of the effective metrics
\begin{equation}
\nabla^\pm_\mu \la J^\mu_\pm \ra =  \mp \frac{1}{48 (4\pi)^2} \sqrt{g_\pm}
\epsilon_{\alpha\beta\gamma\delta} R_{\mu \nu}{}^{\alpha\beta}R^{\mu \nu \gamma\delta} (g_\pm)\;.
\label{chiral-an} 
\end{equation}

The expressions \eqref{tracepm} and  \eqref{chiral-an} are our final results for the anomalies of a Dirac fermion on a MAT background. All other symmetries are anomaly free.
We have evaluated these anomalies using traces with chiral projectors of the heat kernel 
coefficient $a_2(-\Dslash^2)$, associated to the covariant square of the Dirac operator 
in a background metric $g_{\mu\nu}$. For completeness, we list this coefficient and related traces 
\begin{align} 
a_{2}(-\Dslash^2)  = &\ \frac{1}{180} \left(R_{\mu \nu \rho\sigma }R^{\mu \nu \rho\sigma }
-R_{\mu \nu}R^{\mu \nu}\right)+  \frac{1}{288}R^2  -\frac{1}{120} \square  R  +\frac{1}{192} 
{\cal R}_{\mu\nu} {\cal R}^{\mu\nu} \,,
\label{hkc}
\\
 \tr [P_\pm a_{2}(-\Dslash^2)]  = 
& -\frac{1}{720} \left( 7 R_{\mu \nu \rho\sigma } R^{\mu \nu\rho\sigma } +8 R_{\mu \nu} R^{\mu \nu } -5 R^2 +12 \square R\right) 
\cr &
\pm \frac{i}{96} \sqrt{g} \epsilon_{\alpha\beta\gamma\delta} R_{\mu \nu}{}^{\alpha\beta}R^{\mu \nu \gamma\delta} \,,
\end{align} 
where ${\cal R}_{\mu\nu} = R_{\mu\nu a b }\gamma^{ab}$.
The coefficient $a_2$ is deduced from~\eqref{hkc2} in the same way outlined in app.~\ref{appA} (see also~\cite{DeWitt:1964mxt, Vassilevich:2003xt});
these expressions have appeared in the anomaly context already in~\cite{Christensen:1978gi,Christensen:1978md}. 

\subsection{Limits of the MAT background}\label{sec:limits_mat}

We now discuss the limits on the MAT background to recover the theories of Dirac and Weyl 
fermions in a curved spacetime and their anomalies.

Setting $f_{\mu\nu}=0$ reproduces the standard coupling of a massless Dirac fermion to a curved 
background and
corresponds to identifying the two effective metrics $g^+_{\mu\nu}= g^-_{\mu\nu}$. 
The final stress tensor becomes the sum of the two chiral stress tensors, and acquires 
the sum of the two trace anomalies in \eqref{tracepm}.
Thus, we recover the usual trace anomaly of a Dirac field~\cite{Bastianelli:2016nuf}
\begin{align}\label{trace-dirac}
\la T^\mu{}_\mu \ra &=
\frac{1}{360 (4 \pi)^2} \left( 7 R_{\mu \nu \rho\sigma } R^{\mu \nu \rho\sigma } +8 R_{\mu \nu} R^{\mu \nu } -5 R^2 +12 \square R\right) \nonumber\\
&= \frac{1}{180 (4 \pi)^2} \left( -\frac{11}{2} E_4 + 9 C^2 + 6 \square R\right) \,.
\end{align}
Similarly, for the two $U(1)$ symmetries, we obtain
 \begin{equation}
\nabla_\mu \la J^\mu_\pm \ra =  \mp \frac{1}{48 (4\pi)^2} \sqrt{g}
\epsilon_{\alpha\beta\gamma\delta} R_{\mu \nu}{}^{\alpha\beta}R^{\mu \nu \gamma\delta} \,, 
\end{equation}
 which gets translated into the  covariant conservation of the vector current 
 $J^\mu_{_V} = J^\mu_+ +J^\mu_-$,  together with
the anomalous conservation of the axial current $J^\mu_{_A} = J^\mu_+ - J^\mu_-$,
 with the well-known Pontryagin contribution \cite{Kimura:1969iwz, Delbourgo:1972xb}
 \begin{equation}
 \nabla_\mu \la J^\mu_{_V} \ra =  0 \,, \qquad 
 \nabla_\mu \la J^\mu_{_A} \ra =  - \frac{1}{24 (4\pi)^2} \sqrt{g}
\epsilon_{\alpha\beta\gamma\delta} R_{\mu \nu}{}^{\alpha\beta}R^{\mu \nu \gamma\delta}  \,.
\label{chiral-an-dirac}
\end{equation}

Let us now study the case of the Weyl fermion $\lambda$.
This is obtained by taking the collapsing limit in which the effective metric  $g^-_{\mu\nu}$ becomes flat 
($g^-_{\mu\nu}=\eta_{\mu\nu}$ and $g^+_{\mu\nu}=g_{\mu\nu}$),  so that 
the independent right-handed fermion $\rho$ decouples  completely from the background.
Therefore, only the chiral left-handed part contributes to the stress tensor, producing
for the trace anomaly half of the result above. 
 Similarly, we find the anomalous conservation of the $U(1)$ current $J^\mu_+$, 
the only one that remains coupled to the curved background, 
 with the expected Pontryagin contribution.
 To summarize, we find for a left-handed Weyl fermion the following anomalies
\begin{align}
  \la T^\mu{}_\mu \ra &=
\frac{1}{720 (4 \pi)^2} \left( 7 R_{\mu \nu \rho\sigma } R^{\mu \nu \rho\sigma } +8 R_{\mu \nu} R^{\mu \nu } -5 R^2 +12 \square R \right) \nonumber\\
&= \frac{1}{180 (4 \pi)^2} \left( -\frac{11}{4} E_4 + \frac92 C^2 + 3 \square R\right)\,,\\
\nabla_\mu \la J^\mu_+ \ra &=  - \frac{1}{48 (4\pi)^2} \sqrt{g}
\epsilon_{\alpha\beta\gamma\delta} R_{\mu \nu}{}^{\alpha\beta}R^{\mu \nu \gamma\delta} \,.
\end{align}
These results confirm the absence of a Pontryagin term in the trace anomaly of a Weyl fermion. 
The Pontryagin term sits only in the chiral anomaly.
\chapter{Irrelevant operators and their holographic anomalies}\label{chap:holography}

In sec.~\ref{sec:holography} we introduced the AdS/CFT correspondence as an example of a holographic duality, and we showed that the holographic computation of the CFT trace anomaly provides a test of the correspondence.
In particular, we introduced a subgroup of bulk diffeomorphisms, called PBH diffeomorphisms, that at the boundary reduce to a Weyl transformation of the boundary metric.
Then, evaluating a bulk action on the solution of the PBH equations yielded the trace anomaly of the CFT.

As we discussed in sec.~\ref{sec:dictionary}, bulk scalar fields are dual to scalar operators in the CFT.
Thus, the addition of scalar fields in the bulk allows to study the contributions of scalar operators to the trace anomaly of the CFT.
Indeed, this has been done shortly after the first holographic computation of the Weyl anomaly (see e.g.~\cite{deHaro:2000vlm}) for scalar fields dual to relevant or marginal operators.
Scalars dual to irrelevant operators are considered in~\cite{vanRees:hr_irr_op} and their contribution to the conformal anomaly is computed in~\cite{vanRees:CS_eq-w_anomalies} in the case of a flat background metric at the boundary.
Nonetheless, when formulated on a curved background, CFTs with irrelevant operators of integer conformal dimension manifest peculiar properties~\cite{Schwimmer:2019efk}, namely the Weyl transformation of the metric acquires a beta-function which in turn deforms the trace anomaly.

In this chapter, we will provide a holographic description of the four-dimensional CFT studied in~\cite{Schwimmer:2019efk}.
To do so, in sec.~\ref{sec:pbhwscalar} we modify the PBH transformations to include a massive scalar field in the bulk, that sources an integer-dimensional irrelevant operator on the boundary.
As we will show, the scalar field modifies the usual Weyl transformation of the boundary metric and we derive the modified Weyl anomaly of the four-dimensional boundary CFT in sec.~\ref{sec:pbhwscalar_ano}.
More details and explicit results are reported in app.~\ref{app:holo_ano}.
These results provide an additional test of the AdS/CFT correspondence and have been first published in~\cite{Broccoli:2021icm}, on which we base the following presentation.

\section{Massive scalar field and modified PBH transformations}\label{sec:pbhwscalar}
We start by adding a massive scalar field $\Phi$ in the bulk and couple it to the metric, and extend the PBH transformations discussed in sec.~\ref{sec:pbh} to describe this system.
The PBH transformation of scalar fields is also discussed in~\cite{Bianchi:2001de,Bianchi:2001kw,Erdmenger:2001ja}.

From the standard holographic dictionary (see sec.~\ref{sec:dictionary}) we known that a bulk scalar field of mass $m$ is dual to a scalar operator on the boundary theory with dimension $\Delta$, related to the mass by $m^2 = \Delta(\Delta-d)$.\footnote{We adopt here and in the following the same conventions employed in sec.~\ref{sec:pbh}.}
Close to the boundary, we consider the following expansion 
\begin{equation}\label{phiansatz}
    \Phi(z,x) = z^{(d-\Delta)/2} \phi(z,x) \, , \quad \phi(z,x) = \sum _{n=0}^\infty \phi_{(n)}(x) z^n  \,,
\end{equation}
with $\phi_{(0)}$ being the source of the boundary operator.\footnote{As for the metric expansion~\eqref{gansatz}, there are logarithmic terms in the expansion of $\phi(z,x)$ for even integer dimension $d$.
We assume we do not need to include them in the present discussion.}
Requiring that the bulk scalar is indeed a scalar under diffeomorphisms,
\begin{equation}
    \Phi'(z',x') = \Phi(z,x) \, ,
\end{equation}
and choosing the diffeomorphism to be a PBH~\eqref{a}, we obtain the PBH transformation for the field $\phi$ as
\begin{equation}\label{phi}
    \delta \phi = -2\sigma \bigg( \frac{d-\Delta}{2} +z\partial_z \bigg) \phi + a^i \partial_i \phi \, .
\end{equation}
With the expansion in~\eqref{phiansatz}, we get
\begin{equation}
    \delta \phi_{(n)} = -\sigma (d + 2n -\Delta ) + \sum_{m=0}^{n-1} \left( a^i_{(n-m)} \partial_i \phi_{(m)} \right) \, ,
\end{equation}
that to lowest order yields
\begin{equation}\label{dphi0}
    \delta \phi_{(0)} =  \sigma \left( \Delta - d \right) \phi_{(0)} \, ,
\end{equation}
namely the correct transformation of the source of a dimension $\Delta$ operator under Weyl transformation.
Eventually, we want to make contact with the four-dimensional CFT analysed in~\cite{Schwimmer:2019efk}.
Thus, we choose $\Delta= d+1$, and therefore the scalar field is sourcing an irrelevant operator in the CFT.

When we couple the bulk scalar to gravity, the dynamical background will backreact and the metric in FG form will be as follows
\begin{equation}\label{modfg}
    ds^2  = \frac{dz^2}{4z^2} + \frac{1}{z} \left( g_{ij}(z,x) + h_{ij}(z,x) \right) dx^i dx^j \, ,
\end{equation}
where $h_{ij}$ is the backreaction, which depends explicitly on $g_{ij}(z,x)$ and $\Phi(z,x)$.
We are essentially allowing for perturbations of the metric $g_{ij}$ due to the presence of the scalar field $\Phi$, and the metric $g_{ij}$ is then treated as a background, unperturbed, metric.
To first non-trivial order, the backreaction is quadratic in the scalar field, and from now on we will work to order $\mathcal{O}(\sigma,\phi^2)$.
We impose a boundary condition for the backreaction (following \cite{vanRees:hr_irr_op}), namely that the backreaction does not change the boundary metric.
In other words, $g^{(0)}_{ij}$ is still the boundary metric even in the presence of the backreaction.
We will see the effect of this boundary condition later.

We now derive the modifications of the PBH transformations due to the presence of the backreaction by studying the behaviour of~\eqref{modfg} under diffeomorphisms.
The FG form of the bulk metric in~\eqref{modfg} is invariant under a general coordinate transformation $X^\mu = X'^\mu + \xi^\mu (X')$ for 
\begin{equation}\label{modxi}
    \xi^z = -2 z \sigma(x) \, , \quad \xi^i = a^i(z,x) + b^i(z,x) = \frac12 \partial_j \sigma(x) \int_\epsilon^z dz' \left( g^{ij}(z',x) - h^{ij}(z',x) \right) \, ,
\end{equation}
where $b^i$ contains the scalar field corrections brought about by the backreaction and is therefore of order~$\mathcal{O}(\sigma,\phi^2)$, while $a^i$ is still of order~$\mathcal{O}(\sigma,\phi^0)$.
Notice that, contrary to eq.~\eqref{a}, now we are restricting the radial integration to the region $z \geq \epsilon > 0$.
This is necessary to avoid divergences in the integration.
Indeed, since $\Delta=d+1$ and $h_{ij}$ is quadratic in the scalar field, it follows that the backreaction goes as $1/z$ about the boundary, thus making the above integration divergent at $z=0$ and requiring that we integrate over the region $z \geq \epsilon > 0$.
This effect is reminiscent of~\cite{McGough:2016lol,Hartman:2018tkw}: the scalar field is causing the boundary to move into the bulk.
Finally, from form invariance of $G_{ij}$ we find to $\mathcal{O}(\sigma,\phi^2)$
\begin{align}\label{gmodrev}
    \delta g_{ij} + \delta h_{ij} &= 2\sigma (1-z\partial_z) \left( g_{ij} + h_{ij}\right)  \nonumber\\
    & \phantom{=} + \nabla_i a_j + \nabla_j a_i  + \nabla_i b_j + \nabla_j b_i + h_{ik} \nabla_j a^k + h_{jk} \nabla_i a^k + a^k \nabla_k h_{ij} \, ,
\end{align}
where indices are lowered with (and derivatives are covariant w.r.t.) $g_{ij}$.
We refer to eqs.~\eqref{modxi}, \eqref{gmodrev} as the modified PBH transformations.

Given the leading asymptotic behaviour of the metric and the backreaction, we make the following Ansätze for the radial expansion of $a^i$ and $b^i$
\begin{align}
    a^i(z,x) &= \overline{a}^i(\epsilon,x) + \sum_{n=1}^\infty z^n \, a^i_{(n)}(x) \,,\\
    b^i (z,x) &= \overline{b}^i(\epsilon,x) +  \log z \, \Tilde{b}^i_{(1)}(x) + \sum_{n=2}^{\infty}\left[ z^{n-1} \left( \log z \, \Tilde{b}^i_{(n)}(x) + b^i_{(n)}(x) \right) \right] \, , \label{bansatz-d+1}
\end{align}
where $\overline{a}^i$ and $\overline{b}^i$ are constant terms in $z$ and their appearance is due to the lower end of the integration in~\eqref{modxi}.
For the metric and backreaction we assume
\begin{align}
    g_{ij}(z,x) &= \sum _{n=0}^\infty z^n \, g^{(n)}_{ij}(x) \,,\\
    h_{ij}(z,x) &= \frac{1}{z} h^{(0)}_{ij}(x) + \log z \, \Tilde{h}^{(1)}_{ij}(x) + 0 + \sum_{n=2}^\infty \left[ z^{n-1} \left( \log z \, \Tilde{h}^{(n)}_{ij}(x) + h^{(n)}_{ij}(x) \right) \right] \, , \label{hansatz-d+1}
\end{align}
where we stress that there is no term at order $z^0$.
This implements the boundary condition that we anticipated above, namely that the boundary metric is still given by~$g^{(0)}_{ij}$ even in the presence of the backreaction~\cite{vanRees:hr_irr_op}.
The appearance of logarithmic terms is a consequence of the particular choice for the dimension of the scalar field $\Delta=d+1$, and we are thus generalising the analysis of~\cite{vanRees:hr_irr_op} as advocated in~\cite{vanRees:CS_eq-w_anomalies}.

Using the above expansions, from the modified PBH equations we find for the first few terms (the $a^i_{(n)}$ are as before)
\begin{align}
    \Tilde{b}^i_{(1)} =& - \frac{1}{2} g_{(0)}^{im}g_{(0)}^{jn} h^{(0)}_{mn} \partial_j \sigma \label{b1tilde} \,,\\
    \Tilde{b}^i_{(2)} =& -\frac12 g_{(0)}^{im}g_{(0)}^{jn} \Tilde{h}^{(1)}_{mn} \partial_j \sigma \label{b2tilde} \,,\\
    b^i_{(2)} =& \;  \frac{1}{2} \big[ \big( g_{(0)}^{im}g_{(1)}^{jn}  + g_{(1)}^{im}g_{(0)}^{jn} \big) h^{(0)}_{mn} + g_{(0)}^{im}g_{(0)}^{jn} \Tilde{h}^{(1)}_{mn}  \big] \partial_j \sigma \label{b2 d+1} \,,\\
    \Tilde{b}^i_{(3)} =& \; \frac14 \big[  \big( g_{(0)}^{im}g_{(1)}^{jn}  + g_{(1)}^{im}g_{(0)}^{jn} \big) \Tilde{h}^{(1)}_{mn} - g_{(0)}^{im}g_{(0)}^{jn} \Tilde{h}^{(2)}_{mn}  \big] \partial_j \sigma \label{b3tilde} \,,\\
    b^i_{(3)} =& -\frac{1}{4} \bigg[ h_{(2)}^{ij} -\frac12 \Tilde{h}_{(2)}^{ij} + \frac12 \big(  g_{(1)}^{ik} \Tilde{h}^{(1)j}{}_{k} + g_{(1)}^{jk} \Tilde{h}^{(1)i}{}_k \big) + g_{(1)}^{ik} g_{(1)}^{jl} h^{(0)}_{kl} + g_{(1)}^{ik} g^{(1)l}{}_k h^{(0)j}{}_l \nonumber \\
    & + g_{(1)}^{jk} g^{(1)l}{}_k  h^{(0)i}{}_l - g_{(2)}^{ik} h^{(0)j}{}_k - g_{(2)}^{jk} h^{(0)i}{}_k  \bigg] \partial_j \sigma \, , \label{b3 d+1}
\end{align}
where indices are lowered (raised) with the (inverse of) $g^{(0)}_{ij}$; curvatures and covariant derivatives will be w.r.t.~$g^{(0)}_{ij}$.
For the metric and backreaction variation\footnote{The expressions that follow are written up to boundary diffeomorphisms generated by $\bar{a}^i$ and $\bar{b}^i$. Since their presence does not affect the solution of the PBH equations, we discard them for simplicity of notation.}
\begingroup
\allowdisplaybreaks[1]
\begin{align}
    \delta h^{(0)}_{ij} =& \; 4\sigma h^{(0)}_{ij} \label{deltah0 d+1} \,,\\
    \delta \Tilde{h}^{(1)}_{ij} =& \; 2\sigma \Tilde{h}^{(1)}_{ij} + \overset{(0)}{\nabla}_i \Tilde{b}_{(1)j} + \overset{(0)}{\nabla}_j \Tilde{b}_{(1)i}  \label{deltah1tilde} \,,\\
    \delta g^{(0)}_{ij} =& \; 2 \sigma g^{(0)}_{ij} - 2 \sigma \Tilde{h}^{(1)}_{ij} + h^{(0)}_{ik} \overset{(0)}{\nabla}_j a^k_{(1)} + h^{(0)}_{jk} \overset{(0)}{\nabla}_i a^k_{(1)} + a^k_{(1)} \overset{(0)}{\nabla}_k h^{(0)}_{ij} \label{gbeta} \,,\\
    \vdots & \nonumber\\
    \delta \Tilde{h}^{(n+1)}_{ij} =& \; 2\sigma (1-n) \Tilde{h}^{(n+1)}_{ij} \nonumber\\*
    & + \sum_{m=1}^{n+1} \Big( g^{(n-m+1)}_{ik} \partial_j \tilde{b}^k_{(m)} + g^{(n-m+1)}_{jk} \partial_i \tilde{b}^k_{(m)} + \tilde{b}^k_{(m)} \partial_k g^{(n-m+1)}_{ij} \nonumber\\*
    & + \tilde{h}^{(n-m+1)}_{ik} \partial_j a^k_{(m)} + \tilde{h}^{(n-m+1)}_{jk} \partial_i a^k_{(m)} + a^k_{(m)} \partial_k \tilde{h}^{(n-m+1)}_{ij} \Big)\label{deltahntilde} \,,\\
    \delta g^{(n)}_{ij} + \delta h^{(n+1)}_{ij} =& \, 2\sigma (1-n) \left( g^{(n)}_{ij} + h^{(n+1)}_{ij} \right) - 2\sigma \Tilde{h}^{(n+1)}_{ij} \nonumber\\*
    & + \sum_{m=1}^{n+1} \Big( g^{(n-m+1)}_{ik} \partial_j {b}^k_{(m)} + g^{(n-m+1)}_{jk} \partial_i {b}^k_{(m)} + {b}^k_{(m)} \partial_k g^{(n-m+1)}_{ij} \nonumber\\*
    & + {h}^{(n-m+1)}_{ik} \partial_j a^k_{(m)} + {h}^{(n-m+1)}_{jk} \partial_i a^k_{(m)} + a^k_{(m)} \partial_k {h}^{(n-m+1)}_{ij} \Big) \, . \label{deltag+hn}
\end{align}
\endgroup
A few comments are in order here.
The above equation for the metric and the backreaction can be solved in the same spirit outlined in sec.~\ref{sec:pbh}.
The term $h^{(n)}_{ij}$ (and similarly $\tilde{h}^{(n)}_{ij}$) is a symmetric and covariant tensor built from $g^{(0)}_{ij}$ and quadratic in $\phi_{(0)}$ with $2n$ derivatives.
Once the most general expression for $g^{(n)}_{ij} + h^{(n+1)}_{ij}$ (or $\tilde{h}^{(n)}_{ij}$) is written down, it is enough to vary it according to~\eqref{gbeta} and~\eqref{dphi0} up to $\mathcal{O}(\sigma,\phi^2)$ and impose the variation is a PBH to find the sought for expression.
The solution of the backreacted Einstein's equations of motion (as in~\cite{vanRees:hr_irr_op} but with $\Delta = d+1$) will also satisfy the above equations.

As in the pure gravity case, the modified PBH equations fix the expression of the backreaction only to some extent.
For instance, the first term in the expansion is
\begin{equation}\label{h0 d+1}
     h^{(0)}_{ij} = h_0 \, g^{(0)}_{ij} \phi_{(0)}^2
\end{equation}
for some coefficient $h_0$, not fixed by the PBH equation~\eqref{deltah0 d+1}.
The higher order terms in the radial expansion will have more and more free coefficients, that are fixed on-shell given an action.\footnote{\label{foot:on-shell}For instance, given the action of a free massive scalar field coupled to a dynamical metric, then on-shell $h_0$ is proportional to the coefficient of the lowest order term in the radial expansion of the scalar field action and it is thus non-vanishing on-shell.}

Notice that the backreaction modifies the usual Weyl transformation of the boundary metric in~\eqref{gbeta}.
However, unlike in the pure gravity case, as it stands the modified PBH transformation does not reduce to a Weyl transformation of $g^{(0)}_{ij}$.
Indeed, when the diffeomorphism is a modified PBH transformation, from~\eqref{xihat} and~\eqref{modxi} we find that (up to $\mathcal O(\sigma,\phi^2)$)
\begin{equation}
    \hat\xi^z = 0 \, , \quad \hat\xi^i = \epsilon \big( g^{ij}(\epsilon,x) - h^{ij}(\epsilon,x) \big) \big( \xi^z_2 \partial_j \xi^z_1 - \xi^z_1 \partial_j \xi^z_2 \big) \, ,
\end{equation}
so that now we are left with a residual diffeomorphism
\begin{equation}\label{modPBH_comm}
    [\delta_2,\delta_1] g^{(0)}_{ij} = g^{(0)}_{jk}\overset{(0)}{\nabla}_i \hat\xi^k + g^{(0)}_{ik}\overset{(0)}{\nabla}_j \hat\xi^k \, .
\end{equation}
Before solving the modified PBH equation for the backreaction, we thus have to address the issue of the $z=\epsilon$ cutoff.

The holographic dual of the gravitational theory discussed so far is a CFT deformed by an irrelevant operator.
However, since we want to make contact with the unperturbed CFT presented in~\cite{Schwimmer:2019efk}, we have to move the cutoff surface back to the AdS boundary.
Given that 
\begin{equation}
    \bar{b}^i(\epsilon,x) = \frac12 h_0 \log \epsilon \, g_{(0)}^{ij} \phi_{(0)}^2 \partial_j \sigma + \mathcal{O}(\epsilon) \, ,
\end{equation}
the way to move the cutoff back to the boundary, without setting the source to zero, is to take $h_0 = 0$ and then $\epsilon = 0$.
In this limit we can describe a boundary CFT in the presence of an irrelevant operator, avoiding the prescription $\phi_{(0)}^2=0$ advocated in~\cite{Witten:1998qj,deHaro:2000vlm}.
As a bonus, the commutator in~\eqref{modPBH_comm} vanishes and the modified PBH transformation reduces to a Weyl transformation at the boundary.
However, the price to pay is that the solutions of the modified PBH equations are not on-shell anymore (see footnote~\ref{foot:on-shell}).

Now, we proceed by solving the modified PBH equations for the backreaction with $h_0 = 0$.
For $\tilde{h}_{(1)}$ we make the Ansatz
\begin{align}\label{h1tildeansatz}
    \tilde{h}^{(1)}_{ij} =& \;  h_1 \, \overset{(0)}{R}_{ij} \phi_{(0)}^2  + h_2 \, g^{(0)}_{ij} \overset{(0)}{R} \phi_{(0)}^2 + h_3 \, \overset{(0)}{\nabla}_i \phi_{(0)} \overset{(0)}{\nabla}_j \phi_{(0)} + h_4 \, \phi_{(0)} \overset{(0)}{\nabla}_i \overset{(0)}{\nabla}_j \phi_{(0)} \nonumber\\
    &\phantom{=}+ h_5 \, g^{(0)}_{ij} \phi_{(0)} \overset{(0)}{\Box} \phi_{(0)} + h_6 \, g^{(0)}_{ij} (\overset{(0)}{\nabla} \phi_{(0)})^2 \,,
\end{align}
and a solution of~\eqref{deltah1tilde} is given by
\begin{align}
    \Tilde{h}_{(1)ij} =& \; \, h_1 \left( \overset{(0)}{R}_{ij} \phi_{(0)}^2 + (d-2) \phi_{(0)} \overset{(0)}{\nabla}_i \overset{(0)}{\nabla}_j \phi_{(0)} + g^{(0)}_{ij} \phi_{(0)} \overset{(0)}{\Box} \phi_{(0)} - (d-1) g^{(0)}_{ij} ( \overset{(0)}{\nabla} \phi_{(0)} )^2  \right) \nonumber\\
    & + \, h_2 \, g^{(0)}_{ij} \left( \overset{(0)}{R} \phi_{(0)}^2 + 2(d-1) \phi_{(0)} \overset{(0)}{\Box} \phi_{(0)} -d(d-1) ( \overset{(0)}{\nabla} \phi_{(0)} )^2 \right) \, . \label{h1tilde}
\end{align}
Notice that the solutions parametrised by $h_1$ and $h_2$ are proportional to $\hat R_{ij}$ and $\hat R$ respectively, which are the curvature tensors computed from the Weyl invariant metric $g^{(0)}_{ij}/\phi^2_{(0)}$.
Then, the variation of the metric in~\eqref{gbeta} reads
\begin{equation}\label{gbetanoh0}
    \delta g^{(0)}_{ij} = 2\sigma g^{(0)}_{ij} -2 \sigma \beta_{ij} \,, \quad \text{with} \quad \beta_{ij} = h_1 \, \hat R_{ij} + h_2 \, g^{(0)}_{ij} \hat R \,,
\end{equation}
and we interpret it as a modification of the usual Weyl transformation of the boundary metric due to the presence of the scalar field.
The holographic beta-function that we find
is in agreement with~\cite{Schwimmer:2019efk}.
Notice that for $h_1=0$ we could perform a field dependent redefinition of $\sigma$, so that the metric transformation is still a usual Weyl transformation:
\begin{equation}\label{sigma_redef}
    \sigma \to \hat\sigma = \sigma \left(1-h_2 \, \hat R \right) \quad \text{s.t.} \quad \delta g^{(0)}_{ij} = 2\hat\sigma g^{(0)}_{ij} \,.
\end{equation}
We will comment on the effect of this redefinition to the holographic anomalies later.

Similarly, we also solve~\eqref{deltahntilde} for~$\tilde{h}_{(2)}$ and~\eqref{deltag+hn} for~$g_{(1)} + h_{(2)}$.
The Ansatz for~$h_{(2)}$ has thirty-five terms and the modified PBH equation leaves six out of the thirty-five coefficients free, while $\tilde{h}_{(2)}$ is determined in terms of $h_1$ and $h_2$.
We also solve for the trace of $\tilde{h}_{(3)}$ and the trace of $h_{(3)}$ that we will need in the following.
The Ansatz for the trace of~$h_{(3)}$ has sixty-six terms and the modified PBH equation leaves nine coefficients free, while the trace of $\tilde{h}_{(3)}$ is determined in terms of $h_1,h_2$ and the $c_1,c_2$ appearing in~\eqref{g2}.
The expressions of $g_{(1)}$ and $g_{(2)}$ are not modified by the presence of the backreaction and are still given by~\eqref{g1} and~\eqref{g2} respectively.
We provide more details in app.~\ref{app:results}.


\section{Effective boundary action and modified Weyl anomalies}\label{sec:pbhwscalar_ano}
To derive the holographic dual of the modified Weyl anomaly found in~\cite{Schwimmer:2019efk}, we extend the method outlined in sec.~\ref{sec:PBHanomaly} to include the effect of the backreaction.

Consider an action
\begin{equation}\label{gravity+matter}
    S = \int_M d^{d+1} X \sqrt{G} f\left(R(G),\Phi\right) \, ,
\end{equation}
where $f$ is a local function of the curvature and its covariant derivatives and contains also matter field $\Phi$.
We could think of $S$ as gravitational action with a scalar field coupled to the metric and we require that $f\left(R(G), \Phi=0 \right)$ is such that $\text{AdS}_{d+1}$ is a solution of the equations of motion.
Then, we write the action in FG coordinates with backreaction~\eqref{modfg} and, expanding in powers of $z$, we obtain the holographic anomalies by evaluating the corresponding expressions on the solutions of the modified PBH equations.
However, the off-shell solution that we discussed so far with $h_0 = 0$ sets the scalar field action to zero (see footnote~\ref{foot:on-shell}), so that eventually the scalar field contributions to the anomalies is only due to the backreaction.

Defining now $\mathcal{L}=\sqrt{G} f\left(R(G),\Phi\right)$, in the FG coordinates~\eqref{modfg} we write
\begin{gather}
    \mathcal{L} = \sqrt{g_{(0)}} z^{-d/2-1} \left( \mathcal{L}_{g} + \mathcal{L}_{h} \right) \label{Og+Oh} \,,\\
    \mathcal{L}_{g} = \sum_{n=0}^\infty \mathcal{L}^{(n)}_g (x) z^n \, , \quad \mathcal{L}_{h} = \sum_{n=0}^\infty \left[ \log z \, \tilde{\mathcal{L}}_h^{(n)} + \mathcal{L}_h^{(n)} \right] \, ,
\end{gather}
where $\mathcal{L}_h$ contains the backreaction and it is thus quadratic in the scalar field; in the second line we use the expansions~\eqref{hansatz-d+1} with $h_0 =0$ and $\mathcal{L}_g$ is as in~\eqref{lg_FG}.
Following the reasoning of sec.~\ref{sec:PBHanomaly}, we can show that $[\delta_2,\delta_1]\mathcal{L}=0$ upon using the PBH group property for $h_0=0$ and thus $\mathcal{L}_{g}+\mathcal{L}_{h}$ satisfies the following WZ condition
\begin{equation}\label{WZ_wbeta}
   \delta_1 \int_{\partial M} d^d x \sqrt{g_{(0)}} \, \sigma_2(x)  (\mathcal{L}_{g}+\mathcal{L}_{h}) - \delta_2 \int_{\partial M} d^d x \sqrt{g_{(0)}} \, \sigma_1(x) (\mathcal{L}_{g}+\mathcal{L}_{h}) = 0 \, ,
\end{equation}
where $g^{(0)}_{ij}$ now transforms with the beta-function~\eqref{gbetanoh0}.
From~\eqref{deltaSbdy} we interpret $\mathcal{L}_g^{(n)} + \mathcal{L}_h^{(n)}$ as the trace anomaly of the $d=2n$ dimensional CFT at the boundary.\footnote{As in sec.~\ref{sec:PBHanomaly}, we neglect divergent terms that are cancelled by counterterms.
In~\cite{deHaro:2000vlm} the counterterms considered only cancel negative powers of the radial coordinate; with irrelevant operators there is also need of logarithmic counterterms, as considered (in flat space) in~\cite{vanRees:CS_eq-w_anomalies}.}

Considering again the $f(R(G))$ in~\eqref{EHwriem}, we now write it in the FG coordinates with backreaction~\eqref{modfg} and find the corresponding expression for $\mathcal{L}_h$ that we report in~\eqref{fwbackre} in app.~\ref{app:action}; $\mathcal{L}_g$ is still given by~\eqref{Og}.
Expanding the metric and backreaction in~\eqref{Og+Oh} according to~\eqref{hansatz-d+1} with $h_0 =0 $, we identify for instance
\begingroup
\allowdisplaybreaks[1]
\begin{align}
    \mathcal{L}_h^{(1)} &= 4\gamma \overset{(0)}{R}{}^{ij} \tilde{h}^{(1)}_{ij} + \frac12(2  - d) (1+4\gamma) h^{(2)i}{}_{i} - (2 - d) (1+8\gamma) g^{(1)ij} \tilde{h}^{(1)}{}_{ij} \nonumber\\*
    &\phantom{=} + \frac{1}{2} (1 +12\gamma - d(1+4\gamma)) g^{(1)i}{}_{i} \tilde{h}^{(1)j}{}_{j} + (3 - d) (1+4\gamma) \tilde{h}^{(2)i}{}_{i} \label{Oh1} \,,\\
    \mathcal{L}_h^{(2)} &= - \frac{1}{4} (1 + \gamma ) \overset{(0)}{R} h^{(2)i}{}_{i} + 2 (1 + 6 \gamma - \frac{3}{8} d (1 + 4 \gamma )) g^{(1)j}{}_{j} h^{(2)i}{}_{i} + \frac{1}{2}(1 + 12 \gamma ) \overset{(0)}{R}{}_{ij} h^{(2)ij} \nonumber\\
    &\phantom{=} - (5 +28 \gamma  - \frac12 d (3 + 20 \gamma )) g^{(1)}{}_{ij} h^{(2)ij} - \frac{3}{2} (d-4) (1 + 4 \gamma ) h^{(3)i}{}_{i}  \nonumber\\ 
    &\phantom{=} - (5  +68 \gamma  -d(1+ 12\gamma) ) g^{(2)ij} \tilde{h}^{(1)}{}_{ij} -8 \gamma  \overset{(0)}{R}{}^{ij} g^{(1)}{}_{i}{}^{k} \tilde{h}^{(1)}{}_{jk} + 8 \gamma  \nabla_{k}\tilde{h}^{(1)}{}_{ij} \nabla^{k}g^{(1)ij} \nonumber\\
    &\phantom{=} + (5 + 56 \gamma   - d (1 + 12 \gamma )) g^{(1)}{}_{i}{}^{k} g^{(1)ij} \tilde{h}^{(1)}{}_{jk} -8 \gamma  \overset{(0)}{\nabla}_j\tilde{h}^{(1)}{}_{ik}\overset{(0)}{\nabla}{}^{k}g^{(1)ij}  \nonumber\\
    &\phantom{=} - \frac{1}{2} (4  +48 \gamma   -d(1+ 8 \gamma )) g^{(1)i}{}_{i} g^{(1)jk} \tilde{h}^{(1)}{}_{jk} + \frac{1}{2} (3 + 28 \gamma  - d (1 + 4 \gamma )) g^{(2)i}{}_{i} \tilde{h}^{(1)j}{}_{j} \nonumber\\
    &\phantom{=} - \frac{1}{4} (3 +44 \gamma  -d(1+ 4 \gamma) ) g^{(1)}{}_{ij} g^{(1)ij} \tilde{h}^{(1)k}{}_{k} +\frac{1}{2} ( 1+ 4 \gamma ) \overset{(0)}{\Box}h^{(2)i}{}_{i}  \nonumber\\
    &\phantom{=} + \frac{1}{8} (3 + 28 \gamma  - d (1 + 4 \gamma )) g^{(1)i}{}_{i} g^{(1)j}{}_{j} \tilde{h}^{(1)k}{}_{k} -4 \gamma  \overset{(0)}{R}_{ikjl} g^{(1)ij} \tilde{h}^{(1)kl} + 4 \gamma  \overset{(0)}{R}{}^{ij} \tilde{h}^{(2)}{}_{ij} \nonumber\\
    &\phantom{=} - (6  +32 \gamma  +d(1+ 8 \gamma) ) g^{(1)ij} \tilde{h}^{(2)}{}_{ij} + \frac{1}{2} (5 + 28 \gamma  - d (1 + 4 \gamma )) g^{(1)i}{}_{i} \tilde{h}^{(2)j}{}_{j} \nonumber\\
    &\phantom{=}  -2 \gamma  \tilde{h}^{(1)ij} \overset{(0)}{\nabla}_{j}\overset{(0)}{\nabla}_{i}g^{(1)k}{}_{k}  + 4 \gamma  \tilde{h}^{(1)ij} \nabla_{k}\nabla_{j}g^{(1)}{}_{i}{}^{k} - \frac{1}{2} (1 +4 \gamma ) \overset{(0)}{\nabla}_{j}\overset{(0)}{\nabla}_{i}h^{(2)ij} \nonumber\\
    &\phantom{=}- (d-7) (1 + 4 \gamma ) \tilde{h}^{(3)i}{}_{i} + 2 \gamma  \overset{(0)}{R}{}^{ij} g^{(1)k}{}_{k} \tilde{h}^{(1)}{}_{ij} -2 \gamma  \tilde{h}^{(1)ij} \overset{(0)}{\Box}g^{(1)}{}_{ij}\label{Oh2} \,,
\end{align}
\endgroup
which should contribute to the anomaly in $d=2,4$ respectively.
Focusing on the $d=4$ case, the candidate anomaly is $\mathcal{L}_g^{(2)} + \mathcal{L}_h^{(2)}$, which on the PBH solutions yields indeed a solution of the WZ condition.
Notice that $g^{(n)}$ appears in $\mathcal{L}_h^{(n)}$ and together with~$\tilde{h}^{(3)i}{}_{i}$ it causes factors of $(d-4)^{-1}$ to appear in $\mathcal{L}_h^{(2)}$.
Thus, at first sight $\mathcal{L}_g^{(2)} + \mathcal{L}_h^{(2)}$ is singular in $d=4$.
Nonetheless, it is possible to renormalise the free coefficients of the PBH solution for the backreaction (see app.~\ref{app:coeffren}) so that eventually $\mathcal{L}_g^{(2)} + \mathcal{L}_h^{(2)}$ is regular in $d=4$ and can be identified with the holographic anomaly of the boundary CFT.
We thus define $\mathcal{A}^{4d} = (\mathcal{L}_g^{(2)} + \mathcal{L}_h^{(2)})|_{\text{reg}}$ as the regularised, four dimensional holographic anomaly, which of course still satisfies the WZ condition.
The term quadratic in the curvature in~\eqref{EHwriem} allows us to separate again the pure gravity type A and B anomalies, which now receive contributions also from the scalar field.
In $\mathcal{A}^{4d}$ we then identify
\begin{equation}\label{full_ano}
    \mathcal{A}^{4d} = \mathcal{A}^{4d}_{\text{B}} + \mathcal{A}^{4d}_{\text{A}} \,,
\end{equation}
with
\begin{align}
    \mathcal{A}^{4d}_{\text{B}} &= \mathcal{I}^{\text{pg}}_\text{B} + \mathcal{I}_1(h_1) + \mathcal{I}_2 (c_1,h_1,h_2) + \mathcal{I}_3(c_2,h_1,h_2) \label{full_ano_B} \,,\\
    \mathcal{A}^{4d}_{\text{A}} &= \mathcal{I}^{\text{pg}}_\text{A} + \mathcal{I}_4(h_1) + \mathcal{I}_5(h_2) + \mathcal{I}_6(h_2) + \mathcal{I}_7(c_1,h_1,h_2) + \mathcal{I}_8(c_2,h_1,h_2) \nonumber\\
    &\phantom{=} + \mathcal{I}_9(h_{21}) + \mathcal{I}_{10}(h_{22}) + \mathcal{I}_{11}(h_{23})+ \mathcal{I}_{12}(h_{24}) + \mathcal{I}_{13}(h_{25}) + \mathcal{I}_{14}(h_{26}) \label{full_ano_A} \,,
\end{align}
where $\mathcal{I}^{\text{pg}}_\text{B} = - c\, C^2 /16 $ and $\mathcal{I}^{\text{pg}}_\text{A} = a\, E_4 /16$ are the type B and A pure gravity anomaly and the $\mathcal{I}_i$ are tensorial structures quadratic in the scalar field and to sixth order in derivatives, parametrised by the (backreaction) coefficients given in parenthesis.
In particular, $\mathcal{I}_i$ for $i=1,2,3$ contribute to the type B pure gravity anomaly, while $\mathcal{I}_i$ for $i=4,\ldots,14$ contribute to the type A pure gravity anomaly.
We present all the explicit expressions in appendix~\ref{app:full_ano};
here, we discuss the general properties of the individual tensorial structures and make contact with the results of~\cite{Schwimmer:2019efk}.

As we already remarked, $\mathcal{A}^{4d}$ satisfies the WZ condition~\eqref{WZ_wbeta} to $\mathcal{O}(\phi^2)$ when the metric transforms with the beta-function~\eqref{gbetanoh0}.
However, some of the tensorial structures in~$\mathcal{A}^{4d}$ are solutions of the WZ condition without need of the beta-function.
Amongst these solutions, we can distinguish between trivial solutions (i.e.~terms which in the field theory can be cancelled with counterterms) and solutions which corresponds to `ordinary' anomaly in the CFT that cannot be cancelled with counterterms.
The $\mathcal{I}_i$ for $i=9,\ldots,14$ are trivial solutions, while $\mathcal{I}_i$ for $i=2,3,6,7,8$ correspond to ordinary anomalies.
Notice that, even though the latter depend on the metric beta-function coefficients, they are solutions of the WZ condition without need of the beta-function, so that from the field theory point of view they do not correspond to modifications of the type A and B anomalies induced by the beta-function.
The structures $\mathcal{I}_i$ for $i=2,3,7,8$ transform under Weyl transformations as $\delta \mathcal{I}_i = -4\sigma \mathcal{I}_i$, so that they lead to Weyl invariant anomaly integrals; nonetheless, $\mathcal{I}_7$ and $\mathcal{I}_8$ fall into the scalar field contributions to the type A pure gravity anomaly and thus we deduce that the bulk action defined through~\eqref{EHwriem} does not have enough parameters to distinguish between the type A and B terms quadratic in the scalar field.
To do so, it might be necessary to include additional higher derivative terms in the bulk action, which would complicate the computation of the anomalies and for simplicity we didn't consider.

The remaining structures $\mathcal{I}_i$ for $i=1,4,5$ together with the pure gravity terms are solutions of the WZ condition only when the beta-function is taken into account.
However, as we mentioned above, for $h_1 =0$ we can perform a redefinition of $\sigma$ so that the transformation~\eqref{gbetanoh0} is a usual Weyl transformation~\eqref{sigma_redef}.
This suggests that from the field theory point of view the anomaly seeded by the beta-function parametrised by $h_2$ might not be a new anomaly, but rather a deformation of the usual pure gravity $\mathcal{I}^{\text{pg}}_\text{A}$ anomaly.
Indeed, from our holographic computation we find that
\begin{equation}
    \mathcal{I}_5(h_2) = -\frac{a}{16} h_2 \hat R \, E_4 \,,
\end{equation}
so that 
\begin{equation}
    \sigma \left( \mathcal{I}^{\text{pg}}_\text{A} + \mathcal{I}_5(h_2) \right) = \frac{a}{16} \hat \sigma E_4 \,,
\end{equation}
which is a solution of the WZ condition when the metric transforms in the usual way with a redefined $\sigma$ as in~\eqref{sigma_redef}.

The tensorial structures $\mathcal{I}_1$ and $\mathcal{I}_4$, together with the pure gravity terms, are the only terms which survive in~\eqref{full_ano} when all the free coefficients appearing in the anomaly, but $h_1$, are set to zero.
We now focus on these terms to make contact with the anomalies in~\cite{Schwimmer:2019efk}.
Setting to zero all the free coefficients in the anomaly, but $h_1$, we are then left with the following expression for the pure gravity type B anomaly and its scalar field contributions ($g_{ij} \equiv g^{(0)}_{ij}$ and $\phi \equiv \phi_{(0)}$):
\begingroup
\allowdisplaybreaks[1]
\begin{align}
    \mathcal{A}^{4d}_{\text{B}} &= \mathcal{I}^{\text{pg}}_\text{B} + \mathcal{I}_1(h_1) \nonumber\\
    &= -\frac{1}{16} c \, C^2 + c\, h_1 \Bigg( \frac{3}{4} \nabla^{j}\nabla^{i}\phi \Box\nabla_{j}\nabla_{i}\phi -  \frac{3}{16} \Box\phi \Box^2 \phi + \frac{1}{2} \nabla_{k}\nabla_{j}\nabla_{i}\phi \nabla^{k}\nabla^{j}\nabla^{i}\phi \nonumber\\
    &\phantom{=} - \frac{89}{144} R_{i}{}^{k} R^{ij} R_{jk} \phi^2 + \frac{97}{144} R_{ij} R^{ij} R \phi^2 -  \frac{7}{81} R^3 \phi^2 -  \frac{11}{16} R^{ij} R^{kl} R_{ikjl} \phi^2 \nonumber\\
    &\phantom{=} -  \frac{5}{384} R R_{ijkl} R^{ijkl} \phi^2 -  \frac{1}{18} R_{i}{}^{m}{}_{k}{}^{n} R^{ijkl} R_{jmln} \phi^2 + \frac{53}{576} R_{ij}{}^{mn} R^{ijkl} R_{klmn} \phi^2 \nonumber\\
    &\phantom{=} -  \frac{7}{192} R \phi^2 \Box R + \frac{3}{64} R^2 \phi \Box \phi + \frac{1}{6} R_{jklm} R^{jklm} \phi \Box \phi -  \frac{19}{288} \phi^2 \nabla_{i}R \nabla^{i}R \nonumber\\
    &\phantom{=} -  \frac{65}{288} R \phi \nabla_{i}\phi \nabla^{i}R + R^{jk} \phi \nabla_{i}R_{jk} \nabla^{i}\phi + \frac{1}{16} R^{jklm} \phi \nabla_{i}R_{jklm} \nabla^{i}\phi \nonumber\\
    &\phantom{=} + \frac{19}{24} R_{jk} R^{jk} \nabla_{i}\phi \nabla^{i}\phi -  \frac{5}{36} R^2 \nabla_{i}\phi \nabla^{i}\phi -  \frac{1}{3} R_{jklm} R^{jklm} \nabla_{i}\phi \nabla^{i}\phi \nonumber\\
    &\phantom{=} -  \frac{1}{36} R^{ij} \phi^2 \nabla_{j}\nabla_{i}R -  \frac{11}{72} R^{ij} R \phi \nabla_{j}\nabla_{i}\phi -  \frac{35}{288} \phi \Box R \Box \phi -  \frac{13}{48} \nabla_{i}\phi \nabla^{i}R \Box \phi \nonumber\\
    &\phantom{=} -  \frac{1}{12} \phi \nabla^{i}R \Box \nabla_{i}\phi -  \frac{1}{6} R \nabla^{i}\phi \Box \nabla_{i}\phi -  \frac{3}{32} R \phi \Box^2\phi + \frac{1}{3} R_{ij} \phi \nabla^{i}R \nabla^{j}\phi \nonumber\\
    &\phantom{=} + \frac{1}{12} R_{ij} R \nabla^{i}\phi \nabla^{j}\phi -  \frac{1}{2} R^{kl} R_{ikjl} \nabla^{i}\phi \nabla^{j}\phi + \frac{1}{12} \nabla^{i}R \nabla_{j}\nabla_{i}\phi \nabla^{j}\phi \nonumber\\
    &\phantom{=} -  \frac{1}{18} \phi \nabla_{j}\nabla_{i}\phi \nabla^{j}\nabla^{i}R -  \frac{3}{16} R \nabla_{j}\nabla_{i}\phi \nabla^{j}\nabla^{i}\phi -  \frac{1}{2} R^{jk} \phi \nabla^{i}\phi \nabla_{k}R_{ij} \nonumber\\
    &\phantom{=} -  \frac{1}{8} R_{i}{}^{k} R^{ij} \phi \nabla_{k}\nabla_{j}\phi + \frac{1}{2} R^{jk} \nabla^{i}\phi \nabla_{k}\nabla_{j}\nabla_{i}\phi + \frac{1}{8} R^{ij} \phi^2 \Box R_{ij} + \frac{13}{24} \phi \nabla^{j}\nabla^{i}\phi \Box R_{ij} \nonumber\\
    &\phantom{=} -  \frac{1}{6} R_{ij} R^{ij} \phi \Box \phi + \frac{1}{2} R_{i}{}^{j} \nabla^{i}\phi \Box \nabla_{j}\phi + \frac{3}{8} R^{ij} \phi \Box \nabla_{j}\nabla_{i}\phi -  \frac{9}{16} \phi^2 \nabla_{j}R_{ik} \nabla^{k}R^{ij} \nonumber\\
    &\phantom{=} + \frac{61}{96} \phi^2 \nabla_{k}R_{ij} \nabla^{k}R^{ij} + \frac{1}{2} \phi \nabla_{k}\nabla_{j}\nabla_{i}\phi \nabla^{k}R^{ij} + \frac{3}{8} R^{ij} \nabla_{k}\nabla_{j}\phi \nabla^{k}\nabla_{i}\phi \nonumber\\
    &\phantom{=} + \frac{5}{4} \nabla_{i}R_{jk} \nabla^{i}\phi \nabla^{k}\nabla^{j}\phi -  \frac{1}{2} \nabla^{i}\phi \nabla_{k}R_{ij} \nabla^{k}\nabla^{j}\phi + \frac{1}{2} R_{ijkl} \phi \nabla^{i}\phi \nabla^{l}R^{jk} \nonumber\\
    &\phantom{=} + \frac{7}{24} R_{ikjl} \phi^2 \nabla^{l}\nabla^{k}R^{ij} + \frac{5}{24} R^{ij} R_{ikjl} \phi \nabla^{l}\nabla^{k}\phi + \frac{3}{8} R_{ikjl} \nabla^{j}\nabla^{i}\phi \nabla^{l}\nabla^{k}\phi \nonumber\\
    &\phantom{=} + \frac{5}{192} \phi^2 \nabla_{m}R_{ijkl} \nabla^{m}R^{ijkl} \Bigg) \, , \label{modetypeB_h1}
\end{align}
\endgroup
while the pure gravity type A anomaly and the scalar field contributions read (again $g_{ij} \equiv g^{(0)}_{ij}$ and $\phi \equiv \phi_{(0)}$):
\begingroup
\allowdisplaybreaks[1]
\begin{align}
    \mathcal{A}^{4d}_{\text{A}} &= \mathcal{I}^{\text{pg}}_\text{A} + \mathcal{I}_4(h_1) \nonumber\\
    &= \frac{1}{16} a\, E_4 + a\, h_1 \Bigg(    \frac{2}{3} \nabla^{j}\Box\phi \Box\nabla_{j}\phi -  \frac{43}{24} \nabla^{j}\nabla^{i}\phi \Box \nabla_{j}\nabla_{i}\phi + \frac{31}{96} \Box\phi \Box^2\phi \nonumber\\
    &\phantom{=} -  \frac{1}{4} \nabla^{i}\phi \Box^2\nabla_{i}\phi + \frac{7}{32} \phi \Box^3\phi -  \frac{5}{3} \nabla_{k}\nabla_{j}\nabla_{i}\phi \nabla^{k}\nabla^{j}\nabla^{i}\phi + \frac{43}{96} R_{i}{}^{k} R^{ij} R_{jk} \phi^2 \nonumber\\
    &\phantom{=} + \frac{1}{12} R_{ij} R^{ij} R \phi^2 -  \frac{23}{768} R^3 \phi^2 -  \frac{53}{96} R^{ij} R^{kl} R_{ikjl} \phi^2 -  \frac{29}{768} R R_{ijkl} R^{ijkl} \phi^2 \nonumber\\
    &\phantom{=} + \frac{41}{24} R_{i}{}^{m}{}_{k}{}^{n} R^{ijkl} R_{jmln} \phi^2 -  \frac{61}{384} R_{ij}{}^{mn} R^{ijkl} R_{klmn} \phi^2 + \frac{1}{288} R \phi^2 \Box R -  \frac{85}{144} R^2 \phi \Box \phi \nonumber\\
    &\phantom{=} -  \frac{19}{48} R_{jklm} R^{jklm} \phi \Box \phi -  \frac{7}{576} \phi^2 \nabla_{i}R \nabla^{i}R -  \frac{19}{36} R \phi \nabla_{i}\phi \nabla^{i}R + 4 R^{jk} \phi \nabla_{i}R_{jk} \nabla^{i}\phi \nonumber\\
    &\phantom{=} -  \frac{169}{48} R^{jklm} \phi \nabla_{i}R_{jklm} \nabla^{i}\phi + \frac{155}{32} R_{jk} R^{jk} \nabla_{i}\phi \nabla^{i}\phi -  \frac{295}{576} R^2 \nabla_{i}\phi \nabla^{i}\phi \nonumber\\
    &\phantom{=} -  \frac{535}{192} R_{jklm} R^{jklm} \nabla_{i}\phi \nabla^{i}\phi + \frac{5}{8} R^{ij} \phi^2 \nabla_{j}\nabla_{i}R + \frac{77}{48} R^{ij} R \phi \nabla_{j}\nabla_{i}\phi \nonumber\\
    &\phantom{=} -  \frac{1}{16} \nabla_{i}\phi \nabla^{i}\phi \Box R + \frac{55}{96} \phi \Box R \Box\phi + \frac{17}{48} R \Box\phi \Box\phi + \frac{23}{96} \nabla_{i}\phi \nabla^{i}R \Box\phi \nonumber\\
    &\phantom{=} + \frac{1}{4} \phi \nabla^{i}\phi \Box \nabla_{i}R + \frac{55}{96} \phi \nabla^{i}R \Box\nabla_{i}\phi -  \frac{5}{6} R \nabla^{i}\phi \Box \nabla_{i}\phi + \frac{7}{192} \phi^2 \Box^2 R \nonumber\\
    &\phantom{=} + \frac{13}{32} R \phi \Box^2 \phi -  \frac{49}{24} R_{ij} \phi \nabla^{i}R \nabla^{j}\phi  + \frac{13}{6} R_{i}{}^{k} R_{jk} \nabla^{i}\phi \nabla^{j}\phi - 2 R_{ij} R \nabla^{i}\phi \nabla^{j}\phi \nonumber\\
    &\phantom{=} + \frac{13}{6} R^{kl} R_{ikjl} \nabla^{i}\phi \nabla^{j}\phi -  \frac{7}{24} \nabla^{i}\phi \nabla_{j}\nabla_{i}R \nabla^{j}\phi + \frac{67}{48} \phi \nabla_{j}\nabla_{i}\phi \nabla^{j}\nabla^{i}R \nonumber\\
    &\phantom{=} -  \frac{51}{32} R \nabla_{j}\nabla_{i}\phi \nabla^{j}\nabla^{i}\phi + \frac{69}{16} R^{jk} \phi \nabla^{i}\phi \nabla_{k}R_{ij} + \frac{17}{16} R_{i}{}^{k} R^{ij} \phi \nabla_{k}\nabla_{j}\phi \nonumber\\
    &\phantom{=} + \frac{37}{12} R^{jk} \nabla^{i}\phi \nabla_{k}\nabla_{j}\nabla_{i}\phi + \frac{5}{16} R^{ij} \phi^2 \Box R_{ij} + \frac{3}{8} \nabla^{i}\phi \nabla^{j}\phi \Box R_{ij} -  \frac{139}{48} \phi \nabla^{j}\nabla^{i}\phi \Box R_{ij} \nonumber\\
    &\phantom{=} + \frac{21}{8} R_{ij} R^{ij} \phi \Box \phi -  \frac{5}{24} R^{ij} \nabla_{j}\nabla_{i}\phi \Box \phi -  \frac{29}{12} R_{i}{}^{j} \nabla^{i}\phi \Box \nabla_{j}\phi -  \frac{13}{48} R^{ij} \phi \Box \nabla_{j}\nabla_{i}\phi \nonumber\\
    &\phantom{=} + \frac{185}{64} \phi^2 \nabla_{j}R_{ik} \nabla^{k}R^{ij} -  \frac{153}{128} \phi^2 \nabla_{k}R_{ij} \nabla^{k}R^{ij} + \frac{1}{12} \phi \nabla_{k}\nabla_{j}\nabla_{i}\phi \nabla^{k}R^{ij} \nonumber\\
    &\phantom{=} + \frac{27}{16} R^{ij} \nabla_{k}\nabla_{j}\phi \nabla^{k}\nabla_{i}\phi -  \frac{31}{24} \nabla_{i}R_{jk} \nabla^{i}\phi \nabla^{k}\nabla^{j}\phi -  \nabla^{i}\phi \nabla_{k}R_{ij} \nabla^{k}\nabla^{j}\phi \nonumber\\
    &\phantom{=} -  \frac{41}{8} R_{ijkl} \phi \nabla^{i}\phi \nabla^{l}R^{jk} -  \frac{223}{48} R_{ikjl} \phi^2 \nabla^{l}\nabla^{k}R^{ij} -  \frac{101}{16} R^{ij} R_{ikjl} \phi \nabla^{l}\nabla^{k}\phi \nonumber\\
    &\phantom{=} -  \frac{13}{16} R_{ikjl} \nabla^{j}\nabla^{i}\phi \nabla^{l}\nabla^{k}\phi -  \frac{643}{768} \phi^2 \nabla_{m}R_{ijkl} \nabla^{m}R^{ijkl}  \Bigg) \,. \label{modtypeA_h1}
\end{align}
\endgroup
The expressions~\eqref{modetypeB_h1} and~\eqref{modtypeA_h1} satisfy the WZ condition~\eqref{WZ_wbeta} and, since they are parametrised by $a,c$ and $h_1$ only, we can compare them to the anomalies presented in~\cite{Schwimmer:2019efk}.
At first sight, they look different from the one obtained in~\cite{Schwimmer:2019efk}.
However, we checked that the expressions are the same,\footnote{Up to a factor of 2, which is missing in the normalisation of the metric beta-function in~\cite{Schwimmer:2019efk}.\label{foot:normalisation}} up to variation of local counterterms in the field theory and addition of ordinary Weyl invariant anomalies.
These terms are present in~\eqref{modetypeB_h1} and~\eqref{modtypeA_h1} and the bulk action defined through~\eqref{EHwriem} does not have enough parameters to distinguish them  from the modifications of the pure gravity anomalies which are solutions of~\eqref{WZ_wbeta} only when the beta-function is taken into account.

Notice the interplay between the $a$, $c$ coefficients and the metric beta-function coefficient $h_1$, as already observed in~\cite{Schwimmer:2019efk}.
It is this feature that makes the anomalies~\eqref{modetypeB_h1} and~\eqref{modtypeA_h1} solutions of the consistency condition~\eqref{WZ_wbeta} in the presence of the metric beta-function.
We remark that the other tensorial structures which appear in the anomaly solve the WZ condition without need of the metric beta-function (apart from $\mathcal{I}_5$, whose peculiarities we analysed above).

As a final comment, we discuss the flat space limit of~\eqref{full_ano}.
From the explicit expressions in app.~\ref{app:full_ano}, it appears that in flat boundary space and for constant $\sigma\to 1$, the anomaly reduces to (after integration by parts)
\begin{equation}\label{flat}
    \int d^4 x \sqrt{g_{(0)}} \, \sigma(x) \mathcal{A}^{4d} \to \frac{1}{16} c \, h_1 \, \int d^4 x \, \phi_{(0)} \Box^3 \phi_{(0)} \, ,
\end{equation}
i.e.~the type A anomaly vanishes in flat space and the result is proportional to $h_1$.
This contribution comes entirely from~$\mathcal{I}_1$.
The conformal anomaly associated with free massive scalar fields in flat boundary space and its coefficient are computed holographically in~\cite{deHaro:2000vlm}.
Comparing with our result, we can fix the coefficient $h_1$ as
\begin{equation}
    h_1 = \frac{16}{c} \frac{1}{128} \, .
\end{equation}
Notice that this coefficient does not renormalise~\cite{Petkou:1999fv}.
The flat space limit~\eqref{flat} and the value of $h_1$ are in agreement with the results in~\cite{Schwimmer:2019efk}.
Indeed, comparing with~\eqref{gbetanoh0}, from~\cite{Schwimmer:2019efk} we identify (with footnote~\ref{foot:normalisation} in mind)
\begin{equation}
    h_1 = \frac{5}{72} \frac{N}{c_T}
\end{equation}
where $N$ is the normalisation of the two-point function of scalar operators which according to~\cite{Freedman:1998tz} is $N= \pi^{-d/2}(2\Delta-d)\Gamma(\Delta)/\Gamma(\Delta-d/2)$ and from the normalisation of the type B Weyl anomaly we find $c_T = 40 c / \pi^2$.
For $d=4$ and $\Delta =5$, the value of $h_1$ computed in the CFT matches the one found holographically.
\chapter{Conclusions}\label{chap:conclusions}

In this thesis we analysed various aspects of CFTs in two and four dimensions,
in the latter case also in the context of holography.
We introduced and motivated the topics of our interest in chap.~\ref{chap:intro} and we devoted chap.~\ref{chap:intro_tech} to review basic notions which constitute the foundations of this thesis.
In particular, we defined conformal transformations and discussed the implications of conformal symmetry in field theory, with a particular emphasis on two-dimensional CFT.
We then defined Weyl transformations and anomalies, presenting a regularisation scheme that allows to compute anomalies with heat kernel formulae.
We motivated the AdS/CFT correspondence and showed how to compute the trace anomaly of the CFT by studying diffeomorphisms in the bulk.
In the next chapters we presented our contributions to these topics, and we now review our results discussing possible outlooks.

In chap.~\ref{chap:desc_corr} we showed how to systematically compute the R\'enyi entanglement entropy, the sandwiched R\'enyi divergence and the trace square distance of generic descendant states reduced to a single interval subsystem in a CFT. 
In practice the computations can be performed with the help of computer algebra programs and with the implementation of a recursive function that computes any correlator of descendants as a (differential) operator acting on the correlator of the respective primaries.
We explicitly computed the aforementioned quantum measures for the first few descendant states in the vacuum module and for descendants of primaries in the Ising model and in the three-state Potts model. 

From the results in the vacuum module we saw that degenerate descendant states only show equal behaviour for small subsystem sizes. At large central charge any of the above quantities behaved very different for the degenerate states we considered.
This may be a hint that in general the holographic R\'enyi entanglement entropy can be very different between degenerate descendant states.

We also checked explicitly if predictions from the generalized version of QNEC \cite{Lashkari:2018nsl,Moosa:2020jwt} are true for descendant states, namely that the sandwiched R\'enyi divergence is a convex function of subsystem size. 
In the Ising and Potts models the SRD is a convex function in all the cases we checked. Nonetheless, we could show that the SRD of descendant states becomes non-convex for small but positive central charge.

Many of the analytic expressions that we obtained are too large to show them explicitly.
However, showing the results in the small subsystem size limit is possible and they are always in agreement with the expectations from taking the respective limits in the operator product expansion.
Our results show that in this limit the measures we considered do not allow to distinguish between different degenerate states.
Only at larger subsystem size it is possible to distinguish between different degenerate states with these measures, as it is evident from some of the plots we presented.

In chap.~\ref{chap:weyl_fermions} we studied the anomalies of Weyl fermions in different backgrounds.
We employed a Pauli-Villars regularisation scheme, which casts the anomalies in a form that looks similar to the regulated Jacobian used by Fujikawa.
In the scheme we adopted, the anomalies are due to the possible non-invariance of the mass term of the PV field and, once this is specified, the regulators and the Jacobians of the Fujikawa's scheme are determined.
The anomalies are then computed by using heat kernel formulae.
This procedure has already been applied successfully to several contexts in the past, e.g.~in the case of two-dimensional $b$-$c$ systems~\cite{Bastianelli:1990xn} and in the model of chiral bosons~\cite{Bastianelli:1990ev}.

To begin with, we computed the chiral and trace anomaly in the Bardeen model of a Dirac fermion coupled to non-abelian vector and axial gauge fields.
We derived the famous result for the chiral anomaly as a check on our scheme and we found the trace anomaly, which was not known before;
by taking a suitable limit on the background fields, we obtained the chiral and trace anomalies for a left-handed Weyl fermion coupled to non-abelian gauge fields.
Then, we studied the anomalies of a Dirac fermion coupled to the MAT background, which allowed us to rederive the anomalies of a Weyl fermion coupled to a curved spacetime in a set-up similar to Bardeen model. 

In the case of the gauge background, we verified that the consistent trace anomalies for the Dirac and Weyl fermions acquire a gauge invariant form by adding suitable local counterterms to the effetive action.
In particular, the trace anomaly for the Weyl fermion is half the trace anomaly of the Dirac fermion, confirming the finding of~\cite{Bastianelli:2018osv} in the limiting case of an abelian gauge background.
In the case of the MAT background, we reproduced the known trace anomaly for the Dirac fermion, and the one for the Weyl fermion again is found to be half the anomaly for the Dirac fermion~\cite{Bastianelli:2016nuf}.
In both examples we verified the absence of parity-odd terms proportional to the (Chern-)Pontryagin density of the background fields.

In chap.~\ref{chap:holography} we holographically computed the Weyl anomaly of a four-dimensional CFT in the presence of a source for an irrelevant operator of conformal dimension $\Delta=5$.
In the bulk, the massive scalar field induces a backreaction onto the metric, and, since the scalar field sources an irrelevant operator on the boundary, it changes the leading asymptotic behaviour of the metric which becomes more singular as the boundary is approached.
This causes logarithmic terms to appear in the radial expansion of the backreaction which affect the Weyl transformation of the boundary metric.
A metric beta-function appears, which in turn modifies the holographic Weyl anomaly.
The metric beta-function and the holographic Weyl anomaly are in agreement with the known CFT results of~\cite{Schwimmer:2019efk}.

To obtain the holographic results, we studied diffeomorphisms in the bulk and used the equivalence between bulk diffeomorphisms and Weyl transformation at the boundary.
We saw that the presence of a scalar field, which sources an irrelevant operator at the boundary, introduces a radial cutoff, which reflects the fact that the boundary field theory is being deformed by the irrelevant operator.
Eliminating the cutoff, we were able to describe the underformed CFT and the modified PBH transformations reduce to a boundary Weyl transformation.
However, this requires that the backreaction is put off-shell.

The holographic anomaly is obtained as the finite piece of a bulk action, evaluated on the PBH solutions.
The scalar field bulk action does not contribute to the holographic anomaly, which comes from the gravitational action alone evaluated on the PBH solution for the metric including the backreaction due to the scalar field.
In the resulting expression, in addition to the usual pure gravity Weyl anomaly there are contributions which depend on the scalar field and we calculated them to lowest non-trivial order, which is quadratic in the source of the boundary operator.
Some of them correspond to anomalous terms seeded by the metric beta-function, while others correspond to solutions of the Wess-Zumino consistency condition to quadratic order in the scalar field but not seeded by the backreaction.
The holographic anomaly is non-vanishing in flat space, where it reduces to the expected result, which in $d$ boundary dimensions for a scalar operator of dimension $\Delta$ with source $\phi_{(0)}$ is proportional to $\phi_{(0)} \Box^n \phi_{(0)}$ with integer $n = \Delta - d/2$~\cite{vanRees:CS_eq-w_anomalies}.
Thus, our analysis provides an additional test of the AdS/CFT correspondence.

\section*{Further discussions and outlooks}
Our code to compute correlation functions and quantum measures of descendant states offers new possibilities to analyse the holographic behaviour of descendant states.
Indeed, as we already mentioned, the quantum measures we considered did not show the same behaviour for different degenerate states in the limit of large central charge.
Since black hole microstates in $\text{AdS}_3$ are believed to be dual to typical high conformal dimension states, it would be interesting to apply our code to test the distinguishability of black hole microstates.

The code that led to our results is openly accessible, for instance from the preprint version of~\cite{Brehm:2020zri}, and can be used to compute the quantum measures discussed in this work for more descendant states or in different models.
One could for example consider quasi-primary states and check if they behave special compared to generic descendant states.
Other interesting states to study might be those that correspond to currents of the KdV charges~\cite{Sasaki:1987mm,Brehm:2019fyy}. 
The code can also be modified easily to compute other (quantum information theoretical) quantities as long as it is possible to express them in terms of correlation functions. There is e.g.~a so-called R\'enyi relative entropy~\cite{Sarosi:2016oks} that could be computed with the methods presented here.
The code can also be modified to compute correlators on different geometries, e.g.~on the torus~\cite{Brehm:2021wev}.

Various directions can also be exploited to improve the code, e.g. the possibility to use symmetries that might speed up the computations significantly. A faster and more efficient code allows to compute higher R\'enyi indices or higher descendants within reasonable time and without too much memory consumption.

As for the trace anomalies of Weyl fermions, our main aim was to compute them explicitly to verify that they do not contain any parity-odd term proportional the (Chern-)Pontryagin density.
These findings however are at odds with the original claim of~\cite{Bonora:2014qla}, reconfirmed also in~\cite{Bonora:2017gzz,Bonora:2018obr}.
There, the trace anomaly in the coupling to gravity is computed with different techniques and regularisations than the one we employed, including dimensional and $\zeta$-function regularisations.
Let us comment more on this clash.

Although the Pontryagin term satisfies the Wess-Zumino consistency condition for trace anomalies, it is known that CFTs do not support parity-odd terms in the correlation function of three stress tensors~\cite{Stanev:2012nq,Zhiboedov:2012bm}, thus hinting at the absence of such a contribution in the trace anomaly.
Our explicit calculation within the MAT background shows indeed that such terms are absent, thereby confirming the findings of \cite{Bastianelli:2016nuf}, which employed the same regularization scheme we used here.
The analogous case of a Weyl fermion in a gauge background provides additional support for these results.

Retracing our calculation of the trace anomaly, and observing the formulae in \eqref{trace-an}, one may notice that an imaginary  term proportional to the Pontryagin density would indeed arise in the trace anomalies if the contribution from the regulators of the charge conjugated fields were neglected. However, there is no justification for dropping those terms. 
This is consistent with four-dimensional CPT invariance, that requires the two Weyl fermion's helicities (described by the Weyl fermion and its hermitian conjugate) to be present in a massless relativistic QFT.
Thus, both helicities circulate in the loop that produces the anomalies, and their contributions cannot be split in any legal way.

Our computation has been questioned and in~\cite{Liu:2022jxz} the regularisation scheme that we also adopted has been employed to claim that the trace anomaly of chiral fermions has a Pontryagin term.
We find the paper unsound and the claims unjustified.
Indeed, the starting point is an action with a Dirac mass term which is not hermitian and that contains both chiralities, even though one lacks a kinetic term and it is not clarified what its role is in the quantum computations.
As the starting point is unsound, all further manipulations and interpretations look unjustified, including the final result.
We thus reject the criticisms.

Additional support from the results we presented here comes from independent analyses~\cite{Frob:2019dgf,Abdallah:2021eii,Abdallah:2022okt}.
In~\cite{Frob:2019dgf} the trace anomaly for Weyl fermions is computed using Hadamard subtraction and it is shown that there is no parity-odd contribution if the divergence of the stress tensor vanishes.
A critical assessment of the regularizations employed in~\cite{Bonora:2014qla,Bonora:2017gzz,Bonora:2018obr} is also performed, and possible sources of disagreement are pointed out.
The chiral trace anomaly has been then computed in~\cite{Abdallah:2021eii} with dimensional regularisation and revised in~\cite{Abdallah:2022okt}, confirming the absence of the Pontryagin density.
Dimensional regularisation has been then employed also to compute the trace anomaly of a Weyl fermion in a gauge background, confirming the absence of parity-odd terms~\cite{Bastianelli:2022hmu}.

As for the holographic anomalies, it would be interesting to extend the analysis we performed to the case where the backreaction is not put off-shell.
If the backreaction is on-shell, then the dual theory is a CFT deformed by irrelevant operators.
The metric beta-function would then indicate that the geometry is subject to an RG flow, and it would be interesting to provide its holographic description.

Our analysis could be generalised to higher dimensional theories, where the calculations will be more complicated but similar to the one we presented here.
However, it does not provide the correct dual description of a two-dimensional boundary CFT.
Indeed, for $d=2$ and $\Delta=3$ the candidate holographic anomaly should be proportional to $\mathcal{L}_g^{(1)} + \mathcal{L}_h^{(1)}$ as given by~\eqref{Og1} and~\eqref{Oh1}.
On the PBH solutions it yields an expression which satisfies the Wess-Zumino condition~\eqref{WZ_wbeta}, but it vanishes in flat space and therefore does not reproduce the expected result which should be proportional to $\phi_{(0)} \Box^2 \phi_{(0)}$.
This happens because for $d=2$ the solutions parametrised by $h_1$ and $h_2$ in~\eqref{h1tilde} are no longer independent.

The breakdown of the present description in $d=2$ is expected, as the CFT analysis in two dimensions requires the addition of a $\Delta=4$ operator, which is eventually identified with the $T \bar T$ operator, together with an operator of dimension $\Delta =3$~\cite{Schwimmer:2019efk}.
We checked that including an additional scalar field in the bulk which sources a $\Delta =4$ operator on the boundary and then demanding that the boundary CFT is not deformed by the irrelevant operators still does not reproduce the $d=2$ metric beta-function found in~\cite{Schwimmer:2019efk}.
This may be because the standard holographic dictionary tells that the $T \bar T$ operator is dual to a sharp cut-off in the bulk.
However, we saw that requiring that the dual CFT is undeformed translated into eliminating the cut-off induced by the irrelevant operator. 
It would thus be interesting to extend the present analysis to the two dimensional case and see how the CFT results are reproduced by holographic computations.

Relating our holographic analysis to the ones in~\cite{Skenderis:2009nt,Guica:2010sw} would also be desirable.
There, it is shown that CFTs with certain irrelevant operators in $d$ dimensions can be viewed as dual to $(d+1)$-dimensional massive vector theories or to topologically massive gravity for $d=2$.
Topologically massive gravity is known to have a higher derivative generalisation which is a third-way consistent theory~\cite{Bergshoeff:2014pca}, meaning that the equations of motion are consistent only on-shell.
Examples of third-way consistent theories in dimensions higher than three exist for interacting $p$-form theories~\cite{Broccoli:2021pvv}, but finding third-way consistent theories of gravity in higher dimensions is still an open challenge.
We speculate that the holographic description of irrelevant operators might provide insights into the construction of new third-way consistent theories of gravity.

\appendix
\pagestyle{appendices}
\chapter{More on transformation and correlators of descendant states}\label{app:descendant}

\section{The action of conformal transformations}

\subsection{Mathematica code to obtain the v's}\label{app:matv}

We provide here a Mathematica code to compute the relation between the coefficients $v_j$ and $a_k$ as discussed in sec.~\ref{sec:trafo}.

\begin{verbatim}
Poly[m_, f_] := Sum[v[j] t^(j + 1) , {j, 1, m - 1}] D[f, t]

PolyToPowerNN[NN_, m_, f_] := 
    If[NN == 1, Poly[m, f], PolyToPowerNN[NN - 1, m, Poly[m, f]]]
    
lhs[MM_] := 
    v[0] t + v[0] Sum[1/i! PolyToPowerNN[i, MM, t], {i, 1, MM}]
    
Equ[NN_] := 
    Block[{tmp},
        tmp=CoefficientList[lhs[NN] 
        - Sum[a[i] t^i,{i,1,NN}],t][[2;;NN+1]];
        {Table[tmp[[i]] == 0, {i, NN}], Table[v[i], {i,0,NN-1}]}]
    
ListOfVs[NN_] := 
    Block[{tmp}, tmp = Equ[NN];
        Table[v[n],{n,0,NN-1}]/.Solve[tmp[[1]],tmp[[2]]][[1]]]
\end{verbatim}

\subsubsection{Example: Coefficients up to $j=5$}
Up to $j=5$, the relations between $v_j$ and $a_k$ are the following:

\begingroup
\allowdisplaybreaks[1]
\begin{align}
    v_0 & = a_1\,,\\
    v_1 & = \frac{a_2}{a_1}\,,\\
    v_2 & = \frac{a_1 a_3 -a_2^2}{a_1^2}\,,\\
    v_3 & = \frac{3a_2^3 - 5 a_1 a_2 a_3 +2a_1^2a_4}{2a_1^3}\,,\\
    v_4 & = -\frac{16a_2^4 -37a_1a_2^2a_3 + 9 a_1^2a_3^2 +18a_1^2a_2 a_4 -6 a_1^3a_5}{6a_1^4}\,,\\
    v_5 & = \frac{31 a_2^5 -92 a_1 a_2^3a_3 +48 a_1^2a_2^2a_4 -7 a_1^2a_2 \left(-7a_3^2+3a_1a_5\right) +3a_1^3\left(-7a_3a_4 +2a_1a_6\right)}{6a_1^5}\,.
\end{align}
\endgroup

\subsection{Local action for the uniformization map}\label{app:uniformization}

For the local action of the uniformization map
\begin{equation}
    w(z) = \left(\frac{z e^{ -i\pi \frac{l}{L}} - 1}{z -  e^{-i\pi\frac{l}{L}} }\right)^{\frac1n}
\end{equation}
we need the transformation of local coordinates. We choose the standard local coordinates on the $k$th sheet around a point $z_k$
\begin{equation}
   \varphi_{z_k}(\rho) = \rho + z_k\,,
\end{equation}
which are mapped to
\begin{equation}
    \tilde{\beta}_{z_k}(\rho) = \left(\frac{\left(\rho+z_k\right)e^{ -\frac{i\pi l}{L}} - 1}{\rho+z_k -  e^{ -\frac{i\pi l}{L}} }\right)^{\frac1n}\,.
\end{equation}
    
\noindent 
The standard local coordinates on the plane around $w(z_k)$ are simply
\begin{equation}
    \beta_{w(z_k)}(\rho) = \rho + w(z_k)\,.
\end{equation}
    
\noindent
Now, the local coordinate change $\eta_{z_k}$ should satisfy
\begin{equation}
    \tilde{\beta}_{z_k}(\rho) = \beta_{w(z_k)}(\eta_{z_k}(\rho))\,
\end{equation}
and, hence, 
\begin{equation}
    \eta_{z_k}(\rho) = \beta_{w(z_k)}^{-1} \!\left(\tilde{\beta}_{z_k}(\rho)\right)\,.
\end{equation}

\noindent
Since we deal with the standard local coordinates this is straightforward to compute
\begin{equation}
    \eta_{z_k}(\rho) = \tilde{\beta}_{z_k}(\rho) - w(z_k) =  \left(\frac{\left(\rho+z_k\right)e^{ -\frac{i \pi l}{L}} - 1}{\rho+z_k -  e^{ -\frac{i \pi l}{L}} }\right)^{\frac1n} -  \left(\frac{z_k e^{ -\frac{i \pi l}{L}} - 1}{z_k -  e^{ -\frac{i \pi l}{L}} }\right)^{\frac1n}\,,
\end{equation}
and hence for the actual insertion points $z_k = 0_k$ we get
\begin{equation}
    \eta_{0_k}(\rho) = e^{\frac{2\pi i(k-1)}{n} } \left(\left(\frac{\rho \,e^{\frac{i \pi l}{L}} - 1}{\rho -  e^{ \frac{i \pi l}{L}} }\right)^{\frac1n} - e^{\frac{i \pi l}{nL}}\right)\,.
\end{equation}

\noindent 
Expanding this around $\rho$ allows us to solve for the coefficients $v_j$ appearing in the local action $\Gamma_{w(0,k)} \equiv \Gamma_{k,l}$. Up to $j=5$ they are given by
\begin{align}
    v_0 &= \frac{ e^{\frac{2 \pi i (k-1)}{n} + \frac{\pi i (1-n) l }{n L}}}{n} \left(e^{2\pi i \frac{l}{L}} -1\right)\,,\\
    v_1 &= \cos\left(\frac{\pi l}{L}\right) + i \frac{\sin\left(\frac{\pi l}{L}\right)}{n}\,,\\
    v_2 &= \frac{1-n^2}{3n^2} \sin^2\!\left(\frac{\pi l}{L}\right)\,,\\
    v_3 &= \frac{v_2}{2n} \left(n \cos\left(\frac{\pi l}{L}\right) -i \sin\left(\frac{\pi l}{L}\right)\right)\,,\\
    v_4 &= \frac{v_2}{30n^2} \left(n^2-4 + 4(n^2+1) \cos\left(\frac{2\pi l}{L}\right) - 10 i\, n \sin\left(\frac{2 \pi l}{L}\right)\right)\,,\\
    v_5 &= -\frac{v_2 \sin\left(\frac{\pi l}{L}\right) }{18 n^3} \bigg((11n^3+61n) \sin\left(\frac{2\pi l}{L}\right) + i (61 n^2 + 11) \cos\left(\frac{2\pi l}{L}\right) \nonumber\\
    & \phantom{=} + i (29n^2-11) \bigg)\,.
\end{align}

\noindent
Note that for the dual fields we basically have to take the composition of the uniformization map with the inversion, i.e.~we have to repeat the latter arguments for $w(1/z)$. Let us denote the local coordinate change by $\theta_{0_k}(\rho)$. It is given by 
\begin{equation}
    \theta_{0_k}(\rho) = \eta_{0_k}(\rho)|_{l\to-l}\,,
\end{equation}
so that the respective local action is given by $\Gamma_{w(1/z)} = \Gamma_{w(z)} \Gamma_{1/z} \equiv \Gamma_{k,-l}$\,.

\section{Code to compute correlation functions of descendants}
We report here the Mathematica codes with which we implemented the recursive formula discussed in~\ref{sec:rec_formula}.
A working Mathematica notebook with the present codes is also available in the ancillary files of the arXiv version of~\cite{Brehm:2020zri}.
We use the Mathematica implementation of the Virasoro algebra by M.~Headrick that can be downloaded from \url{http://people.brandeis.edu/~headrick/Mathematica/}.  

\subsection{Any \texorpdfstring{$N$}{N}-point function of vacuum descendants}\label{app:VacDesCorr}

The Mathematica code to compute any correlator of descendants of the vacuum is

\begin{verbatim}
VacNptFct[stat_] := Module[{states, ntrivial, TMP, tmp0, tmp1, tmp2},
  
  (*reorders the states s.t. the descendants with more Virasoro 
  generators are mostleft. This makes the recursion faster:*)
  states = Sort[stat, Length[
    GetSequence[#1[[1]]]]<Length[GetSequence[#2[[1]]]] &];
  
  (*Checks at which position there are non-trivial descendants:*)
  ntrivial = Position[Sign[level[states[[All, 1]]]], 1] // Flatten; 
  
  (*When there are no non-trivial descendants the function returns 1. 
  For only one descendant it returns 0 due to translation invariance. 
  In any other case it uses the recursion:*)
  Which[Length[ntrivial] == 0, 1, Length[ntrivial] == 1, 0, True,
   TMP = Table[
     RecStep[states[[ntrivial[[1]], 1]], states[[ntrivial[[i]], 1]], 
      states[[ntrivial[[1]], 2]], states[[ntrivial[[i]], 2]]], {i, 2, 
      Length[ntrivial]}];
   Sum[Sum[
     TMP[[j - 1, nn, 1]] VacNptFct[
       Table[Which[
         i == ntrivial[[1]], {CutFirst[states[[i, 1]]], 
          states[[i, 2]]}, 
         i == ntrivial[[j]], {TMP[[j - 1, nn, 2]], states[[i, 2]]}, 
         True, states[[i]]], {i, Length[states]}]], {nn, 
      Length[TMP[[j - 1]]]}], {j, 2, Length[ntrivial]}]
   ]
  ]
\end{verbatim}
where we define the functions
\begin{verbatim}
RecStep[ L[n1__], L[n2__], z1_, z2_] := 
 Module[{tmp1, test}, 
  tmp1 = List @@ 
     Expand[Sum[- Coeff[-{n1}[[1]], n, z1, z2] L[n - 1] ** L[n2] ** 
         vac, {n, 0, -Plus[n2] + 1}]] ** vac;
  tmp1 = ReArr[tmp1];
  test = 1;
  While[test == 1, test = 0;
   For[nnn = 1, nnn < Length[tmp1], nnn++, 
    If[GetSequence[tmp1[[nnn, 2]]]==GetSequence[tmp1[[nnn + 1, 2]]],
      test = 1; tmp1[[nnn, 1]] = tmp1[[nnn, 1]] + tmp1[[nnn + 1, 1]]; 
     tmp1 = Delete[tmp1, nnn + 1]; Break;]]];
  tmp1 = Simplify[tmp1]
  ]

ReArr[a_ L[m__]] := {a, L[m]}
ReArr[ L[m__]] := {1, L[m]}
ReArr[a_ ] := {a, 1}
SetAttributes[ReArr, Listable];

GetSequence[L[m__]] := {m}
GetSequence[a] := {}
SetAttributes[GetSequence, Listable];

Coeff[m_, n_, zi_, zj_] := 
    (-1)^n Binomial[n + m - 2, n] (zj - zi)^(1 - m - n)
  
CutFirst[ L[n__]] :=  L[{n}[[2 ;; Length[{n}]]] /. List -> Sequence]
\end{verbatim}

\noindent 
The function \verb+VacNptFct+ takes as arguments a list of descendants together with their coordinates. The descendants are given in the form $\verb+L+[-n_1,...,-n_k]$, where $n_i \ge n_{i+1}$, $n_i\in \mathbb{N}$. The coordinates can either be variables or specific values. For example 
\begin{verbatim}
    VacNptFct[{{L[-2],z},{L[-2],w}}]
\end{verbatim}
gives the result for the two-point function of the energy momentum tensor, $\frac{c/2}{(z-w)^4}$.

\subsection{Any \texorpdfstring{$N$}{N}-point function of non-vacuum descendants}\label{app:PrimDesCorr}

Given a correlator of descendants of non-vacuum primaries, we compute the differential operator acting on the correlator of primaries with the function \verb|NPtFct|:
\begin{verbatim}
NPtFct[stat_] :=
 Which[
 
  (* checks the input is given in the correct form :*)
  And @@ Table[Length[stat[[i]]] != 2, {i, Length[stat]}], 
  "The number of fields and coordinates do not match!",
  
  (* If there is only one descendant then it returns 0 due to 
     translational invariance *)
  Length[stat] == 1, 0,
  True, Module[{states, virpos, derivative, tmp, rec, noone, pr},
   
   (*reorders the states s.t. 
     the descendants with more Virasoro generators are most left. 
     This makes the recursion faster :*)
   states = 
    Sort[stat, 
     Length[GetSequence[#1[[1]]]] <= Length[GetSequence[#2[[1]]]] &];
   pr = FindPermutation[stat, states];
   virpos = Position[Sign[level[states[[All, 1]]]], 1] // Flatten;
   derivative = 
    Table[Length[GetSequence[states[[virpos[[i]], 1]]]] == 
      level[states[[virpos[[i]], 1]]], {i, 1, Length[virpos]}];
   
   (*When there are no non-
     trivial descendants the function returns corrp[...]. 
     If the descendants are only level 1 descendants, 
     it returns the appropriate derivatives acting on corrp[...]. 
     In any other case it uses the recursion :*)
   Which[
    Length[virpos] == 0, corrp[stat[[All, 2]] /. List -> Sequence],
    And @@ derivative, 
    Derivative[level[stat[[All, 1]]] /. List -> Sequence][corrp][
     stat[[All, 2]] /. List -> Sequence],
    True, noone = Position[derivative, False] // Flatten;
    rec = 
     Drop[Table[i, {i, 1, Length[states]}], {virpos[[noone[[1]]]]}];
      tmp = 
     Table[RecStep[states[[virpos[[noone[[1]]]], 1]], states[[i, 1]], 
       states[[virpos[[noone[[1]]]], 2]], states[[i, 2]]], {i, rec}];
    Sum[tmp[[i, j, 1]] NPtFct[
       Permute[
        ReplacePart[
         states, {{virpos[[noone[[1]]]], 1} -> 
           CutFirst[states[[virpos[[noone[[1]]]], 1]]], {rec[[i]], 
            1} -> tmp[[i, j, 2]]}], Ordering[PermutationList[pr]]]
       ], {i, 1, Length[tmp]}, {j, 1, Length[tmp[[i]]]}]]]]
\end{verbatim}
where we define the functions:
\begin{verbatim}
RecStep[ L[n1__] ** prim[p1_], L[n2__] ** prim[p2_], z1_, z2_] := 
 Module[{tmp1, test}, 
  tmp1 = List @@ 
    Expand[Sum[-Coeff[-{n1}[[1]],n,z1,z2] L[n-1]**L[n2]**prim[p2],
            {n, 0, -Plus[n2] + 1}]];
  tmp1 = ReArr[tmp1];
  test = 1;
  While[test == 1, test = 0;
   For[nnn = 1, nnn < Length[tmp1], nnn++, 
    If[GetSequence[tmp1[[nnn, 2]]]==GetSequence[tmp1[[nnn + 1, 2]]],
      test = 1; tmp1[[nnn, 1]] = tmp1[[nnn, 1]] + tmp1[[nnn + 1, 1]]; 
     tmp1 = Delete[tmp1, nnn + 1]; Break;]]];
  tmp1 = Simplify[tmp1]
  ]

RecStep[ L[n1__] ** prim[p1_], prim[p2_], z1_, z2_] := 
 Module[{tmp1, test}, 
  tmp1 = List @@ 
    Expand[-Sum[
       Coeff[-{n1}[[1]], n, z1, z2] L[n - 1]**prim[p2], {n, 0, 1}]];
  tmp1 = ReArr[tmp1];
  test = 1;
  While[test == 1, test = 0;
   For[nnn = 1, nnn < Length[tmp1], nnn++, 
    If[GetSequence[tmp1[[nnn, 2]]]==GetSequence[tmp1[[nnn + 1, 2]]],
      test = 1; tmp1[[nnn, 1]] = tmp1[[nnn, 1]] + tmp1[[nnn + 1, 1]]; 
     tmp1 = Delete[tmp1, nnn + 1]; Break;]]];
  tmp1 = Simplify[tmp1]
  ]
  
ReArr[a_ L[m__] ** prim[p_]] := {a, L[m] ** prim[p]}
ReArr[L[m__] ** prim[p_]] := {1, L[m] ** prim[p]}
ReArr[a_ prim[p_]] := {a, prim[p]}
ReArr[prim[p_]] := {1, prim[p]}
SetAttributes[ReArr, Listable];

CutFirst[ L[n__] ** prim[p_]] :=  
 L[{n}[[2 ;; Length[{n}]]] /. List -> Sequence] ** prim[p]
 
GetSequence[L[m__] ** prim[p_]] := {m}
GetSequence[prim[p_]] := {}
SetAttributes[GetSequence, Listable];
\end{verbatim}

\noindent
The function \verb|NPtFct| takes as arguments a list of $N$ lists, where in the innermost lists the first entry is the descendant and the second entry is the coordinate.
The descendants are given as $\verb|L[|-n_1,...,-n_k\verb|]**prim[p]|$, where again $n_i \ge n_{i+1}$, $n_i\in \mathbb{N}$ and \verb|prim[p]| denotes the primary state.
For instance, 
\begin{verbatim}
    tp = NPtFct[{ {L[-2] ** prim[p], z}, {L[-1, -1] ** prim[p], w} }]
\end{verbatim}
produces the output
\begin{align*}
&\frac{6 \, \text{corrp}[z,w] h[p]}{(w-z)^4}-\frac{2 (1+2 h[p]) \text{corrp}^{(0,1)}[z,w]}{(w-z)^3} \\
& +\frac{(2+h[p]) \text{corrp}^{(0,2)}[z,w]}{(w-z)^2}+\frac{\text{corrp}^{(0,3)}[z,w]}{-w+z}
\end{align*}
where \verb|h[p]| is the conformal dimension of \verb|prim[p]| and the function \verb|corrp|, which is a function of the insertion points, denotes the correlator of primaries. The derivatives acting on it are displayed in the Mathematica language.
If we know the explicit expression of \verb|corrp|, we can further simplify the output; in our example we can for instance write:
\begin{verbatim}
    corrp[z1_, z2_] := 1/(z1 - z2)^(2 h[p])
    tp // Simplify
    Clear[corrp]
\end{verbatim}
to get the explicit result
\begin{equation*}
    6 (-w+z)^{-2 (2+h[p])} h[p] \left(3+5 h[p]+2 h[p]^2\right)
\end{equation*}

\section{Explicit results}

We collect here some more results in addition to the ones already presented in chap.~\ref{chap:desc_corr} for the R\'enyi entanglement entropy, sandwiched R\'enyi divergence and trace square distance of (non-)vacuum descendants.
Henceforth we write $x=l/L$.

\subsection{R\'enyi entanglement entropy}

\subsubsection{Vacuum module}\label{app:REresultsVac}

The second R\'enyi entanglement entropy for $L_{-3}\ket{0}$, $L_{-4}\ket{0}$, and $L_{-5}\ket{0}$ are:

\begingroup
\allowdisplaybreaks[1]
\begin{align}\label{eq:RFE[-3]}
    F^{(2)}_{L_{-3}\ket{0}} &= \frac{c^2\sin ^8(\pi x) \cos ^4(\pi x)}{64} \nonumber\\
    &\phantom{=}+\frac{c \sin ^4(\pi x) \cos ^2(\pi x) (255 \cos (2 \pi x)+90 \cos (4 \pi x)+17 \cos (6 \pi x)+1686)}{8192} \nonumber\\
   &\phantom{=}+\frac{\sin ^4(\pi x) (8391 \cos (2 \pi x)+1890 \cos (4 \pi x)+361 \cos (6 \pi x)+7790)}{16384 c}\nonumber\\
   &\phantom{=}+\frac{1}{8388608}(3032808 \cos (2 \pi x)+819919 \cos (4 \pi x)-27612 \cos (6 \pi x) \nonumber\\
   &\phantom{=} + 386 \cos (8 \pi x)+8436 \cos (10 \pi x)+289
   \cos (12 \pi x)+4554382)\,,
\end{align}
\endgroup

\begingroup
\allowdisplaybreaks[1]
\begin{align}\label{eq:RFE[-4]}
    F^{(2)}_{L_{-4}\ket{0}} &= \frac{c^2 \sin ^8(\pi x) (3 \cos (2 \pi x)+2)^4}{6400} \nonumber\\
    &\phantom{=}+\frac{ c \sin ^4(\pi x) (3 \cos (2 \pi x)+2)^2}{6553600} (12760 \cos (2 \pi x)+7396 \cos (4 \pi x) \nonumber\\
    &\phantom{=} + 2152 \cos (6 \pi x)) +1263 \cos (8 \pi x)+140269) \nonumber\\
    &\phantom{=}+\frac{\sin ^4(\pi x)}{6553600 c} (5444642 \cos (2 \pi x)+2684168 \cos (4 \pi x)+913973 \cos (6 \pi x) \nonumber\\
    &\phantom{=} + 286934 \cos (8 \pi x)+59049 \cos (10 \pi x) ) \nonumber\\
    &\phantom{=}+\frac{1}{53687091200} ( 16641312784 \cos (2 \pi x)+4954285000 \cos (4 \pi  x) \nonumber\\
    &\phantom{=} +1976400688 \cos (6 \pi x)  -121298020 \cos (8 \pi x)-5870960 \cos (10 \pi x) \nonumber\\
    &\phantom{=} +13794296 \cos (12 \pi x) + 18977328 \cos (14 \pi x)+1595169 \cos (16 \pi  x) \nonumber\\
    &\phantom{=} +30207894915) \, ,
\end{align}
\endgroup

\begingroup
\allowdisplaybreaks[1]
\begin{align}\label{eq:RFE[-5]}
   F^{(2)}_{L_{-5}\ket{0}} &= \frac{1}{3435973836800} \big[c^2 (-16777216 \cos (2 \pi  x)-51380224 \cos (4 \pi  x) \nonumber \\
   &\phantom{=} -79691776 \cos (6 \pi  x)) -98566144 \cos (8 \pi  x)+121634816 \cos (10 \pi x) \nonumber \\
   &\phantom{=} +17825792 \cos (12 \pi  x) +8388608 \cos (14 \pi  x)-8388608 \cos (16 \pi  x) \nonumber \\
   &\phantom{=} -33554432 \cos (18 \pi  x) +16777216 \cos (20 \pi x)+123731968) \nonumber \\
   &\phantom{=} + c (-4076806144 \cos (2 \pi  x)-9140649984 \cos (4 \pi  x) \nonumber\\
   &\phantom{=} -14113284096 \cos (6 \pi  x)-18862759936 \cos (8 \pi  x) \nonumber\\
   &\phantom{=} + 18117304320 \cos (10 \pi  x)  -27205632 \cos (12 \pi  x)+101367808 \cos (14 \pi  x) \nonumber\\
   &\phantom{=} + 114972672 \cos (16 \pi  x) -28581888 \cos (18 \pi x) + 74997760 \cos (20 \pi  x) \nonumber\\
   &\phantom{=}+27840645120) + c^{-1}(-19727178304 \cos (2 \pi  x)-27011932672 \cos (4 \pi  x) \nonumber\\
   &\phantom{=} -22303010688 \cos (6\pi  x) -10523886336 \cos (8 \pi  x) \nonumber\\
   &\phantom{=} +8735760000 \cos (10 \pi  x) +2720936448 \cos (12 \pi  x) \nonumber\\
   &\phantom{=} + 1016710944 \cos (14 \pi  x)+879348416 \cos (16\pi  x) +919004192 \cos (18 \pi  x) \nonumber\\
   &\phantom{=} +65294248000) + 967144492584 \cos (2 \pi  x)+295129895330 \cos (4 \pi  x) \nonumber\\
   &\phantom{=} +135116995760 \cos (6 \pi  x) +61785824936 \cos (8 \pi  x) \nonumber\\
   &\phantom{=} -5774059280 \cos (10 \pi  x) -260030763 \cos (12 \pi  x) + 179443820 \cos (14 \pi  x) \nonumber\\
   &\phantom{=} +597122990 \cos(16 \pi  x) +506168268 \cos (18 \pi  x) + 83814025 \cos (20 \pi  x) \nonumber\\
   &\phantom{=} +1981464169130 \big] \, .
\end{align}
\endgroup

\subsubsection{Ising model}\label{app:REEresultsIsing}
Up to level 3 descendants of the energy density operator we find the following results for the $n=2$ R\'enyi entanglement entropy:
\begin{align}
    F^{(2)}_{L_{-1}\ket{\varepsilon}} &= \frac{(\cos (2 \pi  x)+7) (1558 + 439 \cos (2 \pi  x)+26 \cos (4 \pi  x)+25 \cos (6 \pi  x))}{16384} \, , \\
    \nonumber\\
    F^{(2)}_{L_{-2}\ket{\varepsilon}} &= \frac{\cos (2 \pi  x)+7}{67108864} (6085442 + 1693410 \cos (2 \pi  x)+514952 \cos (4 \pi  x) \nonumber\\
    &\phantom{ = } +49813 \cos (6 \pi  x)+9270 \cos (8 \pi  x)+35721 \cos (10 \pi  x)) \\
    &= F^{(2)}_{L_{-1}^2\ket{\varepsilon}} \,, \nonumber\\
    \nonumber\\
    F^{(2)}_{L_{-3}\ket{\varepsilon}} &= \frac{\cos (2 \pi  x)+7}{17179869184} (1523423468 + 432147835 \cos (2 \pi  x)+111740030 \cos (4 \pi  x) \nonumber\\ 
    & \phantom{= } +65921129 \cos (6 \pi  x)+7438836 \cos (8 \pi  x)+1584475 \cos (10 \pi  x) \nonumber\\ 
    & \phantom{= } +626850 \cos
   (12 \pi  x)+4601025 \cos (14 \pi  x)) \\
   &= F^{(2)}_{L_{-1}^3\ket{\varepsilon}} = F^{(2)}_{L_{-2}L_{-1}\ket{\varepsilon}}\,, \nonumber
\end{align}
where the common prefactor is due to the factorization of the holomorfic and antiholomorfic parts of the correlator.
Even though $\mathcal{D}^{F(2)}_{L_{-1}^2} \neq \mathcal{D}^{F(2)}_{L_{-2}}$, at level~2 we find the same entanglement entropy for the different descendants and the same happens at level~3.
This reflects the existence of only one physical state at level~2 and~3.

For $\sigma$ descendants:
\begingroup
\allowdisplaybreaks[1]
\begin{align}
    F^{(2)}_{L_{-1}\ket{\sigma}} &= \frac{435 + 60 \cos (2 \pi  x)+17 \cos (4 \pi  x)}{512}\,, \\
    \nonumber\\
    F^{(2)}_{L_{-2}\ket{\sigma}} &= \frac{1}{2097152}(1560707 + 438088 \cos (2 \pi  x)+75420 \cos (4 \pi  x) \nonumber\\
    &\phantom{=} + 8312 \cos (6 \pi  x)+14625 \cos (8 \pi  x) ) \\
    &= F^{(2)}_{L_{-1}^2\ket{\sigma}}\,, \nonumber\\
    \nonumber\\
    F^{(2)}_{L_{-3}\ket{\sigma}} &= \frac{1}{63438848}( 42511910 + 16535144 \cos (2 \pi  x)+2825131 \cos (4 \pi  x) \nonumber\\
    &\phantom{=} +1123684 \cos (6 \pi  x) + 179114 \cos (8 \pi  x)+141364 \cos (10 \pi  x) \nonumber\\
    &\phantom{=}+122501 \cos (12 \pi x) )\,, \\
    \nonumber\\
   F^{(2)}_{L_{-1}^3\ket{\sigma}} &= \frac{1}{5585604968448} (3968670881070 + 1175831066472 \cos (2 \pi  x)\nonumber\\
   &\phantom{=} +306581016863 \cos (4 \pi  x) + 102222772068 \cos (6 \pi  x) \nonumber\\
   &\phantom{=}+11235770850 \cos (8 \pi  x) + 5235592500 \cos (10 \pi  x)\nonumber\\
   &\phantom{=}+15827868625 \cos (12 \pi  x))\,, \\
   \nonumber\\
   F^{(2)}_{L_{-2}L_{-1}\ket{\sigma}} &= \frac{1}{173946175488} (120071187054 + 40660967528 \cos (2 \pi  x)\nonumber\\
   &\phantom{=} +9353937343 \cos (4 \pi  x) + 2427785700 \cos (6 \pi  x)\nonumber\\
   &\phantom{=}+860494498 \cos (8 \pi  x) + 152046004 \cos (10 \pi  x)+419757361 \cos (12 \pi x))\,.
\end{align}
\endgroup
In this case we have one physical state at level~2, while two physical states at level~3 and we thus find different expressions for the REEs for degenerate states at level~3.

\subsubsection{Three-state Potts model}\label{app:REEresultsPotts}
For the first descendant of the energy density in the three-states Potts model we find: 
\begingroup
\allowdisplaybreaks[1]
\begin{align}
    F^{(2)}_{L_{-1}\ket{\varepsilon}} &= \tfrac{1}{64} \, F\!\left(-\tfrac{8}{5},-\tfrac{1}{5};-\tfrac{2}{5};\eta \right){}^2 (\cos (2 \pi  x)+7)^2 \nonumber\\
    &\phantom{=} +\tfrac{1}{166400}\big\{ F\left(-\tfrac{8}{5},-\tfrac{1}{5};-\tfrac{2}{5};\eta \right) \sin ^4(\pi  x) \big[9 \sin (\pi  x) \left(49 F\!\left(\tfrac{12}{5},\tfrac{19}{5};\tfrac{18}{5};\eta \right) \sin ^3(\pi  x) \right.  \nonumber\\
    &\phantom{=} \left. +260 F\! \left(\tfrac{7}{5},\tfrac{14}{5};\tfrac{13}{5};\eta \right) \sin (2 \pi 
   x)\right)+5200 F\!\left(-\tfrac{3}{5},\tfrac{4}{5};\tfrac{3}{5};\eta \right) \cos (\pi  x) \nonumber\\
   & \phantom{=}  +520 F\!\left(\tfrac{2}{5},\tfrac{9}{5};\tfrac{8}{5};\eta \right)
   (31 \cos (2 \pi  x)+41)\big] \big\} \nonumber\\
   &\phantom{=} -\left(111411200\ 2^{2/5} \Gamma \left(-\tfrac{8}{5}\right) \Gamma \left(\tfrac{17}{10}\right) \Gamma
   \left(\tfrac{12}{5}\right)\right)^{-1} \nonumber\\
   & \phantom{=} \times \bigg\{\Gamma \left(-\tfrac{2}{5}\right) \Gamma \left(\tfrac{3}{10}\right) \Gamma \left(\tfrac{13}{5}\right) F\! \left(\tfrac{6}{5},\tfrac{13}{5};\tfrac{12}{5};\eta \right) \sin ^{\tfrac{28}{5}}(\pi  x) \nonumber\\
   &\phantom{=}  \times \big[ 34 F\! \left(\tfrac{6}{5},\tfrac{13}{5};\tfrac{12}{5};\eta
   \right) (14196 \cos (2 \pi  x)+15129 \cos (4 \pi  x)+7667) \nonumber\\
   &\phantom{=} +13 \sin ^2(\pi  x) \big(9016 F\!\left(\tfrac{26}{5},\tfrac{33}{5};\tfrac{32}{5};\eta \right) \sin
   ^6(\pi  x)  \nonumber\\
   &\phantom{=} +17 F\! \left(\tfrac{11}{5},\tfrac{18}{5};\tfrac{17}{5};\eta \right) (8977 \cos (\pi  x)+6479 \cos (3 \pi  x))  \nonumber\\
   &\phantom{=}  +79488 F\! \left(\tfrac{21}{5},\tfrac{28}{5};\tfrac{27}{5};\eta \right) \sin ^4(\pi  x) \cos (\pi  x) \nonumber\\
   &\phantom{=}  +99 F\! \left(\tfrac{16}{5},\tfrac{23}{5};\tfrac{22}{5};\eta \right)
   \sin ^2(\pi  x) (1373 \cos (2 \pi  x)+1003)\big)\big] \bigg\} \,,
\end{align}
\endgroup
where $F \equiv \; _2F_1 $ is the hypergeometric function and $\eta = \sin^2 \left( \frac{\pi x}{2} \right)$.
For higher-level descendants the expressions are more involved, and we limit ourselves to show this simplest example.

\subsection{Sandwiched R\'enyi divergence}

\subsubsection{Vacuum module}\label{app:SRDresultsvac}

Some explicit expressions for the SRD between the vaccum and light states are:

\begingroup
\allowdisplaybreaks[1]
\begin{align}
    \mathcal{F}^{(2)}_{L_{-3}\ket{0}} &= \frac{e^{12 i \pi x}}{2048 \left(1+e^{2 i \pi x}\right)^{12}} \big[ c^{-1} \big(-202752 \cos (2 \pi x)-3775424 \cos (4 \pi x)\nonumber\\
    &\phantom{=} +356352 \cos (6 \pi x) + 888064 \cos (8 \pi x)-184320 \cos (10 \pi x)-137\cos (12 \pi x)\nonumber\\
    &\phantom{=} +27648 \cos (14 \pi x) + 14272 \cos (16 \pi x)+3072 \cos (18 \pi x)+288 \cos (20 \pi x) \nonumber\\
    &\phantom{=} +2874176 \big) -86237760 \cos (2 \pi x)+49130280 \cos (4 \pi x) \nonumber\\
    &\phantom{=} -14301120 \cos (6 \pi x) + 2900567 \cos (8 \pi x)-122592\cos (10 \pi x) \nonumber\\
    &\phantom{=} +12228 \cos (12 \pi x) -1568 \cos (14 \pi x)-2062 \cos (16 \pi x)-416 \cos (18 \pi x) \nonumber\\
    &\phantom{=} + 276 \cos (20 \pi x)+160 \cos (22 \pi x) +25 \cos (24 \pi x)+57010590 \big] \,,
\end{align}
\endgroup

\begingroup
\allowdisplaybreaks[1]
\begin{align}
    \mathcal{F}^{(2)}_{L_{-4}\ket{0}} &= \frac{e^{16 i \pi x}}{163840 \left(1+e^{2 i \pi x}\right)^{16}} \big[ c^{-1} \big( -2183086080 \cos (2 \pi x)-14065619072 \cos (4 \pi x) \nonumber\\
    &\phantom{=} + 3329505280 \cos (6 \pi x)+3373021952 \cos (8 \pi x)-1144576000 \cos(10 \pi x) \nonumber\\
    &\phantom{=} -7513472 \cos (12 \pi x)-26378240 \cos (14 \pi x)+22635008 \cos (16 \pi x) \nonumber\\
    &\phantom{=} + 21217280 \cos (18 \pi x)+10149760 \cos (20 \pi x) +3225600 \cos (22 \pi x) \nonumber\\
    &\phantom{=} + 693504 \cos (24 \pi x)+92160 \cos (26 \pi x)+5760 \cos (28 \pi x)+10666626560 \big) \nonumber\\
    &\phantom{=} -1927558100400 \cos (2 \pi x)+1107347224880 \cos (4 \pi x) \nonumber\\
    &\phantom{=} -420523178000 \cos (6 \pi x)+102577128040 \cos (8 \pi x) \nonumber\\
    &\phantom{=} -14129186800 \cos (10 \pi x)+1083586960 \cos (12 \pi x) -18763600 \cos (14 \pi x) \nonumber\\
    &\phantom{=} +2105260 \cos (16 \pi x)-2374000 \cos (18 \pi x) -2296400 \cos (20 \pi x) \nonumber\\
    &\phantom{=} -722000 \cos (22 \pi x)+205400 \cos (24 \pi x) + 286800 \cos (26 \pi x) \nonumber\\
    &\phantom{=} +120720 \cos (28 \pi x)+25200 \cos (30 \pi x) + 2205 \cos (32 \pi x) \nonumber\\
    &\phantom{=} +1161961353975 \big] \,.
\end{align}
\endgroup

\subsubsection{Ising model}\label{app:SRDising}
We present here some of the correlation functions related to the SRD computation for non-vacuum descendants.
For simplicity we show only the results for $\varepsilon$ descendants in the Ising model:

\begin{align}
    \mathcal{F}^{(2)}_{L_{-1}\ket{\varepsilon}} &= \frac{\cos (4 \pi  x)+7}{4096 \cos ^8(\pi x)} (954 -776 \cos (2 \pi  x)+319 \cos (4 \pi  x) +4 \cos (6 \pi  x) \nonumber\\
    &\phantom{=} +6 \cos (8 \pi  x)+4 \cos (10 \pi  x)+\cos (12 \pi  x)) \,, \\
    \nonumber\\
   \mathcal{F}^{(2)}_{L_{-2}\ket{\varepsilon}} &= \frac{\cos (4 \pi  x)+7}{1048576  \cos ^{12}(\pi  x)} (1546754 -2155216 \cos (2 \pi  x)+864034 \cos (4 \pi  x) \nonumber\\
   &\phantom{=} -139296 \cos (6 \pi  x)+13320 \cos (8 \pi  x)+224 \cos (10 \pi  x) +469 \cos (12 \pi  x) \nonumber\\
   &\phantom{=} +456 \cos (14 \pi  x) +246 \cos (16 \pi  x) +72 \cos (18 \pi  x)+9 \cos (20 \pi  x)) \\
   & = \mathcal{F}^{(2)}_{L_{-1}^2\ket{\varepsilon}} \,, \nonumber\\
   \nonumber\\
   \mathcal{F}^{(2)}_{L_{-3}\ket{\varepsilon}} &= \frac{\cos (4 \pi  x)+7 }{67108864\cos ^{16}(\pi  x)} (938450676-1469899184 \cos (2 \pi  x) \nonumber\\
   &\phantom{=} +710758371 \cos (4 \pi  x) -201143980 \cos (6 \pi  x)+32581122 \cos (8 \pi  x) \nonumber\\
   &\phantom{=} -2510220 \cos (10 \pi  x)+99537 \cos (12 \pi  x)+5080 \cos (14 \pi  x) \nonumber\\
   &\phantom{=} +13484 \cos (16 \pi  x)+15448 \cos (18 \pi  x)+11059 \cos (20 \pi  x) +5260 \cos (22 \pi  x) \nonumber\\
   &\phantom{=} +1630 \cos (24 \pi  x)+300 \cos  (26 \pi  x) +25 \cos (28 \pi  x)) \\
   & = \mathcal{F}^{(2)}_{L_{-1}^3\ket{\varepsilon}} = \mathcal{F}^{(2)}_{L_{-2}L_{-1}\ket{\varepsilon}}\,.\nonumber
\end{align}

\subsection{Trace square distance}

\subsubsection{Vacuum module}\label{app:TSDvacResults}

Here are some explicit expressions for the TSD between light states:
\begingroup
\allowdisplaybreaks[1]
\begin{align}\label{eq:TSD31}
     T^{(2)}_{L_{-3}\ket{0},\ket{0}} &=\frac{1}{64} c^2 \sin ^8(\pi x) \cos ^4(\pi x)\nonumber\\
     &\phantom{=}-\frac{c \sin ^6(\pi x) \cos ^2(\pi x) (124 \cos (2 \pi x)+17 \cos (4 \pi x)+243)}{2048}\nonumber\\
     &\phantom{=}+\frac{\sin^4(\pi x) (8391 \cos (2 \pi x)+1890 \cos (4 \pi x)+361 \cos (6 \pi x)+7790)}{16384 c}\nonumber\\
    &\phantom{=}+\frac{1}{8388608}(-7864320 \cos (\pi x)+1984232 \cos (2 \pi x)-524288
   \cos (3 \pi x) \nonumber\\
    &\phantom{=} + 295631 \cos (4 \pi x)-27612 \cos (6 \pi x)+386 \cos (8 \pi x) + 8436 \cos (10 \pi x) \nonumber\\
    &\phantom{=} +289 \cos (12 \pi x)+6127246) \,,
\end{align}
\endgroup
%
\begingroup
\allowdisplaybreaks[1]
\begin{align}
    T^{(2)}_{L_{-4}\ket{0},\ket{0}} &= \frac{c^2 \sin ^8(\pi x) (3 \cos (2 \pi x)+2)^4}{6400} \nonumber\\
    &\phantom{=} -\frac{c \sin ^6(\pi x) (3 \cos (2 \pi x)+2)^2 }{1638400} (15489 \cos (2 \pi x)+4678 \cos (4 \pi x) \nonumber\\
    &\phantom{=} + 1263 \cos (6 \pi x)+19530)\nonumber \\
    &\phantom{=} + \frac{\sin ^4(\pi x)}{6553600 c} (5444642 \cos (2 \pi x)+2684168 \cos (4 \pi x) + 913973 \cos (6 \pi x) \nonumber\\
    &\phantom{=} +286934 \cos (8 \pi x)+59049 \cos (10 \pi x)+3718434)\nonumber\\
    &\phantom{=} +\frac{1}{53687091200}(-48486154240 \cos (\pi x)+12614780944 \cos (2 \pi x) \nonumber\\
    &\phantom{=} -4445962240 \cos (3 \pi x)  + 2269930440 \cos(4 \pi x)-754974720 \cos (5 \pi x) \nonumber\\
    &\phantom{=} +634223408 \cos (6 \pi x)  -121298020 \cos (8 \pi x)-5870960 \cos (10 \pi x) \nonumber\\
    &\phantom{=} +13794296 \cos (12 \pi x)  + 18977328 \cos (14 \pi x)+1595169 \cos (16 \pi x) \nonumber\\
    &\phantom{=} +38260958595) \,.
\end{align}
\endgroup

\subsubsection{Ising model}\label{app:TSDising}
Here are some results for $\varepsilon$ descendants in the Ising models:
\begingroup
\allowdisplaybreaks[1]
\begin{align}
    T^{(2)}_{L_{-1}\ket{\varepsilon},\ket{\varepsilon}} &= \frac{ \cos (2 \pi  x)+7}{1024} \sin ^4\left(\frac{\pi  x}{2}\right) (11 + 92 \cos (\pi  x)+148 \cos (2 \pi  x) \nonumber\\
    & \phantom{=} + 100 \cos (3 \pi  x)+ 25 \cos (4 \pi  x)) \,,\\
    \nonumber\\
    T^{(2)}_{L_{-2}\ket{\varepsilon},\ket{\varepsilon}} &= \frac{ \cos (2 \pi  x)+7}{4194304} \sin ^4\left(\frac{\pi  x}{2}\right) (2347723 + 2430412 \cos (\pi  x) \nonumber\\
    &\phantom{=} +1872304 \cos (2 \pi  x)  +1393796 \cos (3 \pi  x)+1144940 \cos (4 \pi  x) \nonumber\\
    &\phantom{=} +751500 \cos (5 \pi  x) +366480 \cos (6 \pi  x)+142884 \cos (7 \pi  x) \nonumber\\
    &\phantom{=} +35721 \cos (8 \pi  x)) \\
   & = T^{(2)}_{L_{-1}^2\ket{\varepsilon},\ket{\varepsilon}} \,, \nonumber\\
   \nonumber\\
   T^{(2)}_{L_{-3}\ket{\varepsilon},\ket{\varepsilon}} &= \frac{ (\cos (2 \pi  x)+7)}{1073741824} \sin ^4\left(\frac{\pi  x}{2}\right) ( 912429118 + 1218353112 \cos (\pi  x) \nonumber\\
   &\phantom{=} +998414040 \cos (2 \pi  x)+780711528 \cos (3 \pi  x)  +629316847 \cos (4 \pi x) \nonumber\\
   &\phantom{=} +497152212 \cos (5 \pi  x) +393829628 \cos (6 \pi  x)+276532300 \cos (7 \pi  x) \nonumber\\
   &\phantom{=} +168888850 \cos (8 \pi  x)+94527900 \cos (9 \pi  x) +46637100 \cos (10 \pi x) \nonumber\\
   &\phantom{=} +18404100 \cos (11 \pi  x) +4601025 \cos (12 \pi  x)) \\
   &= T^{(2)}_{L_{-1}^3\ket{\varepsilon},\ket{\varepsilon}} =T^{(2)}_{L_{-2}L_{-1}\ket{\varepsilon},\ket{\varepsilon}} \,. \nonumber
\end{align}
\endgroup

For $\sigma$ descendants:
\begingroup
\allowdisplaybreaks[1]
\begin{align}
    T^{(2)}_{L_{-1}\ket{\sigma},\ket{\sigma}} &= -\frac{\cos ^2\left(\frac{\pi  x}{2}\right) \sqrt{\csc \left(\frac{\pi  x}{2}\right)}}{32768 \sin ^{\frac{7}{2}}\left(\frac{\pi  x}{2}\right) (\cos (\pi  x)+1)^{5/4}} \Bigg\{66752\ 2^{3/4} \sqrt{\sin ^7\left(\frac{\pi  x}{2}\right) \sin (\pi 
   x)} \nonumber\\
   &\phantom{=}+\sqrt[4]{\cos (\pi  x)+1} \big[ 46488 \cos (\pi  x)-16137 \cos (2 \pi  x)+3612 \cos (3 \pi  x)  \nonumber\\
   &\phantom{=} -1006 \cos (4 \pi  x)+76 \cos (5 \pi  x)-7 \cos (6 \pi  x)\big]-33026
   \sqrt[4]{\cos (\pi  x)+1}  \nonumber\\
   &\phantom{=} -16 \sqrt[4]{2} \sin ^4\left(\frac{\pi  x}{2}\right) \sqrt{\cos \left(\frac{\pi  x}{2}\right)} \big[-3888 \cos (\pi  x)-804 \cos (2 \pi 
   x) \nonumber\\
   &\phantom{=}  +48 \cos (3 \pi  x)+129 \cos (4 \pi  x)+7099\big]\Bigg\} \,, \\
   \nonumber\\
   T^{(2)}_{L_{-2}\ket{\sigma},\ket{\sigma}} &= \left(33554432\ 2^{3/4} \sin
   ^{\frac{5}{4}}(\pi  x) \sqrt{\csc \left(\frac{\pi  x}{2}\right)}\right)^{-1}\Bigg\{ 2 \big[ 51995040 \cos (\pi  x) \nonumber\\
   &\phantom{=}  +6726368 \cos (2 \pi  x)-50083264 \cos (3 \pi  x)-27705396 \cos (4 \pi  x) \nonumber\\
   &\phantom{=} -23115584 \cos (5 \pi  x)+3517116 \cos (6 \pi  x)-1600624
   \cos (7 \pi  x) \nonumber\\
   &\phantom{=} +688029 \cos (8 \pi  x)-18480 \cos (9 \pi  x)+36 \cos (10 \pi  x)+93860567\big] \nonumber\\
   &\phantom{=} \times \sqrt[4]{\sin ^3(\pi  x) (\cos (\pi  x)+1)}+\big[-124036846 \sin (\pi x) \nonumber\\
   &\phantom{=}-131438432 \sin (2 \pi  x) -31447124 \sin (3 \pi  x)+24870528 \sin (4 \pi  x) \nonumber\\
   &\phantom{=} +31182128 \sin (5 \pi  x)+21514960 \sin (6 \pi  x)-2930095 \sin (7 \pi  x) \nonumber\\
   &\phantom{=} +1582144 \sin (8 \pi  x)-453993 \sin (9 \pi  x)+18480 \sin (10 \pi  x)\nonumber\\
   &\phantom{=} -36 \sin (11 \pi  x)\big] \sqrt[4]{\cot \left(\frac{\pi  x}{2}\right)}\Bigg\} \\
   & = T^{(2)}_{L_{-1}^2\ket{\sigma},\ket{\sigma}} \,, \nonumber\\
   \nonumber\\
   T^{(2)}_{L_{-3}\ket{\sigma},\ket{\sigma}} &= \left(64961380352\ 2^{3/4} \sin^{\frac{5}{4}}(\pi  x) \sqrt{\csc \left(\frac{\pi  x}{2}\right)}\right)^{-1} \nonumber\\
   &\phantom{=} \times \Bigg\{ \big[-804063279604 \sin (\pi  x)-679262872576 \sin (2 \pi  x) \nonumber\\
   &\phantom{=} -676379752341 \sin (3 \pi  x)-151873435008 \sin (4 \pi  x) \nonumber\\
   &\phantom{=} +78179526785 \sin (5 \pi  x)  +162043727616 \sin (6 \pi  x) \nonumber\\
   &\phantom{=} +203623756022 \sin (7 \pi  x)+116161263104 \sin (8 \pi  x) \nonumber\\
   &\phantom{=} +68995216314 \sin (9 \pi  x)+27584342272 \sin (10 \pi  x) \nonumber\\
   &\phantom{=} -1884177607 \sin (11 \pi x)  +2161088640 \sin (12 \pi  x) \nonumber\\
   &\phantom{=} -1097984829 \sin (13 \pi  x)+30000 (1568 \sin (14 \pi  x) \nonumber\\
   &\phantom{=} +93 \sin (15 \pi  x))\big] \sqrt[4]{\cot \left(\frac{\pi  x}{2}\right)} -2\big[-463898496256 \cos (\pi  x) \nonumber\\
   &\phantom{=} -339771582456 \cos (2 \pi  x)+161014585088 \cos (3 \pi  x) \nonumber\\
   &\phantom{=} +347509883805 \cos (4 \pi  x)+308827933824 \cos (5 \pi  x) \nonumber\\
   &\phantom{=} +269411694364 \cos(6 \pi  x) +145953734016 \cos (7 \pi  x) \nonumber\\
   &\phantom{=} +65832431142 \cos (8 \pi  x)+29792470912 \cos (9 \pi  x) \nonumber\\
   &\phantom{=} -3124129172 \cos (10 \pi  x)+2208128640 \cos (11 \pi  x) \nonumber\\
   &\phantom{=} -1220635853 \cos (12 \pi  x) +47040000 \cos (13 \pi  x)+2790000 \cos (14 \pi  x) \nonumber\\
   &\phantom{=} -619071617270\big] \sqrt[4]{\sin ^3(\pi  x) (\cos (\pi  x)+1)}\Bigg\} \,.
\end{align}
\endgroup
For degenerate states at level~3 the expressions are different, but we report here only one for simplicity.

\subsubsection{Three-states Potts model}\label{app:TSDpotts}
We conclude this appendix with the TSD between the first-level descendant of $\varepsilon$ and the primary state itself in the Potts model:
\begingroup
\allowdisplaybreaks[1]
\begin{align}
    T^{(2)}_{L_{-1}\ket{\varepsilon},\ket{\varepsilon}} &= \frac{1}{8} F\!\left(-\tfrac{8}{5},-\tfrac{1}{5};-\tfrac{2}{5};\eta \right){}^2 \sin^4\!\left(\frac{\pi  x}{2}\right) (4 \cos (\pi  x)+\cos (2 \pi  x)+19) \nonumber\\
    &\phantom{=}+\frac{1}{5200}\Bigg\{(F\!\left(-\tfrac{8}{5},-\tfrac{1}{5};-\tfrac{2}{5};\eta \right) \sin^{10}\!\left(\frac{\pi  x}{2}\right) \cos^8\!\left(\frac{\pi  x}{2}\right) \csc ^6(\pi  x)\nonumber\\
    &\phantom{=}  \times\bigg[(41600 F\!\left(-\tfrac{3}{5},\tfrac{4}{5};\tfrac{3}{5};\eta \right) (2 \cos (\pi  x)+\cos (2 \pi  x)-15)+32 \cos^2\!\left(\frac{\pi  x}{2}\right) \nonumber\\
    &\phantom{=}\times\bigg(9
   \sin (\pi  x) \left(49 F\!\left(\tfrac{12}{5},\tfrac{19}{5};\tfrac{18}{5};\eta \right) \sin ^3(\pi  x)+260 F\!\left(\tfrac{7}{5},\tfrac{14}{5};\tfrac{13}{5};\eta \right) \sin (2 \pi  x)\right)\nonumber\\
   &\phantom{=}  +520 F\!\left(\tfrac{2}{5},\tfrac{9}{5};\tfrac{8}{5};\eta \right) (31 \cos
   (2 \pi  x)+9)\bigg)\bigg]\Bigg\} \nonumber\\
   &\phantom{=}-\left(1782579200\ 2^{2/5} \Gamma
   \left(-\frac{8}{5}\right) \Gamma \left(\frac{17}{10}\right) \Gamma \left(\frac{12}{5}\right)\right)^{-1}\nonumber\\
   &\phantom{=}\times\Bigg\{\Gamma \left(-\frac{2}{5}\right) \Gamma \left(\frac{3}{10}\right) \Gamma \left(\frac{13}{5}\right) F\!\left(\tfrac{6}{5},\tfrac{13}{5};\tfrac{12}{5};\eta \right) \sin ^{\frac{28}{5}}(\pi  x) \nonumber\\
   &\phantom{=} \times\bigg[ 544 F\!\left(\tfrac{6}{5},\tfrac{13}{5};\tfrac{12}{5};\eta
   \right) (9600 \cos (\pi  x)-17164 \cos (2 \pi  x) +15129 \cos (4 \pi  x) \nonumber\\
   &\phantom{=} -1293) +208 \sin ^2(\pi  x) \bigg( 9016 F\!\left(\tfrac{26}{5},\tfrac{33}{5};\tfrac{32}{5};\eta \right) \sin ^6(\pi  x)+17 F\!\left(\tfrac{11}{5},\tfrac{18}{5};\tfrac{17}{5};\eta \right) \nonumber\\
   &\phantom{=}\times(3697 \cos
   (\pi  x)+6479 \cos (3 \pi  x)+800)+79488 F\!\left(\tfrac{21}{5},\tfrac{28}{5};\tfrac{27}{5};\eta \right) \sin ^4(\pi  x)  \nonumber\\
   &\phantom{=} \times \cos (\pi  x) +99 F\!\left(\tfrac{16}{5},\tfrac{23}{5};\tfrac{22}{5};\eta \right) \sin ^2(\pi  x) (1373 \cos (2 \pi  x)+843)\bigg)\bigg]\Bigg\} \,,
\end{align}
\endgroup
where $F \equiv \; _2F_1 $ is the hypergeometric function and $\eta = \sin^2 \left( \frac{\pi x}{2} \right)$.
We computed also TSDs for higher-level descendants, but the expressions are more complicated and we won't show them here.
\chapter{Conventions and further details on the fermionic models}\label{app:weyl_fermion}

\section{Conventions on fermions and Dirac matrices}\label{app:fermion_conventions} 
We collect here our conventions on Dirac matrices and Dirac and Weyl fermions.
They are employed in chap.~\ref{chap:weyl_fermions} and were also used in~\cite{Bastianelli:2018osv,Bastianelli:2019fot,Bastianelli:2019zrq}.

In a four-dimensional Minkowski spacetime we use a flat metric $\eta_{ab}$.
The Dirac matrices $\gamma^a$ satisfy 
\begin{equation}
\{ \gamma^a, \gamma^b \} =2 \eta^{ab} \,,
\end{equation}
and the conjugate Dirac spinor $\bpsi$ is defined using $\beta= i\gamma^0$ by
\begin{equation} 
\bpsi = \psi^\dagger \beta \;.
\end{equation}
The hermitian chiral matrix $\gamma^5$ is given by
\begin{equation} 
\gamma^5=- i \gamma^0 \gamma^1\gamma^2 \gamma^3 \,,
\end{equation}
and used to define the chiral projectors
\begin{equation}
P_L = \frac{\mathbb{I}+\gamma_5}{2} \,, \qquad P_R = \frac{\mathbb{I}-\gamma_5}{2} \,,
\end{equation}
that split a Dirac spinor $\psi$ into its left and right  Weyl components
\begin{equation}
\psi = \lambda +\rho \;, \qquad \lambda=P_L \psi \;,  \quad \rho=P_R \psi  \;.
\end{equation}
The charge conjugation matrix $C$ satisfies
\begin{equation}
C\gamma^{a} C^{-1} = -\gamma^{a T }  \;,
\label{C-matrix}
\end{equation}
it is antisymmetric and used to define the charge conjugation of  the spinor $\psi$ by
\begin{equation}
\psi_c = C^{-1} \bpsi^T \,,
\end{equation}
for which the roles of particle and antiparticle get interchanged.
Note that a chiral spinor  $\lambda$ has its charge conjugated field $\lambda_c$  of opposite chirality.
A Majorana spinor $\mu$ is a spinor  that equals its charged conjugated spinor
\begin{equation}
\mu =\mu_c \;.
\label{Majc}
\end{equation}
This constraint is incompatible with the chiral constraint, and Majorana-Weyl  spinors do not exist in 4 dimensions.

We find it convenient, as a check on our formulae, to use the  chiral representation of the gamma matrices.
In terms of $2\times 2$  blocks they are given by
\begin{equation}
\gamma^0= -i
\left( \begin{array}{cc}
 0 & \mathbb{I} \\ 
\mathbb{I} & 0
\end{array} \right) \,, \qquad 
\gamma^i = -i  \left( \begin{array}{cc}
0 &  \sigma^i \\ 
- \sigma^i & 0
\end{array} \right) \,,
\end{equation}
where $\sigma^i$ are the Pauli matrices, so that 
 \begin{equation}
\gamma^5=
\left( \begin{array}{cc}
\mathbb{I} & 0 \\ 
0 & - \mathbb{I}
\end{array} \right) 
\;, \qquad
\beta = i\gamma^0 = 
\left( \begin{array}{cc}
0 & \mathbb{I} \\ 
\mathbb{I} & 0
\end{array} \right) \;.
\end{equation}
The chiral representation makes evident that the Lorentz generators in the spinor space 
$M^{ab}=\frac{1}{4} [\gamma^a,\gamma^b] = \frac12 \gamma^{ab}  $
take a block diagonal form
\begin{equation}\label{lor_gen}
M^{0i} = \frac12 
\left( \begin{array}{cc}
\sigma^i & 0\\ 
0&-\sigma^i  
\end{array} \right) \;, \qquad 
M^{ij} = \frac{i}{2} \epsilon^{ijk} 
\left( \begin{array}{cc}
\sigma^k & 0 \\ 
 0 & \sigma^k
\end{array} \right)  \,,
\end{equation}
and do not mix the chiral components of  a Dirac spinor (as $\gamma^5$ is also block diagonal). 
The usual two-dimensional Weyl spinors appear inside a four-dimensional Dirac spinor as follows 
\begin{equation}
\psi = 
\left( \begin{array}{c}
l \\ 
r 
\end{array} \right)  \;, \qquad
\lambda = 
\left( \begin{array}{c}
l \\ 
0
\end{array} \right)  \;, \qquad
\rho = 
\left( \begin{array}{c}
0 \\ 
r 
\end{array} \right)  
\label{A.12}
\end{equation}
where  $l$ and $r$ indicate  two-dimensional independent spinors of opposite chirality.
In the chiral representation one may take the charge conjugation matrix $C$ to be given by
\begin{equation}
C = \gamma^2 \beta = -i 
\left( \begin{array}{cc}
\sigma^2 & 0\\ 
0&-\sigma^2  
\end{array} \right)  \,,
\end{equation}
and satisfies 
\begin{equation}
C=-C^T=-C^{-1}=-C^\dagger =C^* \,,
\end{equation}
and we remark that some of these relations are  representation dependent.
In the chiral representation the Majorana constraint \eqref{Majc} takes the form 
\begin{equation}
\mu=\mu_c    \qquad \to  \qquad  
\left( \begin{array}{c} l \\ r \end{array} \right)  =
\left( \begin{array}{c} i \sigma ^2 r^* \\ - i \sigma ^2 l^* \end{array} \right)  \,,
\end{equation}
which shows that the two-dimensional spinors $l$ and $r$ cannot be independent. 
The Majorana condition can be solved in terms of the single two-dimensional left-handed spinor $l$ as 
\begin{equation}
\mu = \left( \begin{array}{c} l \\  -i \sigma^2 l^* \end{array} \right) \,,
\end{equation}
which, evidently, contains the four-dimensional chiral spinors $\lambda$ and $\lambda_c$ defined by
\begin{equation}
\lambda = \left( \begin{array}{c} l \\  0 \end{array} \right) 
\;, \qquad
\lambda_c = \left( \begin{array}{c} 0 \\  -i \sigma^2 l^* \end{array} \right)  \;.
\end{equation}
In a four-dimensional spinors notation one can write
\begin{equation}
\mu =\lambda +\lambda_c \;.
\end{equation}
Alternatively, the Majorana condition can be solved in terms of the two-dimensional right-handed spinor $r$ as
\begin{equation}
\mu = \left( \begin{array}{c} i \sigma^2 r^* \\  r \end{array} \right ) \,,
\end{equation}
which contains the four-dimensional chiral spinors $\rho$ and $\rho_c$
\begin{equation}
\rho = \left( \begin{array}{c} 0  \\    r  \end{array} \right) 
\;, \qquad
\rho_c = \left( \begin{array}{c}  i \sigma^2 r^*  \\  0 \end{array} \right) \,,
\end{equation}
and $\mu = \rho+\rho_c$.  This solution is of course the same as the previous one, 
as one may identify $\lambda=\rho_c$.

The explicit dictionary between Weyl and Majorana spinors shows that the field theory 
of a Weyl spinor is equivalent to that of a Majorana spinor, as Lorentz symmetry fixes uniquely their actions, which are bound to be identical.

Finally, we normalize our $\epsilon$ symbols by $\epsilon_{0123}=-1$ and $\epsilon^{0123}=1$, so that
\begin{equation}
\frac14
\tr(\gamma^5 \gamma^a\gamma^b\gamma^c\gamma^d) = i \epsilon^{abcd} \;.
\end{equation}


\section{Metric, vielbein and related connections}\label{app:vielbein}
In a Lorentzian spacetime we use a metric $g_{\mu\nu}$ which is covariantly constant
\begin{equation}
    \nabla_\lambda g_{\mu\nu} = \partial_\lambda g_{\mu\nu} - \Gamma^\rho_{\lambda\mu} g_{\nu\rho} - \Gamma^\rho_{\lambda\nu}g_{\mu\rho} = 0 \,,
\end{equation}
so that the Levi-Civita connection $\Gamma^\lambda_{\mu\nu}$ is given by
\begin{equation}
    \Gamma^\lambda_{\mu\nu} = \frac12 g^{\lambda\rho} \left( \partial_\mu g_{\nu\rho} + \partial_\nu g_{\mu\rho} - \partial_{\rho} g_{\mu\nu} \right) \,.
\end{equation}
The covariant derivative on a vector field acts as
\begin{equation}
    \nabla_\mu V^\lambda = \partial_\mu V^\lambda + \Gamma_{\mu\nu}^\lambda V^\nu \,,
\end{equation}
and we remind our conventions for the curvature tensor
\begin{equation}
[\nabla_\mu, \nabla_\nu] V^\lambda = R_{\mu\nu}{}^\lambda{}_{\rho} V^\rho \,, \quad R_{\mu\nu} = R_{\mu\lambda\nu}{}^\lambda \, , \quad R = R^\mu{}_\mu \, .
\end{equation}

Introducing the vielbein $e_\mu{}^a$ by setting
\begin{equation}\label{vielbein}
    g_{\mu\nu} =  e_\mu{}^a \, e_\nu{}^b \, \eta_{ab} \, ,
\end{equation}
we gain a new gauge symmetry, namely the local Lorentz transformations of the tangent space; in this way we are able to couple a fermionic theory to gravity.
Latin letters are used for indices of the tangent space, and $\eta_{ab}$ is a Minkowski metric.
In four dimensions, we refer to $e_\mu{}^a$ as the vierbein.
The covariant derivative now needs a corresponding connection, i.e.~the spin connection, which in the absence of torsion is defined by requiring the vielbein to be covariantly constant
\begin{equation}
    \nabla_\mu e_\nu{}^a = \partial_\mu e_\nu{}^a - \Gamma^\lambda_{\mu\nu} e_\lambda{}^a + \omega_\mu{}^a{}_b e_\nu{}^b =0 \, .
\end{equation}
This condition can be solved by
\begin{equation}
    \omega_\mu{}^{ab} = e^{b\nu} \left( \Gamma^\lambda_{\mu\nu} e_\lambda{}^a - \partial_\mu e_\nu{}^a \right) \,,
\end{equation}
or equivalently
\begin{align}
    \omega_\mu{}^{ab} &= \frac12 e^{a\lambda} e^{b\nu} e_{c\mu} \left( \partial_\nu e_\lambda{}^c - \partial_\lambda e_\nu{}^c \right) + \frac12 e^{a\nu} \left( \partial_\mu e_\nu{}^b - \partial_\nu e_\mu{}^b \right) -\frac12 e^{b\nu} \left( \partial_\mu e_\nu{}^a - \partial_\nu e_\mu{}^a \right)
\end{align}
in which it is manifest that $\omega_\mu{}^{ab} + \omega_\mu{}^{ba} =0$.
Then, the covariant derivative acts on a spinor $\psi$ as
\begin{equation}
    \nabla_\mu \psi = \partial_\mu \psi + \frac12 \omega_{\mu a b} M^{ab} \psi \,,
\end{equation}
where $M^{ab}$ are the generators of the Lorentz group, which in our conventions are made explicit in eq.~\eqref{lor_gen}.


\section{The stress tensor of the Bardeen model}
\label{appB} 

In chap.~\ref{chap:weyl_fermions} we have defined the stress tensor of the Bardeen model as
\begin{equation}
T^\mu{}_a = \frac{1}{e}\frac{\delta S}{\delta e_\mu{}^a} \,,
\end{equation}
where $S=\int\, d^4 x {\cal L}$ is the action associated to the lagrangian  
\eqref{lag2}.
We review here its classical properties, which in sec.~\ref{sec:weyl_fermions_gauge} we only need in the flat space limit. 
The model depends on the background fields $e_\mu{}^a, A_\mu, B_\mu$, and 
statisfies various background gauge symmetries, responsible for the properties of the gauge currents 
and stress tensor.
Let us discuss the latter.

The infinitesimal background symmetries associated to general coordinate invariance
(with infinitesimal local parameters $\xi^\mu$), local Lorentz invariance  
(with infinitesimal local parameters $\omega_{ab}$), and Weyl invarance 
(with infinitesimal local parameter $\sigma$), take the form
\begin{equation} 
\begin{aligned}
\delta e_\mu{}^a  & = \xi^\nu \partial_\nu  e_\mu{}^a + (\partial_\mu \xi^\nu) e_\nu{}^a +\omega^a{}_b e_\mu{}^b +\sigma e_\mu{}^a \,,\cr
\delta A_\mu  & = \xi^\nu \partial_\nu  A_\mu + (\partial_\mu \xi^\nu) A_\nu \,,\cr
\delta B_\mu  & = \xi^\nu \partial_\nu  B_\mu + (\partial_\mu \xi^\nu) B_\nu \,,\cr
\delta \psi  & = \epsilon^\mu \partial_\mu  \psi
+ \frac14 \omega_{ab} \gamma^{ab}\psi -\frac{3}{2} \sigma \psi \,.
\end{aligned} 
\label{transf}
\end{equation}

Under the Weyl symmetry $\delta_\sigma$ with local parameter $\sigma$, the gauge fields do not transform, 
and the invariance of the action implies 
\begin{equation}
\begin{aligned}
\delta_\sigma S &= \int d^4 x \left ( \frac{\delta S}{\delta e_\mu{}^a} \delta_\sigma e_\mu{}^a +
\frac{\delta_R S}{\delta \psi} \delta_\sigma \psi + \delta_\sigma \bpsi 
\frac{\delta_L S}{\delta \bpsi} 
\right ) \cr
&=  \int d^4 x  e\,  T^\mu{}_a \delta_\sigma e_\mu{}^a = 
 \int d^4 x  e\,  T^a{}_a \sigma = 0 \,,
  \end{aligned}
\end{equation}
where we have used left and right derivatives for the Grassmann valued fields and in the second line we have implemented the equations of motion of the spinor fields.
Thus local Weyl invariance implies tracelessness of the stress tensor, with the trace computed 
with the vierbein as $T^a{}_a = T^\mu{}_a  e_\mu{}^a $. 
Thus, the stress tensor is traceless at the classical level. 

Similarly, the Lorentz symmetry $\delta_\omega$ with local parameters $\omega_{ab}$ implies
\begin{equation}
\begin{aligned}
\delta_\omega S &= \int d^4 x \left ( \frac{\delta S}{\delta e_\mu{}^a} \delta_\omega e_\mu{}^a +
\frac{\delta_R S}{\delta \psi} \delta_\omega \psi + \delta_\omega \bpsi 
\frac{\delta_L S}{\delta \bpsi} 
\right ) \cr
&=  \int d^4 x  e\,  T^\mu{}_a \delta_\omega e_\mu{}^a =
 \int d^4 x  e\,  T^\mu{}_a \omega^a{}_b e_\mu{}^b 
=  \int d^4 x  e\,  T^{ba}\omega_{ab} = 0 \,,
  \end{aligned}
\end{equation}
and constrains the antisymmetric part of the stress tensor to vanish on-shell.
Again, we have used the fermionic equations of motion and the fact that the gauge fields 
$A_\mu$ and $B_\mu$ do not transform under local Lorentz transformations. Considering the
arbitrariness of the local parameters and their antisymmetry, $\omega_{ab} = - \omega_{ba}$,  
we recognize that the stress tensor with flat indices is symmetric,  $T^{ab} = T^{ba}$.
At the quantum level, our PV regularization preserves this symmetry, and no anomalies can arise 
in the local Lorentz sector.

Finally, a suitable conservation law of the stress tensor arises as a consequence of the infinitesimal 
diffeomorphism invariance $\delta_\xi$.  It is actually useful to combine it with 
additional local Lorentz and gauge symmetries (with composite parameters),
 so to obtain a conservation in the following way
\begin{equation}
\begin{aligned}
\delta_\xi S &= \int d^4 x \biggl ( \frac{\delta S}{\delta e_\mu{}^a} \delta_\xi e_\mu{}^a +
\frac{\delta_R S}{\delta \psi} \delta_\xi \psi + 
\delta_\xi  \bpsi \frac{\delta_L S}{\delta \bpsi} 
+\frac{\delta S}{\delta A_\mu^\alpha } \delta_\xi A_\mu^\alpha  
+\frac{\delta S}{\delta B_\mu^\alpha } \delta_\xi B_\mu^\alpha  
\biggr) \cr
&= \int d^4 x  e\,  \biggl ( T^\mu{}_a {\cal L}_\xi e_\mu{}^a 
+ J^{\mu \alpha}  {\cal L}_\xi A_\mu^\alpha
+ J_5^{\mu \alpha}  {\cal L}_\xi B_\mu^\alpha
\biggr)
\cr 
& =  \int d^4 x  e\, \biggl(  T^\mu{}_a \nabla_\mu \xi^a 
+ J^{\mu \alpha}  \xi^\nu  \hat F_{\nu\mu}^\alpha 
+ J_5^{\mu \alpha}  \xi^\nu \hat G_{\nu\mu}^\alpha \biggr)
\cr 
&=  - \int d^4 x  e\,  \xi^a  \biggl (
 \nabla_\mu 
T^\mu{}_a  - J^{b \alpha} \hat F_{ab}^\alpha -J_5^{b \alpha} \hat G_{ab}^\alpha
\biggr) 
= 0 \,,
  \end{aligned}
\end{equation}
 where in the second line we have used the fermion equations of motion
 (and denoted with Lie derivatives  ${\cal L}_\xi$ the transformation rules under diffeomorphisms), 
 and in the third line
 added for free to the Lie derivative of the vierbein 
 a spin connection term (it amounts to a local Lorentz transformation with composite parameter, 
 and it drops out once the  stress tensor is symmetric),
 and to the Lie derivatives of the gauge fields suitable gauge transformations 
 (which also drop out after partial integration as the corresponding currents are covariantly conserved, 
 recall eqs.~\eqref{gauge-conservation}  and \eqref{uno}
 which we use here in their curved space version), and then integrated by parts.
 The arbitrariness of the local parameters $\xi^a(x)$ allows to derive
 a covariant conservation law, which contains a contribution from the gauge fields (that would vanish 
once the gauge fields are made dynamical, which allows to use their equations of motion). 
It reads
\begin{equation}
 \nabla_\mu 
T^\mu{}_a  = J^{b \alpha} \hat F_{ab}^\alpha +J_5^{b \alpha} \hat G_{ab}^\alpha \;.
\end{equation}
Our PV regularization preserves diffeomorphism invariance,  
and thus no anomaly may appear in the Ward identities arising from this symmetry.

We end this appendix by presenting the explicit expression of the stress 
tensor for the Bardeen model in flat spacetime:
\begin{equation}
    T_{ab} = \frac14 \bpsi \Big( \gamma_a \overset{\leftrightarrow}{D}_b (A,B) + \gamma_b \overset{\leftrightarrow}{D}_a (A,B) \Big) \psi \,,
\end{equation}
where $\overset{\leftrightarrow}{D}_a (A,B) = D_a (A,B) - \overset{\leftarrow}{D}_a (A,B) $,
with the second derivative meaning here $\overset{\leftarrow}{D}_a (A,B) =
\overset{\leftarrow}{\partial}_a  -A_a -B_a \gamma^5$.
One can verify explicitly all the statements about the classical 
background symmetry derived above, namely
\begin{equation}
\partial_a T^{ab} = \bpsi \gamma_a (\hat F^{ab} +\hat G^{ab} \gamma^5)\psi \; , 
\qquad T^{ab} =  T^{ba} \;, 
\qquad   T^a{}_a =0  \;.
\label{cons-flat}
\end{equation}
The PV regularization used in the main text 
preserves the corresponding quantum Ward identities except the last one,
as the mass term in curved space is not Weyl invariant, and a trace anomaly develops.


\section{Heat kernel traces}
\label{appA} 
We outline here how to compute the anomalies in the Bardeen model and present some intermediate results.
Similarly, one also proceeds in the case of the MAT background.

In Bardeen model, the role of the operator $H$ introduced in sec.~\ref{sec:heat_kernel} is played by the regulators $R_\psi$ and $R_{\psi_c}$. 
Focusing on the regulator ${R}_\psi   = -  \DDslash^2(A,B) $, we expand it as
\begin{align}\label{reg_bardeen}
{R}_\psi   &=  -D^a(A) D_a(A) + B^2 - \gamma^5 \left(D^a(A) B_a\right) \nonumber\\
&\phantom{=} -\frac12 \gamma^{ab} \left[ \hat F_{ab} - 4B_a B_b +\gamma^5 \left( \hat G_{ab} - 4 B_a D_b(A) \right) \right] \,,
\end{align}
where $\hat F_{ab} $ and $ \hat G_{ab}$ are the Bardeen curvatures given in \eqref{Bardeen-curvatures}, the covariant derivative is $D_a(A) =\partial_a+A_a$, the covariant divergence of $B_a$ is $D^a(A) B_a = (\partial^a B_a) +[ A^a, B_a]$ and $B^2 = B^a B_a$.
We compare this operator to the generic operator $H$ in~\eqref{opH}, where in this case we consider it in flat space and with only a gauge covariant derivative with connection $W_a$, i.e.~$H$ takes the form
\begin{equation}\label{Hbardeen}
    H = - \Box +V = -\partial^a\partial_a - 2W^a\partial_a -(\partial_a W^a) - W^2 +V \,.
\end{equation}
Comparing with~\eqref{reg_bardeen} we identify
\begin{align}
W_a &= A_a +\gamma_{ab}\gamma^5 B^b \,,\\
V &=  -2 B^2 -\gamma^5 \left(D^a(A) B_a\right) -\frac12 \gamma^{ab} \hat F_{ab} \,,\\
  {\cal F}_{ab} &= F_{ab}(A) + \left(\gamma_{ac}\gamma_{bd}- \gamma_{bc}\gamma_{ad}\right) B^c B^d 
+ \gamma^5 \left(\gamma_{ca} D_b(A) B^c -\gamma_{cb} D_a(A) B^c\right) \,,
\end{align}
with
\begin{equation}
    {\cal F}_{ab} = \partial_a W_b-\partial_b W_a + [W_a, W_b] \,.
\end{equation}

Now the coefficient $a_2(R_\psi)$, which enters in the anomaly calculations in four dimensions, can be made explicit using~\eqref{hkc2} in flat space.
In this limit, which is the one of interest for the Bardeen model, it reduces to
\begin{equation}\label{a2_flat}
    a_2(x,H) = \frac12 V^2  -\frac{1}{6} \Box V + \frac{1}{12} {\cal F}_{ab}^2 \,,
\end{equation}
and the Dirac traces can be directly computed.
We list here some intermediate results, grouping together the contributions from the same terms in~\eqref{a2_flat}.
We find (with $D_a = D_a(A) $, $DB=D_a B^a$, $B^2=B^a B_a$, and so on):

$i)$ from  $a_2 =\frac12  V^2$ 
\begin{align}
\tr \left[ \gamma^5 T^\alpha a_{2} \left({R}_\psi\right)\right] &= \tr_{_{\!\! YM}}  T^\alpha \bigg [ \frac{i}{2} \epsilon^{abcd} \hat F_{ab}   \hat F_{cd}  + 4 \{B^2, DB \}  \bigg ] \,,\\
\tr \left[T^\alpha a_{2} \left({R}_\psi\right)\right] &=  \tr_{_{\!\! YM}}  T^\alpha \left[ -\hat F^{ab}  \hat F_{ab} + 8 B^4  +  2 DB DB\right ] \,,\\
\tr \left[ a_{2} \left({R}_\psi\right)\right] &=   \tr_{_{\!\! YM}} \left[ -\hat F^{ab}  \hat F_{ab} + 8 B^4  +  2 DB DB\right ] \,.
\end{align}

$ii)$ from  $a_2 =-\frac{1}{6} \nabla^2 V$ 
\begin{align}
\tr \left[ \gamma^5 T^\alpha a_{2} \left({R}_\psi\right)\right] &= \tr_{_{\!\! YM}}  T^\alpha \bigg [ -\frac23 i \epsilon^{abcd} \left(  B_a B_b \hat F_{cd}  + 2 B_a \hat F_{bc}  B_d +\hat F_{ab}  B_c B_d \right)  \nonumber\\
&\phantom{=} +\frac23 D^2 D B +\frac13 [\hat F^{ab},\hat G_{ab}]
+\frac43 [D^a \hat F_{ab}, B^b]  - 2 \{ B^2,  DB \} \nonumber\\
&\phantom{=} + 4 B_a DB B^a \bigg ] \,,\\
\tr \left[T^\alpha a_{2} \left({R}_\psi\right)\right] &= \tr_{_{\!\! YM}}  T^\alpha \bigg[ \frac{i}{6} \epsilon^{abcd} \left( [ \hat G_{ab}, \hat F_{cd}] -4 [ B_a,  D_b \hat F_{cd}]  \right) +8 \left(B_aB^2 B^a- B^4\right) \nonumber\\
&\phantom{=} +\frac43 D^2 B^2 +\frac43 \left(\hat F^{ab} B_a B_b+ 2 B_a\hat F^{ab} B_b+B_a B_b \hat F^{ab} \right)  \bigg ] \,,\\
\tr \left[ a_{2} \left({R}_\psi\right)\right] &= \tr_{_{\!\! YM}}  \bigg[ \frac43 D^2 B^2 \bigg ] \,.
\end{align}

$iii)$ from $a_2 = \frac{1}{12} {\cal F}_{ab}^2 $ 
\begin{align}
\tr \left[ \gamma^5 T^\alpha a_{2} \left({R}_\psi\right)\right] &= \tr_{_{\!\! YM}}  T^\alpha \bigg [  i \epsilon^{abcd} \left(\frac16 \hat G_{ab}  \hat G_{cd} 
-\frac23 \{ \hat F_{ab} , B_c B_d\}  +\frac{16 }{3}  B_a B_b  B_c B_d \right)\nonumber\\
 &\phantom{=} -\frac83 \{ B^2,  DB\} +\frac43  \{ \{ B^a,B^b\}, D_a B_b \} \bigg ] \,, \\
 \tr \left[ T^\alpha a_{2} \left({R}_\psi\right)\right]  &=  \tr_{_{\!\! YM}}  T^\alpha \bigg [  \frac13 \hat F^{ab}\hat F_{ab} -\frac43 \{ \hat F_{ab}, B^a B^b \} +\frac83\left( B_a B_b B^a B^b  - B^4\right) \nonumber \\
&\phantom{=} - 8 B_a B^2 B^a  -\frac43 D_a B_b D^a B^b  -\frac23 DB DB  \bigg ] \,, \\
\tr \left[ a_{2} \left({R}_\psi\right)\right] &=  \tr_{_{\!\! YM}} \bigg [  \frac13 \hat F^{ab}\hat F_{ab} -\frac83 \hat F_{ab}B^a B^b  +\frac83 B_a B_b B^a B^b  - \frac{32}{3}B^4 \nonumber\\
&\phantom{=}  -\frac43 D_a B_b D^a B^b  -\frac23 DB DB  \bigg ] \,.
\end{align}

The analogous results for $a_2(R_{\psi_c})$  are obtained by replacing $A\to -A^T$ and $B\to B^T$ 
(and also $T^\alpha \to T^{\alpha T}  $ for the explicit $T^\alpha$ appearing in the traces). 
Their effect is just to double the contribution from   $a_2(R_{\psi})$ in the chiral and trace anomalies.

\section{Dirac mass in MAT background}\label{app:dirac_mass}
We give here a  brief description of a different regularization that could be employed for a Dirac fermion coupled to the MAT backgound discussed in sec.~\ref{sec:weyl_mat}.
There, we used a Majorana mass for the PV fields, arguing that it was most convenient as it breaks the least number of symmetries.
We give support here for that claim, by showing how a Dirac mass for the PV fields could be coupled to the MAT background. 

In a flat background the Dirac mass is given by
\begin{equation}
 - M\bar{\psi}\psi =  \frac12 M \left(\psi_c^T C \psi + \psi^T C \psi_c\right)\;.
\end{equation}
This term
breaks the $U(1)_A$ axial symmetry while maintaining the $U(1)_V$ vector symmetry.
This continues to be the case also when one tries to MAT-covariantize it. 
There are various options to couple the Dirac mass to the MAT geometry.
We may choose to use only  the metric $g_{\mu\nu}$ in the mass term, without any axial extension,
so that 
\begin{equation}
  \Delta_{\scriptscriptstyle D} \mathcal{L} = - \sqrt{g} M\bar{\psi}\psi 
  = \frac{\sqrt{g}}{2} M (\psi^T_c C \psi + \psi^T C \psi_c )
  \label{dir-mass}
\end{equation}
has the virtue of preserving the vector-like diffeomorphisms and  vector-like local Lorentz transformations
on top of the $U(1)_V$ symmetry, while breaking all of their axial extensions.
It also breaks both vector and axial Weyl symmetries, 
which are then expected to be anomalous as well. 
Counterterms should eventually be introduced to achieve the equivalence 
with the results in sec.~\ref{sec:limits_mat}.
Other choices are also possible, e.g.~$\sqrt{g} \to \tfrac12( \sqrt{g_+} + \sqrt{g_-})$,
which shares the same property of 
 preserving the vector-like diffeomorphisms  and local Lorentz transformations.

In the following we limit ourselves to derive the regulators to be used  for computing 
the anomalies in this scheme.
  We add to the lagrangian \eqref{lag_axial} written in a symmetric form 
\begin{equation}
 \mathcal{L}  =  
  \frac12 \psi_c^T C  \sqrt{\bar{\hat{g}}} \hat{\Dslash}  \psi
+ \frac12 \psi^T C  \sqrt{\hat{g}} \bar{\hat{\Dslash}} \psi_c
 \label{lag-2}
\end{equation}
the mass term \eqref{dir-mass}, and comparing it with \eqref{PV-l}  we find,
choosing the field basis  
$
    \phi = \left(
    \begin{array}{c}
    \psi\\
    \psi_c
    \end{array}
    \right) 
    $,
\begin{equation}
    T\mathcal{O} = \left(
    \begin{array}{cc}
        0 & C \sqrt{\hat{g}}  \bar{\hat{\Dslash}} \\
        C \sqrt{\bar{\hat{g}}}  {\hat{\Dslash}} & 0
    \end{array}
    \right) 
 \, , \quad\quad
    T = \left(
    \begin{array}{cc}
        0 & \sqrt{g}  C \\
   \sqrt{g}         C & 0
    \end{array}
    \right) \, , 
\end{equation}
and thus
\begin{equation}
    \mathcal{O} =\left(
    \begin{array}{cc}
    \sqrt{\frac{\bar{\hat{g}}}{g}} \, {\hat{\Dslash}}
         & 0 \\
        0 & 
        \sqrt{\frac{\hat{g}}{g}} \, \bar{\hat{\Dslash}}
    \end{array}
    \right) \,, \quad\quad
    \mathcal{O}^2 = \left(
    \begin{array}{cc}
        \sqrt{\frac{\bar{\hat{g}}}{g}} \, {\hat{\Dslash}} \sqrt{\frac{\bar{\hat{g}}}{g}} \, {\hat{\Dslash}}
        & 0 \\
        0 &   \sqrt{\frac{\hat{g}}{g}} \, \bar{\hat{\Dslash}}  \sqrt{\frac{\hat{g}}{g}} \, \bar{\hat{\Dslash}}
    \end{array}
    \right) \,,
\end{equation}
with the differential operators acting on everything placed on their right hand side. 
This regulator  $\mathcal{R}=-\mathcal{O}^2$  is difficult to work with, but it has the virtue of being
covariant under vector diffeomorphisms.
Its structure is again more transparent when splitting the Dirac fermion into its  chiral parts, so that on the basis 
\begin{equation}
    \phi = \left(
    \begin{array}{c}
    \lambda\\
    \rho \\
    \rho_c\\
    \lambda_c
    \end{array}
    \right) \,,
\end{equation}
we find a block diagonal regulator 
\begin{equation}
    \mathcal{R} 
    = \left(
    \begin{array}{cccc} R_\lambda &0  &0 &0 \\
        0& R_\rho   &0  &0 \\
  0 & 0& R_{\rho_c}
  &0 \\  0& 0 &0 & R_{\lambda_c}
      \end{array}
    \right) \,,
\end{equation}
with entries

\begin{equation}
\begin{aligned}
R_\lambda &= - \sqrt{\frac{g_-}{g}}\Dslash_- \sqrt{\frac{g_+}{g}}\Dslash_+ P_+ 
\;, \qquad
R_\rho = - \sqrt{\frac{g_+}{g}}\Dslash_+  \sqrt{\frac{g_-}{g}}\Dslash_- P_-  \\
R_{\lambda_c}  &= - \sqrt{\frac{g_-}{g}}\Dslash_- \sqrt{\frac{g_+}{g}}\Dslash_+ P_- 
\;, \qquad
R_{\rho_c}  = - \sqrt{\frac{g_+}{g}}\Dslash_+  \sqrt{\frac{g_-}{g}}\Dslash_- P_+  \;.
  \end{aligned}
\end{equation}
The functions $\sqrt{\frac{g_\mp}{g}}$ are scalar functions under 
the vector-like diffeomorphism, so that these regulators are covariant under that symmetry.
The projectors $P_\pm$ take just the unit value on the  corresponding chiral spinor space,
but we have kept them to remind on which space the different regulators act.
A systematic analysis of all the anomalies, including the axial gravitational anomaly, 
may be feasible in this scheme, at least  when treating
$f_{\mu\nu}$ as a perturbation.
The final results have to agree with the anomalies presented in sec.~\ref{sec:limits_mat} by adding suitable counterterms to the effective action.
\chapter{Further details on the holographic anomalies with scalar fields}\label{app:holo_ano}

\section{Ansätze for the backreaction}\label{app:results}

In this appendix we collect the Ans\"atze that we make to solve eqs.~\eqref{deltahntilde},~\eqref{deltag+hn} for $n=1,2$.
The explicit results can be found in the Mathematica notebook provided in the supplementary material of~\cite{Broccoli:2021icm}.\footnote{The results were obtained with the help of the xAct collection of Mathematica packages~\cite{Martin-Garcia:2007bqa,Martin-Garcia:2008yei,MARTINGARCIA2008597,Brizuela:2008ra,Nutma:2013zea}.}
In the following, we assume that $g_{ij} \equiv g^{(0)}_{ij}$ and curvatures and covariant derivatives are w.r.t.~$g^{(0)}_{ij}$, and $\phi \equiv \phi_{(0)}$.
We also choose $h_0 =0$, as explained in the text.

The solution of the PBH equation for $g^{(1)}_{ij}$ is given in~\eqref{g1} and it is not modified by the presence of the backreaction.
The most generic Ansatz for $h^{(2)}_{ij}$ has thirty-five terms and six coefficients will not be fixed by the modified PBH equation.
We write the Ansatz as
\begingroup
\allowdisplaybreaks[1]
\begin{align}
   h^{(2)}_{ij} =& \;  + h_{21}^{} g_{ij} R_{kl} R^{kl} \phi^2 + h_{22}^{} R_{ij} R \phi^2 + h_{23}^{} R^{kl} R_{ikjl} \phi^2 + h_{24}^{} \phi^2 \nabla_{j}\nabla_{i}R  + h_{25}^{} R_{ij} \phi \Box \phi \nonumber\\
   &\phantom{=} + h_{26}^{} g_{ij} R \phi \Box \phi  + \beta_{1}^{} R_{i}{}^{k} R_{jk} \phi^2  + \beta_{2}^{} g_{ij} R^2 \phi^2  + \beta_{3}^{} R_{i}{}^{klm} R_{jklm} \phi^2 \nonumber\\
   &\phantom{=} + \beta_{4}^{} g_{ij} R_{klmn} R^{klmn} \phi^2 + \beta_{6}^{} R \nabla_{i}\phi \nabla_{j}\phi + \beta_{5}^{} (\phi \nabla_{i}\phi \nabla_{j}R + \phi \nabla_{i}R \nabla_{j}\phi) \nonumber\\
   &\phantom{=} + \beta_{7}^{} R \phi \nabla_{j}\nabla_{i}\phi + \beta_{8}^{} (R_{j}{}^{k} \phi \nabla_{k}\nabla_{i}\phi + R_{i}{}^{k} \phi \nabla_{k}\nabla_{j}\phi) + \beta_{9}^{} \phi^2 \Box R_{ij} + \beta_{10}^{} g_{ij} \phi^2 \Box R \nonumber\\
   &\phantom{=} + \beta_{11}^{} \nabla_{j}\nabla_{i}\phi \Box \phi + \beta_{12}^{} (\nabla_{j}\phi \Box \nabla_{i}\phi + \nabla_{i}\phi \Box \nabla_{j}\phi) + \beta_{13}^{} \phi \Box\nabla_{j}\nabla_{i}\phi \nonumber\\
   &\phantom{=} + \beta_{14}^{} g_{ij} \phi \nabla_{k}\phi \nabla^{k}R + \beta_{15}^{} (\phi \nabla_{i}R_{jk} \nabla^{k}\phi + \phi \nabla_{j}R_{ik} \nabla^{k}\phi) \nonumber\\
   &\phantom{=} + \beta_{16}^{} (R_{jk} \nabla_{i}\phi \nabla^{k}\phi + R_{ik} \nabla_{j}\phi \nabla^{k}\phi) + \beta_{17}^{} \phi \nabla_{k}R_{ij} \nabla^{k}\phi + \beta_{18}^{} R_{ij} \nabla_{k}\phi \nabla^{k}\phi \nonumber\\
   &\phantom{=} + \beta_{19}^{} g_{ij} R \nabla_{k}\phi \nabla^{k}\phi + \beta_{20}^{} \nabla_{k}\nabla_{j}\nabla_{i}\phi \nabla^{k}\phi + \beta_{21}^{} \nabla_{k}\nabla_{j}\phi \nabla^{k}\nabla_{i}\phi \nonumber\\
   &\phantom{=} + \beta_{22}^{} g_{ij} R^{kl} \phi \nabla_{l}\nabla_{k}\phi  + \beta_{23}^{} g_{ij} \Box \phi \Box \phi + \beta_{24}^{} g_{ij} \nabla^{k}\phi \Box \nabla_{k}\phi + \beta_{25}^{} g_{ij} \phi \Box^2 \phi \nonumber\\
   &\phantom{=} + \beta_{26}^{} g_{ij} R_{kl} \nabla^{k}\phi \nabla^{l}\phi + \beta_{27}^{} R_{ikjl} \nabla^{k}\phi \nabla^{l}\phi + \beta_{28}^{} R_{ikjl} \phi \nabla^{l}\nabla^{k}\phi \nonumber\\
   &\phantom{=} + \beta_{29}^{} g_{ij} \nabla_{l}\nabla_{k}\phi \nabla^{l}\nabla^{k}\phi \,, 
\end{align}
\endgroup
and leave the coefficients $h_{21},\ldots,h_{26}$ free.
The full solution is in the aforementioned Mathematica notebook.

A similar Ansatz is made for $\tilde{h}^{(2)}_{ij}$ and its coefficients are fixed in terms of $h_1$ and $h_2$ by the modified PBH equation.

The solution for $g^{(2)}_{ij}$ is given in~\eqref{g2} and it is not modified by the presence of the backreaction.
Thus, here we focus on ${h}^{(3)}_{ij}$, which appears in eq.~\eqref{Oh2} only as ${h}^{(3)i}{}_{i}$ and we then do not compute ${h}^{(3)}_{ij}$ but rather its trace that satisfies the following PBH equation 
\begin{equation}
    \delta g^{(2)i}{}_{i} + \delta {h}^{(3)i}{}_{i} = -2\sigma g^{(2)i}{}_{i} + g^{(0)ij} \delta g^{(2)}{}_{ij} +2\sigma \beta^{ij} g^{(2)}_{ij} -2\sigma {h}^{(3)i}{}_{i} + g^{(0)ij} \delta {h}^{(3)}_{ij} \, .
\end{equation}
The most generic Ansatz for ${h}^{(3)i}{}_{i}$ has sixty-six terms and nine coefficients will not be fixed by the modified PBH equation.
We write the Ansatz as
\begingroup
\allowdisplaybreaks[1]
\begin{align}
    {h}^{(3)i}{}_{i} =& \; h_{31}^{} R^{jk} \phi \nabla_{i}R_{jk} \nabla^{i}\phi + h_{32}^{} R^{ij} \phi^2 \nabla_{j}\nabla_{i}R + h_{33}^{} R_{ij} R^{ij} R \phi^2 + h_{34}^{} R^{ij} R \phi \nabla_{j}\nabla_{i}\phi \nonumber\\
    &\phantom{=} + h_{35}^{} R_{ij} \phi \nabla^{i}R \nabla^{j}\phi + h_{36}^{} R^{ij} R^{kl} R_{ikjl} \phi^2 + h_{37}^{} R^{ij} \phi^2 \Box R_{ij} + h_{38}^{} R_{ij} R^{ij} \phi \Box \phi \nonumber\\
    &\phantom{=} + h_{39}^{} R^{ij} \nabla_{j}\nabla_{i}\phi \Box \phi + \gamma_{1}^{} R_{i}{}^{k} R^{ij} R_{jk} \phi^2  + \gamma_{2}^{} R^3 \phi^2  + \gamma_{3}^{} R R_{ijkl} R^{ijkl} \phi^2 \nonumber\\
    &\phantom{=} + \gamma_{4}^{} R^{ij} R_{i}{}^{klm} R_{jklm} \phi^2 + \gamma_{5}^{} R_{i}{}^{m}{}_{k}{}^{n} R^{ijkl} R_{jmln} \phi^2 + \gamma_{6}^{} R_{ij}{}^{mn} R^{ijkl} R_{klmn} \phi^2 \nonumber\\
    &\phantom{=} + \gamma_{7}^{} R \phi^2 \Box R + \gamma_{8}^{} R^2 \phi \Box \phi + \gamma_{9}^{} R_{jklm} R^{jklm} \phi \Box \phi + \gamma_{10}^{} \phi^2 \nabla_{i}R \nabla^{i}R \nonumber\\
    &\phantom{=} + \gamma_{11}^{} R \phi \nabla_{i}\phi \nabla^{i}R  + \gamma_{12}^{} R^{jklm} \phi \nabla_{i}R_{jklm} \nabla^{i}\phi + \gamma_{13}^{} R_{jk} R^{jk} \nabla_{i}\phi \nabla^{i}\phi \nonumber\\
    &\phantom{=} + \gamma_{14}^{} R^2 \nabla_{i}\phi \nabla^{i}\phi + \gamma_{15}^{} R_{jklm} R^{jklm} \nabla_{i}\phi \nabla^{i}\phi   + \gamma_{16}^{} \nabla_{i}\phi \nabla^{i}\phi \Box R + \gamma_{17}^{} \phi \Box R \Box \phi \nonumber\\
    &\phantom{=} + \gamma_{18}^{} R \Box \phi \Box \phi + \gamma_{19}^{} \nabla_{i}\phi \nabla^{i}R \Box\phi + \gamma_{20}^{} \phi \nabla^{i}\phi \Box \nabla_{i}R + \gamma_{21}^{} \phi \nabla^{i}R \Box \nabla_{i}\phi \nonumber\\
    &\phantom{=} + \gamma_{22}^{} R \nabla^{i}\phi \Box \nabla_{i}\phi + \gamma_{23}^{} \phi^2 \Box^2 R + \gamma_{24}^{} R \phi \Box^2\phi  + \gamma_{25}^{} R_{i}{}^{k} R_{jk} \nabla^{i}\phi \nabla^{j}\phi \nonumber\\
    &\phantom{=} + \gamma_{26}^{} R_{ij} R \nabla^{i}\phi \nabla^{j}\phi + \gamma_{27}^{} R^{kl} R_{ikjl} \nabla^{i}\phi \nabla^{j}\phi + \gamma_{28}^{} R_{i}{}^{klm} R_{jklm} \nabla^{i}\phi \nabla^{j}\phi \nonumber\\
    &\phantom{=} + \gamma_{29}^{} \nabla^{i}\phi \nabla_{j}\nabla_{i}R \nabla^{j}\phi + \gamma_{30}^{} \nabla^{i}R \nabla_{j}\nabla_{i}\phi \nabla^{j}\phi + \gamma_{31}^{} \phi \nabla_{j}\nabla_{i}\phi \nabla^{j}\nabla^{i}R \nonumber\\
    &\phantom{=} + \gamma_{32}^{} R_{i}{}^{klm} R_{jklm} \phi \nabla^{j}\nabla^{i}\phi + \gamma_{33}^{} R \nabla_{j}\nabla_{i}\phi \nabla^{j}\nabla^{i}\phi + \gamma_{34}^{} R^{jk} \phi \nabla^{i}\phi \nabla_{k}R_{ij} \nonumber\\
    &\phantom{=} + \gamma_{35}^{} R_{i}{}^{k} R^{ij} \phi \nabla_{k}\nabla_{j}\phi + \gamma_{36}^{} R^{jk} \nabla^{i}\phi \nabla_{k}\nabla_{j}\nabla_{i}\phi  + \gamma_{37}^{} \nabla^{i}\phi \nabla^{j}\phi \Box R_{ij} \nonumber\\
    &\phantom{=} + \gamma_{38}^{} \phi \nabla^{j}\nabla^{i}\phi \Box R_{ij}  + \gamma_{39}^{} R_{i}{}^{j} \nabla^{i}\phi \Box \nabla_{j}\phi + \gamma_{40}^{} \nabla^{j}\Box \phi \Box \nabla_{j}\phi \nonumber\\
    &\phantom{=} + \gamma_{41}^{} R^{ij} \phi \Box \nabla_{j}\nabla_{i}\phi  + \gamma_{42}^{} \nabla^{j}\nabla^{i}\phi \Box\nabla_{j}\nabla_{i}\phi + \gamma_{43}^{} \Box\phi \Box^2\phi + \gamma_{44}^{} \nabla^{i}\phi \Box^2\nabla_{i}\phi \nonumber\\
    &\phantom{=} + \gamma_{45}^{} \phi \Box^3\phi  + \gamma_{46}^{} \phi^2 \nabla_{j}R_{ik} \nabla^{k}R^{ij} + \gamma_{47}^{} \phi^2 \nabla_{k}R_{ij} \nabla^{k}R^{ij} \nonumber\\
    &\phantom{=} + \gamma_{48}^{} \phi \nabla_{k}\nabla_{j}\nabla_{i}\phi \nabla^{k}R^{ij}  + \gamma_{49}^{} R^{ij} \nabla_{k}\nabla_{j}\phi \nabla^{k}\nabla_{i}\phi + \gamma_{50}^{} \nabla_{i}R_{jk} \nabla^{i}\phi \nabla^{k}\nabla^{j}\phi \nonumber\\
    &\phantom{=} + \gamma_{51}^{} \nabla^{i}\phi \nabla_{k}R_{ij} \nabla^{k}\nabla^{j}\phi  + \gamma_{52}^{} \nabla_{k}\nabla_{j}\nabla_{i}\phi \nabla^{k}\nabla^{j}\nabla^{i}\phi + \gamma_{53}^{} R_{ijkl} \phi \nabla^{i}\phi \nabla^{l}R^{jk} \nonumber\\
    &\phantom{=} + \gamma_{54}^{} R_{ikjl} \phi^2 \nabla^{l}\nabla^{k}R^{ij}  + \gamma_{55}^{} R^{ij} R_{ikjl} \phi \nabla^{l}\nabla^{k}\phi + \gamma_{56}^{} R_{ikjl} \nabla^{j}\nabla^{i}\phi \nabla^{l}\nabla^{k}\phi \nonumber\\
    &\phantom{=} + \gamma_{57}^{} \phi^2 \nabla_{m}R_{ijkl} \nabla^{m}R^{ijkl} \,,\label{h3ansatz}
\end{align}
\endgroup
and leave the coefficients $h_{31},\ldots,h_{39}$ free.
The full solution is in the aforementioned Mathematica notebook.

Similarly, since $\tilde{h}^{(3)}$ appears in eq.~\eqref{Oh2} only as $\tilde{h}^{(3)i}{}_{i}$, we do not compute $\tilde{h}^{(3)}_{ij}$ but rather its trace that satisfies the following PBH equation 
\begin{equation}
    \delta \tilde{h}^{(3)i}{}_{i} = -2\sigma \tilde{h}^{(3)i}{}_{i} + g^{(0)ij} \delta \tilde{h}^{(3)}_{ij} \, .
\end{equation}
A similar Ansatz to~\eqref{h3ansatz} is made for $\tilde{h}^{(3)i}{}_{i}$ and its coefficients are fixed in terms of $h_1,h_2,c_1,c_2$ by the modified PBH equation.

\section{Gravitational action with backreaction}\label{app:action}
From~\eqref{EHwriem} in FG coordinates with backreaction~\eqref{modfg} we identify:
\begingroup
\allowdisplaybreaks[1]
\begin{align}
     \mathcal{L}_h &= \tfrac12 d(1 + 4 \gamma ) h^{i}{}_{i} \nonumber\\
     & + \tfrac12 z (1+4\gamma)\big[ 2 (d-1) g'_{ij} h^{ij} - (d-1) g^{ij}g'_{ij} h^{k}{}_{k} \nonumber\\
     &\phantom{=} -2 (d-1) g^{ij}h'_{ij} + h^{ij} R_{ij} - \tfrac{1}{2} h^{i}{}_{i} R -\nabla^{j}\nabla^{i}h_{ij}  + \Box h^{i}{}_{i}\big] \nonumber\\
     & + z^2 \big[ -2 (1 + 4\gamma ) g''_{ij} h^{ij} + (3  -4 (d-5) \gamma ) g^{im}g^{jn}g'_{ik} g'_{mn} h^{jk} \nonumber\\
     &\phantom{=}- (1 + 8 \gamma ) g^{ij} g'_{ij} g'_{kl} h^{kl} + (1 + 4 \gamma ) g^{ij} g''_{ij} h^{k}{}_{k} -2 \gamma  R_{ikjl} \nabla^{l}\nabla^{k}h^{ij} \nonumber\\
     &\phantom{=}- (\tfrac{3}{4} -  (d-5) \gamma ) g^{im}g^{jn}g'_{ij} g'_{mn} h^{k}{}_{k} + (\tfrac{1}{4} + 2 \gamma ) (g^{ij}g'_{ij})^2 h^{k}{}_{k} \nonumber\\
     &\phantom{=}- (3 - 4 (d-5) \gamma ) g^{ik}g^{jl}g'_{ij} h'_{kl} + 2(1 + 4 \gamma ) g^{ij} h''_{ij} + 2 \gamma  g'_{ij}  R^{ij} h^{k}{}_{k} \nonumber\\
     &\phantom{=} + (1 + 8 \gamma ) g^{ij} g^{kl} g'{ij} h'_{kl} + 4 \gamma h'_{ij} R^{ij} -8 \gamma  g'_{ij} h^{i}{}_{k} R^{jk} - \gamma  h^{ij} R_{i}{}^{klp} R_{jklp} \nonumber\\
     &\phantom{=} + \tfrac{1}{4} \gamma  h^{i}{}_{i} R_{jklp} R^{jklp} -2 \gamma  g'_{ij} \nabla^{j}\nabla^{i}h^{k}{}_{k} + 4 \gamma  g'_{ij} \nabla_{k}\nabla^{j}h^{ik} -2 \gamma  g'_ {ij} \Box h^{ij} \big] \nonumber\\
     & + \gamma z^3 \big[ -12 g^{im}g^{jn} g'_{ik} g'_{mn} g'_{jl} h^{kl} + 8 g^{ij} g^{mn} g'_{ij} g'_{lm} g'_{mk} h^{kl} \nonumber\\
     &\phantom{=}+ 4 g^{im}g^{jn} g'_{ij} g'_{mn} g'_{kl} h^{kl} + 2 g^{im}g^{jn}g^{kl} g'_{ik}g'_{mn} g'_{jl} h^{p}{}_{p} \nonumber\\
     &\phantom{=} -2 g^{ij} g^{km} g^{nl} g'_{ij} g'_{kl} g'_{mn} h^{p}{}_{p}  + 12 g^{im}g^{jn} g^{kl} g'_{ik} g'_{mn} h'_{jl} \nonumber\\
     &\phantom{=}-8 g^{ij} g^{km} g^{ln} g'_{ij} g'_{kl} h'_{mn} -4 g^{ik} g^{jl} g^{mn} g'_{ij} g'_{kl} h'_{mn}  - g'_{ij} g'_{kl} h^{p}{}_{p} R^{ikjl} \nonumber\\
     &\phantom{=} -4 g'_{ij} h'_{kl} R^{ikjl} -6 g'_{ij} g'_{kl} h^{i}{}_{p} R^{jklp} -4 g^{km} g^{ln} h^{ij} \nabla_{i}g'_{kl} \nabla_{j}g'_{mn} \nonumber\\
     &\phantom{=} -4 g'_{ij} \nabla^{i}g'_{kl} \nabla^{j}h^{kl} -8 g^{il} \nabla^{j}h'_{ik} \nabla^{k}g'_{jl} + 8 g^{ik}g^{jl}\nabla_{k}h'_{ij} \nabla^{k}g'_{kl} \nonumber\\
     &\phantom{=} + 8 g'_{ij} \nabla^{i}g'_{kl} \nabla^{l}h^{jk} -2  g'_{ij} g'_{kl} \nabla^{l}\nabla^{j}h^{ik} + 2 g'^{ij} g'^{kl} \nabla^{l}\nabla^{k}h^{ij}  \nonumber\\
     &\phantom{=} + 8 g^{kn} h^{ij} \nabla_{j}g'_{kl} \nabla^{l}g'_{in} + 4 g^{in} g'_{jn} \nabla^{j}h^{kl} \nabla_{l}g'_{ik} + 4 h^{ij} \nabla^{k}g'_{jl} \nabla^{l}g'_{ik} \nonumber\\
     &\phantom{=} -4 g^{in} g'_{ij} \nabla^{k}h^{jl} \nabla_{l}g'_{kn} -8 g^{kn} h^{ij} \nabla_{l}g'_{jk} \nabla^{l}g'_{in} -4 g^{in} g'_{ij} \nabla_{l}h^{jk} \nabla^{l}g'_{kn} \nonumber\\
     &\phantom{=} -2 g^{jn} h^{i}{}_{i} \nabla^{k}g'_{jl} \nabla^{l}g'_{kn} + 2 g^{jm}g^{km} h^{i}{}_{i} \nabla_{l}g'_{jk} \nabla^{l}g'_{mn}\big] \nonumber\\
     & +\gamma z^4 \big[ 8 g^{im}g^{kn}g'_{ij} g'_{kl} g''_{mn} h^{jl} -16 g^{il}g''_{ik} g''_{jl} h^{jk}  + 16 g^{im} g^{jn} g'_{ik} g'_{mn} g''_{jl} h^{kl} \nonumber\\
     &\phantom{=} + 4 g^{ik} g^{jl}  g''_{ij} g''_{kl} h^{k}{}_{k} -4  g^{kq} g^{ir} g^{js} g'_{ik} g'_{rs} g'_{jl} g'_{pq} h^{lp} -8 g^{im}g^{jn}g^{kl} g'_{ik}g'_{mn} h''_{jl} \nonumber\\
     &\phantom{=} -4  g^{ik}g^{jl} g^{mr}g^{ns}g^{pq} g'_{ij} g'_{kl} g'_{mn} g'_{rq} h^{ps} -4 g^{im}g^{jn}g^{kl} g'_{ik}g'_{mn} g''_{jl} h^{p}{}_{p} \nonumber\\
     &\phantom{=} + \tfrac12 g^{im}g^{jn} g^{kr} g^{ls} g'_{ik} g'_{mn} g'_{jl} g'_{rs} h^{p}{}_{p}  + \tfrac12 (g^{ik}g^{jl}g'_{ij} g'_{kl})^2 h^{p}{}_{p} -16 g^{im}g^{jn} g^{kl} g'_{mn} g''_{ik} h'_{jl}\nonumber\\
     &\phantom{=} + 4 g^{im}g^{jn}g^{kr}g^{ls} g'_{ik} g'_{mn} g'_{jl} h'_{rs} + 4 g^{im}g^{jn} g^{kr} g^{ls} g'_{ij} g'_{mn} g'_{kl} h'_{rs} + 16 g^{ik}g^{jl}g''_{ij} h''_{kl} \big] \,,\label{fwbackre}
\end{align}
\endgroup
while $\mathcal{L}_g$ is still given by~\eqref{Og}.
Notice that also terms of $\mathcal{O}(z^4)$ contribute to the anomaly in $d=4$, since $h''_{ij} = - z^{-2}\tilde{h}^{(1)}_{ij} + \mathcal{O}(z^{-1})$.

\section{Coefficients renormalisation}\label{app:coeffren}
Evaluating~\eqref{Og2} and~\eqref{Oh2} on the PBH solutions, we find that $\mathcal{L}_g^{(2)} + \mathcal{L}_h^{(2)}$ satisfies the WZ condition~\eqref{WZ_wbeta}, even though it is singular in $d=4$.
In particular, expanding around $d=4$ we find that
\begin{equation}\label{og2oh2series}
    \mathcal{L}_g^{(2)} + \mathcal{L}_h^{(2)} = \frac{1}{d-4} I^{(-1)} + I^{(0)} + (d-4) I^{(1)} + (d-4)^2 I^{(2)} + \ldots \,,
\end{equation}
with ($g_{ij} \equiv g^{(0)}_{ij}$ and $\phi \equiv \phi_{(0)}$)
\begin{align}
    I^{(-1)} &= c \, h_{1} \Bigg( \frac{7}{24}  R_{ij} R^{ij} R \phi^2 -  \frac{1}{24} R^3 \phi^2 -  \frac{1}{2} R^{ij} R^{kl} R_{ikjl} \phi^2 + \frac{1}{24} R \phi^2 \Box R  -  \frac{1}{12} R^2 \phi \Box\phi \nonumber\\
    &\phantom{=}+ \frac{1}{12} R^{ij} \phi^2 \nabla_{j}\nabla_{i}R + \frac{1}{3} R^{ij} R \phi \nabla_{j}\nabla_{i}\phi + \frac{1}{12} \phi \Box R \Box\phi  + \frac{1}{6} \phi \nabla_{j}\nabla_{i}\phi \nabla^{j}\nabla^{i}R \nonumber\\
    &\phantom{=} -  \frac{1}{4}  R^{ij} \phi^2 \Box R_{ij} -  \frac{1}{2}  \phi \nabla^{j}\nabla^{i}\phi \Box R_{ij} + \frac{1}{4}  R_{ij} R^{ij} \phi \Box\phi -  R^{ij} R_{ikjl} \phi \nabla^{l}\nabla^{k}\phi \Bigg) \, .
\end{align}
Notice that $I^{(-1)}$ is entirely type B and $\delta I^{(-1)} = -4\sigma I^{(-1)}$ under Weyl transformation up to $\mathcal{O}(\phi^2)$.
However, $\mathcal{L}_g^{(2)} + \mathcal{L}_h^{(2)}$ contains also several free coefficients, which are not fixed by the PBH equations, i.e.~$\{c_1,c_2,h_1,h_{21},\ldots,h_{26},h_{31},\ldots,h_{39}\}$, and it is possible to regularise~\eqref{og2oh2series} by renormalising some of these coefficients.
In particular, there is a unique shift which cancels $I^{(-1)}$ and it consists of shifting some of the free coefficients which appear in $I^{(2)}$ as follows:
\begin{alignat*}{4}
    & h_{32} \to h_{32} + \frac{c \, h_1}{18 a (d-4)^2} \,, \quad && h_{33} \to h_{33} + \frac{7 c \, h_1}{36 a (d-4)^2} \,, \quad && h_{34} \to h_{34} + \frac{2 c\, h_1}{9a (d-4)^2} \,,\\
    & h_{36} \to h_{36} - \frac{c\, h_1}{3 a (d-4)^2} \,, \quad && h_{37} \to h_{37} - \frac{c\, h_1}{6 a (d-4)^2} \,, \quad && h_{38} \to h_{38} + \frac{c\, h_1}{6a (d-4)^2} \, .
\end{alignat*}
Since the shifted coefficients only appear in $I^{(2)}$, this shift does not introduce new singularities, and of course the regularised expression satisfies the Wess-Zumino condition.

\section{Complete expression of the anomaly}\label{app:full_ano}
We provide here the complete expression of the holographic anomaly $\mathcal{A}^{4d}$ as in~\eqref{full_ano}, recalling that in the type A and B pure gravity anomalies with the respective scalar field contributions we identify the following terms
\begin{align}
    \mathcal{A}^{4d}_{\text{B}} &= \mathcal{I}^{\text{pg}}_\text{B} + \mathcal{I}_1(h_1) + \mathcal{I}_2 (c_1,h_1,h_2) + \mathcal{I}_3(c_2,h_1,h_2) \,, \label{full_ano_B_app}\\
    \mathcal{A}^{4d}_{\text{A}} &= \mathcal{I}^{\text{pg}}_\text{A} + \mathcal{I}_4(h_1) + \mathcal{I}_5(h_2) + \mathcal{I}_6(h_2) + \mathcal{I}_7(c_1,h_1,h_2) + \mathcal{I}_8(c_2,h_1,h_2) \nonumber\\
    &\phantom{=} + \mathcal{I}_9(h_{21}) + \mathcal{I}_{10}(h_{22}) + \mathcal{I}_{11}(h_{23})+ \mathcal{I}_{12}(h_{24}) + \mathcal{I}_{13}(h_{25}) + \mathcal{I}_{14}(h_{26}) \, . \label{full_ano_A_app}
\end{align}
We proceed by presenting and analysing them individually.
Again, we assume that $g_{ij} \equiv g^{(0)}_{ij}$ and curvatures and covariant derivatives are w.r.t.~$g^{(0)}_{ij}$, and $\phi \equiv \phi_{(0)}$.
The interested reader can also find the expressions of $\mathcal{A}^{4d}_{\text{B}}$ and $\mathcal{A}^{4d}_{\text{A}}$ in the Mathematica notebook provided in the supplementary material of~\cite{Broccoli:2021icm}.

As for $\mathcal{A}^{4d}_{\text{B}}$, we already presented  $\mathcal{I}^{\text{pg}}_\text{B}$ and $\mathcal{I}_1$ in~\eqref{modetypeB_h1}.
We move then to $\mathcal{I}_2$, which is given by
\begin{equation}
    \mathcal{I}_2 (c_1,h_1,h_2) = 4 c \, c_1 (h_1+4 h_2) \hat R \, C^2 \, .
\end{equation}
This tensorial structure leads to a Weyl invariant anomaly integrand, as already pointed out in~\cite{Schwimmer:2019efk}.
Thus, in the CFT it contributes to the anomaly as an `ordinary' anomaly, which satisfy the WZ condition without need of the beta-function.
Henceforth we will refer to this kind of structures as `ordinary anomaly' for short.
Notice that $\mathcal{I}_2$ vanishes in flat space.
A similar discussion follows for $\mathcal{I}_3$ since it is related to $\mathcal{I}_2$ by
\begin{equation}
    \mathcal{I}_3 (c_2,h_1,h_2) = \mathcal{I}_2\left(\frac{c_2}{4},h_1,h_2\right) \, .
\end{equation}

As for $\mathcal{A}^{4d}_{\text{A}}$, the tensorial structures $\mathcal{I}^{\text{pg}}_\text{A}$ and $\mathcal{I}_4$ are given in~\eqref{modtypeA_h1} and $\mathcal{I}_5$ is also already presented and discussed in the main text.
Then, we move to $\mathcal{I}_6$ which corresponds to ordinary anomalies and thus satisfies the WZ condition without need of the metric beta-function.
Explicitly, it reads
\begingroup
\allowdisplaybreaks[1]
\begin{align}
    \mathcal{I}_6(h_2) &= a \, h_2 \Big( 3 \nabla^{j}\Box\phi \Box\nabla_{j}\phi -  \tfrac{15}{2} \nabla^{j}\nabla^{i}\phi \Box\nabla_{j}\nabla_{i}\phi + \tfrac{15}{8} \Box\phi \Box^2\phi + \tfrac{9}{8} \phi \Box^2\phi \nonumber\\
    &\phantom{=}-  \tfrac{15}{2} \nabla_{k}\nabla_{j}\nabla_{i}\phi \nabla^{k}\nabla^{j}\nabla^{i}\phi - \tfrac{3}{8} R_{i}{}^{k} R^{ij} R_{jk} \phi^2 + \tfrac{5}{2} R_{ij} R^{ij} R \phi^2 -  \tfrac{23}{64} R^3 \phi^2 \nonumber\\
    &\phantom{=} -  \tfrac{39}{8} R^{ij} R^{kl} R_{ikjl} \phi^2 -  \tfrac{17}{64} R R_{ijkl} R^{ijkl} \phi^2 + \tfrac{33}{4} R_{i}{}^{m}{}_{k}{}^{n} R^{ijkl} R_{jmln} \phi^2 \nonumber\\
    &\phantom{=} -  \tfrac{9}{32} R_{ij}{}^{mn} R^{ijkl} R_{klmn} \phi^2 -  \tfrac{3}{16} R \phi^2 \Box R -  \tfrac{17}{8} R^2 \phi \Box \phi -  \tfrac{15}{8} R_{jklm} R^{jklm} \phi \Box \phi \nonumber\\
    &\phantom{=} -  \tfrac{1}{2} \phi^2 \nabla_{i}R \nabla^{i}R -  \tfrac{15}{4} R \phi \nabla_{i}\phi \nabla^{i}R + \tfrac{99}{4} R^{jk} \phi \nabla_{i}R_{jk} \nabla^{i}\phi \nonumber\\
    &\phantom{=} -  \tfrac{129}{8} R^{jklm} \phi \nabla_{i}R_{jklm} \nabla^{i}\phi + \tfrac{189}{8} R_{jk} R^{jk} \nabla_{i}\phi \nabla^{i}\phi -  \tfrac{23}{8} R^2 \nabla_{i}\phi \nabla^{i}\phi \nonumber\\
    &\phantom{=} -  \tfrac{51}{4} R_{jklm} R^{jklm} \nabla_{i}\phi \nabla^{i}\phi + \tfrac{19}{8} R^{ij} \phi^2 \nabla_{j}\nabla_{i}R + 6 R^{ij} R \phi \nabla_{j}\nabla_{i}\phi + \tfrac{17}{8} \phi \Box R \Box \phi \nonumber\\
    &\phantom{=} + \tfrac{17}{8} R \Box \phi \Box \phi + \tfrac{1}{8} \nabla_{i}\phi \nabla^{i}R \Box \phi + \tfrac{3}{2} \phi \nabla^{i}\phi \Box\nabla_{i}R + \tfrac{21}{8} \phi \nabla^{i}R \Box\nabla_{i}\phi \nonumber\\
    &\phantom{=} -  \tfrac{15}{4} R \nabla^{i}\phi \Box\nabla_{i}\phi + \tfrac{3}{16} \phi^2 \Box^2 R + \tfrac{3}{2} R \phi \Box^2 \phi -  \tfrac{27}{4} R_{ij} \phi \nabla^{i}R \nabla^{j}\phi \nonumber\\
    &\phantom{=} + \tfrac{27}{2} R_{i}{}^{k} R_{jk} \nabla^{i}\phi \nabla^{j}\phi -  \tfrac{37}{4} R_{ij} R \nabla^{i}\phi \nabla^{j}\phi + \tfrac{15}{2} R^{kl} R_{ikjl} \nabla^{i}\phi \nabla^{j}\phi \nonumber\\
    &\phantom{=} + \tfrac{7}{4} \nabla^{i}R \nabla_{j}\nabla_{i}\phi \nabla^{j}\phi + \tfrac{23}{4} \phi \nabla_{j}\nabla_{i}\phi \nabla^{j}\nabla^{i}R -  \tfrac{59}{8} R \nabla_{j}\nabla_{i}\phi \nabla^{j}\nabla^{i}\phi \nonumber\\
    &\phantom{=} + \tfrac{57}{4} R^{jk} \phi \nabla^{i}\phi \nabla_{k}R_{ij} + \tfrac{15}{4} R_{i}{}^{k} R^{ij} \phi \nabla_{k}\nabla_{j}\phi + \tfrac{27}{2} R^{jk} \nabla^{i}\phi \nabla_{k}\nabla_{j}\nabla_{i}\phi \nonumber\\
    &\phantom{=} + \tfrac{9}{4} R^{ij} \phi^2 \Box R_{ij} -  \tfrac{39}{4} \phi \nabla^{j}\nabla^{i}\phi \Box R_{ij} + \tfrac{21}{2} R_{ij} R^{ij} \phi \Box\phi -  \tfrac{15}{4} R^{ij} \nabla_{j}\nabla_{i}\phi \Box\phi \nonumber\\
    &\phantom{=} - 9 R_{i}{}^{j} \nabla^{i}\phi \Box\nabla_{j}\phi + \tfrac{129}{16} \phi^2 \nabla_{j}R_{ik} \nabla^{k}R^{ij} -  \tfrac{9}{32} \phi^2 \nabla_{k}R_{ij} \nabla^{k}R^{ij} \nonumber\\
    &\phantom{=} + \tfrac{3}{2} \phi \nabla_{k}\nabla_{j}\nabla_{i}\phi \nabla^{k}R^{ij} + \tfrac{45}{4} R^{ij} \nabla_{k}\nabla_{j}\phi \nabla^{k}\nabla_{i}\phi + 3 \nabla_{i}R_{jk} \nabla^{i}\phi \nabla^{k}\nabla^{j}\phi \nonumber\\
    &\phantom{=} -  \tfrac{21}{2} \nabla^{i}\phi \nabla_{k}R_{ij} \nabla^{k}\nabla^{j}\phi - 18 R_{ijkl} \phi \nabla^{i}\phi \nabla^{l}R^{jk} -  \tfrac{75}{4} R_{ikjl} \phi^2 \nabla^{l}\nabla^{k}R^{ij} \nonumber\\
    &\phantom{=} -  \tfrac{99}{4} R^{ij} R_{ikjl} \phi \nabla^{l}\nabla^{k}\phi -  \tfrac{3}{4} R_{ikjl} \nabla^{j}\nabla^{i}\phi \nabla^{l}\nabla^{k}\phi -  \tfrac{237}{64} \phi^2 \nabla_{m}R_{ijkl} \nabla^{m}R^{ijkl} \Big) \, .
\end{align}
\endgroup
Notice that $\mathcal{I}_6$ vanishes in flat space after integration by parts.

The tensorial structures in $\mathcal{I}_7$ correspond to Weyl invariant ordinary anomalies and they are thus solutions of the WZ condition without need of the metric beta-function.
Explicitly, we have
\begingroup
\allowdisplaybreaks[1]
\begin{align}
    \mathcal{I}_7(c_1,h_1,h_2) &= a\, c_1 \Big( -48 h_{1}^{} R_{i}{}^{k} R^{ij} R_{jk} \phi^2 + 4 (9 h_{1}^{} + 8 h_{2}^{}) R_{ij} R^{ij} R \phi^2 \nonumber\\
    &\phantom{=} -  \tfrac{16}{3} (h_{1}^{} + h_{2}^{}) R^3 \phi^2 - 4 (h_{1}^{} + 4 h_{2}^{}) R R_{ijkl} R^{ijkl} \phi^2 - 4 h_{1}^{} R \phi^2 \Box R \nonumber\\
    &\phantom{=} - 4 (5 h_{1}^{} + 8 h_{2}^{}) R^2 \phi \Box \phi - 12 (h_{1}^{} + 8 h_{2}^{}) R_{jklm} R^{jklm} \phi \Box \phi \nonumber\\
    &\phantom{=} - 8 h_{1}^{} R \phi \nabla_{i}\phi \nabla^{i}R + 48 h_{1}^{} R^{jk} \phi \nabla_{i}R_{jk} \nabla^{i}\phi + 24 h_{1}^{} R^{jklm} \phi \nabla_{i}R_{jklm} \nabla^{i}\phi \nonumber\\
    &\phantom{=} - 48 (3 h_{1}^{} + 8 h_{2}^{}) R_{jk} R^{jk} \nabla_{i}\phi \nabla^{i}\phi + 8 (3 h_{1}^{} + 8 h_{2}^{}) R^2 \nabla_{i}\phi \nabla^{i}\phi \nonumber\\
    &\phantom{=} + 24 (3 h_{1}^{} + 8 h_{2}^{}) R_{jklm} R^{jklm} \nabla_{i}\phi \nabla^{i}\phi + 64 h_{1}^{} R^{ij} R \phi \nabla_{j}\nabla_{i}\phi\nonumber\\
    &\phantom{=} - 16 h_{1}^{} R \Box\phi \Box\phi + 48 h_{1}^{} R_{ij} \phi \nabla^{i}R \nabla^{j}\phi + 16 h_{1}^{} R \nabla_{j}\nabla_{i}\phi \nabla^{j}\nabla^{i}\phi \nonumber\\
    &\phantom{=} - 96 h_{1}^{} R^{jk} \phi \nabla^{i}\phi \nabla_{k}R_{ij} - 96 h_{1}^{} R_{i}{}^{k} R^{ij} \phi \nabla_{k}\nabla_{j}\phi + 24 h_{1}^{} R^{ij} \phi^2 \Box R_{ij} \nonumber\\
    &\phantom{=} + 24 (3 h_{1}^{} + 8 h_{2}^{}) R_{ij} R^{ij} \phi \Box\phi + 96 h_{1}^{} R^{ij} \nabla_{j}\nabla_{i}\phi \Box\phi \nonumber\\
    &\phantom{=} - 96 h_{1}^{} R^{ij} \nabla_{k}\nabla_{j}\phi \nabla^{k}\nabla_{i}\phi + 192 h_{1}^{} R_{ijkl} \phi \nabla^{i}\phi \nabla^{l}R^{jk} \nonumber\\
    &\phantom{=} - 48 h_{1}^{} R_{ikjl} \phi^2 \nabla^{l}\nabla^{k}R^{ij} - 96 h_{1}^{} R^{ij} R_{ikjl} \phi \nabla^{l}\nabla^{k}\phi \nonumber\\
    &\phantom{=} - 96 h_{1}^{} R_{ikjl} \nabla^{j}\nabla^{i}\phi \nabla^{l}\nabla^{k}\phi  \Big) \, .
\end{align}
\endgroup
Notice that $\mathcal{I}_7$ vanishes in flat space.
Similar properties hold for $\mathcal{I}_8$ since it is related to $\mathcal{I}_7$ by
\begin{equation}
    \mathcal{I}_8(c_2,h_1,h_2) = \mathcal{I}_7\left(\frac{c_2}{4},h_1,h_2\right) \, .
\end{equation}

The remaining tensorial structures $\mathcal{I}_9,\ldots,\mathcal{I}_{14}$ are parametrised by the free coefficients which appear in~$h^{(2)}_{ij}$.
Each of them solves the WZ condition without need of the metric beta-function and in the field theory they can be eliminated with counterterms.
Thus, they correspond to trivial solutions of the WZ condition.
Explicitly, they read:
\begingroup
\allowdisplaybreaks[1]
\begin{align}
    \mathcal{I}_9(h_{21}) &= a \, h_{21} \Big( -6 R_{i}{}^{k} R^{ij} R_{jk} \phi^2 + 6 R_{ij} R^{ij} R \phi^2 \nonumber\\
    &\phantom{=} -  \tfrac{3}{4} R^3 \phi^2 - 6 R^{ij} R^{kl} R_{ikjl} \phi^2 -  \tfrac{3}{4} R R_{ijkl} R^{ijkl} \phi^2 + 6 R_{i}{}^{m}{}_{k}{}^{n} R^{ijkl} R_{jmln} \phi^2 \nonumber\\
    &\phantom{=} + \tfrac{3}{2} R_{ij}{}^{mn} R^{ijkl} R_{klmn} \phi^2 -  \tfrac{1}{2} R \phi^2 \Box R -  \tfrac{1}{2} R^2 \phi \Box \phi -  \tfrac{3}{2} R_{jklm} R^{jklm} \phi \Box \phi \nonumber\\
    &\phantom{=} -  \tfrac{1}{2} \phi^2 \nabla_{i}R \nabla^{i}R - 2 R \phi \nabla_{i}\phi \nabla^{i}R + 12 R^{jk} \phi \nabla_{i}R_{jk} \nabla^{i}\phi - 6 R^{jklm} \phi \nabla_{i}R_{jklm} \nabla^{i}\phi \nonumber\\
    &\phantom{=} + 3 R_{jk} R^{jk} \nabla_{i}\phi \nabla^{i}\phi -  \tfrac{1}{2} R^2 \nabla_{i}\phi \nabla^{i}\phi -  \tfrac{3}{2} R_{jklm} R^{jklm} \nabla_{i}\phi \nabla^{i}\phi + 3 R^{ij} \phi^2 \Box R_{ij} \nonumber\\
    &\phantom{=} + 3 R_{ij} R^{ij} \phi \Box \phi + 3 \phi^2 \nabla_{k}R_{ij} \nabla^{k}R^{ij} - 6 R_{ikjl} \phi^2 \nabla^{l}\nabla^{k}R^{ij}\nonumber\\
    &\phantom{=} -  \tfrac{3}{2} \phi^2 \nabla_{m}R_{ijkl} \nabla^{m}R^{ijkl} \Big) \,,
\end{align}
\endgroup

\begingroup
\allowdisplaybreaks[1]
\begin{align}
    \mathcal{I}_{10}(h_{22}) &= a \, h_{22} \Big( 3 \nabla^{j}\Box\phi \Box \nabla_{j}\phi - 3 \nabla_{k}\nabla_{j}\nabla_{i}\phi \nabla^{k}\nabla^{j}\nabla^{i}\phi + 3 R_{i}{}^{k} R^{ij} R_{jk} \phi^2 \nonumber\\
    &\phantom{=} -  \tfrac{3}{2} R_{ij} R^{ij} R \phi^2 + \tfrac{3}{16} R^3 \phi^2 + \tfrac{3}{16} R R_{ijkl} R^{ijkl} \phi^2 -  \tfrac{3}{2} R_{i}{}^{m}{}_{k}{}^{n} R^{ijkl} R_{jmln} \phi^2 \nonumber\\
    &\phantom{=} -  \tfrac{3}{8} R_{ij}{}^{mn} R^{ijkl} R_{klmn} \phi^2 + \tfrac{1}{2} R \phi^2 \Box R + \tfrac{3}{4} R^2 \phi \Box \phi + \tfrac{3}{4} R_{jklm} R^{jklm} \phi \Box \phi \nonumber\\
    &\phantom{=} + \tfrac{5}{8} \phi^2 \nabla_{i}R \nabla^{i}R + \tfrac{3}{2} R \phi \nabla_{i}\phi \nabla^{i}R - 9 R^{jk} \phi \nabla_{i}R_{jk} \nabla^{i}\phi + \tfrac{9}{4} R^{jklm} \phi \nabla_{i}R_{jklm} \nabla^{i}\phi \nonumber\\
    &\phantom{=} -  \tfrac{3}{2} R_{jk} R^{jk} \nabla_{i}\phi \nabla^{i}\phi + \tfrac{3}{8} R^2 \nabla_{i}\phi \nabla^{i}\phi + \tfrac{3}{8} R_{jklm} R^{jklm} \nabla_{i}\phi \nabla^{i}\phi + R^{ij} \phi^2 \nabla_{j}\nabla_{i}R \nonumber\\
    &\phantom{=} - 2 R^{ij} R \phi \nabla_{j}\nabla_{i}\phi + \phi \Box R \Box \phi + R \Box \phi \Box \phi + \tfrac{1}{2} \nabla_{i}\phi \nabla^{i}R \Box \phi + \tfrac{3}{2} \phi \nabla^{i}R \Box \nabla_{i}\phi \nonumber\\
    &\phantom{=} -  R_{ij} \phi \nabla^{i}R \nabla^{j}\phi + 3 R_{i}{}^{k} R_{jk} \nabla^{i}\phi \nabla^{j}\phi -  \tfrac{3}{2} R_{ij} R \nabla^{i}\phi \nabla^{j}\phi + 3 R^{kl} R_{ikjl} \nabla^{i}\phi \nabla^{j}\phi \nonumber\\
    &\phantom{=} -  \tfrac{1}{2} \nabla^{i}R \nabla_{j}\nabla_{i}\phi \nabla^{j}\phi + 2 \phi \nabla_{j}\nabla_{i}\phi \nabla^{j}\nabla^{i}R -  R \nabla_{j}\nabla_{i}\phi \nabla^{j}\nabla^{i}\phi \nonumber\\
    &\phantom{=} + 3 R^{jk} \phi \nabla^{i}\phi \nabla_{k}R_{ij} + 6 R_{i}{}^{k} R^{ij} \phi \nabla_{k}\nabla_{j}\phi - 3 R^{jk} \nabla^{i}\phi \nabla_{k}\nabla_{j}\nabla_{i}\phi \nonumber\\
    &\phantom{=} - 3 R^{ij} \phi^2 \Box R_{ij}  - 6 \phi \nabla^{j}\nabla^{i}\phi \Box R_{ij} - 3 R_{ij} R^{ij} \phi \Box \phi - 6 R^{ij} \nabla_{j}\nabla_{i}\phi \Box \phi \nonumber\\
    &\phantom{=} + 6 R_{i}{}^{j} \nabla^{i}\phi \Box \nabla_{j}\phi  + 3 \phi^2 \nabla_{j}R_{ik} \nabla^{k}R^{ij} -  \tfrac{9}{2} \phi^2 \nabla_{k}R_{ij} \nabla^{k}R^{ij} \nonumber\\
    &\phantom{=} - 3 \phi \nabla_{k}\nabla_{j}\nabla_{i}\phi \nabla^{k}R^{ij}  + 3 R^{ij} \nabla_{k}\nabla_{j}\phi \nabla^{k}\nabla_{i}\phi - 3 \nabla_{i}R_{jk} \nabla^{i}\phi \nabla^{k}\nabla^{j}\phi \nonumber\\
    &\phantom{=} + 3 \nabla^{i}\phi \nabla_{k}R_{ij} \nabla^{k}\nabla^{j}\phi  - 3 R_{ijkl} \phi \nabla^{i}\phi \nabla^{l}R^{jk} + 9 R_{ikjl} \nabla^{j}\nabla^{i}\phi \nabla^{l}\nabla^{k}\phi \nonumber\\
    &\phantom{=} + \tfrac{3}{8} \phi^2 \nabla_{m}R_{ijkl} \nabla^{m}R^{ijkl} \Big) \,,
\end{align}
\endgroup

\begingroup
\allowdisplaybreaks[1]
\begin{align}
    \mathcal{I}_{11}(h_{23}) &= a \, h_{23} \Big( \tfrac{3}{2} \nabla^{j}\Box \phi \Box\nabla_{j}\phi -  \tfrac{3}{2} \nabla_{k}\nabla_{j}\nabla_{i}\phi \nabla^{k}\nabla^{j}\nabla^{i}\phi - \tfrac{3}{2} R_{i}{}^{k} R^{ij} R_{jk} \phi^2 \nonumber\\
    &\phantom{=} + \tfrac{9}{4} R_{ij} R^{ij} R \phi^2 -  \tfrac{9}{32} R^3 \phi^2 - 3 R^{ij} R^{kl} R_{ikjl} \phi^2 -  \tfrac{9}{32} R R_{ijkl} R^{ijkl} \phi^2 \nonumber\\
    &\phantom{=} + \tfrac{9}{4} R_{i}{}^{m}{}_{k}{}^{n} R^{ijkl} R_{jmln} \phi^2 + \tfrac{9}{16} R_{ij}{}^{mn} R^{ijkl} R_{klmn} \phi^2 + \tfrac{1}{8} R^2 \phi \Box \phi \nonumber\\
    &\phantom{=} -  \tfrac{3}{8} R_{jklm} R^{jklm} \phi \Box \phi + \tfrac{1}{16} \phi^2 \nabla_{i}R \nabla^{i}R -  \tfrac{1}{4} R \phi \nabla_{i}\phi \nabla^{i}R + \tfrac{3}{2} R^{jk} \phi \nabla_{i}R_{jk} \nabla^{i}\phi \nonumber\\
    &\phantom{=} -  \tfrac{15}{8} R^{jklm} \phi \nabla_{i}R_{jklm} \nabla^{i}\phi + \tfrac{3}{4} R_{jk} R^{jk} \nabla_{i}\phi \nabla^{i}\phi -  \tfrac{1}{16} R^2 \nabla_{i}\phi \nabla^{i}\phi \nonumber\\
    &\phantom{=} -  \tfrac{9}{16} R_{jklm} R^{jklm} \nabla_{i}\phi \nabla^{i}\phi + \tfrac{1}{2} R^{ij} \phi^2 \nabla_{j}\nabla_{i}R -  R^{ij} R \phi \nabla_{j}\nabla_{i}\phi + \tfrac{1}{2} \phi \Box R \Box \phi \nonumber\\
    &\phantom{=} + \tfrac{1}{2} R \Box \phi \Box \phi + \tfrac{1}{4} \nabla_{i}\phi \nabla^{i}R \Box \phi + \tfrac{3}{4} \phi \nabla^{i}R \Box \nabla_{i}\phi -  \tfrac{1}{2} R_{ij} \phi \nabla^{i}R \nabla^{j}\phi \nonumber\\
    &\phantom{=} + \tfrac{3}{2} R_{i}{}^{k} R_{jk} \nabla^{i}\phi \nabla^{j}\phi -  \tfrac{3}{4} R_{ij} R \nabla^{i}\phi \nabla^{j}\phi + \tfrac{3}{2} R^{kl} R_{ikjl} \nabla^{i}\phi \nabla^{j}\phi \nonumber\\
    &\phantom{=} -  \tfrac{1}{4} \nabla^{i}R \nabla_{j}\nabla_{i}\phi \nabla^{j}\phi + \phi \nabla_{j}\nabla_{i}\phi \nabla^{j}\nabla^{i}R -  \tfrac{1}{2} R \nabla_{j}\nabla_{i}\phi \nabla^{j}\nabla^{i}\phi \nonumber\\
    &\phantom{=} + \tfrac{3}{2} R^{jk} \phi \nabla^{i}\phi \nabla_{k}R_{ij} + 3 R_{i}{}^{k} R^{ij} \phi \nabla_{k}\nabla_{j}\phi -  \tfrac{3}{2} R^{jk} \nabla^{i}\phi \nabla_{k}\nabla_{j}\nabla_{i}\phi \nonumber\\
    &\phantom{=} - 3 \phi \nabla^{j}\nabla^{i}\phi \Box R_{ij} - 3 R^{ij} \nabla_{j}\nabla_{i}\phi \Box \phi + 3 R_{i}{}^{j} \nabla^{i}\phi \Box \nabla_{j}\phi + \tfrac{3}{2} \phi^2 \nabla_{j}R_{ik} \nabla^{k}R^{ij} \nonumber\\
    &\phantom{=} -  \tfrac{3}{4} \phi^2 \nabla_{k}R_{ij} \nabla^{k}R^{ij} -  \tfrac{3}{2} \phi \nabla_{k}\nabla_{j}\nabla_{i}\phi \nabla^{k}R^{ij} + \tfrac{3}{2} R^{ij} \nabla_{k}\nabla_{j}\phi \nabla^{k}\nabla_{i}\phi \nonumber\\
    &\phantom{=} -  \tfrac{3}{2} \nabla_{i}R_{jk} \nabla^{i}\phi \nabla^{k}\nabla^{j}\phi + \tfrac{3}{2} \nabla^{i}\phi \nabla_{k}R_{ij} \nabla^{k}\nabla^{j}\phi -  \tfrac{3}{2} R_{ijkl} \phi \nabla^{i}\phi \nabla^{l}R^{jk} \nonumber\\
    &\phantom{=} - 3 R_{ikjl} \phi^2 \nabla^{l}\nabla^{k}R^{ij} + \tfrac{9}{2} R_{ikjl} \nabla^{j}\nabla^{i}\phi \nabla^{l}\nabla^{k}\phi -  \tfrac{9}{16} \phi^2 \nabla_{m}R_{ijkl} \nabla^{m}R^{ijkl} \Big) \,,
\end{align}
\endgroup

\begingroup
\allowdisplaybreaks[1]
\begin{align}
    \mathcal{I}_{12}(h_{24}) &= a \, h_{24} \Big( 3 \nabla^{j}\Box\phi \Box\nabla_{j}\phi - 3 \nabla_{k}\nabla_{j}\nabla_{i}\phi \nabla^{k}\nabla^{j}\nabla^{i}\phi -6 R_{i}{}^{k} R^{ij} R_{jk} \phi^2 \nonumber\\
    &\phantom{=} + \tfrac{15}{2} R_{ij} R^{ij} R \phi^2 -  \tfrac{15}{16} R^3 \phi^2 - 9 R^{ij} R^{kl} R_{ikjl} \phi^2 -  \tfrac{15}{16} R R_{ijkl} R^{ijkl} \phi^2 \nonumber\\
    &\phantom{=} + \tfrac{15}{2} R_{i}{}^{m}{}_{k}{}^{n} R^{ijkl} R_{jmln} \phi^2 + \tfrac{15}{8} R_{ij}{}^{mn} R^{ijkl} R_{klmn} \phi^2 -  \tfrac{1}{4} R \phi^2 \Box R + \tfrac{1}{2} R^2 \phi \Box \phi \nonumber\\
    &\phantom{=} -  \tfrac{3}{2} R_{jklm} R^{jklm} \phi \Box \phi -  \tfrac{1}{8} \phi^2 \nabla_{i}R \nabla^{i}R -  \tfrac{3}{2} R \phi \nabla_{i}\phi \nabla^{i}R + 9 R^{jk} \phi \nabla_{i}R_{jk} \nabla^{i}\phi \nonumber\\
    &\phantom{=} -  \tfrac{27}{4} R^{jklm} \phi \nabla_{i}R_{jklm} \nabla^{i}\phi + \tfrac{3}{2} R_{jk} R^{jk} \nabla_{i}\phi \nabla^{i}\phi + \tfrac{1}{8} R^2 \nabla_{i}\phi \nabla^{i}\phi \nonumber\\
    &\phantom{=} -  \tfrac{15}{8} R_{jklm} R^{jklm} \nabla_{i}\phi \nabla^{i}\phi + R^{ij} \phi^2 \nabla_{j}\nabla_{i}R - 4 R^{ij} R \phi \nabla_{j}\nabla_{i}\phi -  \tfrac{1}{2} \nabla_{i}\phi \nabla^{i}\phi \Box R \nonumber\\
    &\phantom{=} + \tfrac{1}{2} \phi \Box R \Box \phi + R \Box\phi \Box\phi + \tfrac{1}{2} \nabla_{i}\phi \nabla^{i}R \Box \phi + \tfrac{3}{2} \phi \nabla^{i}R \Box\nabla_{i}\phi -  R_{ij} \phi \nabla^{i}R \nabla^{j}\phi \nonumber\\
    &\phantom{=} + 3 R_{i}{}^{k} R_{jk} \nabla^{i}\phi \nabla^{j}\phi -  \tfrac{7}{2} R_{ij} R \nabla^{i}\phi \nabla^{j}\phi + 9 R^{kl} R_{ikjl} \nabla^{i}\phi \nabla^{j}\phi \nonumber\\
    &\phantom{=} -  \nabla^{i}\phi \nabla_{j}\nabla_{i}R \nabla^{j}\phi -  \tfrac{1}{2} \nabla^{i}R \nabla_{j}\nabla_{i}\phi \nabla^{j}\phi + \phi \nabla_{j}\nabla_{i}\phi \nabla^{j}\nabla^{i}R \nonumber\\
    &\phantom{=} -  R \nabla_{j}\nabla_{i}\phi \nabla^{j}\nabla^{i}\phi + 3 R^{jk} \phi \nabla^{i}\phi \nabla_{k}R_{ij} + 6 R_{i}{}^{k} R^{ij} \phi \nabla_{k}\nabla_{j}\phi \nonumber\\
    &\phantom{=} - 3 R^{jk} \nabla^{i}\phi \nabla_{k}\nabla_{j}\nabla_{i}\phi + \tfrac{3}{2} R^{ij} \phi^2 \Box R_{ij} + 3 \nabla^{i}\phi \nabla^{j}\phi \Box R_{ij} - 3 \phi \nabla^{j}\nabla^{i}\phi \Box R_{ij} \nonumber\\
    &\phantom{=} - 6 R^{ij} \nabla_{j}\nabla_{i}\phi \Box \phi + 6 R_{i}{}^{j} \nabla^{i}\phi \Box \nabla_{j}\phi + 3 \phi^2 \nabla_{j}R_{ik} \nabla^{k}R^{ij} \nonumber\\
    &\phantom{=} - 3 \phi \nabla_{k}\nabla_{j}\nabla_{i}\phi \nabla^{k}R^{ij} + 3 R^{ij} \nabla_{k}\nabla_{j}\phi \nabla^{k}\nabla_{i}\phi - 3 \nabla_{i}R_{jk} \nabla^{i}\phi \nabla^{k}\nabla^{j}\phi \nonumber\\
    &\phantom{=} + 3 \nabla^{i}\phi \nabla_{k}R_{ij} \nabla^{k}\nabla^{j}\phi - 3 R_{ijkl} \phi \nabla^{i}\phi \nabla^{l}R^{jk} - 9 R_{ikjl} \phi^2 \nabla^{l}\nabla^{k}R^{ij} \nonumber\\
    &\phantom{=} + 6 R^{ij} R_{ikjl} \phi \nabla^{l}\nabla^{k}\phi + 9 R_{ikjl} \nabla^{j}\nabla^{i}\phi \nabla^{l}\nabla^{k}\phi -  \tfrac{15}{8} \phi^2 \nabla_{m}R_{ijkl} \nabla^{m}R^{ijkl} \Big) \,,
\end{align}
\endgroup

\begingroup
\allowdisplaybreaks[1]
\begin{align}
    \mathcal{I}_{13}(h_{25}) &= a \, h_{25} \Big( \tfrac{1}{2} \nabla^{j}\Box \phi \Box \nabla_{j}\phi - 4 \nabla^{j}\nabla^{i}\phi \Box \nabla_{j}\nabla_{i}\phi + \Box\phi \Box^2\phi \nonumber\\
    &\phantom{=} -  \tfrac{7}{2} \nabla_{k}\nabla_{j}\nabla_{i}\phi \nabla^{k}\nabla^{j}\nabla^{i}\phi - \tfrac{5}{2} R_{i}{}^{k} R^{ij} R_{jk} \phi^2 + \tfrac{9}{4} R_{ij} R^{ij} R \phi^2 -  \tfrac{9}{32} R^3 \phi^2 \nonumber\\
    &\phantom{=} - 2 R^{ij} R^{kl} R_{ikjl} \phi^2 -  \tfrac{9}{32} R R_{ijkl} R^{ijkl} \phi^2 + \tfrac{9}{4} R_{i}{}^{m}{}_{k}{}^{n} R^{ijkl} R_{jmln} \phi^2 \nonumber\\
    &\phantom{=} + \tfrac{9}{16} R_{ij}{}^{mn} R^{ijkl} R_{klmn} \phi^2 -  \tfrac{3}{4} R_{jklm} R^{jklm} \phi \Box\phi -  \tfrac{1}{16} \phi^2 \nabla_{i}R \nabla^{i}R \nonumber\\
    &\phantom{=} + 2 R^{jk} \phi \nabla_{i}R_{jk} \nabla^{i}\phi -  \tfrac{21}{8} R^{jklm} \phi \nabla_{i}R_{jklm} \nabla^{i}\phi + \tfrac{3}{4} R_{jk} R^{jk} \nabla_{i}\phi \nabla^{i}\phi \nonumber\\
    &\phantom{=} -  \tfrac{1}{16} R^2 \nabla_{i}\phi \nabla^{i}\phi -  \tfrac{13}{16} R_{jklm} R^{jklm} \nabla_{i}\phi \nabla^{i}\phi -  \tfrac{1}{4} R^{ij} \phi^2 \nabla_{j}\nabla_{i}R + \tfrac{1}{4} \phi \Box R \Box \phi \nonumber\\
    &\phantom{=} + \tfrac{1}{2} R \Box\phi \Box \phi + \tfrac{3}{4} \nabla_{i}\phi \nabla^{i}R \Box \phi + \tfrac{3}{4} \phi \nabla^{i}R \Box \nabla_{i}\phi + R \nabla^{i}\phi \Box \nabla_{i}\phi + \tfrac{1}{2} R \phi \Box^2 \phi \nonumber\\
    &\phantom{=} -  R_{ij} \phi \nabla^{i}R \nabla^{j}\phi + \tfrac{1}{2} R_{i}{}^{k} R_{jk} \nabla^{i}\phi \nabla^{j}\phi -  \tfrac{3}{4} R_{ij} R \nabla^{i}\phi \nabla^{j}\phi + \tfrac{7}{2} R^{kl} R_{ikjl} \nabla^{i}\phi \nabla^{j}\phi \nonumber\\
    &\phantom{=} + \tfrac{1}{4} \nabla^{i}R \nabla_{j}\nabla_{i}\phi \nabla^{j}\phi -  \tfrac{1}{2} \phi \nabla_{j}\nabla_{i}\phi \nabla^{j}\nabla^{i}R -  \tfrac{1}{2} R^{jk} \phi \nabla^{i}\phi \nabla_{k}R_{ij} \nonumber\\
    &\phantom{=} -  R_{i}{}^{k} R^{ij} \phi \nabla_{k}\nabla_{j}\phi -  \tfrac{7}{2} R^{jk} \nabla^{i}\phi \nabla_{k}\nabla_{j}\nabla_{i}\phi + \tfrac{1}{2} R^{ij} \phi^2 \Box R_{ij} -  \tfrac{1}{2} \phi \nabla^{j}\nabla^{i}\phi \Box R_{ij} \nonumber\\
    &\phantom{=} + \tfrac{1}{2} R_{ij} R^{ij} \phi \Box \phi - 2 R^{ij} \nabla_{j}\nabla_{i}\phi \Box \phi - 2 R_{i}{}^{j} \nabla^{i}\phi \Box \nabla_{j}\phi - 2 R^{ij} \phi \Box\nabla_{j}\nabla_{i}\phi \nonumber\\
    &\phantom{=} -  \phi^2 \nabla_{j}R_{ik} \nabla^{k}R^{ij} + \tfrac{5}{4} \phi^2 \nabla_{k}R_{ij} \nabla^{k}R^{ij} -  \tfrac{7}{2} \phi \nabla_{k}\nabla_{j}\nabla_{i}\phi \nabla^{k}R^{ij} \nonumber\\
    &\phantom{=} + \tfrac{1}{2} R^{ij} \nabla_{k}\nabla_{j}\phi \nabla^{k}\nabla_{i}\phi -  \tfrac{5}{2} \nabla_{i}R_{jk} \nabla^{i}\phi \nabla^{k}\nabla^{j}\phi -  \tfrac{3}{2} \nabla^{i}\phi \nabla_{k}R_{ij} \nabla^{k}\nabla^{j}\phi \nonumber\\
    &\phantom{=} -  \tfrac{1}{2} R_{ijkl} \phi \nabla^{i}\phi \nabla^{l}R^{jk} -  \tfrac{3}{2} R_{ikjl} \phi^2 \nabla^{l}\nabla^{k}R^{ij} + R^{ij} R_{ikjl} \phi \nabla^{l}\nabla^{k}\phi \nonumber\\
    &\phantom{=} -  \tfrac{1}{2} R_{ikjl} \nabla^{j}\nabla^{i}\phi \nabla^{l}\nabla^{k}\phi -  \tfrac{9}{16} \phi^2 \nabla_{m}R_{ijkl} \nabla^{m}R^{ijkl} \Big) \,,
\end{align}
\endgroup

\begingroup
\allowdisplaybreaks[1]
\begin{align}
    \mathcal{I}_{14}(h_{26}) &= a \, h_{26} \Big( 3 \nabla^{j}\Box\phi \Box\nabla_{j}\phi - 12 \nabla^{j}\nabla^{i}\phi \Box\nabla_{j}\nabla_{i}\phi + 3 \Box\phi \Box^2\phi \nonumber\\
    &\phantom{=} - 12 \nabla_{k}\nabla_{j}\nabla_{i}\phi \nabla^{k}\nabla^{j}\nabla^{i}\phi -6 R_{i}{}^{k} R^{ij} R_{jk} \phi^2 + 6 R_{ij} R^{ij} R \phi^2 -  \tfrac{3}{4} R^3 \phi^2 \nonumber\\
    &\phantom{=} - 6 R^{ij} R^{kl} R_{ikjl} \phi^2 -  \tfrac{3}{4} R R_{ijkl} R^{ijkl} \phi^2 + 6 R_{i}{}^{m}{}_{k}{}^{n} R^{ijkl} R_{jmln} \phi^2 \nonumber\\
    &\phantom{=} + \tfrac{3}{2} R_{ij}{}^{mn} R^{ijkl} R_{klmn} \phi^2 + \tfrac{1}{4} R \phi^2 \Box R + \tfrac{1}{4} R^2 \phi \Box \phi -  \tfrac{3}{2} R_{jklm} R^{jklm} \phi \Box \phi \nonumber\\
    &\phantom{=} + \tfrac{1}{4} \phi^2 \nabla_{i}R \nabla^{i}R + R \phi \nabla_{i}\phi \nabla^{i}R - 6 R^{jklm} \phi \nabla_{i}R_{jklm} \nabla^{i}\phi + \tfrac{1}{4} R^2 \nabla_{i}\phi \nabla^{i}\phi \nonumber\\
    &\phantom{=} -  \tfrac{3}{2} R_{jklm} R^{jklm} \nabla_{i}\phi \nabla^{i}\phi + \tfrac{3}{2} \phi \Box R \Box \phi + \tfrac{3}{2} R \Box \phi \Box \phi + 3 \nabla_{i}\phi \nabla^{i}R \Box \phi \nonumber\\
    &\phantom{=} + 3 \phi \nabla^{i}R \Box \nabla_{i}\phi + 3 R \nabla^{i}\phi \Box \nabla_{i}\phi + \tfrac{3}{2} R \phi \Box^2\phi - 3 R_{ij} \phi \nabla^{i}R \nabla^{j}\phi \nonumber\\
    &\phantom{=} + 3 R_{i}{}^{k} R_{jk} \nabla^{i}\phi \nabla^{j}\phi - 3 R_{ij} R \nabla^{i}\phi \nabla^{j}\phi + 12 R^{kl} R_{ikjl} \nabla^{i}\phi \nabla^{j}\phi \nonumber\\
    &\phantom{=} - 12 R^{jk} \nabla^{i}\phi \nabla_{k}\nabla_{j}\nabla_{i}\phi - 6 \phi \nabla^{j}\nabla^{i}\phi \Box R_{ij} - 6 R^{ij} \nabla_{j}\nabla_{i}\phi \Box\phi \nonumber\\
    &\phantom{=} - 3 R_{i}{}^{j} \nabla^{i}\phi \Box \nabla_{j}\phi - 6 R^{ij} \phi \Box \nabla_{j}\nabla_{i}\phi - 12 \phi \nabla_{k}\nabla_{j}\nabla_{i}\phi \nabla^{k}R^{ij} \nonumber\\
    &\phantom{=} - 12 \nabla_{i}R_{jk} \nabla^{i}\phi \nabla^{k}\nabla^{j}\phi - 6 R_{ikjl} \phi^2 \nabla^{l}\nabla^{k}R^{ij} -  \tfrac{3}{2} \phi^2 \nabla_{m}R_{ijkl} \nabla^{m}R^{ijkl}  \Big) \,.
\end{align}
\endgroup
Notice that $\mathcal{I}_9$ vanishes in flat space and similarly  $\mathcal{I}_{10},\ldots,\mathcal{I}_{14}$ after integration by parts.

\backmatter
\newpage
\phantomsection
\addcontentsline{toc}{chapter}{Bibliography}

\sloppy
\fancyhead[LE,RO]{\textsc{Bibliography}}
\printbibliography

\end{document}